\documentclass[letterpaper,3p,times,11pt]{elsarticle}
\usepackage{amsfonts,amsthm,amsmath,amssymb,slashed}
\usepackage{txfonts,setspace,hypernat,graphicx,epstopdf}

\usepackage{color}
\definecolor{MyDarkBlue}{rgb}{0.15,0.15,0.45}

\usepackage[linktocpage=true]{hyperref}
\hypersetup{
colorlinks=true,
citecolor=MyDarkBlue,
linkcolor=MyDarkBlue,
urlcolor=MyDarkBlue,
pdfauthor={Jonathan Bagger, Neil Lambert, Sunil Mukhi 
and Constantinos Papageorgakis},
pdftitle={Multiple Membranes in M-theory},
pdfsubject={hep-th}}

\def\one{{\,\hbox{1\kern-.8mm l}}} \newcommand{\Sp}{\mathrm{Sp}}
\newcommand{\Dslash}{\not{\hbox{\kern-4pt $D$}}}
\newcommand{\pdslash}{\not{\hbox{\kern-2pt $\partial$}}}
\newcommand{\SO}{\mathrm{SO}} \newcommand{\SL}{\mathrm{SL}}
\newcommand{\SU}{\mathrm{SU}} \newcommand{\U}{\mathrm{U}}
\newcommand{\cL}{\mathcal{L}} 
\newcommand{\cN}{\mathcal{N}} \newcommand{\cO}{\mathcal{O}}
 \newcommand{\cR}{\mathcal{R}}
\newcommand{\eps}{\epsilon} \newcommand{\veps}{\varepsilon}
\newcommand{\Tr}{\mathrm{Tr}}

 \newcommand{\STr}{\mathrm{STr}}

\newcommand{\twobytwo}[4]{\left(\begin{array}{cc}
      #1&#2\\#3&#4\end{array}\right)} \newcommand{\ie}{\emph{i.e.}\:}
\newcommand{\eg}{\emph{e.g.}\;} \newcommand{\pd}{\partial}

\newcommand{\Comment}[1]{{}}
\newcommand{\bth}{{\bar\theta\,}}
\newcommand{\kap}{\kappa}
\newcommand{\Pislash}{{\Pi\!\!\!\! /\,\,}}

\def\IZ{{\mathbb Z}}

\def\IR{{\mathbb R}}

\def\bZ            {{\bar Z}}

\newcommand{\uf}{ {\mathfrak{u}}}
\newcommand{\su}{{\mathfrak{su}}}
\newcommand{\cp}{\Cset \mathrm{P}}
\newcommand{\Cset}{{\,\,{{{^{_{\pmb{\mid}}}}\kern-.45em{\mathrm C}}}}}
\newcommand{\boldX}{\hbox{\boldmath $X$}}
\newcommand{\boldA}{\hbox{\boldmath $A$}}
\newcommand{\boldF}{\hbox{\boldmath $F$}}
\newcommand{\boldG}{\hbox{\boldmath $G$}}
\newcommand{\boldH}{\hbox{\boldmath $H$}}
\newcommand{\boldD}{\hbox{\boldmath $D$}}
\newcommand{\boldpsi}{\hbox{\boldmath $\psi$}}

\newcommand{\boldsigma}{\hbox{\boldmath $\sigma$}}
\newcommand{\boldalpha}{\hbox{\boldmath $\alpha$}}
\newcommand\boldB{\hbox{\boldmath $B$}}
\newcommand\boldM{\hbox{\boldmath $M$}}
\newcommand\boldphi{\hbox{\boldmath $\phi$}}
\newcommand\boldchi{\hbox{\boldmath $\chi$}}

\newcommand{\alp}{{2\pi\alpha'}}
\newcommand{\gYM}{{g_{\!\it YM}}}\newcommand{\gYMs}{{g^2_{\!\it YM}}}

\newcommand{\ra}{\rightarrow}
\newcommand{\bc}{\begin{center}}
\newcommand{\ec}{\end{center}}\newcommand{\ba}{\begin{array}}
\newcommand{\ea}{\end{array}}\newcommand{\beq}{\begin{equation}}
\newcommand{\eeq}{\end{equation}}\newcommand{\bea}{\begin{eqnarray}}
\newcommand{\eea}{\end{eqnarray}}\newcommand{\bmx}{\begin{pmatrix}}
\newcommand{\emx}{\end{pmatrix}}\newcommand{\nn}{\nonumber}
\newcommand{\be}{\begin{equation}}
\newcommand{\ee}{\end{equation}}

\newcommand{\vep}{\varepsilon}

\newcommand{\s}{\sigma}
\newcommand{\x}{\xi}

\newcommand{\del}{\partial}
\newcommand{\half}{\frac{1}{2}}
\newcommand{\bpsi}{{\overline \psi}}

\newcommand{\tD}{{\tilde D}}
\newcommand{\tQ}{{\tilde Q}}

\newcommand{\eref}[1]{Eq.~(\ref{#1})}

\newcommand{\hA}{{\hat{A}}}
\newcommand{\hX}{{\hat{X}}}\newcommand{\hlambda}{{\hat{\lambda}}}
\newcommand{\hD}{{\hat{D}}}\newcommand{\hB}{{\hat{B}}}

\newcommand{\boldf}{{\,\bf f}}
\newcommand{\tA}{{\tilde A}}

\newcommand{\hG}{{\hat{G}}}
\newcommand{\cG}{{\cal G}}

\newcommand{\spn}{{\it spin}}

\def\IB{\relax{\rm I\kern-.18em B}}
\def\IC{{\relax\hbox{\kern.3em{\cmss I}$\kern-.4em{\rm C}$}}}
\def\ID{\relax{\rm I\kern-.18em D}}\def\IE{\relax{\rm I\kern-.18em E}}
\def\IF{\relax{\rm I\kern-.18em F}}\def\II{\relax{\rm I\kern-.18em I}}
\def\IZ{\relax{\sf Z\kern-.35em Z}}\def\Id{\relax{1\kern-.32em 1}}
\def\IG{\relax\hbox{$\inbar\kern-.3em{\rm G}$}}
\def\IR{\relax{\rm I\kern-.18em R}}
\newcommand\sfrac[2]{{\textstyle\frac{#1}{#2}}}

\newcommand\shalf{{\textstyle\frac12}}

\normalsize

\makeatletter
\@addtoreset{equation}{subsection}
\makeatother

\journal{Physics Reports}

\begin{document}

\rightline{CERN-PH-TH-2012-94}
\rightline{RUNHETC-2012-05}
\rightline{TIFR/TH/12-06}
\rightline{NI-12-013}

\begin{frontmatter}
\title{{\huge Multiple Membranes in M-theory}}

\author[jhu]{Jonathan Bagger}
\ead{bagger@jhu.edu}

\author[cern,kcl,INI]{Neil Lambert}
\ead{neil.lambert@cern.ch}

\author[tifr,INI]{Sunil Mukhi}
\ead{mukhi@tifr.res.in}

\author[rutgers,INI]{Constantinos Papageorgakis}
\ead{papageorgakis@physics.rutgers.edu}

\address[jhu]{Department of Physics and Astronomy,
Johns Hopkins University\\ 3400 North Charles Street,
Baltimore, MD 21218, USA\\$\,$}
\address[cern]{Theory Division, CERN\\ 1211 Geneva 23, Switzerland \\ $\,$}
\address[kcl]{Department of Mathematics, King's College London\\ London WC2R 2LS, UK \\ $\,$}
\address[tifr]{Tata Institute of Fundamental Research\\ Homi Bhabha Road,
  Mumbai 400 005, India\\ $\,$}
\address[rutgers]{NHETC and Department of Physics and Astronomy, Rutgers University\\
 126 Frelinghuysen Road, Piscataway, NJ 08854-8019, USA\\ $\,$}
\address[INI]{Isaac Newton Institute for Mathematical Sciences \\
20 Clarkson Road, Cambridge, CB3 OEH, UK\\ $\,$}

\begin{abstract}
We review developments in the theory of multiple, parallel
membranes in M-theory. After discussing the inherent difficulties
with constructing a maximally supersymmetric lagrangian 
with the appropriate field content and symmetries, we introduce
3-algebras and show how they allow for such a description.
Different choices of 3-algebras lead to distinct classes of 2+1
dimensional theories with varying degrees of supersymmetry.
We then demonstrate that these theories are equivalent to
conventional superconformal Chern-Simons gauge theories
at level $k$, but
with bifundamental matter.  Analysing the physical properties of
these theories leads to the identification of a certain subclass of
models with configurations of M2-branes on $\mathbb Z_k$
orbifolds.  These models give rise to a whole new gauge/gravity
duality in the form of an $\textrm{AdS}_4/\textrm{CFT}_3$
correspondence.  We also discuss mass
deformations, higher derivative corrections, and the
possibility of extracting information about M5-brane physics.
\end{abstract}

\begin{keyword}
String theory\sep M-theory\sep Branes
\end{keyword}

\end{frontmatter}

\newpage

{\small
 \begin{spacing}{1}
\tableofcontents
\end{spacing}
}

\parskip 6pt

\newcommand\bepsilon{{\bar \epsilon}}
\newcommand\homega{{\hat\omega}}
\newcommand\tilder{{\tilde r}}

\section[M-theory and M-branes: a brief review]{{\Large{\bf M-theory and M-branes: a brief review}}}\label{chapter1}

M-theory is a proposed interacting quantum theory involving fields and
extended objects (``branes'') propagating in 11 spacetime
dimensions. Its existence has been inferred from the properties of its
massless fields, including the spacetime metric, which couple to each
other via a specific classical lagrangian known as ``11d
supergravity,'' to be described in more detail below. In addition to
the massless fields, M-theory possesses two kinds of
stable branes, namely 2-branes (equivalently referred to as
``membranes''), and 5-branes. These are dynamical objects that extend in
two and five spatial dimensions respectively (as well as time) and
possess a characteristic tension and charge.

11d supergravity \cite{Cremmer:1978km} is a locally supersymmetric
lagrangian field theory involving massless bosons and fermions. It is
special in that eleven is the highest spacetime
dimension in which a consistent supersymmetric
theory can be written down that has spins $\le 2$ \cite{Nahm:1977tg}. The
theory has one 32-component spinor supercharge in eleven dimensions
and, if we allow no more than two-derivative interactions, its lagrangian is
unique.  However, given the non-renormalisability of gravity in any dimension
greater than or equal to four, it is not obvious how to extend this lagrangian
to an ultraviolet-complete quantum theory.  For this reason, the role of the 11d
supergravity lagrangian in quantum physics remained unclear for many
years.

The situation for supergravity theories in 10 spacetime dimensions is
superficially similar. Type II supergravities have two spacetime
supersymmetry charges, which in turn singles out ten as the maximum
allowed spacetime dimension. With only two-derivative interactions and
this amount of supersymmetry there is not one unique lagrangian, but
rather two possible lagrangians with different field contents. These
are referred to as type IIA and type IIB supergravity.  Again, because of non-renormalisability, 
these lagrangians by themselves do not define an ultraviolet-complete
quantum theory.

However, in these cases, an ultraviolet completion is known.
Type IIA and IIB supergravities are
the low-energy limits of corresponding superstring theories.  In
particular, this is how type IIB supergravity was originally
discovered \cite{Green:1981yb} and subsequently constructed 
\cite{Schwarz:1983wa,Howe:1983sra}. As shown in Ref.~\cite{Green:1981yb},
there are two superstring theories in 10 dimensions, type IIA and IIB.
Quantisation of these strings reveals, in particular, a spectrum of
massless particles. Computations of string scattering amplitudes for
these modes can be used to read off their low-energy lagrangians, and
the resulting theories are type IIA and type IIB supergravity.\footnote{Type IIA
supergravity also arises by compactifying 11d supergravity on a circle 
\cite{Giani:1984wc,Huq:1983im,Campbell:1984zc}, as we will discuss in more detail later.}  
This fact, together with considerable evidence that string theory is
ultraviolet finite, encourages us to think of superstring theory as
the ultraviolet completion of type II supergravity. The full theory includes
not just the massless modes of supergravity but also extended objects,
specifically strings. It was later understood that the
spectrum also includes extended branes. 
These have been studied from a variety of complementary points of view:
in terms of worldvolume field theories of the degrees of freedom bound
to them, as charged extended soliton-like solutions in supergravity,
and as one-dimensional matrix models.

This relation between superstring theory and 10d supergravity provides a basis to conjecture the existence of a theory that similarly completes 11d supergravity in the ultraviolet (UV). Indeed, it was long expected that fundamental membranes play a role analogous to the one that strings play in completing ten-dimensional supergravities (see for example Refs.\cite{Bergshoeff:1987cm,Duff:1987cs}).  This idea was further stimulated by the discovery that when compactifying 11d supergravity, wrapped membranes naturally turn into the fundamental strings of type IIA superstring theory \cite{Duff:1987bx}. While it has not actually proved possible to quantise fundamental membranes and derive 11d supergravity from them, it was argued via duality \cite{Townsend:1995kk,Witten:1995ex,Schwarz:1995jq} that there is a consistent UV completion of 11d supergravity and that stable membranes are an important part of this theory.

The details of this conjectured theory, called ``M-theory,'' will be described below. In addition, for previous reviews on M-theory, its duality properties, compactifications, as well as complementary aspects of membrane dynamics, we refer the reader to \cite{Townsend:1996xj,Obers:1998fb,Sorokin:1999jx,Acharya:2004qe,Berman:2007bv,Simon:2011rw}. As we will see, 11d supergravity has no scalar fields and no dimensionless couplings. Therefore in particular it has no tunable coupling constant. The same must therefore be true of the hypothetical M-theory whose low-energy action is postulated to be 11d supergravity. It follows that unlike string theory, M-theory has no perturbative expansion. This makes it considerably harder to study than string theory. We believe in the existence of M-theory only because of many different properties and relationships that have been uncovered in the last three decades. These together provide convincing evidence for the existence of an elegant and internally consistent structure. In this chapter and the next, we will attempt to exhibit this structure. A key feature will be the presence of supersymmetric membranes and 5-branes.

The fact that 10d type IIA supergravity arises by dimensional
reduction of 11d supergravity strongly suggests that the ultraviolet completions of the
two theories (explicitly known in the former case and conjectural in
the latter) are related. Indeed it has been convincingly
argued \cite{Townsend:1995kk,Witten:1995ex,Schwarz:1995jq} that upon
starting with type IIA string theory and allowing the string coupling
to become very strong, the resulting theory reveals a hidden eleventh
dimension and should be thought of as M-theory. Conversely, upon
compactifying the underlying eleven-dimensional spacetime on a spatial
circle, M-theory reduces to type IIA string theory with the string
coupling being related to the radius of the circle. This
comparison is more subtle and rich than comparison of merely the
low-energy effective actions. Indeed, here one keeps the Kaluza-Klein
states \cite{Huq:1983im} arising on compactification, as well as states
arising from wrapped or unwrapped branes, and (as we explain below) a
perfect match ensues.  Thus M-theory is in fact a limit of string
theory: more precisely, a novel and unexpected description of the
dynamics of string theory in a strong-coupling region where the
familiar string formalism (specifically perturbation theory) is not
applicable.

A major puzzle in M-theory has been to understand which, if any, of
its degrees of freedom plays a ``fundamental'' role analogous to that
of the fundamental string in string theory. This is at least partially
answered \cite{Duff:1987bx} by noticing that when M-theory turns into
type IIA string theory upon compactifying a spatial dimension, the
membrane of M-theory wrapped on this dimension can be identified with
the fundamental string. In this sense the membrane appears to be the
most fundamental object in M-theory, providing renewed justification
for earlier attempts to treat it thus. Indeed it is presumably the
origin of the letter ``M'' in ``M-theory.'' One must however be very
careful about this interpretation because while quantisation and
scattering are well-understood (perturbatively) for the fundamental
string, there is no simple analogue for the membrane in M-theory.

A key feature of modern string theory is the dynamics of multiple
D-branes \cite{Polchinski:1995mt}, which are described by the end
points of open strings. This description provides a great deal of
insight into the worldvolume dynamics of the branes, which is
described by familiar classes of gauge theories augmented by
higher-derivative corrections. It should not come as a surprise that
the dynamics of multiple membranes (and also of multiple 5-branes) is
more complex than that of multiple D-branes. In parallel with our
limited understanding of everything else about M-theory, relatively
little has been known about the degrees of freedom localised on
membranes and 5-branes. In the last few years, however, considerable
progress has been made in understanding the interacting field theory
on multiple membranes in M-theory. This constitutes the subject of the
present review.

We note that there have been various attempts to directly define an eleven-dimensional quantum theory of gravity involving membranes. The first, well before the name M-theory was coined, aimed to quantise membranes as one does for strings. However this was later found to be fraught with difficulties (see \eg \cite{deWit:1988ig}). A later definition involved reducing the degrees of freedom to those of matrices living on the worldvolume of D0-branes in the so-called infinite momentum frame \cite{Banks:1996vh}.
That both of these approaches involve fundamental
degrees of freedom that begin with the letter ``M'' is surely one of the
reasons for the current name: M-theory.  There is a great deal of literature concerning a single, quantum,  membrane in eleven-dimensional supergravity; for example see the pioneering works \cite{Hoppe:1982,Hoppe:1988gk}. This review cannot claim to do justice to this topic; rather we aim to give a review of recent results concerning the infrared quantum description of multiple M2-branes in terms of novel highly supersymmetric gauge theories, analogous to the role of Yang-Mills gauge theories on D-branes.

In the remaining part of this chapter we provide a pedagogical discussion of 11d supergravity and its relationship to type IIA supergravity in 10d, along with a survey of the stable branes of M-theory. For the latter, we start in historical order with their worldvolume descriptions and go on to describe their appearance as stable classical solutions of 11d supergravity. We then show how M-theory and type IIA string theory branes are related.

The rest of this review is organised as follows. In Chapter~\ref{chapter2} we make precise the definition of multiple membrane field theory and discuss the Basu-Harvey proposal, involving a triple-bracket, for the structure of such a theory. We also review the basics of supersymmetric Chern-Simons theories which are the foundation on which multiple membrane theories are built. In Chapter~\ref{chapter3} we develop the mathematical structure of superconformal Chern-Simons field theories based on ``3-algebras.'' This includes the construction of the Bagger-Lambert-Gustavsson (BLG) model, which was the first example of a maximally supersymmetric lagrangian that was not a Yang-Mills theory, and the first description of multiple (albeit only two) M2-branes. All these developments come together in Chapter~\ref{chapter4} where the Aharony-Bergman-Jafferis-Maldacena (ABJM) action, the effective description for multiple membranes in M-theory placed at an orbifold singularity, is presented. In Chapter~\ref{chapter5} we begin the analysis of various features of the ABJM theory, including its relation to super-Yang-Mills via the novel Higgs mechanism as well as its connection to the BLG models. In Chapter~\ref{chapter6} we continue with some more advanced topics, covering the description of M-theory momentum by monopole operators, as well as an extension of the theory through a mass deformation and its subsequent spacetime interpretation in terms of dielectric membranes. In Chapter~\ref{chapter7} we consider more general superconformal Chern-Simons theories with reduced supersymmetry ($\mathcal N = 5,4$) and give their 3-algebra description. In Chapter~\ref{chapter8} a few other potentially interesting directions are presented where 3-algebra-based theories play a role, including Lorentzian 3-algebras, higher derivative corrections and applications to M5-branes. Some closing remarks are presented in Chapter~\ref{chapter9}. For a less technical review, the reader is referred to \cite{Lambert:2012wr}.

\subsection{Eleven-dimensional supergravity}

A massless field with spin equal to 2 in four dimensions (and its
analogues in higher dimensions) can be consistently coupled in a field
theory only if the couplings obey general coordinate invariance. The
spin-2 field is then identified with the metric of spacetime and upon
quantisation becomes a ``graviton,'' the mediator of the gravitational
force. Moreover, interacting massless fields with spin greater than 2 are
believed to be inconsistent unless there are infinitely many of
them. Therefore in trying to construct a supergravity theory we should
look for a supermultiplet of bosonic and fermionic fields for which
the highest spin is 2. 

Indeed, just assuming supersymmetry and a spin-2 field one deduces
that the supersymmetry must be {\it local} in spacetime. This arises
from the fact that the anticommutator of two supersymmetries is a
translation generator -- in gravity, translations are promoted to
local (general coordinate) transformations and the supersymmetry
generators accordingly must generate transformations that are local in
spacetime.  Theories of this sort are called ``supergravities.''

The metric or graviton field is denoted $G_{MN}$ with
${M,N=0,1,\cdots,D-1}$. In $D$ spacetime dimensions this has
$\frac{1}{2}(D-1)(D-2)-1$ on-shell degrees of freedom.  This counting
comes from the fact that the little group is $\SO(D-2)$, and the
on-shell graviton transforms in the symmetric traceless representation
of this group.

Supersymmetry requires that there be a superpartner for the graviton,
known as the ``gravitino'' $\Psi_{M,\,\alpha}$. The gravitino is
the gauge particle of local supersymmetry (just as a Yang-Mills field
is the gauge particle of usual gauge invariance). It is a fermion
with both a vector and a spinor index, $M=0,1,\cdots,D-1~{\rm
and}~\alpha=1,2,\cdots,\tD$.  Here $\tD$ is the dimension of the
irreducible spinor representation of the little group, which depends
in a complicated way on $D$. The gravitino $\Psi_{M,\,\alpha}$ has
$\frac{1}{2}(D-3)\tD$ on-shell degrees of freedom. To see this, note
that a simple spinor of $\tD$ components has $\frac{1}{2}\tD$
components on-shell while a $D$-component vector has $D-2$ components
on-shell. Thus a gravitino apparently has $\frac{1}{2}(D-2)\tD$ degrees
of freedom. However due to the well-known properties of
$\Gamma$-matrices the ``$\Gamma$-trace'' of the gravitino field,
defined as $(\Gamma^M \Psi_{M})_\alpha\equiv \Gamma^M_{\alpha\beta}
\Psi_{M,\,\beta}$, is clearly an irreducible representation
by itself. Therefore to get an irreducible representation one must
remove this part by imposing ``$\Gamma$-tracelessness''
\be
(\Gamma^M \Psi_M)_\alpha=0\;,
\ee
which subtracts $\frac{1}{2}\tD$ on-shell degrees of freedom, leaving
the number quoted above.

To find a supermultiplet one can now compare the number of physical
degrees of freedom of a graviton and a gravitino and try to account
for the difference -- if any -- by introducing additional fields. We
look for supermultiplets with the minimal amount of supersymmetry in a
given dimension. From the discussion above, it follows that there will
be a single gravitino. We exhibit the degrees-of-freedom count for
various spacetime dimensions in the following table

\begin{center}
{
\begin{tabular}{c|c|c|c||c|}
Spacetime dimension & Spinor dimension & Graviton & Gravitino & Difference\\
$D$ & $\tD$ & $\frac{1}{2}(D-1)(D-2)-1$ & $\frac{1}{2}(D-3)\tD$ & \\[2mm]
\hline\hline
4 & 4 & 2 & 2 & 0\\
\hline
5& 8 & 5 & 8 &  3\\ 
\hline
6&  8 & 9&  12&  3\\ 
\hline
7&  16 & 14&  32& 18 \\ 
\hline
8&  16 & 20&  40&  20\\ 
\hline
9&  16 & 27&  48&  21\\ 
\hline
10&  16 & 35& 56 & 21 \\ 
\hline
11&  32 & 44&  128& 84 \\ 
\hline
12&  64 & 54&  288 & 234 \\ 
\hline
\end{tabular}
}
\end{center}

\bigskip

The deficit can be made up by adding new bosons to the
theory. However, once we reach $D > 11$ there are so many bosons
needed that we inevitably encounter ``spin $>2$'' fields (we have not
proved this here but it is the content of a theorem to which we
referred earlier).  For $D=11$, we need to add bosonic fields with 84
on-shell degrees of freedom to obtain a matching of on-shell degrees
of freedom, a necessary condition for supersymmetry. Fortunately there
is an irreducible representation of the little group $\SO(9)$ that has
precisely this dimension. It is the antisymmetric 3-form
$C_{MNP}$. In general this has $\frac{1}{6}(D-2)(D-3)(D-4)$ on-shell
degrees of freedom, and for $D=11$ this is precisely 84!

Thus we may hope to find an 11d supergravity theory containing the massless
fields 
\be
G_{MN},~ C_{MNP},~ \Psi_{M,\alpha}\;.
\ee
Indeed, it was shown by Cremmer, 
Julia and Scherk \cite{Cremmer:1978km} that the following action
is supersymmetric
\bea
S_{11d} &=& \frac{1}{16\pi {\cal G}_{(11)}}
\Bigg[\int\!\! d^{11}\!x~ \sqrt{||G||}~\Bigg(R -
\frac{1}{2} |G|^2 + \frac16\int C\wedge G\wedge G
-\frac{i}{2}\,{\bar\psi}_M
\Gamma^{MNP}D_N\,\Big(\frac{\omega + \homega}{2}\Big)\,\psi_P\nn\\
&& \hspace{15mm} -\frac{i}{384}\Big(\bpsi_M\Gamma^{MNABCD}\psi_N + 12
\bpsi^A\Gamma^{BC}\psi^D\Big)\Big(G_{ABCD}+\hG_{ABCD}\Big)
\Bigg)  
\Bigg]\;.
\eea
Here, ${\cal G}_{(11)}$ is the Newton constant in 11 dimensions. It has dimensions of
[length]$^9$ and is often written in terms of 
$\ell_p$, the 11d Planck length, via
\be
\label{ellpdef}
16\pi {\cal G}_{(11)}=
\frac{(2\pi\ell_p)^9}{2\pi}\;.
\ee
The other quantities appearing in the above action are defined as follows.
$R$ is the Ricci scalar and
$D_M(\omega)$ is the covariant derivative
\be
D_M(\omega)\psi_N\equiv \del_M\psi_N
-\frac14\,\omega_{MAB}\Gamma^{AB}\psi_N \;.
\ee
The spin connections $\omega$ and $\homega$ are defined in terms
of the vielbein $E_M^A$ (defined by $E_M^A
E_N^A=G_{MN}$) and the gravitino $\psi_M$ as
\bea
\omega_{MAB}&=&\omega^{(0)}_{MAB}(E) +
\frac{i}{16}\,\bigg[\bpsi_N \Gamma_{MAB}^{\phantom{MAB}NP}\psi_P
- 2\big(\bpsi_M\Gamma_B\psi_A - \bpsi_M\Gamma_A\psi_B
 +\bpsi_B\Gamma_M\psi_A\big)\bigg]\nn\\
\homega_{MAB}&=&\omega_{MAB} - \frac{i}{16}\,
\bpsi_N \Gamma_{MAB}^{\phantom{MAB}NP}\psi_P\;.
\eea
with $\omega^{(0)}_{MAB}(E)$ being the usual Levi-Civita spin
connection associated to $E_M^A$. $\homega$ has the property of being supercovariant -- its supersymmetry variation does not contain derivatives of the supersymmetry parameter. Finally, the field strengths $G$ and
$\hG$ are defined as
\bea
G_{LMNP}&\equiv& 4\,\del_{[L} C_{MNP]}\nn\\ 
\hG_{LMNP}&\equiv& G_{LMNP}+\frac32 i\, \bpsi_L \Gamma_{MN}\psi_P\;.
\eea
In form notation $G=dC$ and
\beq
|G|^2 = |dC|^2 \equiv \frac{1}{4!}G_{LMNP}G^{LMNP}\;.
\eeq

The supersymmetry transformations, labelled by an
arbitrary spacetime-dependent infinitesimal spinor parameter
$\epsilon(x)$, are as follows
\bea
\delta E^A_M &=& \frac{i}{2}\bepsilon\, \Gamma^A\Psi_M\nn\\[2mm]
\delta C_{MNP} &=& -\frac32 i\,\bepsilon\, \Gamma_{[MN}\Psi_{P]}\nn\\[2mm]
\delta \Psi_M &=& 2D_M(\homega)\,\epsilon + \frac{1}{144}
\Big(\Gamma^{PQRS}_{\phantom{PQRS}M}
+8 \Gamma^{QRS}\delta_M^{\phantom{M}P}\Big)\,\hG_{PQRS}\,\epsilon\;,
\eea
where the antisymmetrised gamma-matrices
are defined as\footnote{The 
action and supersymmetry transformations given above follow 
from those of Ref.~\cite{Cremmer:1978km} by rescaling $\psi_M$ 
and $A_{MNP}$ such that a common factor of $(16\pi\cG)^{-1}$
appears in front of all terms, and then sending 
$\Gamma_\mu\to i\Gamma_\mu$, $\Gamma^\mu\to i\Gamma^\mu$, 
$\del_\mu\to \del_\mu$, $\del^\mu\to -\del^\mu$, ${\bar\psi}\to i{\bar\psi}$ to convert from the ``mostly minus'' metric
there to the ``mostly plus'' one used in the present review.}
$\Gamma^{P_1\cdots P_n}
\equiv\Gamma^{[P_1}\Gamma^{P_2}\cdots\Gamma^{P_n]}$.

The above action is general-coordinate-invariant (and actually, since
it involves fermions, also local-Lorentz-invariant), as one
would expect of any action involving gravity. Additionally it is
invariant up to a total derivative under the ``2-form gauge symmetry''
\be
\delta C = d\Lambda\;,
\ee
where $\Lambda$ is an arbitrary infinitesimal spacetime-dependent
2-form.

\subsection{Relation to string theory}

So far we have not exhibited any relationship between M-theory and
string theory. A relationship between them is strongly suggested
by the compactification of 11d supergravity to 10 dimensions. For
this, we assume the eleventh dimension $x^{10}$ is compactified on a
circle with periodicity  $x^{10}\equiv x^{10}+2\pi R_{10}$ and find the massless fields in
ten dimensions by taking the eleven-dimensional fields to be
independent of $x^{10}$. We must also decompose tensors and spinors
into irreducible representations of the ten-dimensional Lorentz
algebra. This is part of the process called Kaluza-Klein reduction
(which additionally involves massive fields in the lower dimension
arising from non-constant modes over the compact space).\footnote{For a review of Kaluza-Klein supergravity see \cite{Duff:1986hr}.}

Let us now use the indices $\mu,\nu,\cdots$ to denote 10d spacetime
indices and $a,b,\cdots$ for 10d tangent-space indices. Then the 11d
metric and 3-form reduce as follows 
\cite{Huq:1983im,Campbell:1984zc,Witten:1995ex}
\bea
&& G^{(11)}_{\mu\nu}=G^{(10)}_{\mu\nu} + e^{2\gamma}A_\mu A_\nu,\quad 
G^{(11)}_{\mu\,10}=e^{2\gamma}A_\mu,\quad 
G^{(11)}_{10\,10}=e^{2\gamma}\nn\\
&&
C^{(11)}_{\mu\nu\rho}=C^{(10)}_{\mu\nu\rho},
\quad C^{(11)}_{\mu\nu\,10} = B_{\mu\nu}\;.
\eea
The quantities on the right hand side of the equalities are all
ten-dimensional  (both in the sense that they are representations of
the 10d Lorentz group and that they depend on the 10d
coordinates). This has not been denoted explicitly by a label except where
confusion may occur.\footnote{Note that $\gamma$ is a scalar field. The 
decomposition above anticipates that an exponential parametrisation
for the scalar will be natural.}

The decomposition of the metric in the first line was chosen in part
so that, using the standard identity for block matrix determinants
\bea
\Bigg|\Bigg|\,\begin{matrix}
A&B\\ C&D
\end{matrix}\,\Bigg|\Bigg| =
\big|\big|\,A-B^TD^{-1}C\,\big|\big|~~\big|\big|\,D\,\big|\big| \ ,
\eea
we have 
\bea 
\sqrt{\big|\big| G^{(11)}\big|\big|} =
\sqrt{\big|\big|G^{(10)}\big|\big|}~e^\gamma\;.  
\eea

The curvature term of the 11d action then reduces as
\bea
\frac{2\pi}{(2\pi\ell_p)^9}\int
d^{11}x\sqrt{\big|\big|G^{(11)}\big|\big|}
~R\to
\frac{(2\pi)^2 R_{10}}{(2\pi\ell_p)^9}
\int d^{10}x\sqrt{\big|\big|G^{(10)}\big|\big|}
~\left(e^\gamma\big(R - \half|d\gamma|^2\big)
-\half e^{3\gamma}|dA|^2\right)\;.
\eea
Similarly the 3-form-dependent terms of the 11d
action reduce as
\bea
-\frac{2\pi}{(2\pi\ell_p)^9}
\frac1{2}\int d^{11}x\sqrt{\big|\big|G^{(11)}\big|\big|}\,|G|^2
&\to &
-\frac{(2\pi)^2 R_{10}}{(2\pi\ell_p)^9}\half\int d^{10}x
\sqrt{\big|\big|G^{(10)}\big|\big|}\,\left(
e^{\gamma}\,|dC^{(10)}|^2+ e^{-\gamma}|dB|^2\right)+\cdots \nn\\
-\frac{2\pi}{(2\pi\ell_p)^9}
\frac1{6}\int C\wedge G\wedge G &\to &
-\frac{(2\pi)^2 R_{10}}{(2\pi\ell_p)^9}
\frac1{2}\int B\wedge dC^{(10)}\wedge dC^{(10)}\;,
\eea
where we are being schematic and have omitted terms that involve powers of $A$. We notice that the bosonic fields of the dimensionally reduced theory, that is
a metric $G_{\mu\nu}$, a scalar $\gamma$ and a 1-form, 2-form and
3-form $A,B,C$ (we drop the superscript $(10)$ from now on), are in
one-to-one correspondence with the fields of type IIA supergravity in
10 dimensions. The latter has a metric, a scalar $\Phi$ called the
dilaton and a 2-form $B$, all coming from the Neveu-Schwarz-Neveu-Schwarz sector, and 1-form
and 3-form Ramond-Ramond potentials $A$ and $C$. 

The bosonic part of the type IIA supergravity action is 
\bea 
S_{IIA} & = &
\frac{2\pi}{(2\pi\ell_s)^8} \int
d^{10}x\sqrt{\big|\big|G^{(10)}\big|\big|}\, \left[e^{-2\Phi}\Big(R+
  |d\Phi|^2 - \half|dB|^2\Big)
  - \Big(\half |dA|^2 + \half |dC|^2\Big)\right]\nn\\
&&-\frac{2\pi}{(2\pi\ell_s)^8}\frac1{2} \int B\wedge dC\wedge dC +\cdots\ ,
\eea
where $\ell_s$ is the ``string length'' associated to type IIA string
theory, of which this 10d supergravity is the low-energy limit.

To match the two sides we may perform a Weyl transformation on the
metric and also a rescaling of $\gamma$. However we are not allowed to
absorb powers of $e^\gamma$ in the gauge potentials $A,B,C$ as these
will lead to derivative couplings with the dilaton which are not
present in type IIA supergravity in the frame in which we are
working. It is now easy to see that with
\beq
G_{\mu\nu}\to e^{-\gamma} G_{\mu\nu}, \quad \Phi=\frac32 \gamma\ ,
\eeq
the two actions match perfectly up to the overall constants in front
of the integrals.

To match these constants, we first note that the 10d and 11d Planck
lengths are related by virtue of the relation between 10d and 11d
metrics
\beq
G^{(11)}_{\mu\nu} = e^{-\gamma}G^{(10)}_{\mu\nu}
= e^{-\frac{2}{3}\Phi}G^{(10)}_{\mu\nu}\;.
\eeq
This tells us that a given physical distance $L$, when measured in
units of $\ell_p$, is related to the same distance as measured in
units of $\ell_s$ by
\be
\frac{L}{\ell_p}=e^{-\half\gamma}\frac{L}{\ell_s}=e^{-\frac{1}{3}\Phi}\frac{L}{\ell_s}\ .
\ee
From the dilaton dependence of the type IIA action above we can read
off that the constant part or VEV of the dilaton defines the string
coupling via
\beq
e^{\langle\Phi\rangle}=g_s\;.
\eeq
It follows that
\beq
\ell_p = g_s^{\frac{1}{3}}\ell_s\;.
\label{ellsellp}
\eeq
With this identification, we can match the coefficients if we set
\beq
\frac{(2\pi)^2 R_{10}}{(2\pi\ell_p)^9}
= \frac{2\pi}{(2\pi\ell_s)^8}\frac{1}{g_s^2}\;,
\label{matchcoef}
\eeq
where on the RHS we have extracted the VEV of $e^{-2\Phi}$ from the
integral. Substituting \eref{ellsellp} in  \eref{matchcoef} we
immediately find
\beq
R_{10} = g_s\ell_s\;.
\eeq

To summarise, we have seen in this section that 11d
supergravity, when compactified on a circle to 10d, is identical to
type IIA supergravity. There is a definite relationship between the
Planck lengths of the two theories, and also between the radius of
compactification of the 11d theory (a parameter absent in the 10d
description) and the coupling constant of the 10d theory (absent in
the 11d description). At small radius or weak coupling
the type IIA description is more appropriate, while at large radius or
strong coupling it is the 11d description that is more appropriate.
As we remarked earlier, since type IIA supergravity in 10d has a
consistent ultraviolet completion in the form of type IIA string
theory, this strongly suggests that 11d supergravity also has a
consistent UV completion, which corresponds to the strongly coupled
limit of type IIA string theory. It is this hypothetical completion
that bears the name ``M-theory.''

\subsection{Motivations to study extended objects}

There are two distinct kinds of limitations in our understanding of
M-theory. One is that we have formulated it in a fixed spacetime
background\footnote{While here we have only chosen flat Minkowski
spacetime, many other noncompact and partially compactified
backgrounds are known and have been investigated.} and it is not clear
how to study M-theory in a background-independent way. Of course an
analogous problem holds also in the existing formalisms of string
theory. The other limitation is that there is no direct way to {\em
prove} the existence of a consistent ultraviolet completion of 11d
supergravity. In contrast, it can be quite convincingly demonstrated
using the string perturbation expansion that superstring theories in
10 dimensions are ultraviolet finite, so at least in perturbation
theory we can be sure they provide consistent UV completions of their
low-energy supergravity theories. This cannot be repeated in M-theory
due to the absence of a coupling constant. 

However, given that in string theory it is the string size that cuts
off possible ultraviolet infinities, one might suspect that something
similar holds in M-theory, namely that it is a theory of not just point
particles but also extended objects, one or more of which
somehow provides an ultraviolet cutoff. This provides an important
motivation to study extended objects or {\em branes} in 11d
supergravity, to which we turn our attention in the following section.

Another related motivation to study extended objects in 11d supergravity is that the spectrum of type IIA string theory contains, besides the fundamental string, a profusion of other stable supersymmetric extended objects. The latter include both Dirichlet branes (``D-branes'') that exhibit unusual and striking features, as well as other more
conventional branes. If the
relationship that we have discussed above between 11d supergravity
and type IIA supergravity in 10d lifts to a relationship between the
hypothetical M-theory and the UV-complete type IIA string theory,
there must be a precise relationship between the stable branes in the
two theories. With this motivation in mind we construct branes of
M-theory in the next section from two points of view: as extended
worldvolume field theories and as soliton-like extended solutions of
classical 11d supergravity. Then we go on to discuss their
relationship with branes of type IIA string theory in 10d. As the
title of this review indicates, the M-theory brane that will be of
greatest interest to us is the 2-brane or membrane.

\subsection{Worldvolume actions for M-theory branes}
\label{worldvol}

Worldvolume actions for particles or extended objects determine
(after quantisation) the quantum mechanical behaviour of these
objects. Typically they are made up of kinetic terms and couplings to
gauge fields under which the object is charged. For M-branes,
the worldvolume action crucially includes couplings to
the 3-form gauge field. 

To understand the origin of such couplings, recall the
well-known coupling of a particle to a gauge field $A_M$, which is
\beq
\int A_M \,dX^{M}\ ,
\eeq
integrated along the worldline of a particle. The worldline itself
is given by some function $X^M(\tau)$ where $\tau$ is 
a parameter. Then the above
coupling can be better written as
\beq
\int A_M\Big(X\,(\tau)\Big)\, 
\frac{dX^M}{d\tau}\,d\tau=\int A_\tau\,d\tau\;,
\eeq
where 
\beq
A_\tau\equiv A_M\,\frac{dX^M}{d\tau}\ ,
\eeq
is the ``pull-back'' of the gauge field onto the worldline of the particle.

In string theory we encounter a generalisation of this where the
particle worldline $X^M(\tau)$ is replaced by the string
worldsheet $X^M(\sigma,\tau)$ where $\sigma$ labels points
along the string. The analogous coupling of the
string is to a 2-form field $B_{MN}$
\beq
\int B_{MN} ~dX^M\wedge dX^N
= \int B_{\mu\nu} ~d\xi^{\,\mu}\wedge d\xi^{\,\mu}\ ,
\eeq
where $\xi^\mu=(\xi^0,\xi^1)=(\tau,\sigma)$ and
\beq
B_{\mu\nu} \equiv B_{MN}\,\frac{dX^M}{d\xi^{\,\mu}}
\frac{dX^N}{d\xi^{\,\nu}}\ ,
\eeq
is the pull-back of the $B$-field to the string worldsheet.

In general, the rank $r$ of the gauge potential is related to the
spatial dimension $p$ of the charged object by $r=p+1$. In the
examples above, we see that point particles ($p=0$) are electrically
charged under 1-form potentials, as is familiar in electromagnetism,
while strings ($p=1$) are ``electrically'' charged under 2-forms. Now
as long as all dimensions are noncompact, the only gauge field in 11d
supergravity is the 3-form $C_{MNP}$. It follows that the only
possible electrically charged objects in this theory are 
2-branes, or membranes, whose charge is manifested via the 
worldvolume coupling 
\beq \int C_{MNP}
~dX^M\wedge dX^N\wedge dX^P = \int C_{\mu\nu\lambda}
~d\xi^{\,\mu}\wedge d\xi^{\,\nu}\wedge d\xi^{\,\lambda} =\frac16\int
d^3\xi\,\epsilon^{\,\mu\nu\lambda}C_{\mu\nu\lambda} \;.
\eeq
Here $(\xi^0,\xi^1,\xi^2)$ are
the worldvolume coordinates, with the first one being worldvolume
time and the last two labelling points on
the membrane, while
\beq 
C_{\mu\nu\lambda} \equiv
C_{MNP}\,\frac{dX^{\,M}}{d\xi^\mu}
\frac{dX^{\,N}}{d\xi^\nu}\frac{dX^{\,P}}{d\xi^\lambda} \ ,
\eeq 
is the pull-back of the $C$-field to the 2-brane worldvolume.

While the above must be a term in the 2-brane action in M-theory, it
cannot of course be the whole story. As mentioned above we need to add
kinetic terms. In addition, as we will explain later, the stable
2-branes in M-theory are actually supersymmetric. Therefore we have to
supersymmetrise the worldvolume action.

We first present the bosonic part of the M2-brane action. It contains
11 scalar fields $X^M(\xi)$ representing the brane coordinates, and a
worldvolume metric $g_{\mu\nu}$ that is treated as an independent
field. The 11d supergravity fields $G_{MN}$ and $C_{MNP}$ are treated
as fixed backgrounds and the action is \cite{Bergshoeff:1987cm}
\beq
S_{M2}^{bosonic}=\int d^3\xi \left(\half \sqrt{|g|}\,
g^{\mu\nu}G_{MN}\,\del_\mu X^{\,M}\del_\nu X^{\,N}
-\half \sqrt{|g|} + \frac16\epsilon^{\,\mu\nu\lambda}
C_{MNP}\,\del_\mu X^M\del_\nu
X^N\del_\lambda X^P\right)\ .
\eeq
This is rather similar to the well-known action for a string
worldsheet. Note however that while the worldvolume metric decouples
for that case (in the critical dimension), here it remains a dynamical
degree of freedom. Moreover the cosmological term in the worldvolume
metric sets it equal, via the equations of motion, to the pull-back of
the spacetime metric onto the brane
\beq
g_{\mu\nu} = G_{MN}\,\del_\mu X^M\del_\nu X^N\;.
\eeq

Supersymmetrising this action is most effectively done in superspace.
To avoid going into all the complexities of the superspace
construction, we restrict ourselves at present to a flat target
spacetime, $G_{MN}=\eta_{MN}$ with vanishing 3-form $C_{MNP}$, which
will provide sufficient insight. In this case, the superspace action
is easily reduced to an action for the bosonic coordinates $X^{\,M}$
and a set of fermionic coordinates $\theta^\alpha,
\alpha=1,2,\cdots,32$. The latter are spinors in spacetime and
scalars on the brane worldvolume. Although the number of bosonic and
fermionic coordinates is not equal, we will soon see that both of them
are effectively reduced to 8 degrees of freedom
thanks to various symmetries. The discussion that follows is based
largely on Ref.~\cite{Bergshoeff:1987qx}.

Let us define the quantity
\beq
\Pi_\mu^M \equiv \del_\mu X^M-i\bth\Gamma^{\,M}\del_\mu\theta\;.
\label{pidef}
\eeq
The supersymmetric M2-brane action in flat spacetime with a vanishing
3-form gauge field is then
\bea
S_{M2}^{susy} &=& \int d^3\xi~ \Bigg(\half \sqrt{|g|}\,
g^{\mu\nu}\Pi_\mu^{\,M}\Pi_\nu^{\,N}
-\half \sqrt{|g|}\nn\\
&&+~\frac{i}{2}\epsilon^{\,\mu\nu\lambda}\,
\bth\Gamma_{MN}\del_\mu\theta\,\bigg[\Pi_\nu^{\,M} 
\Pi_\lambda^{\,N}+ i\,\Pi_\nu^{\,M}~\bth\Gamma^{\,N}\!\del_\lambda\theta 
-\frac13\,\bth\Gamma^M\!\del_\nu\theta~\,
\bth\Gamma^{\,N}\!\del_\lambda\theta\bigg]~\Bigg)\;.
\label{mtwosusy}
\eea 
Note that even though the background 3-form $C_{MNP}$ has been
set to zero, the last term in the above action resembles a
3-form coupling -- in particular, it is independent of
the worldvolume metric and therefore topological. Indeed, it arises
from a 3-form coupling in superspace.

The symmetries of this action under spacetime translations and Lorentz
transformations, as well as under local worldvolume reparametrisations, are
manifest. That leaves the  fermionic symmetries, which are of two types. One
is a rigid supersymmetry transformation with a constant parameter
$\veps^\alpha$, which is a spacetime spinor and a worldvolume
scalar. This transformation is
\bea
\delta X^{\,M} \!\!\!&=&\!\!\! -i\bth\Gamma^M\veps\nn\\
\delta\theta \!\!\!&=&\!\!\!\veps\nn\\
\delta g_{\mu\nu} \!\!\!&=&\!\!\! 0\;.
\label{rigidsusy}
\eea
We see that the worldvolume metric is neutral under this
rigid spacetime supersymmetry. 
The other is a {\em local} fermionic symmetry, called
$\kappa$-{\em symmetry}, with an arbitrary worldvolume
coordinate-dependent parameter $\kappa^\alpha(\xi)$ that, like
$\veps^\alpha$, is a spacetime spinor and worldvolume scalar. The
worldvolume metric transforms non-trivially under the
$\kappa$-symmetry transformations.
It is convenient to define the quantities
\bea
\Pislash_\mu &\equiv& \Pi_\mu^{\,M}\,\Gamma_{M}\nn\\
\tau^{\,\mu} &\equiv & \frac{1}{2\sqrt{|g|}}\,\epsilon^{\,\mu\nu\lambda}\,
\Pislash_\nu\,\Pislash_\lambda\nn\\
\Gamma&\equiv& \frac{1}{6\sqrt{|g|}}\, \eps^{\,\mu\nu\lambda}\,
\Pislash_\mu\,\Pislash_\nu\,\Pislash_\lambda\,\;.
\label{usequan}
\eea

The $\kappa$-symmetry transformations are then given 
by\footnote{We assume the membrane is closed
and has no boundary.}
\bea
\delta X^{\,M} &=& i\,\bth\Gamma^{\,M}(1+\Gamma)\,\kap\nn\\
\delta\theta &=& (1+\Gamma)\,\kap\nn\\
\delta\, \Big(\!\sqrt{|g|}\,g^{\mu\nu}\Big) &=&
i\,g^{\sigma(\mu}\eps^{\,\nu)\lambda\rho}\, {\bar\kap}\,(1+\Gamma)\,
\del_\sigma\theta\,\Pislash_\lambda\,\Pislash_\rho
\\
&&+~ \frac{2i}{3\sqrt{|g|}}\,\eps^{\,\sigma\tau(\mu}\eps^{\,\nu)\lambda\rho}~
{\bar\kap}\,\Pislash^\alpha\del_\alpha\theta~
\bigg(\Pi_\sigma^{\,M}\Pi_{\lambda\,M}\,\Pi_\tau^{\,N}\Pi_{\rho\,N}+
\Pi_\sigma^{\,M}\Pi_{\lambda\,M}\,g_{\tau \rho}+
g_{\sigma \lambda}\,g_{\tau \rho}\bigg)\;.\nn
\label{kappasym}
\eea
We will return shortly to the question of gauge-fixing this local symmetry.
It is useful to note at this stage that the $\kappa$-symmetry
variation of $\Pi_\mu^M$, defined in \eref{pidef}, vanishes. As a
consequence all the quantities in \eref{usequan} are
$\kappa$-invariant.\footnote{The factors of $|g|$ cancel out against
implicit powers in the $\eps$ symbol.}

Let us now examine the equations of motion following from the action
\eref{mtwosusy}. As already indicated above in a bosonic context, the
equation of motion for the worldvolume metric sets it equal to the
pull-back of the spacetime metric. In the present case the spacetime
is flat but since we are dealing with a supersymmetric theory, we find
from \eref{pidef} that the pull-back is implemented via the
super-covariant quantity $\Pi_\mu^{\,M}$
\beq
g_{\mu\nu} = \Pi_{\mu\,M}\Pi_\nu^{\,M}\;.
\label{pback}
\eeq
This equation ensures the useful relations
\bea
\Gamma^2\!\!\!\! &=&\!\!\!\! 1\nn\\
\tau^{\,\mu} \!\!\!\! &=&\!\!\!\! 
g^{\mu\nu}\,\Pislash_\nu\,\Gamma=g^{\mu\nu}\,\Gamma\,\Pislash_\nu\nn\\
\{\tau^{\,\mu},\tau^\nu\}\!\!\!\!  &=& \!\!\!\! 2g^{\mu\nu}\;.
\label{userel}
\eea

The equation of motion for the bosonic coordinates $X^M$ is
\beq
A^M\equiv 
\del_\mu\left\{\,
\!\sqrt{|g|}\,g^{\mu\nu}\Pi_\nu^{\,M} - i\eps^{\,\mu\nu\lambda}\,
\left(\bth\,\Gamma^{MN}\del_\nu\theta\right)\,
\bigg(\Pi_{\lambda\,N} + \frac{i}{2}\bth\,
\Gamma_{N}\del_\lambda\theta\bigg)~\,\right\}
=0\;,
\label{boseeq}
\eeq
while the equation for the fermionic coordinates is found to be
\beq
(1-\Gamma)\,g^{\mu\nu}\,\Pi_\mu^{\,M}\,\Gamma_M\,\del_\nu\theta=0\;.
\eeq
The latter equation can be rewritten, using both relations in
\eref{userel}, as
\beq
B= (1-\Gamma)\,\tau^{\,\mu}\,\del_\mu\theta=0\ .
\label{fermieq}
\eeq
Note that both the above equations are invariant under the rigid
supersymmetry transformations as given in \eref{rigidsusy}.

Using the equations of motion we can finally analyse the on-shell
degrees of freedom of the super-membrane. The reason we have given
names to the LHS of the above equations is that
it becomes easy to display three relations among them 
\beq
\Pi_\mu^{\,M} A_M = -2i\sqrt{|g|}~\del_\mu\bth B \;.
\eeq 
Since these
equations involve only the canonical momenta $\Pi_\mu^{\,M}$ of the
bosonic coordinates $X^M$, without any time derivatives of the
momenta, they are not dynamical evolution equations. Instead, they
amount to constraints. In this way the 11 bosonic coordinates are
reduced to 8 independent coordinates.

For the fermions, we started with $\theta^\alpha$ which has 32
components. By virtue of the last equation of \eref{userel}, $\tau^\mu$
acts like a gamma-matrix and therefore \eref{fermieq} is like a Dirac
equation. However it differs from a conventional Dirac equation by
having the projection operator $(1-\Gamma)$ in front. Indeed this is
what ensures $\kappa$-symmetry, which acts by a shift in $\theta$
preceded by the orthogonal projector $(1+\Gamma)$ (the remaining
quantities are already $\kappa$-invariant as we have noted.) This
allows us to remove half the degrees of freedom of $\theta^\alpha$.
The Dirac equation then has its usual effect of halving the remaining
degrees of freedom, so at the end we are left with 8 on-shell fermionic
coordinates. The matching of on-shell Bose and Fermi degrees of freedom is a
necessary condition for supersymmetry.

To extract the physical degrees of freedom one must choose a
suitable gauge that fixes worldvolume reparametrisations and
$\kappa$-symmetry. A convenient choice is {\em static gauge}, in which
we choose the time and two arbitrary spatial directions in
the target spacetime and identify them with the worldvolume
coordinates. Thus, we first carry out a split and re-labelling
\beq
X^M\rightarrow (X^{\,\mu},X^I),\quad \mu=0,1,2;\quad I=3,4,\cdots,10\ ,
\eeq
and then impose the gauge-fixing conditions
\beq
X^{\,\mu}=\xi^{\,\mu},\quad \mu=0,1,2\;.
\eeq
For our purposes it is sufficient to assume this has been done locally.
Whether these conditions can be imposed globally will depend on the
topology of the membrane.

Once the static gauge has been chosen, we must re-examine the
symmetries of the theory. Those which violate the gauge condition
will, clearly, no longer be symmetries of the gauge-fixed
theory. However some linear combinations of them may preserve the
gauge and these will be genuine symmetries. An example of this is the
combination of general coordinate transformations on the worldvolume
(which can be represented infinitesimally as local worldvolume
translations) and spacetime Lorentz symmetry
\beq
\delta X^M = \eta^{\,\nu}(\xi)\,\del_\nu X^M+\Lambda^M_{~\,N}X^N\;.
\eeq
Choosing $M=\mu$, we see that each of these terms separately violates
the gauge condition. However performing both transformations together
on $X^\mu$, we get
\beq
\delta X^\mu = \eta^{\,\mu} + \Lambda^\mu_{\,\nu}\,\xi^{\,\nu} + 
\Lambda^\mu_{~I}\,X^I\;.
\eeq
This variation vanishes for the special choice
\beq
\eta^{\,\mu} = - \Lambda^\mu_{\,\nu}\,\xi^{\,\nu} 
- \Lambda^\mu_{~I}\,X^I\;.
\label{combpres}
\eeq
It follows that the gauge-fixed theory will be invariant under those
combinations of worldvolume translations and spacetime Lorentz
transformations that satisfy \eref{combpres} above, namely
\beq
\delta X^I = -\left(\Lambda^\mu_{\,\nu}\,\xi^{\,\nu} +
\Lambda^\mu_{~I}X^I\right) \del_\mu X^I + \Lambda^I_{~J}X^J\;.
\label{thosecomb}
\eeq
The first term on the right-hand-side corresponds to a worldvolume Lorentz
transformation for a set of {\em scalars} $X^I$. To see this, note
that under
\beq
\xi^{\,\mu} \to \xi^{\,\mu} + \ell^{\,\mu}_{\,~\nu}\,\xi^\nu\;,
\eeq
where $\ell_{\mu\nu}=-\ell_{\nu\mu}$ is the parameter of worldvolume
Lorentz transformations, a worldvolume scalar $\phi(\xi)$ changes by
\beq
\delta\phi = \ell^{\,\mu}_{\,~\nu}\,\xi^{\,\nu}\del_\mu\phi\;.
\eeq
This tells us that $\Lambda^{\mu}_{\,~\nu}$ is to be identified with
$-\ell^{\,\mu}_{\,~\nu}$ and the $\SO(2,1)$ subgroup of the spacetime Lorentz
group $\SO(10,1)$ is thereby identified with the $\SO(2,1)$ worldvolume
Lorentz group.

The last term on the RHS of \eref{thosecomb} shows that
the $X^I$ are vectors under rigid $\SO(8)$ rotations of the spacetime
transverse to the membrane worldvolume, generated by the parameters
$\Lambda^I_{~J}$. Finally, the second term on the RHS of
\eref{thosecomb} is a non-linear transformation that parametrises the
coset $\SO(10,1)/\SO(2,1)\times\SO(8)$.

The same combination of worldvolume general coordinate
transformations and spacetime Lorentz transformations on the
fermionic coordinate $\theta$ (which is a spacetime spinor and
worldvolume scalar) becomes, in the static gauge
\beq
\delta\theta = -\Lambda^\mu_{~\nu}\,\xi^{\,\nu}\del_\mu\,\theta + 
\frac14 \Lambda_{\mu\nu}\Gamma^{\,\mu\nu}\theta + 
\frac14 \Lambda_{IJ}\Gamma^{\,IJ}\theta\;,
\label{thetacomb}
\eeq
where we have written only those terms that depend on the 
$\SO(2,1)\times \SO(8)$ parameters. The first two terms in
\eref{thetacomb} give the transformation laws of a worldvolume {\em
spinor}, while the last term is the transformation law of a spacetime
spinor under transverse $\SO(8)$ rotations in spacetime.

We have only gauge-fixed the worldvolume reparametrisations. It still
remains to fix the local $\kappa$-symmetry on the worldvolume. This
may be achieved by imposing
\beq
(1+\Gamma^*)\,\theta=0\ ,
\eeq
where
\beq
\Gamma^*\equiv \Gamma^1\Gamma^2\cdots \Gamma^8\;.
\eeq
This projects $\theta$ to a chiral spinor with respect to $\SO(8)$.
In what follows we will assume the above steps have been carried out
and the fermionic coordinate is re-labelled
$\psi^A,~A=1,2,\cdots,8$ corresponding to a set of 8 real
two-component worldvolume spinors transforming in the spinor of
$\SO(8)$.

In parallel with the case of bosonic symmetries discussed above, we
now find that the (rigid) spacetime supersymmetry transformations are
not by themselves invariances of the gauge-fixed action, but must be
accompanied by a compensating $\kappa$-symmetry transformation as in
\eref{kappasym}. One can easily show that the static-gauge theory has
maximal or ${\cal N}=8$ global supersymmetry in 2+1 dimensions.

From here on we will always work in static gauge. 
The bosonic part of the action is 
\beq
S_{M2,\,bosonic}^{\,static~gauge} =  -T_{M2}
\int d^3\xi \sqrt{-\det\Big(\eta_{\mu\nu}+ 
\frac{1}{T_{M2}} \del_\mu X^I\del_\nu X^I\Big)}\sim 
-\half\int \del_\mu X^I \del^\mu
X^I + \frac{1}{T_{M2}}\, {\mathcal O}(\del X)^4 + \cdots\ ,
\label{bosstatic}
\eeq
where $T_{M2}  = (2\pi)^{-2}\ell_p^{-3}$, and on
the right hand side we have dropped a constant and
restored the precise dependence on the 11d
Planck length $\ell_p$, as well as constant factors.
We see that the action is an expansion in powers of derivatives,
where the leading term is simply the free kinetic term for 8
worldvolume scalars $X^I$. 

The number 8 coincides with the number of spatial directions
transverse to the M2-brane. This is no coincidence but can be derived
by noticing that in the presence of a 2-brane, spatial translational
invariance of the bulk theory is broken from ten independent
translations to only two (those along the brane). The eight broken
translations correspond to the directions transverse to the brane.
From the worldvolume point of view these appear as spontaneously
broken symmetries, and we therefore expect -- and find -- an equal
number of massless Goldstone bosons -- the scalar fields $X^I$.

The above action (after incorporating the fermion terms) represents a
single M2-brane. The question now is to understand what should be the
action for multiple M2-branes. This is an interesting problem even at
lowest-derivative order, and is the main subject of this
review. Before addressing it directly, we continue by reviewing a
different approach to M2-branes, wherein they are seen as stable
supersymmetric soliton solutions of the bulk 11d supergravity.

\subsection{M-branes as solitons}

In this section we display the stable brane solutions of 11d
supergravity. Their stability will be guaranteed
by supersymmetry through a result of Witten and Olive
\cite{Witten:1978mh}, who showed that for charged configurations in
supersymmetric theories, the charge in appropriate units typically
provides an exact quantum-mechanical lower bound on their mass (or
tension, for extended objects). This bound was originally discovered
(in a classical, non-supersymmetric context) by Bogomolny, Prasad and
Sommerfield \cite{Bogomolny:1975de,Prasad:1975kr} and is known as the
BPS bound.  The simplest BPS branes preserve half of the 32 spacetime
supersymmetries of the supergravity theory. In addition to
guaranteeing stability, this property will provide a relatively simple
method to discover the brane solutions.

In what follows it will be convenient to use the eleven-dimensional
``Planck length'' $\ell_p$, as defined in \eqref{ellpdef}. The
condition for a background $G_{MN}, C_{MNP}, \Psi_{M,\alpha}$ to
preserve supersymmetry is that there should exist some nonzero spinor
or spinors $\epsilon$ such that the supersymmetry variations on the
given background vanish.  Since we only consider bosonic backgrounds
(the fermions are set to zero), the supersymmetry variations of the
bosons vanish identically. Thus we only have to check the
supersymmetry variations of the fermions. Then the requirement for a
supersymmetric solution is
\be
\delta \Psi_M \equiv D_M(\omega)\,\epsilon - \frac{i}{288}
\Big(\Gamma^{PQRS}_{\phantom{PQRS}M}
+8 \Gamma^{QRS}\delta_M^{\phantom{M}P}\Big)\,G_{PQRS}\,\epsilon
= 0 \;,
\ee
where we have dropped the hats on $\omega$ and $G_4$ because the
fermionic terms have been set to zero.
Being first order, these equations are much easier to solve
than the full second-order equations of motion.  Moreover, because of
supersymmetry, the corresponding configurations still satisfy the EOM.  

Charged solutions carry the flux of some (generalised) gauge field.
The only possible flux in uncompactified 11d supergravity comes from
the 3-form $C_3$, whose field strength is the 4-form $G_4=dC_3$
defined above.
The spatial components of this 4-form, $G_{lmnp}$, are analogous
to a magnetic field while the components with one time and three space
indices, $G_{0mnp}$, are analogous to an electric field. Accordingly,
classical solutions will be labelled ``electric'' or ``magnetic''
depending on which of these fluxes they involve.  The electric field
is conveniently studied by dualising it to a 7-form
\begin{equation}\label{fdual}
G_7  = \star G_4
 - \frac{1}{2}C_3\wedge G_4\ .
\end{equation}
and then retaining the spatial components $G_{lmnpqrs}$ of this 7-form. 
The $C_3\wedge G_4$ contribution ensures that on shell, $dG_7=0$
in the presence of the Chern-Simons term.

Let us first find the magnetically charged classical
solution. As discussed above, this will have a nontrivial flux
$G_{lmnp}$. The magnetic charge will be $\int_{S^4}
G_4=Q^{(m)}$.  Here $S^4$ is a 4-sphere that encloses the charged
object. This in turn tells us the dimensionality of the object, for in
$D$ spacetime dimensions (equivalently $D-1$ spatial dimensions), a $d$-sphere encloses a $D-d-2$ dimensional object.\footnote{In 3 space  dimensions this is familiar as the fact that a 2-sphere $S^2$ encloses a point, and a
circle $S^1$ encloses an infinitely extended string.} Since we are now
considering a 4-sphere in 11 dimensions, the above formula tells us
that the charged object must extend along $11-4-2=5$
dimensions. Therefore this is a 5-brane, henceforth referred to as the
{\em $M5$-brane} \cite{Gueven:1992hh}.

By a similar argument involving spatial components of the 7-form flux
defined above, we conclude that an electrically charged object in $11$
dimensions must extend along $11-7-2=2$ dimensions. This is therefore
a 2-brane, called the {\em $M2$-brane} or {\em membrane}. In this case
we will have a nonzero value of the electric charge $\int_{S^7} G_7 =
Q^{(e)}$ where $S^7$ is a 7-sphere enclosing the 
M2-brane \cite{Duff:1990xz}.

There can be more general objects carrying both types of
charges \cite{Papadopoulos:1996uq,Tseytlin:1996bh,Howe:1997ue}. 
These would be interpreted as bound states
of M2- and M5-branes. They will turn out to preserve less
supersymmetry than the individual planar M2- and M5-branes. Note that
in $11$ uncompactified dimensions there are no other gauge fields and
therefore no other types of charges available. As a result we do not
expect to find any other stable, charged solitonic objects in the
theory. In particular, there are no stable strings, which is further
evidence that 11d supergravity is not the low-energy limit of a string
theory.

\subsection{The M2 and M5-brane tension}

Let us now describe the M2-brane solution in some
detail \cite{Duff:1990xz}. We take the coordinates along the brane to
be $y^{\,\mu}=(y^0,y^1,y^2)$ while the coordinates transverse to the
brane are denoted\footnote{Here $x^I$ are
  just coordinates and not functions of $y^\mu$, which is why we
  denote them by lower-case letters.}  $x^I=(x^1,x^2,\cdots,x^8)$.
  A planar 2-brane will have a
symmetry $\SO(2,1)\times\SO(8)$ corresponding to Lorentz
transformations within the brane worldvolume and rotations of the
space transverse to the brane. We also expect to have translational
invariance along the brane, \ie in the $y$-coordinates.

These symmetries determine the M2-brane metric and electric flux to
be of the form
\bea
\label{mtwosol}
ds^2 &=& f_{(1)}(r)\,dy^{\,\mu} dy_\mu + f_{(2)}(r)\, dx^I dx^I\nn\\
G_{012r} &=& f_{(3)}(r) \;,
\eea
where $r$ is the radial distance from the brane
\be
r = \sqrt{(x^1)^2 + (x^2)^2 + \cdots + (x^{8})^2}\ ,
\ee
and $f_{(i)}(r),i=1,2,3$ are functions of $r$ that need to
be determined. 

Imposing the equations of motion of 11d supergravity on the above
ansatz, one finds that the three functions
$f_{(1)}(r)$, $f_{(2)}(r)$ and $f_{(3)}(r)$ are all determined by a
single function
\be
H_{M2}(r) = 1 + \frac{(r_{M2})^6}{r^6}\;,
\ee
where $r_{M2}$ is a constant and $H_{M2}(r)$ is harmonic in
the eight transverse dimensions: $\del_I\del_I H_{M2}(r)=0$. In terms of
this function we have
\bea
\label{fsol}
f_{(1)}(r) &=& H_{M2}(r)^{-\frac23}\nn\\
f_{(2)}(r) &=& H_{M2}(r)^{\frac13}\nn\\
f_{(3)}(r) &=& - \frac{\del}{\del r}\left(H_{M2}(r)^{-1}\right)\;.
\eea

We can evaluate the total charge of the solution by integrating the
appropriate flux. Using \eref{fdual} and inserting the solution for
$G_4$ specified in Eqs.~(\ref{mtwosol}), (\ref{fsol}) we find the dual
7-form flux to be
\be
G_{J_1J_2\cdots J_7} = 6\,(r_{M2})^6\, \epsilon_{I J_1J_2\cdots
J_7}\frac{x^I}{r^8} \; \ .
\ee
In spherical polar coordinates $(r,\theta^{\,i})$ with
$i=1,2,\cdots,7$, $G_7$ has components only in the angular directions
and can be written
\be
G_{\theta_1\theta_2\cdots \theta_7} = 6\,(r_{M2})^6\,
\epsilon_{\theta_1\theta_2\cdots
\theta_7}\ ,
\ee
from which it follows that the electric charge of the M2-brane is
\be
\label{qe}
Q^{(e)} =  6\,(r_{M2})^6\,\Omega_7 = 2\pi^4 (r_{M2})^6\ ,
\ee 
with $\Omega_7=\frac{1}{3}\pi^4$ being the volume of a unit 7-sphere.

By comparing the metric with Newton's law in the weak-field
approximation, we can obtain a relation between the parameter $r_{M2}$
in the solution and the tension of an M2-brane. The basic formula
relates the time-time component of a static $p$-brane metric in $D$
spacetime dimensions to the brane tension. For static, pointlike
sources, Einstein's equations in $D$ spacetime dimensions
\be
R_{\mu\nu} -\half g_{\mu\nu}R = 8\pi \cG_{(D)}T_{\mu\nu}\ ,
\ee
reduce to
\be
\label{rzerozero}
R_{00} = \frac{D-3}{D-2}\, 8\pi \cG_{(D)}\,\rho\;.
\ee
Using $R_{00}= -\half \nabla^2 g_{00}$ to leading
order in the Newtonian approximation, one gets
\be
\nabla^2 g_{00} = -2\,\frac{D-3}{D-2}\,8\pi \cG_{(D)}\,\rho\;.
\ee
Comparing this with Newton's equation
\be
\nabla^2\phi = 4\pi \cG_{(D)}\,\rho\ ,
\ee
we identify
\be
g_{00} = -\left(1+4\,\frac{D-3}{D-2}\,\phi\right)\;.
\ee 
For a pointlike source with $\rho(x) = M \delta^{D-1}(x)$, one has
\be
\label{pointphi}
\phi(x) = -\frac{4\pi \cG_{(D)}M}{(D-3)\,\Omega_{D-2}}\frac{1}{r^{D-3}}\ ,
\ee
and therefore
\be
g_{00} = -\left(1-\frac{16\pi
\cG_{(D)}M}{(D-2)\,\Omega_{D-2}}\frac{1}{r^{D-3}}
\right)\;.
\ee

Since we assumed the source to be pointlike, this formula describes
the Newtonian limit for black holes in arbitrary spacetime
dimensions. It is easily generalised to extended black $p$-branes. In
this case, we label the coordinates $A,B=0,1,\cdots,D-1$ of which the
subset $\mu,\nu=0,1,\cdots, p$ lie along the brane. For a static brane
configuration $T_{\mu\nu}=\eta_{\mu\nu}\,\rho$ and hence
\eref{rzerozero} is modified to
\be
R_{00} = \frac{D-p-3}{D-2}\, 8\pi \cG_{(D)}\,\rho\ ,
\ee
and therefore
\be
g_{00} = -\left(1+4\,\frac{D-p-3}{D-2}\,\phi\right)\;.
\ee
Moreover, \eref{pointphi} changes to
\be
\label{branephi}
\phi(x) = -\frac{4\pi \cG_{(D)}T_p}{(D-p-3)\,
\Omega_{D-p-2}}\frac{1}{r^{D-p-3}}\;,
\ee
where $T_p$ is the tension of the $p$-brane, with dimensions of (mass)$^{p+1}$.
Combining these two results, we have
\be
g_{00} = -\left(1-\frac{16\pi
\cG_{(D)}T_p}{(D-2)\,\Omega_{D-p-2}}\frac{1}{r^{D-p-3}}
\right)\;.
\ee

Applying this formula to M2-branes in 11 dimensions and comparing with
\eref{mtwosol} we find
\be
(r_{M2})^6 = \frac{8\pi}{3} \frac{\cG_{(11)}}{\Omega_7}\,n_2\, T_{M2}\;,
\ee
where $n_2$ is the number of 2-branes and $T_{M2}$ is the tension of
a single M2-brane. Using \eref{ellpdef} we then obtain
\be
\label{rmtwo}
(r_{M2})^6=128\,\pi^4 n_2\,\ell_p^9\,T_{M2}\;.
\ee

For the M5-brane \cite{Gueven:1992hh}, we take the coordinates on the
brane to be $y^{\,\mu}=(y^0,y^1,\cdots,y^5)$, and the coordinates
transverse to the brane to be $x^I=(x^1,x^2,\cdots,x^5)$. By
reasoning similar to the M2-brane case, we assume a symmetry
$\SO(5,1)\times\SO(5)$ and also translational invariance in the
$y$-coordinates. These symmetries fix the metric to be of the form
\be
ds^2 = g_{(1)}(r)\,dy^{\,\mu} dy_\mu + g_{(2)}(r)\, dx^I dx^I\;.
\ee
Here $r$ is the radial distance from the 5-brane
\be
r = \sqrt{(x^1)^2 + (x^2)^2 + \cdots + (x^{5})^2}\;.
\ee
Thus we need to find the functions $g_{(1)}(r)$ and $g_{(2)}(r)$, and
as before these are determined by a single function
\be
H_{M5}(r) = 1 + \frac{(r_{M5})^3}{r^3}\;,
\ee
where $r_{M5}$ is a constant (that will be related to the
magnetic charge of the 5-brane), and $H_{M5}(r)$ is harmonic in
the 5 transverse dimensions: $\del_I\del_I H_{M5}(r)=0$.
In  terms of this function we have
\bea
g_{(1)}(r) &=& H_{M5}(r)^{-\frac13}\nn\\
g_{(2)}(r) &=& H_{M5}(r)^{\frac23}\;.
\eea

Additionally the magnetic flux of the solution is
\be
G_{\theta_1\theta_2\theta_3\theta_4} = 3 (r_{M5})^3
\epsilon_{\theta_1\theta_2\theta_3\theta_4}\;.
\ee
The magnetic charge of the solution is
\be
\label{qm}
Q^{(m)} = \int_{S^4} G = 3\,(r_{M5})^3 \Omega_4= 8\pi^2 (r_{M5})^3\;,
\ee
where $\Omega_4=\frac{8}{3}\pi^2$ is the volume of a unit 4-sphere.

Finally, using the Newtonian approximation once more, we find the relation
\be
\label{rmfive}
(r_{M5})^3 = 32\,\pi^6 n_5\,\ell_p^9 \,T_{M5}\;,
\ee
where $n_5$ is an integer, the number of M5-branes.

Since M-theory has only one
dimensional parameter $\ell_p$, we can predict on dimensional grounds
that
\be
T_{M2}\sim \frac{1}{\ell_p^3},\quad T_{M5}\sim \frac{1}{\ell_p^6}\;.
\ee
Additional information on the actual values can be obtained using the
Dirac quantisation condition, which tells us that
\be
\frac{1}{16\pi G_{(11)}}\,Q^{(e)}Q^{(m)} = 2\pi n\;,
\ee 
where $n$ is an integer \cite{Teitelboim:1985yc}. Choosing single branes and the minimum
quantum, \ie $n_2=n_5=n=1$, and making use of
Eqs.~(\ref{qe}), (\ref{rmtwo}), (\ref{qm}), (\ref{rmfive}) we find
\be
\frac{2\pi}{(2\pi\ell_p)^9}\,Q^{(e)}Q^{(m)} 
= (2\pi)^8 \ell_p^9\,T_{M2}\,T_{M5}\;.
\ee
Setting the RHS equal to $2\pi$ (because of the Dirac quantization condition), we find
\be
T_{M2}\,T_{M5}= \frac{(2\pi)^2}{(2\pi\ell_p)^9}\;.
\ee
In the next section, we will argue that the correct answers are
\be
T_{M2}=\frac{2\pi}{(2\pi\ell_p)^3},\quad T_{M5}
= \frac{2\pi}{(2\pi\ell_p)^6}\;.
\ee

In the above solitonic description of branes it follows, using the
techniques of soliton physics, that the translation symmetries broken
by the brane are collective coordinates. Therefore the brane
worldvolume will support a corresponding number of massless scalar
fields. The M2-brane theory should then have 8 massless scalars, which we have
already encountered in a previous section, while the M5-brane theory
should have 5 massless scalars.

\subsection{Relation to branes in string theory}

If M-theory exists, the brane solitons we have found must be among its stable quantum excitations. The relation to string theory suggested in the previous section then tells us that after compactifying on a circle, the M-theory branes must reduce to one of the branes in type IIA string theory \cite{Witten:1995ex,Townsend:1995kk,Schwarz:1995jq,Aharony:1996xr}.  Indeed one should be able to account for {\it all} stable branes in type IIA string theory from the M-theory perspective. This is potentially a challenge, since type IIA string theory has stable BPS D0, D2, D4 and D6-branes,\footnote{It also has D8-branes but, being domain walls, these change the nature   of the spacetime and render the low-energy theory massive. It has recently been argued   \cite{Aharony:2010af} that massive type IIA theory does not have a strong-coupling, weakly curved limit.} as well as the fundamental string and the NS5-brane, while M-theory only has M2 and M5-branes. At the same time, we have already reproduced all the massless $p$-form gauge fields under which the branes of type IIA string theory are charged; this provides a hint that things should work out properly.

Recall that BPS D$p$-branes in type II string theory have tensions
\beq
 T_p = \frac{1}{g_s}\frac{2\pi}{(2\pi\ell_s)^{p+1}}
\label{dtensions}\;.
\eeq
In addition, there is a stable string (the fundamental string) and its
electric dual, the NS5-brane. The formulae for their
tensions are as follows 
\beq
 T_{F1}=\frac{2\pi}{(2\pi\ell_s)^2},\quad  T_{NS5} =
\frac{1}{g_s^2}\frac{2\pi}{(2\pi\ell_s)^6}
\label{nstensions}\;.
\eeq

We may now try to derive these results starting with M-branes.
However there is a potential problem. The tensions of string theory
branes were calculated at weak coupling. One might expect them to be
renormalised by the time we reach M-theory in the strong coupling
limit. Fortunately here we may rely on the fact that the branes under
discussion are maximally supersymmetric. It can be argued that the
tension of such supersymmetric branes is exact \cite{Dabholkar:1989jt} -- an example of
a non-renormalisation theorem. Therefore we are free to proceed and
compare BPS branes in the two theories.

Now when compactifying on a circle, the M2-brane can be either wrapped
on the circle or transverse to the circle. In the first case it looks
(as $R_{10}\to 0$) like a string or ``1-brane.'' In the second case it
is a 2-brane. Doing the same thing for an M5-brane, we get a 4-brane
when it is wrapped along the circle and a 5-brane when it is transverse
to it.  To match with the branes in string theory, the only
possibilities are that the wrapped M2 becomes the fundamental string
(F1), the transverse M2 becomes the D2-brane, the wrapped M5 becomes
the D4-brane and the transverse M5 becomes the NS5 brane.

This gives rise to a definite set of predictions. Let us start with the
M2-brane. Above, we stated without proof that its tension is
\beq
T_{M2}=\frac{1}{4\pi^2\ell_p^3}\;.
\eeq
Assuming this to be true and wrapping on the circle, the tension of
the resulting brane is
\bea
T_{M2}^{\textrm{wrapped}}&=& T_{M2}\times 2\pi R_{10}\nn\\
&=& \frac{1}{4\pi^2 g_s\ell_s^3}\times 2\pi g_s\ell_s\nn\\
&=& \frac{1}{2\pi\ell_s^2}\;,
\eea
which is correct. This result basically serves to fix the tension
of the M2-brane.

Now consider the transverse M2-brane. Its tension is
\bea
T_{M2} &=& \frac{1}{4\pi^2\ell_p^3}\nn\\
&=& \frac{1}{4\pi^2 (g_s^{\frac{1}{3}}\ell_s)^3}\nn\\
&=& \frac{1}{g_s}\frac{1}{4\pi^2 \ell_s^3}\nn\\
&=& T_{D2}\;.
\eea
This is a remarkable agreement, and a very precise test of the
M-theory conjecture.

For the M5-brane, the story proceeds as follows.
We have previously proposed that its tension is
\beq
T_{M5}=\frac{1}{32\pi^5\ell_p^6}\;.
\eeq
Wrapping on the circle, the tension of the resulting brane is
\bea
T_{M5}^{\textrm{wrapped}}&=& T_{M5}\times 2\pi R_{10}\nn\\
&=& \frac{g_s\ell_s}{16\pi^4 g_s^2 \ell_s^6}\nn\\
&=& \frac{1}{g_s}\frac{2\pi}{(2\pi \ell_s)^5}\nn\\
&=& T_{D4}\;,
\eea
which is correct, but again can be thought of as a
determination of $T_{M5}$.

Finally, the transverse M5-brane gives
\bea
T_{M5} &=& \frac{1}{32\pi^5\ell_p^6}\nn\\[2mm]
&=& \frac{1}{g_s^2}\frac{1}{32\pi^5 \ell_s^6}\nn\\[2mm]
&=& \frac{1}{g_s^2}\frac{2\pi}{(2\pi\ell_s)^6}\nn\\[2mm]
&=& T_{NS5}\;,
\eea
which is again a successful test of the equivalence between
M-theory and type IIA string  theory.

This leaves the D0 and D6 branes.  From \eref{dtensions}, the mass of
a D0 brane is 
\beq 
T_0 = \frac{1}{g_s\ell_s}=\frac{1}{R_{10}}\;.
\eeq
What mode of M-theory can have this mass? A crucial clue comes from
the fact that in string theory, D0-branes are charged under the Ramond-Ramond
1-form gauge potential $A_\mu$. In comparing 11d and 10d
supergravity, we found that $A_\mu$ in the latter arises from
Kaluza-Klein reduction of the metric of the former on the M-theory
circle. This suggests that D0-branes must arise from modes in M-theory
carrying momentum along the M-circle.

Indeed we now argue that a single D0-brane corresponds to the mode of
M-theory with one unit of momentum along the compact direction.
On a compact dimension of length $L$, the momentum is
quantised in integers as 
\beq 
p = \frac{2\pi n}{L} \;.
\eeq 
For massless particles in 11d, we have 
\beq 
E^2 = p_1^2 + \cdots p_9^2 + p_{10}^2 \;.
\eeq 
After compactification, a fixed value of $p_{10}$ will appear as a
mass.  Since $L=2\pi R_{10}$, we have that the 10d mass of states
carrying this momentum is $|p_{10}| = n/R_{10}$. Thus a single
D0-brane ($n=1$) can be identified with an M-theory mode carrying
a single unit of momentum along $x^{10}$.

This leads to a new prediction. From the M-theory point of view there
can be a momentum mode along the compact direction for any integer
$n$. In type IIA string theory, this can only be a bound state of $n$
D0-branes! This is a statement about string theory that we did not
know before the discovery of M-theory. It was subsequently verified
directly within string theory \cite{Sen:1995vr}.

To find D6-branes in M-theory, we first examine D0-branes in a little more detail. As mentioned above, they carry an electric charge under $A_\mu$. This charge is the integral of a suitable differential form over a sphere enclosing the D0-brane. In 10 dimensions, a 0-brane can be enclosed by an 8-sphere, $S^8$, and therefore its charge must be defined as the integral of an 8-form which, in turn, is the Poincar\'e dual of the 2-form field strength $F=dA$ of the Ramond-Ramond 1-form $A_\mu$. As we just saw, from the M-theory point of view $A_\mu$ arises as a Kaluza-Klein gauge field. One expects to find a dual object which can be enclosed by a two-sphere $S^2$ and is a magnetic source for the same field strength.  Such an object will be a 6-brane.\footnote{Indeed, it is known that in type IIA string theory, the D6-brane is the magnetic   dual of the D0-brane.} A magnetically charged object under a Kaluza-Klein gauge field is called a Kaluza-Klein monopole \cite{Sorkin:1983ns,Gross:1983hb}.  We conclude that if the D6-brane of type IIA string theory is to arise in M-theory, it must be a Kaluza-Klein monopole.

Let us first discuss such monopoles abstractly and later embed them
into M-theory.  Consider the metric, known as multi-Taub-NUT, in 4 Euclidean
dimensions \cite{Gibbons:1979zt}
\bea\label{KKinone}
ds_{\hbox{\tiny Taub-NUT}}^2 &=& U({\vec x})\,d{\vec x}\cdot d{\vec x}
+  \frac{1}{U({\vec x})}\Big(dy + {\vec A}\cdot d{\vec x}\Big)^2\;,
\eea
where ${\vec A}$ is the vector potential for a magnetic
monopole in 3 dimensions
\beq
{\vec B} = {\vec\nabla}\times{\vec A}\ ,
\eeq
and $U({\vec x})$ is a harmonic function in 3d determined by
\beq
{\vec\nabla}U = -{\vec B}\;.
\eeq
It can be shown that this metric solves the 4d Euclidean Einstein
equation without sources.

We choose a specific harmonic function $U$ depending on a real
number $R$, namely
\beq
U({\vec x}) = 1 + \frac{R}{2r}\;,
\eeq
where $r=|{\vec x}|$.
Thus the magnetic field is
\beq
{\vec B} =  \frac{R}{2}\frac{{\vec x}}{r^3}\;.
\eeq
As $r\to 0$ the metric written above is apparently singular 
due to the terms
\beq
\frac{R}{2r}\,dr^2 + \frac{2r}{R}\, dy^2\;.
\eeq
The singularity can be avoided as follows. Define
\beq
\tilder=\sqrt{2rR}\;.
\eeq
The dangerous terms then become
\beq
d\tilder^2 + \frac{\tilder^2}{R^2}\,dy^2\;.
\eeq
Now the second term is non-singular if $y$ is an {\em angular
  coordinate} with periodicity precisely $2\pi R$. Being a
non-singular metric with a monopole charge, this is called a
Kaluza-Klein monopole (more precisely it is the spatial metric,
but we can then add $-dt^2$ to make it the describe the 
worldline). The
monopole is located at the core near $r\to 0$, where the Kaluza-Klein
circle shrinks to zero size.

Let us now embed this solution in M-theory by taking the ${\vec x}$
directions to be $x^7,x^8,x^9$ and the Kaluza-Klein direction $y$ to be $x^{10}$
with periodicity $2\pi R_{10}$.  The resulting object is translationally
invariant along $x^1,x^2,\cdots,x^6$ so it is a 6-brane.  And it is
magnetically charged under the Kaluza-Klein gauge field arising from
compactification of $x^{10}$. So we have a candidate object in
compactified M-theory that can be matched with the D6-brane
of type IIA string theory.

To compute the tension, we just integrate the energy density
${\vec\nabla}^2 U$ along the four dimensions in which the monopole is
embedded. Since $U$ is independent of the compact direction, we get
\bea
T_{KK6} &=& \frac{2\pi}{(2\pi\ell_p)^9}\times 2\pi R_{10}\int d^3x \,
{\vec\nabla}^2 U\nn\\
&=& \frac{2\pi}{(2\pi\ell_p)^9}\times (2\pi R_{10})^2\nn\\
&=& \frac{1}{g_s}\frac{2\pi}{(2\pi\ell_s)^7} = T_{D6}\;.
\eea
Thus we have successfully understood the D6-brane as arising from an
object in M-theory. This completes our survey of how D-branes of type
IIA string theory arise from M-theory.

We can now see if type IIB string theory is likewise illuminated by
M-theory. Supersymmetric branes in type IIB string theory can be
obtained from those of type IIA by circle compactification and
T-duality. It is easy to check that this reproduces the tensions of
all the BPS branes of type IIB: D1, D3, D5, D7 as well as F1 and NS5, given in
\eref{dtensions} and \eref{nstensions}. However we get some
additional and highly nontrivial information out of M-theory.

Recall that in type IIB there are two types of strings,
F-strings of tension $1/2\pi\ell_s^2$ and
D-strings of tension $1/2\pi g_s \ell_s^2$.
Based on a continuous symmetry of type IIB supergravity, it has been
argued that type IIB string theory has a discrete S-duality symmetry
group that (for vanishing Ramond-Ramond axion $\chi$) includes the
nonperturbative strong-weak duality
\beq
g_s\to \frac{1}{g_s},\quad \ell_s\to \sqrt{g_s}\,\ell_s\;.
\eeq
Under this ``S-duality'' symmetry, the F-string and D-string are
interchanged. An easy check of the proposal is that the tensions of
these strings get interchanged by the proposed duality. 
Additionally, it has been shown that $p$ F-strings and $q$
D-strings form stable bound states called $(p,q)$
strings, if $p,q$ are co-prime \cite{Witten:1995im}.
These have tension
\beq
T_{p,q} = \frac{1}{2\pi\ell_s^2}\sqrt{p^2 + \frac{q^2}{g_s^2}}\;.
\eeq
The above facts are difficult to prove rigorously because S-duality is
intrinsically nonperturbative in nature, exchanging a weakly coupled
with a strongly coupled theory. We will now see that M-theory explains
and even predicts these results, in a beautifully simple geometric way.

Suppose we compactify M-theory on two circles $x^9,x^{10}$ of radii
$R_{9},R_{10}$ to get type IIA string theory in 9 dimensions. From the
above discussion it should be clear that the 
M2-brane wrapped on $x^{10}$ is the type IIA F-string, while
the M2-brane wrapped on $x^{9}$ is the D2-brane wrapped on $x^{9}$.
Now let us perform a T-duality along $x^9$. 
This duality transformation maps type IIA string theory onto the type
IIB theory. It can be shown that
in the process, fundamental strings are mapped to fundamental strings
and D-branes to D-branes \cite{Polchinski:1995mt}.  The dimension
of branes decreases by one unit if they are initially wrapped on the
T-duality direction, and increases by one unit if they are initially
transverse to this direction. Therefore under this T-duality, the 
type IIA F-string becomes the type IIB F-string, and the
D2-brane wrapped on $x^9$ becomes the type IIB D-string.

It follows that the interchange of the F-string and D-string in type
IIB string theory is just the interchange of the directions
$x^9$ and $x^{10}$ in M-theory!
But the latter is part of Lorentz invariance and is a manifest
geometrical symmetry of M-theory. S-duality can be extended to include
the Ramond-Ramond axion field $\chi$ and then corresponds to the 
group $\mathrm{PSL}(2,\mathbb Z)$.
On the M-theory side, this is realised as the group of modular
transformations on the 2-torus (with the angle between the two sides
being related to the type IIB axion). This is therefore a ``proof'' of
S-duality, though of course it requires us to believe in the
existence of M-theory and the validity of its proposed relationship to
type IIA string theory after compactification, facts which themselves
have not been rigorously proven. Still it is satisfying that a highly
consistent picture emerges using M-theory.

Finally we address $(p,q)$ string bound states. In the proposed
relationship of M-theory to type IIB string theory, it was
shown \cite{Witten:1995ex} that
\beq
g_s \hbox{ (IIB)} = \frac{R_{10}}{R_9},\qquad
\ell_s \hbox{ (IIB)} = \sqrt{\frac{\ell_p^3}{R_{10}}}\;.
\eeq
This follows easily using the Buscher T-duality
rules \cite{Buscher:1987qj}. Next, suppose that in the same
compactification we wrap an M2-brane $p$ times along $x^{10}$ and $q$
times along $x^9$.  The result, after T-dualising on $x^9$, is a
string-like object in type IIB theory that has $p$ units of F-string
charge as well as $q$ units of D-string charge. The tension of the
resulting string will be
\bea
T_{M2}^{\textrm{wrapped}} &=& T_{M2}\sqrt{ p(2\pi R_{10})^2
+ q(2\pi R_9)^2}\nn\\
&=& \frac{1}{2\pi\ell_s^2}\,\sqrt{p^2 + \frac{q^2}{g_s^2}}\ =\ T_{p,q}\;.
\eea
Since the first line is just the total length of the $(p,q)$ string, as
follows from Pythagoras' theorem, we see that M-theory has geometrised
the tension of $(p,q)$ string bound states.


\section[Multiple membranes: background and early 
attempts]{{\Large {\bf Multiple membranes: background 
and early attempts}}}\label{chapter2}

We will now focus our discussion 
on M2-brane worldvolume theories. As we have already
mentioned, the description of multiple M2-branes had been an important
open problem since the discovery of M-theory. In the following
sections we will present various pieces of the relevant background and
early ideas, which led to the modern understanding of these
configurations.

\subsection{M2-branes as strongly coupled D2-branes}\label{abelianM2D2}

Let us return to the bosonic part of the single M2-brane action in
static gauge,\footnote{Recall that the 8 scalars
  $X^I$ are supplemented by a set of fermionic coordinates
  $\psi^A,~A=1,\cdots,8$ with each $\psi^A$ being a complex
  2-component spinor on the worldvolume.} \eref{bosstatic},
\beq
S_{M2}^{bosonic} =  -\frac{1}{(2\pi)^2\ell_p^3}
\int d^3\xi \sqrt{-\det\Big(\eta_{\mu\nu}+ 
(2\pi)^2\ell_p^3\, \del_\mu X^I\del_\nu X^I\Big)}\ ,
\label{mtbos}
\eeq
where $I=1,\cdots,8$.

The above action can be compared with the corresponding
action for a single D2-brane in type IIA string theory. The latter has
seven scalars $X^i,i=1,\cdots,7$ that transform under an SO(7)
symmetry, as well as an abelian worldvolume
gauge field $A_\mu$. The bosonic part of this action in static gauge
is
\be
S_{D2}^{bosonic} = -\frac{1}{(\alp)^2 \gYMs}\int d^3\xi\sqrt{-\det\Big(
  \eta_{\mu\nu} + (\alp)^2 \del_\mu X^i\del_\nu X^i+ \alp
  F_{\mu\nu}\Big)}\;,
\label{dtbos}
\ee
where $F_{\mu\nu}=\del_\mu A_\nu - \del_\nu A_\mu$ and the coupling
constant $\gYM$ is related to the type IIA string coupling $g_s$ by
\be
\gYMs=\frac{g_s}{\sqrt{\alpha'}}\;.
\label{gYMdef}
\ee
Clearly the action is invariant under SO(7), representing rotations
in the space transverse to the membrane.

Because D-branes are loci where open strings end, the above action
can be directly derived using techniques of open-string 
theory \cite{Polchinski:1998rr}.
The factor of $(\gYM)^{-2}\sim g_s^{-1}$ in front of the entire action
reflects the fact that it is a tree-level open-string action.
The coefficients of the $\del X\del X$ and $F$ terms have been
chosen so that upon expanding \eref{dtbos} in powers of $\alpha'$,
the leading terms are of the canonically normalised form 
\be 
\frac{1}{\gYMs}\left(-\half
\del_\mu X^i\del^\mu X^i -\frac{1}{4}F_{\mu\nu}F^{\mu\nu}\right)\;.
\ee

The single-D2-brane and single-M2-brane actions can be transformed into each other \cite{Townsend:1995af,Schmidhuber:1996fy}
in a way that mirrors the duality of the parent string theory and
M-theory. We now demonstrate this explicitly for the bosonic part of
the actions. For this, we start with \eref{dtbos} and manipulate it
using a transformation called {\em abelian duality}, in which it is
replaced by the equivalent action 
\be\label{abelianduality}
\cL = \half
\varepsilon^{\mu\nu\lambda} B_\mu F_{\nu\lambda} -
\frac{1}{(\alp)^2\gYMs} \sqrt{-\det(\eta_{\mu\nu} + (\alp)^2 \del_\mu X^i\del_\nu X^i + (\alp)^2 g_{YM}^4
  B_\mu B_\nu) }\;.  
\ee 
Here $B_\mu$ is a non-dynamical field that
appears in algebraic (rather than derivative) form in the action and
therefore in the equations of motion. It can be integrated out by
solving its own equations of motion and substituting the result back
in the above action. Upon doing this,  one recovers
\eref{dtbos}.

We may instead choose to integrate out the gauge field $A_\mu$. Its
equation of motion tells us that $\del_\mu B_\nu - \del_\nu B_\mu=0$
and therefore $B_\mu$ is the gradient of a scalar, which we write as
\be 
B_\mu \to -\frac{1}{\gYM} \del_\mu X^8\;, 
\label{btoxe}
\ee 
where the
coefficient is chosen so that the eventual kinetic term for $X^8$ is
correctly normalised. Recalling the relation 
$\ell_p^3=g_s\ell_s^3=g_s(\alpha')^{\frac32}$ (see \eref{ellsellp} of
Chapter~\ref{chapter1}) and \eref{gYMdef} above, and 
rescaling $X^i\to \gYM X^i$, 
we end up with the action
\be
\cL = -\frac{1}{(2\pi)^2\ell_p^3}\sqrt{-\det(\eta_{\mu\nu}+ 
(2\pi)^2\ell_p^3\, \del_\mu X^I\del_\nu X^I)}\;,
\ee
where the new scalar $X^8$ defined in \eref{btoxe} 
now appears symmetrically with the seven original scalars $X^i$.

Apparently this action depends solely on $\ell_p$ and has SO(8)
symmetry. However quantisation of flux in the original gauge theory
imposes the periodicity condition 
\be
 X^8 \sim X^8 + 2\pi\gYM\;,
\ee
which violates SO(8) and introduces a dependence on $\gYM$. It is
only in the limit $\gYM\to\infty$ (which is the same as the M-theory
limit $g_s\to\infty$) that the dependence on $\gYM$ disappears and the
field $X^8$ becomes noncompact like the remaining seven scalars. In
this limit we indeed find the correct M2-brane action which depends solely on
$\ell_p$ and has $\SO(8)$ invariance.

These manipulations teach us that the action for a single D2-brane
gets transformed into that for a single M2-brane in the strong
coupling limit in which type IIA string theory transforms into
M-theory.  Moreover, since $\gYM$ is the coupling constant of the
D2-brane theory, it emerges that the M2-brane field theory is the {\em
  strongly-coupled limit} of the D2-brane theory. This is a very
helpful insight, that can be used as follows. Consider the low-energy
limits of the M2-brane and D2-brane worldvolume actions. In the
former, this is achieved by taking $\ell_p\to 0$ while in the latter
it requires $\ell_s=\sqrt{\alpha'}\to 0$. The resulting field theories
are (we reintroduce the fermion terms at this stage)
\bea
S_{M2}^{\textrm{susy}}&\buildrel{\ell_p\to 0}\over{=}& \int d^3\xi~ \Big(
-\half\del_\mu X^I \del^\mu X^I
+\frac{i}{2}\,\bpsi^A \gamma^\mu\del_\mu\psi^A
\Big)\nn\\
S_{D2}^{\textrm{susy}}&\buildrel{\ell_s\to 0}\over{=}& \frac{1}{\gYMs}
\int d^3\xi~
\Big(-\half \del_\mu X^i \del^\mu X^i-\frac{1}{4}F_{\mu\nu}
F^{\mu\nu}
+\frac{i}{2}\,\bpsi^A \gamma^\mu\del_\mu\psi^A
\Big)\;,
\label{lowder}
\eea 
where in the first action, $A=1,2,\cdots,8$ runs over the indices of the
8 real dimensional spinor representation of $\SO(8)$, while in the
second it takes the same values but should be identified with the
spinor representation of $\SO(7)$. In this lowest-derivative limit
we have two free (quadratic) field theories and their equivalence via
abelian duality is simple to check by following the same steps that
were used to go from \eref{dtbos} to \eref{mtbos}.

However, we can now do more. For D-branes in string theory, we know
the low-energy worldvolume action not only for a single brane but for
any number $n$ of branes. In this case we have open strings stretching
from one brane to itself (which give rise to $n$ copies of the
single-brane action) but also (oriented) open strings stretching
between each pair of distinct branes. These add $2\times \half n(n-1)$
degrees of freedom so that altogether one has $\mathcal O(n^2)$
degrees of freedom. These are realised as $n\times n$ Hermitian
matrices and the action, in the limit $\ell_s\to 0$, is that of
$\U(n)$ Yang-Mills theory with seven scalar fields in the adjoint
representation as well as adjoint fermions whose couplings follow from
the ${\cal N}=8$ supersymmetry.

The action in the
second line of \eref{lowder} is thereby generalised to the action of
${\cal N}=8$ supersymmetric Yang-Mills theory in 2+1 dimensions 
\be
S_{n\,D2}\buildrel{\ell_s\to 0}\over{=} \frac{1}{\gYMs}
\int d^3\xi~ \Tr\,\Big(-\half
D_\mu \boldX^i D^\mu \boldX^i +\frac{1}{4}[\boldX^i,\boldX^j]^2
-\frac{1}{4}\boldF_{\mu\nu} \boldF^{\mu\nu}
+\frac{i}{2}{\bar\boldpsi}^A\gamma^\mu D_\mu \boldpsi^A -
{\bar\boldpsi^A}\Gamma^i_{AB}[\boldX^i,\boldpsi^B] \Big)\ , 
\ee 
where $\boldX=X^I T^I, \boldA_\mu=A_\mu^I T^I, \boldpsi=\psi^IT^I$ with
$T^I, I=1,2,\cdots,n^2$ being the generators of the Lie algebra
$\U(n)$,
\bea
D_\mu \boldX^i &=& \del_\mu \boldX^i -i [\boldA_\mu,\boldX^i]\nn\\
\boldF_{\mu\nu}&=& \del_\mu \boldA_\nu - \del_\nu \boldA_\mu -i
[\boldA_\mu,\boldA_\nu]\ , 
\label{covderdef}
\eea 
and $\Gamma^i_{AB}$ are matrices which
convert the product of two SO(7) spinors with indices $A,B$ into an
SO(7) vector with index $i$. These can be derived from 
10-dimensional gamma-matrices.

We are finally in a position to {\em define} the field theory on the
worldvolume of multiple membranes: Simply consider the ${\cal N}=8$
supersymmetric $\U(n)$ Yang-Mills theory with coupling constant $\gYM$ and
take the M-theory limit $g_s\to\infty$, which implies $\gYM\to\infty$.
Because $\gYM$ has dimensions of $({\rm length})^{-\half}$ in 2+1d,
the strong-coupling limit is the same as the long-distance or infrared
(IR) limit of the field theory. If there is to be a nontrivial field
theory of multiple membranes, it must therefore be the (conformally
invariant) IR fixed point of ${\cal N}=8$ supersymmetric $\U(n)$ Yang-Mills
theory. The existence of such a Spin(8)-invariant interacting IR fixed point for three-dimensional SYM was argued in \cite{Sethi:1997sw} based on S-duality.  Our ultimate goal will be a lagrangian description of this
field theory.

Of course this part of the discussion holds for the limit of small
$\ell_s$ or $\ell_p$ in which higher-derivative terms are ignored.
When these terms are included, even the generalisation of ${\cal N}=8$
Yang-Mills theory (which is computable in string perturbation theory)
is not fully known except to the lowest nontrivial order.
Therefore we will concentrate mostly on the $\ell_p\to 0$ limit for
multiple membranes, though in some cases we will also be able to
obtain higher-derivative corrections to lowest nontrivial order in
$\ell_p$.

\subsection{Brane funnels}\label{BH}

As we reviewed briefly above, a D-brane in string theory is
characterised by the fact that open fundamental strings can end on it.
This fact was used to derive the field theory on multiple D-branes.
One may wonder whether an analogous property holds for branes in
M-theory and can be similarly used to learn about M-theory branes. The
analogues are easily constructed by thinking about M-theory as the
strongly coupled limit of type IIA string theory. Starting with a
fundamental string ending on a D2-brane in type IIA, the M-theory
limit converts the D2-brane into an M2-brane and the F-string into
another M2-brane with a different orientation. It is easy to 
establish \cite{Aharony:1996xr} that the two are smoothly 
connected into a single M2-brane. It follows
that multiple M2-branes can be connected to each other not by strings
(which are in any case absent in M-theory) but by M2-branes in such a
way that the entire configuration is a single M2-brane with several
asymptotic regions describing both the initial parallel branes and the
``connecting'' branes.
 
Similarly, one may start with a fundamental string ending on a
D4-brane and take the M-theory limit. At the end one has an M2-brane
ending on an M5-brane, with the common part of their worldvolumes
being a string.  It was shown by Strominger \cite{Strominger:1995ac}
that M2-branes ending on M5-branes satisfy consistency conditions for
the worldvolume couplings and are supersymmetric whenever both sets of
branes are individually parallel and the M2's are normally incident on
the M5's.  This leads us to consider the possibility that worldvolume
field theories (perhaps for both M2 {\em and} M5-branes) could be
reconstructed or guessed using brane intersections.

\subsubsection{D-brane fuzzy funnels}

In fact the M2-M5 relationship is similar to a relationship among
D-branes in type IIB string theory. There, one can use strong-weak
duality (S-duality) to transform the supersymmetric configuration of
an open fundamental string ending on a D3-brane, a configuration known
as a ``BIon'' \cite{Callan:1997kz,Howe:1997ue,Gibbons:1997xz}. The
fundamental string turns into a D-string, while the D3-brane remains
unchanged, so one ends up with a supersymmetric configuration of a
D-string incident normally on a D3-brane. This can be extended to
multiple parallel D-strings ending normally on multiple parallel
D3-branes \cite{Diaconescu:1996rk,Constable:1999ac}. This system
carries very useful information in the form of ``Nahm equations," 
as we will shortly see.

Before we do that, let us use a series of dualities to highlight the
relationship between the intersecting D1$\perp$D3 and M2$\perp$M5
systems. By compactifying an M2$\perp$M5 configuration on a circle
within the M5 but {\em not} within the M2, we obtain a D2$\perp$D4
system. A T-duality along the direction common to both, leads us to
type IIB string theory and the D2- and D4-branes become, respectively,
D-strings and D3-branes; in other words the D1$\perp$D3 system. This
relationship motivated Basu and Harvey \cite{Basu:2004ed} to guess
some aspects of the multiple membrane worldvolume field theory. Their
strategy was to conjecture a generalisation of the Nahm equations that
describe the D1$\perp$D3 system.

Let us first review these equations and their uses. The key point is
that one can understand the D1$\perp$D3 system in terms of the
worldvolume theory of {\em either} the D3-brane {\em or} the D-string.
In the first picture the D1-branes arise as a
soliton ``spike'' in the D3 worldvolume theory, while in the second
picture the D3-brane arises as a ``fuzzy'' or noncommutative sphere in the D-string worldvolume theory. Consider $N$
coincident D1-branes in type IIB string theory oriented along $x^9$
and ending on a single D3-brane spanning the $x^1,x^2,x^3$ directions.
The latter has an abelian worldvolume gauge theory of DBI type,
containing six transverse scalars which we will label
$X^4,X^5,\cdots,X^9$ in addition to a gauge field and
fermions.\footnote{As before, we use upper-case letters 
to denote fields and lower-case letters for the worldvolume 
coordinates.}

This abelian field theory has been shown, see \eg
\cite{Constable:1999ac}, to admit a classical monopole solution
\be 
X^9=\frac{N}{2r}, \qquad
F_{\theta\phi}=-r^2\del_r X^9 \;,
\label{bion}
\ee 
where $r=\sqrt{(x^1)^2+(x^2)^2+(x^3)^2}$ is the radial direction
within the D3-brane. This solution carries a magnetic charge 
\be
\frac{1}{2\pi}\int F = N \ 
\ee 
and has its energy density concentrated along a spike, extending in
the $x^9$ direction and located at $r= 0$. The energy density is found
to be $N/2\pi g_s\alpha'$, which is precisely the tension of $N$ semi-infinite 
D-strings. Thus this classical solution is identified with $N$
D-strings, viewed as excitations of the D3-brane. The solution is
supersymmetric, as one would expect given the geometry of the
configuration.

Seeing the D3-brane in the worldvolume theory of $N$ D-strings is a
little less trivial. This time we use the fact that the
latter theory is non-abelian and has eight $N\times N$ matrix-valued
scalar fields $\boldX^1,\boldX^2,\cdots,\boldX^8$. We choose the gauge
$\boldA_9=0$ and consider solutions for which $\boldX^i=0,~ i\in
4,5,\cdots, 8$. Then the equations of motion can be reduced using
supersymmetry from the usual second-order form to the first-order form
\be
\frac{\del\boldX^i}{\del x^9}
=\pm\frac{i}{2}\epsilon^{ijk}[\boldX^j,\boldX^k],\quad
i,j,k\in 1,2,3\;.
\ee
These are the Nahm equations. In terms of $N\times N$ matrices
$\boldalpha^i$ that form $N$-dimensional representations of $\SU(2)$,
thereby satisfying
\be 
[\boldalpha^i,\boldalpha^j]=2i\epsilon^{ijk}\boldalpha^k\;, 
\ee
a solution is given by 
\be 
\boldX^i = \pm \frac{1}{2x^9}\boldalpha^i,\quad i=1,2,3\;.  
\ee 
In the conventional D-brane interpretation, these worldvolume scalars
parametrise the geometry transverse to the D-string, and in particular
can be thought of as discretised/noncommutative/``fuzzy'' versions of
the usual Euclidean coordinates on a sphere. Hence, the physical
radius of our fuzzy $S^2$ is defined at a fixed value of $x^9$ (a
point on the D-string) as the appropriately normalised sum
\be R^2 = \frac{(2\pi\alpha')^2}{N}
\Tr\sum_{i=1,2,3}(\boldX^i)^2 
\label{physR}\;.
\ee
This may be evaluated using the fact
that the $\boldalpha^i$ have a quadratic Casimir 
\be
\sum_{i=1,2,3}(\boldalpha^i)^2=N^2-1 \ ,
\ee
and we find 
\be\label{physradd1}
R=\frac{\pi\alpha' \sqrt{N^2-1}}{x^9}\;. 
\ee
Therefore the D-string description corresponds to a ``fuzzy funnel,''
the ``mouth'' of which grows towards smaller positive values of $x^9$
and eventually blows up into a D3-brane at zero.

At large $N$, the fuzzy sphere becomes a commutative $S^2$ and
\eref{physradd1} can be equated to the formula in \eref{bion} after
identifying $(R,x^9)$ in the D-string problem with $(r,X^9)$ in the
D3-brane problem. Other properties of the D-string also match between
the two descriptions. Importantly, the ``fuzzy funnel'' picture
is valid even inside the core, unlike the ``BIon.''

\subsubsection{The Basu-Harvey solution}

As advertised, the above intersection can be generalised to the case
of M-theory.  For M2-branes ending on an M5-brane at a string, a
classical solution analogous to \eref{bion} was constructed by
\cite{Howe:1997ue} and is known as the ``self-dual string soliton.''
Instead of a ``spike,'' one now looks for a ``ridge'' solution to the
M5-brane worldvolume theory. The spatial volume of the M5-brane is
oriented along $x^1,x^2,\cdots, x^5$ with all the other coordinates
vanishing.  One takes the self-dual string to lie along $x^5$. The
M2-branes will extend along $x^5,x^{10}$ thereby ending on a string at
$x^{10}=0$ as desired. The soliton of the M5-brane theory has the
profile
\be
X^{10}\sim \frac{N}{r^2}\;,
\ee
where $r=\sqrt{(x^1)^2+(x^2)^2+(x^3)^2+(x^4)^2}$ is the radial 
direction within the M5-brane. The challenge is now to find an analogue
of the Nahm equation, in the worldvolume theory of multiple
M2-branes, which reproduces the above profile.

The idea would be that, since the M2 and M5 branes are codimension 4
objects, this time one has to construct a fuzzy 3-sphere rather than a
2-sphere. The fuzzy 2-sphere was relatively straightforward to realise
using irreducible representations of $\SU(2)$. However, for the
3-sphere it turns out that a more complicated construction is required
\cite{Guralnik:2000pb,Ramgoolam:2001zx,Ramgoolam:2002wb}. In
particular, the $\SO(4)$-covariant matrix construction of the fuzzy
3-sphere gives rise to more degrees of freedom than needed. One can
perform a projection down to the required subset, although this spoils
the associativity of the matrix product,\footnote{A related discussion can be found in Appendix B of
  \cite{Papageorgakis:2006ed}. For a treatment of this
  non-associativity in the context of the M2$\perp$M5 system see
  \cite{Berman:2006eu}.} even for large $N$.
We will see in Section~\ref{dielM2M5} that
the realisation of the 3-sphere as a Hopf fibration is more
appropriately suited for the description of these systems
\cite{Nastase:2009ny, Nastase:2009zu,Nastase:2010uy}, but we can
nevertheless uncover several qualitative aspects of membrane dynamics
with the former approach, which was the one used in
\cite{Basu:2004ed}.

To proceed, consider the decomposition of the 3-sphere isometry
algebra $\spn(4)\simeq \su(2)\oplus \su(2)$.  Its representations are
labelled by $(j_1,j_2)$ with each entry being the spin of the
representation of the corresponding $\su(2)$. The dimension of such a
representation is $(2j_1+1)(2j_2+1)$.  Now choose an odd integer $n$
and define the two representations
\be
\cR^+=\left(\frac{n+1}{4},\frac{n-1}{4}\right),\qquad
\cR^-=\left(\frac{n-1}{4},\frac{n+1}{4}\right)\;.
\ee
The dimension of $\cR^+\oplus\cR^-$ is $N=\frac{1}{2}(n+1)(n+3)$.  With
this construction, the coordinates on the fuzzy 3-sphere are $N\times
N$ matrices $\boldG^i, i=1,2,3,4$ that map $\cR^+\leftrightarrow
\cR^-$.

Let ${\cal P}_{\cR^+},{\cal P}_{\cR^-}$ be the projection operators,
respectively, onto the representations $\cR^+,\cR^-$. Then one can
define a matrix $\boldG_5$ by 
\be 
\boldG_5 \equiv {\cal P}_{\cR^+}- {\cal P}_{\cR^-} \;.
\label{gfive}
\ee
We also need a quantity called the
``Nambu 4-bracket,'' defined by
\be
[A_1,A_2,A_3,A_4]\equiv \sum_{{\rm permutations}~\sigma} 
{\rm sign}(\sigma)\, A_{\sigma(1)}A_{\sigma(2)}A_{\sigma(3)}
A_{\sigma(4)}\;.
\ee
In terms of the above, the Basu-Harvey proposal for the
equation describing an M5-brane in the worldvolume theory of $N$
M2-branes is
\be
\frac{\del\boldX^i}{\del x^{10}} =
\frac{1}{4!}\frac{b}{8\pi\ell_p^3}\epsilon^{ijkl}
[\boldG_5,\boldX^j,\boldX^k,\boldX^l]\;.
\label{basuharv}
\ee
Here, $i,j,k,l\in(1,2,3,4)$ are indices labelling four spatial
directions transverse to the M2-branes. They are interpreted as the
spatial directions of the M5-brane transverse to the string lying
along $x^5$, with $b$ an arbitrary parameter to be determined.
The above equation amounts to a conjecture that will be supported by
finding solutions with the desired properties. These solutions depend
on the fuzzy 3-sphere coordinates $\boldG^i$ referred to above, which
are a set of four $N\times N$ matrices for any integer $N$ equal to
$\frac{1}{2}(n+1)(n+3)$ with $n$ odd. 

We first briefly describe the
construction\footnote{More details can be found
  in \cite{Basu:2004ed,Guralnik:2000pb,Ramgoolam:2001zx,Ramgoolam:2002wb}.} of the $\boldG^i$.
The smallest allowed value of $N$ is 4 (corresponding
to $n=1$), and in this case, in terms of the $4\times
4~\Gamma$-matrices of $\spn(4)$, we have $\boldG^i=\Gamma^i,
\boldG_5=\Gamma_5$.
The case for general $n$ is built up using tensor products involving
the $\Gamma^i$. Define
\be
\rho_s(\Gamma^i) \equiv \one\otimes\cdots\otimes\Gamma^i\otimes
\cdots \otimes\one,\quad s=1,2,\cdots,n\;,
\ee
where we have an $n$-fold product of identity matrices except for a single 
$\Gamma^i$ appearing in the $s$'th place. By summing over $\rho_s$ for
all $s$ from $1$ to $n$, we construct the symmetrised object
\be
\sum_{s=1}^n\rho_s(\Gamma^i) = (\Gamma^i\otimes\one\otimes\cdots\otimes\one) + 
(\one\otimes \Gamma^i\otimes\cdots\otimes\one) +\cdots+
(\one\otimes\cdots\otimes\one\otimes\Gamma^i)\;.
\ee
This matrix has dimension $4^n\times 4^n$. Finally, we define
\be
\boldG^i = {\cal P}_{\cR^+} \sum_{s=1}^n\rho_s(\Gamma^iP_-)\,{\cal
  P}_{\cR^-} ~+~{\cal P}_{\cR^-} \sum_{s=1}^n\rho_s(\Gamma^iP_+)
\,{\cal P}_{\cR^+}\;,
\ee
where $P_\pm =\half(1\pm\Gamma^5)$ and ${\cal P}_{\cR^\pm}$ are the
projection matrices defined above \eref{gfive}. Note that the ${\cal P}_{\cR^\pm}$
project the matrix sandwiched between them from dimension $4^n$ down 
to $N=\frac{1}{2}(n+1)(n+3)$.

Now that we have defined $\boldG^i$, the solution of \eref{basuharv}
proposed by Basu and Harvey takes the form
\be
\boldX^i(x^{10}) = i\, {\hat R}(x^{10})\, \boldG^i\ .
\ee
Inserting this ansatz into \eref{basuharv}, and
using the identity\footnote{This is derived in Appendix A of 
\cite{Basu:2004ed}.}
\be
\epsilon^{ijkl}\boldG_5 \boldG^i \boldG^j \boldG^k=-2(n+2)\boldG^i \;,
\ee
one immediately finds that
\be
{\hat R}(x^{10}) = \sqrt{\frac{2\pi\ell_p^3}{(n+2) b\,x^{10}}}\;.
\label{hatR}
\ee

By analogy with the D1$\perp$D3 case, the physical radius may be defined
as\footnote{An explicit $\ell_p$-dependence, analogous to the
  $\alpha'$-dependence of \eref{physR}, is absent from the definition
  of $R$ here because it is already accounted for in ${\hat R}$.}
\be
R^2 = \frac{1}{N}\, \Big|\Tr \sum_{i=1}^4 (\boldX^i)^2\Big|\;.
\ee
Inserting $\boldX^i$ from the solution above, we find
\be
R = \sqrt{N}\,|{\hat R}|\;.
\ee
Finally, substituting the functional form of ${\hat R}$ from
\eref{hatR} and solving for $x^{10}$ as a function of $R$, we find
\be
x^{10} = \frac{2\pi\ell_p^3 N}{(n+2)b\, R^2}\;.
\ee
This qualitatively has the correct (quadratic) fall-off with distance
$R$ within the M5-brane that is supposed to be described by this
classical solution, since a harmonic function in four spatial
dimensions should go like $R^{-2}$ at large $R$. However, the $N$
dependence does not seem correct. The solution should scale like $N$,
at least for large $N$, representing the fact that it describes $N$
M2-branes intersecting an M5. Since $N\sim n^2$ this scaling does not
hold as long as $b$ is held fixed. This implies that
$b$ should vanish like $1/\sqrt{N}$ for large $N$, or
equivalently $b^2 N$ is held fixed in the large-$N$ limit.

Next Basu and Harvey conjectured a form for the energy functional 
for such configurations. As we will see, their conjecture
inspires one to guess some of the terms in the lagrangian for multiple
M2-branes. The Basu-Harvey functional is
\be
E=T_{M2}\int d^2\sigma~ \Tr\,\Bigg[~
\Bigg(\frac{d\boldX^i}{dx^{10}}+ \frac{b}{8\pi\ell_p^3}
\epsilon^{ijkl}\boldG_5\boldX^j\boldX^k\boldX^l\Bigg)^2 +~
\Bigg(1-\frac{b}{16\pi\ell_p^3}\epsilon^{ijkl}
\Bigg\{\frac{d\boldX^i}{dx^{10}},\boldG_5 \boldX^j
\boldX^k\boldX^k\Bigg\}~\Bigg)^2~\Bigg]^\half\;,
\label{bhenergy}
\ee
where $T_{M2}=2\pi/(2\pi\ell_p)^3$ is the tension of a single
M2-brane, and the 4-bracket $[\boldG_5,\boldX^j,\boldX^k,\boldX^l]$
has been replaced by $4!\,\boldG_5\boldX^j\boldX^k\boldX^l$, to which
it is equal as long as $\boldX^i$ is among the solutions we are
considering.
 
The first term vanishes when the Basu-Harvey equation is
satisfied, and in this case one has
\be
E\,\Big|_{BH} = T_{M2}\int
d^2\sigma~\Bigg(1-\frac{b}{16\pi\ell_p^3}
\epsilon^{ijkl}
\Bigg\{\frac{d\boldX^i}{dx^{10}},\boldG_5 \boldX^j
\boldX^k\boldX^k\Bigg\}~\Bigg)\;.
\ee
The above expression is divergent due to the infinite length of all
the directions in the problem. Recalling that $\sigma$ represents the
two coordinates $x^5$ (along the self-dual string) and $x^{10}$
(transverse to the M5-brane), we can introduce a parameter $L$ to
regulate the length of the self-dual string along the M2-M5
intersection. It can then be shown \cite{Basu:2004ed} that
\be
E= NT_{M2}L\int dx^{10} + T_{M5} L\int 2\pi^2 R^3 dR\;,
\ee 
which is nicely interpreted as the sum of energies of $N$ M2-branes
and one M5-brane (here $T_{M5}=2\pi/(2\pi\ell_p)^6$ is the
M5-brane tension). This result can be considered the best
justification for the ansatz of the analogue Nahm equation
\eref{basuharv} as well as the energy functional \eref{bhenergy}.

The above expression for the energy suggests a set of terms in the
lagrangian of multiple M2-branes. For this we define a {\it triple-product}
\be\label{BHthree}
\boldH^{KLM}\equiv [\boldX^K,\boldX^L,\boldX^M]\equiv 
\{\,[\boldX^K,\boldX^L],\boldX^M\,\}+
\{\,[\boldX^L,\boldX^M],\boldX^K\,\}+
\{\,[\boldX^M,\boldX^K],\boldX^L\,\}\;,
\ee
which is totally antisymmetric in the indices $K,L,M$. The energy
functional then leads to (part of) the action \cite{Basu:2004ed}
\be S=-T_{M2}\int d^3\sigma~\Tr\,\Bigg[1 + (\del_a \boldX^M)^2
-\frac{b^2}{12}\left(\boldH^{KLM}\right)^2
+\frac{b^2}{48}\left[\del_a\boldX^{[K},
  \boldH^{LMN]}\right]^2\Bigg]^\half\;.  
\ee 
We see that the proposed
action contains a sextic scalar self-interaction, while the matrix
$\boldG_5$ no longer appears. Ref.~\cite{Basu:2004ed} also showed that
membrane fluctuations about the classical solution \eref{basuharv}
pass several physical consistency checks. Generalisations of the
Basu-Harvey equations corresponding to M2-branes ending on M5-brane
intersections leading to calibrated geometries were considered in
\cite{Berman:2005re,Copland:2007by}.

We will stop the analysis of the Basu-Harvey equation here, but
various of the features that came up in the above discussion, most
notably a version of the triple-product \eref{BHthree}, will crucially
re-emerge in subsequent chapters and the full description of multiple
M2-branes.

\subsection{Supersymmetric CS theories with $\mathcal N\leq 3$}

We will now switch gears and discuss a set of interacting
three-dimensional supersymmetric field theories. It will soon become
clear how these could be potentially related to the theory of multiple
membranes.

Pure Chern-Simons field theory in 2+1d has the lagrangian 
\be \cL_{CS} =\frac{k}{4\pi}\Tr\,\Big(\boldA\wedge 
d\boldA - \frac{2i}{3}
\boldA\wedge \boldA\wedge \boldA\Big) \ .  
\label{ellcs}
\ee 
Here we use the matrix-valued field $\boldA_\mu$ defined above
\eref{covderdef} and convert it to a differential 1-form via
$\boldA=\boldA_\mu dx^\mu$. Thus the lagrangian is a differential
3-form. We also allow $T^I,I=1,2,\cdots,{\rm dim}\,{\cal G}$ 
to be the Hermitian generators of an arbitrary Lie algebra
$\cG$ in the adjoint representation. Importantly, whenever the
associated gauge group is compact, the ``Chern-Simons level'' $k$
assumes discrete values for the path integral to remain invariant
under global gauge transformations in the quantum theory. We will
discuss this in more detail in Chapter~\ref{chapter3}.

Because the lagrangian is a 3-form, it can be integrated over a
3-manifold without the need to specify a metric. The action obtained
thereby is diffeomorphism-invariant even without coupling to a metric --
in other words, it has topological invariance \cite{Witten:1988hf}.
The gauge field is non-propagating and the only physical observables
are Wilson loop expectation values.  Coupling such a theory to scalar
or fermionic matter destroys the topological invariance, since a
metric is needed to define the matter kinetic terms and couplings.
However, if carefully done it can preserve conformal invariance and/or
any supersymmetry.

It is therefore natural to treat this class of theories as a starting
point to think about the worldvolume field theories on multiple
membranes in M-theory, an effort initiated in
\cite{Schwarz:2004yj,Gaiotto:2007qi}. While our principal goal is the
study of superconformal theories with $\mathcal N\ge 4$, and their
relevance to multiple membranes, this section is devoted to reviewing
status of theories with a modest amount ($\mathcal N\le 3$) of
supersymmetry, with an emphasis on those that are conformal
invariant.\footnote{Though our emphasis here is on supersymmetry,
  quite general non-supersymmetric theories in 2+1d can be non-trivial
  and exactly conformal invariant. For example, this is true when the
  matter consists of minimally coupled fermions \cite{Giombi:2011kc}
  or scalars \cite{Aharony:2011jz} with a suitable choice of coupling
  constants. The argument for conformal invariance
  hinges on the impossibility of a flow to triviality because the
  Chern-Simons coefficient $k$ is quantised.}

\subsubsection{$\mathcal N=1$ supersymmetry}

We start with the simplest supersymmetric Chern-Simons theory. The
$\mathcal N=1$ supersymmetry multiplet in 2+1d consists of a gauge
field $\boldA_\mu$ and a two-component (real) Majorana spinor $\boldchi$. 
The Chern-Simons lagrangian is simply 
\be 
\cL_{\mathcal N=1} =
\frac{k}{4\pi}\Tr\, \Big(\boldA\wedge d\boldA -
\frac{2i}{3}\boldA\wedge\boldA\wedge\boldA-
i{\bar\boldchi}\boldchi \Big) \, ;
\ee 
it is invariant up to a total derivative under the transformations
\bea
\delta \boldA_\mu &=&i{\bar\epsilon}\gamma_\mu\boldchi\nn\\
\delta \boldchi&=&-\half\gamma^{\mu\nu}\boldF_{\mu\nu}\epsilon \;.
\eea
Because the fermion is non-dynamical, this theory has no
propagating modes. One has to couple matter supermultiplets in order
to have propagating modes in the theory.

The $\mathcal N=1$ scalar multiplet consists of a real scalar $\phi$,
a 2-component Majorana spinor $\psi$ and a real auxiliary field $C$. Since
they will all transform in some definite representation of the gauge
group, we assign an index $a=1,2,\cdots,\dim R$ to them. Superspace
techniques can be used \cite{Schwarz:2004yj} to find possible
interaction terms. To maintain scale invariance (at least classically), the
potential must be sixth order in fields, and
to preserve gauge invariance, the coefficient of the superpotential 
must be invariant under the action of the gauge group. 
At the end of the day, one 
finds that the following matter lagrangian is supersymmetric:
\be \cL_{\mathcal N=1}^{\mathrm{matter}} = -\half
\del_\mu\phi^a\del^\mu\phi^a +
\frac{i}{2}\bpsi^a\gamma^\mu\del_\mu\psi^a +\half C^a C^a +
t_{abcd}\phi^a\phi^b \Big(\frac13 \phi^c C^d-\half\bpsi^c\psi^d\Big)\;,
\ee 
where $t_{abcd}$ is real, totally symmetric and invariant under the
gauge group. The auxiliary field
$C^a$ can be eliminated via its own equations of motion. One sees by
inspection that this leads to terms of order $\phi^6$ in addition to
the term $\phi^2\psi^2$ that is already present. Dimensional arguments
tell us that both such terms have canonical dimension 3, because
$[\phi]=\half$ and $[\psi]=1$ in 2+1d. This confirms that the matter
lagrangian above is classically scale invariant. This is not, however, generically preserved at the quantum level.

The supersymmetry transformation laws are
\bea
\delta\phi^a &=& i{\bar \epsilon}\psi^a\nn\\
\delta\psi^a &=& -\Big(\gamma^\mu\del_\mu\phi^a-C^a\Big)\epsilon\nn\\
\delta C^a &=& -i{\bar\epsilon}\gamma^\mu\del_\mu\psi^a\;.
\eea
It is now straightforward to couple a scalar multiplet to the
Chern-Simons vector multiplet. One simply converts the derivatives in
the scalar and fermion kinetic terms to
covariant derivatives
\be
\del_\mu\phi^a \to \del_\mu \phi^a -i A_\mu^I (T^I)_{ab}\phi^b\;,
\ee
where $(T^I)_{ab}$ are the generators of $\cG$ in the representation
of the matter supermultiplet. Additionally, there is a cubic Yukawa coupling
between the gauge fermion, the matter scalar and the matter fermion
\be
\phi^a {\bar \chi}^I T^I_{ab}\psi^b\;.
\ee
The full lagrangian and transformation laws can be found in 
\cite{Schwarz:2004yj}.

\subsubsection{$\mathcal N=2$ supersymmetry}

Chern-Simons gauge theory can be extended to have ${\mathcal N}=2$ supersymmetry \cite{Lee:1990it} by choosing $\boldchi$ to be Dirac instead of Majorana and adding two more scalars, $\boldsigma$ and $\boldD$, with the lagrangian
\be 
\cL_{\mathcal N=2} = \frac{k}{4\pi}\Tr\, \Big(\boldA\wedge d\boldA
- \frac{2i}{3} \boldA\wedge \boldA\wedge \boldA-
i{\bar\boldchi}\boldchi+2\boldD\boldsigma \Big) \;.
\ee 
The supersymmetry transformation rules are now given in terms of a
Dirac spinor $\epsilon$ as follows 
\bea
\delta \boldA_\mu &=& \frac{i}{2}\Big({\bar\epsilon}\gamma_\mu\boldchi - {\bar
  \boldchi}\gamma_\mu\epsilon\Big)\nn\\
\delta\boldsigma &=& -\half\Big({\bar \epsilon}\boldchi 
- {\bar\boldchi}\epsilon\Big)\nn\\
\delta \boldD &=& \half\left({\bar\epsilon}\gamma^\mu D_\mu\boldchi 
+ D_\mu{\bar \boldchi}\gamma^\mu\epsilon\right)
-\half\left({\bar\epsilon}[\chi,\sigma]+
[{\bar\chi},\sigma]\epsilon\right)
\nn\\
\delta\boldchi &=& \Big(-\half\gamma^{\mu\nu}\boldF_{\mu\nu} +
i\gamma^\mu D_\mu\boldsigma -i \boldD\Big)\epsilon \;.
\eea

The ${\cal N}=2$ matter multiplet contains the fields $(\phi^{aA},
\psi^{aA},F^{aA})$ just as in the familiar 3+1d ${\cal N}=1$ chiral
supermultiplet. Here $a$ runs over the dimension of the
representation of the gauge group in which the multiplet transforms,
while $A$ runs over the $N_f$ flavours of this
supermultiplet. The corresponding antichiral multiplet, obtained by
complex conjugation, is denoted $(\phi_{aA},\psi_{aA},F_{aA})$.
Henceforth we will suppress the $a$ index to make the notation more
compact. Then, exactly as in 3+1d, one specifies a holomorphic
superpotential $W$ and writes the lagrangian
\be
-\half \del_\mu\phi_A\del^\mu\phi^{A} +
i{\bar\psi}_A\gamma^\mu\del_\mu\psi^A
+F_AF^A + (F^A W,_A+c.c.).
\label{suplag}
\ee
Here $W,{_A}\equiv \del W(\phi)/\del\phi^A$. The lagrangian is invariant
under the supersymmetry transformations
\bea
\delta \phi^A&=&i{\bar\epsilon}\psi^A\nn\\
\delta\psi^A &=& -\gamma^\mu\del_\mu\phi^A+F^A\epsilon^*\nn\\
\delta F^A &=& -i{\bar\epsilon}^*\gamma^\mu\del_\mu\psi^A
\eea
and also under a $\U(N_f)$ flavour symmetry.
As before, in order to have classical scale invariance, $W$ must be a quartic
function of its argument.

These matter multiplets can be coupled to the gauge supermultiplet by
replacing ordinary derivatives with gauge-covariant derivatives
via minimal coupling. As before, one needs to add some extra terms in
order to achieve full ${\cal N}=2$ supersymmetry for the coupled
system. These are
\be 
-\sigma^I\sigma^J(\phi_A T^IT^J\phi^{A})+
D^I(\phi_A T^I\phi^{A})-
i\sigma^I({\bar\psi}_A T^I\psi^{A})-
\phi_A\,{\bar \chi}^I T^I\psi^{A}
+ \phi^{A}T^I{\bar \psi}_A\,\chi^I \;.
\ee
The full lagrangian and transformation laws can be found in
 \cite{Schwarz:2004yj}. Notice that the lagrangian is linear in $D^I$,
which therefore acts as a Lagrange multiplier determining $\sigma^I$
as bilinears in the $\phi$ fields. This in turn permits the
elimination of $\sigma$ which, from the $\sigma^2\phi^2$ term above,
gives rise to sextic terms in $\phi$. 

In the absence of a superpotential, the final result is
\be
\cL_{N=2, gauged}=\cL_{CS} + \cL_{kinetic} + {\cL}_{scalar-fermion} -V(\phi)\;,
\ee
where $\cL_{CS}$ is given by \eref{ellcs}, $\cL_{kinetic}$ are 
the standard minimally coupled kinetic terms of the scalars and 
fermions, and the remaining pieces are \cite{Gaiotto:2007qi}
\bea
\cL_{scalar-fermion} &=& -\frac{4\pi i}{k}(\phi_AT^I\phi^A)({\bar\psi}_B T^I \psi^B)
-\frac{8\pi i}{k}({\bar\psi}_A T^I\phi^A)(\phi_B T^I \psi^B)\nn\\
V(\phi) &=& \frac{16\pi^2}{k^2}(\phi_A T^I\phi^A)(\phi_B T^J\phi^B)
(\phi_C T^IT^J\phi^C)\;.
\eea

Classical conformal invariance is of course generically violated by quantum corrections. In one higher dimension, lagrangians with the
same amount of supersymmetry ($\cN=1$ in 3+1d) can easily be made
classically conformal invariant (by choosing a cubic superpotential)
but quantum corrections generically induce a nonzero $\beta$-function
and the quantum theory is no longer conformal. But here we encounter a
miracle of 2+1 dimensions: the lagrangian above (with vanishing superpotential) is exactly conformal
even at the quantum level \cite{Gaiotto:2007qi}. A brief sketch of the
argument is as follows. If quantum corrections generated a
superpotential term, this would be holomorphic in the superfield
$\Phi^A$ (which contains $\phi^A,\psi^A$) but such holomorphicity is
inconsistent with the symmetry of the above action under $\Phi^A\to
e^{i\alpha}\Phi^A$. Next, it is well known (with or without
supersymmetry) that the Chern-Simons level $k$ cannot be renormalised
other than by a finite 1-loop shift \cite{Kapustin:1994mt}. This only
leaves the possibility of corrections to the K\"ahler potential of the
theory. However it can be argued that these are either irrelevant in
the infrared or can be absorbed in a rescaling of $\Phi^A$. The reader
is referred to Ref.~\cite{Gaiotto:2007qi} for more details.

One can add a superpotential as in \eref{suplag} (see \eg \cite{Ivanov:1991fn}) but in this case quantum corrections will generically induce a nontrivial $\beta$-function and spoil conformal invariance. However there is a specific way in which this can be done while preserving and even enhancing superconformal symmetry, as we will see in the following section.

\subsubsection{$\mathcal N=3$ supersymmetry}

The amount of supersymmetry present in Chern-Simons-matter theories
can be further enhanced to ${\cal N}=3$ \cite{Kao:1992ig} while maintaining conformal invariance. For this we introduce a pair of
chiral superfields $Q,\tQ$ transforming in conjugate representations of the gauge group, coupled to the ${\cal N}=2$ gauge multiplet described above, and with a quartic superpotential
\be
W(Q,\tQ) = \alpha(\tQ T^I Q)(\tQ T^I Q)\;,
\ee
where $T^I$ are the generators of the gauge group in the chosen representation.  At $\alpha=0$ we have an $\cN=2$ superconformal theory as described above. For any finite value of $\alpha$, as one would generically expect, the theory develops a $\beta$-function for $\alpha$ and conformal invariance is broken (of course ${\cal N}=2$ supersymmetry is maintained). However it has been argued \cite{Gaiotto:2007qi} that the RG flow takes one to an attractive fixed-point at $\alpha=2\pi/k$. At this fixed point it turns out \cite{Kao:1992ig,Gaiotto:2007qi} that the supersymmetry is enhanced to $\cN=3$ and the resulting theory is exactly superconformal.

It was initially thought that ${\cal N}=3$ was the maximum number of supersymmetries allowed for a Chern-Simons-matter gauge theory \cite{Kao:1992ig,Schwarz:2004yj}. However this assumes a simple gauge group.  We will see that, somewhat surprisingly, the construction of Chern-Simons-matter theories with more supersymmetry is possible if the gauge group is not simple. Indeed, unlike supersymmetric Yang-Mills theories where the choice of gauge group is arbitrary, the possible amount of supersymmetry of a Chern-Simons-matter theory is closely linked with the choice of gauge group. A related observation is that $\mathcal N=3$ is the maximum amount of supersymmetry for which one can write down a Lagrangian including both Yang-Mills and Chern-Simons terms \cite{Kao:1993gs}. 


\section[Three-dimensional CS gauge theories based on 3-algebras]{{\Large{\bf Three-dimensional CS gauge theories based on 3-algebras}}}\label{chapter3}

Our task now is to try and construct an effective field theory for the worldvolume dynamics of multiple M2-branes propagating in flat eleven-dimensional spacetime. The solitonic picture of M2-branes and their relation to D2-branes, both discussed in Chapter~\ref{chapter1}, tell us we should look for a theory that preserves half of the 32 spacetime supercharges, leading to ${\cal N}=8$ supersymmetry on the worldvolume. Later we will see that we should also allow for orbifolds which generically break more supersymmetry.

We will do this by searching for field theories with the correct symmetries and therefore our first task is to determine what these symmetries are. We want ${\cal N}=8$ three-dimensional theories with eight dynamical scalars and fermions, but no other dynamical modes. In particular, we do not expect any dynamical gauge fields. One way to see this is to note that the scalars and fermions together make up all the dynamical degrees of freedom of the three-dimensional supermultiplet. However, as we shall see, this does not exclude the possibility of {\it   non-dynamical} gauge modes.\footnote{Here we refer to the degrees of freedom in the classical lagrangian. In the full quantum theory this distinction is somewhat obscure, since in three dimensions a vector is dual to a scalar.} Indeed, we have already seen that in three dimensions there is the possibility of having pure Chern-Simons theories, with or without dynamical scalars, and it was already suggested in \cite{Schwarz:2004yj} that such theories could be suitable candidates for describing multiple M2-branes.

An additional criterion for selecting our candidate theories is that in the limit where gravity is decoupled from the branes, we should  end up with a conformal field theory. Perhaps the simplest reason for this is that M-theory has no parameters and only one scale: the eleven-dimensional Planck scale. The gravity-decoupling limit corresponds to considering vanishingly small energy excitations, or equivalently, taking the eleven-dimensional Planck length to zero, $\ell_p\to 0$. Hence, there is no scale in the decoupled theory. We have already encountered another reason for this in Section~\ref{abelianM2D2}: since M-theory can be thought of as the strong-coupling limit of type IIA string theory, M2-branes are the strong-coupling limit of D2-branes. D2-branes are described by three-dimensional maximally supersymmetric Yang-Mills theories. These theories are super-renormalisable, which means their coupling constant $g_{YM}$ has a positive scale dimension and therefore it increases in the infrared. Thus the strong coupling limit is the same as the IR limit, and the theory must either become free (which is ruled out on physical grounds) or reach an interacting conformal invariant fixed point.

From the geometrical point of view, a stack of M2-branes in eleven dimensions breaks the $\SO(1,10)$ Lorentz group to $\SO(1,2)\times \SO(8)$. While the $\SO(1,2)$ factor becomes the Lorentz group on the worldvolume, the $\SO(8)$ is identified with the R-symmetry and in particular rotates the scalar fields (and acts on the fermions as well). Finally M-theory has a parity symmetry, which M2-branes in a flat background should preserve.

In the rest of this chapter, we will look into the general construction of lagrangians with the above properties and $\mathcal N=8$ or $\mathcal N=6$ supersymmetry. This will involve the introduction of an interesting algebraic structure intimately connected with supersymmetry: 3-algebras, which generalise the notion of conventional Lie algebras. During the course of our discussion we will find that these 3-algebra theories  also admit a conventional Lie algebra formulation in terms of bifundamental matter fields in three dimensions. This provides a connection to the theories of Chapter~\ref{chapter2} and sets the stage for the ABJM theory with $\U(n)\times\U(n)$ gauge symmetry of Chapter~\ref{chapter4}. Superspace constructions of these theories are given in \cite{Mauri:2008ai,Bandos:2008jv,Cederwall:2008vd,Bandos:2008df,Cherkis:2008ha,Bandos:2009dt}.

\subsection{$\mathcal N=8$ 3-algebra theories: BLG}\label{Neq8}

To proceed we simply start from scratch and attempt to construct the theory that we are looking for. This was done in \cite{Bagger:2006sk,Gustavsson:2007vu,Bagger:2007jr,Bagger:2007vi} and is commonly known as BLG theory. The supersymmetries that are preserved by the M2-branes can be taken to satisfy
\begin{equation}
\Gamma_{012}\epsilon=\epsilon\;.
\end{equation}
In this section we work in conventions where our spinors are real
32-component spinors of eleven-dimensional spacetime. This is a somewhat non-standard way to describe a field theory in 2+1d, where irreducible (Majorana) spinors are 2-component. However we use this notation because it greatly helps us relate symmetries on the brane to those in the bulk.

The worldvolume fermions can be thought of as Goldstino modes for the supersymmetry broken by the brane. They therefore satisfy the opposite supersymmetry condition $\Gamma_{012}\Psi=-\Psi$. Let us call the scalar fields $X^I$ and, as for D-branes we assume that they, along with the fermions, take values in some vector space with a basis $T^a$, in other words
\bea
X^I &=& X^I_aT^a\nn\\
\Psi &=&\Psi_a T^a\ .
\eea
Here $I = 1,...,8$ is an R-symmetry index. Despite the notation, we do not require the $T^a$ to generate a Lie algebra; we will shortly see that they do something rather different.

For each value of the index $a$, the scalars have 8 degrees of freedom due to the R-symmetry index. The fermions have 32 degrees of freedom, reduced to 16 by the parity condition above, and further reduced to 8 on-shell. Hence the on-shell bosonic and fermionic degrees of freedom match and, as can be easily checked, the free theory is invariant under the (on-shell) supersymmetry transformations
\begin{eqnarray}
\delta X^I_d &=& i\bar\epsilon\Gamma^I\Psi_d\nonumber\\
\delta \Psi_d &=& \partial_\mu X^I_d\Gamma^\mu\Gamma^I\epsilon\ .
\end{eqnarray}

To introduce interactions we need to include a term in $\delta\Psi$ that is non-linear in the scalar fields. Now $\Psi$ and $\epsilon$ have opposite eigenvalues with respect to $\Gamma_{012}$ and in addition it is easy to see that $[\Gamma_{012},\Gamma_\mu]=0$ but $\{\Gamma_{012},\Gamma_I\}=0$. Thus any term on the right hand side of $\delta\Psi_d$ must have an odd number of $\Gamma^I$ factors. Furthermore we wish to look for a conformal field theory. Since the scaling dimensions of $X^I_a$, $\Psi_a$ and $\epsilon$ are  $\frac{1}{2}$, $1$ and $-\frac{1}{2}$ respectively we see that the interaction term we are looking for should be cubic in $X^I_a$. Thus a natural guess is
\begin{eqnarray}
\delta X^I_d &=& i\bar\epsilon\Gamma^I\Psi_d \nonumber\\
\delta \Psi_d &=& \partial_\mu X^I_d\Gamma^\mu\Gamma^I\epsilon - \frac{1}{3!}X^I_aX^J_bX^K_c f^{abc}{}_d\Gamma^{IJK}\epsilon\;.
\end{eqnarray}
Here we  have introduced coupling constants $f^{abc}{}_d$ which, without loss of generality, are antisymmetric in $a,b,c$.
By analogy with normal Lie algebras, we propose to view them as structure constants for a ``triple product'' or ``3-bracket'' that acts on the  vector space spanned by $T^a$ as
\begin{equation}
[T^a,T^b,T^c] = f^{abc}{}_d T^d\;.
\end{equation}
Thus we can say that the vector space in which the scalars and fermions are valued has a  {\it Lie 3-algebra} structure, namely a totally anti-symmetric triple product (with certain additional properties, as we will see).

Next we must check that this supersymmetry algebra closes. In the more familiar case of D-brane theories this  happens on-shell, up to translations and gauge transformations. Here, if we compute $[\delta_1,\delta_2]\,X^I_a$ we find
\begin{equation}
[\delta_1,\delta_2]\,X^I_d = -2i\bar\epsilon_2 \Gamma^\mu \epsilon_1 \partial_\mu X^I_d - (2i\bar\epsilon_2\Gamma^{JK}\epsilon_1 X^J_aX^K_bf^{abc}{}_d) X^I_c \;.
\end{equation}
The first term is simply a translation, as expected, with parameter $v^\mu = -2i{\bar\epsilon}_2\Gamma^\mu\epsilon_1$. The second term must be interpreted as a new symmetry
\begin{equation}
\delta X^I_d  = \tilde \Lambda^c{}_dX^I_c\;,\qquad \tilde \Lambda^c{}_d = -2i\bar\epsilon_2\Gamma^{JK}\epsilon_1 X^J_aX^K_bf^{abc}{}_d\ .
\end{equation}
This must be a gauge symmetry, since $\tilde \Lambda^c{}_d$ depends explicitly on the $X^J_b$ which in turn depend on $x^\mu$. By multiplying
both sides of the above equation with $T^d$ we can write the above transformation as
\be
\delta X^I =\alpha_{JK} [X^I,X^J,X^K]
\ee
with parameters $\alpha_{JK}=2i\bar\epsilon_2\Gamma^{JK}\epsilon_1$. A general gauge symmetry transformation on an arbitrary vector $X$ in our vector space therefore has the form
\be
\delta X = [X,A,B]
\ee
where $A,B$ are two more vectors in the same space.

In order for this symmetry to hold in the interacting theory, we require that it act as a derivation on the triple product
\begin{eqnarray}
\delta [X,Y,Z] = [\delta X, Y, Z] + [X,\delta Y,Z]+[X,Y,\delta Z]\;,
\end{eqnarray}
which in turn requires that the triple product satisfy the ``fundamental identity''
\begin{equation}\label{FI}
[A,B,[X,Y,Z]] = [[A,B,X],Y,Z] + [X,[A,B,Y],Z] + [X,Y,[A,B,Z]]\ ,
\end{equation}
or equivalently
\begin{equation}
f^{abc}{}_gf^{efg}{}_d = f^{efa}{}_gf^{gbc}{}_d + f^{agc}{}_df^{efb}{}_g + f^{abg}{}_d f^{efc}{}_g\ .
\end{equation}

Next we must introduce a gauge field for this gauge symmetry. Following the standard procedure we define
\begin{equation}
D_\mu X^I_d = \partial_\mu X^I_d - \tilde A_\mu{}^c{}_d X_c^I\ ,
\label{covderblg}
\end{equation}
and similarly for $\Psi_d$. This is gauge covariant provided that
\begin{equation}
\delta \tilde A_\mu{}^c{}_d = \partial_\mu \tilde \Lambda{}^c{}_d + \tilde A_\mu{}^c{}_e \tilde \Lambda{}^e{}_d - \tilde \Lambda{}^c{}_e \tilde A_\mu{}^e{}_d
\end{equation}
under a gauge transformation.
We can also compute the field strength from $[D_\mu,D_\nu]\,X^I_b = \tilde F_{\mu\nu}{}^a{}_bX^I_a$ and find
\begin{equation}
 \tilde F_{\mu\nu}{}^a{}_b = \partial_\nu \tilde A_\mu{}^a{}_b -  \partial_\mu \tilde A_\nu{}^a{}_b - \tilde A_{\mu}{}^a{}_c\tilde A_\nu{}{}^c{}_b +  \tilde A_{\nu}{}^a{}_c\tilde A_\mu{}^c{}_b\;.
\end{equation}
These are familiar expressions from gauge theory and indeed the fundamental identity (\ref{FI}) ensures that the set of all $\tilde\Lambda^a{}_b$ form a closed set under matrix commutation. Thus the 3-algebra defines an ordinary Lie algebra generated by the elements $\tilde \Lambda^a{}_b$ that act naturally on the 3-algebra. The underlying gauge symmetry of the theory is therefore that of an ordinary gauge theory based on a Lie algebra. We will give a more mathematical treatment of 3-algebras below.

It remains to specify the supersymmetry transformation law of the gauge field. This is easily done using index structure and dimensional counting (noting that the above equations determine the canonical dimension of $\tA_\mu{}^a{}_b$ to be $+1$). We are thereby led to postulate the complete set of supersymmetry transformations
 \begin{eqnarray}\label{susygauged2}
\nonumber \delta X^I_a &=& i\bar\epsilon\Gamma^I\Psi_a\\[2mm]
\delta \Psi_a &=& D_\mu X^I_a\Gamma^\mu \Gamma^I\epsilon -\frac{1}{6}
X^I_bX^J_cX^K_d f^{bcd}{}_{a}\Gamma^{IJK}\epsilon \\[2mm]
\nonumber \delta\tilde A_{\mu}{}^b{}_a &=& i\bar\epsilon
\Gamma_\mu\Gamma_IX^I_c\Psi_d f^{cdb}{}_{a}\;.
\end{eqnarray}
A priori it is not at all obvious that these supersymmetries close into translations and gauge transformations on-shell. In fact at this stage we do not even know what ``on-shell'' means since we do not know the equations of motion of the theory we are seeking.

Fortunately, the requirement that the above supersymmetry transformations close is very powerful. It determines the equations of motion for some of the fields and also fixes the normalisation of the supersymmetry variation $\delta {\tilde A}_\mu{}^a{}_b$ above, which could not have been determined by dimensional counting. At the end one finds that
\begin{eqnarray}
  \nonumber [\delta_1,\delta_2]\, X^I_a &=&
  v^\lambda\partial_\lambda X^I_a
  +(
\tilde\Lambda^b{}_a-v^\lambda\tilde A_\lambda{}^b{}_a) X^I_b \\
 {}[\delta_1,\delta_2]\, \Psi_a  &=& v^\lambda \partial_\lambda\Psi_a + (\tilde\Lambda^b{}_a
 -v^\lambda\tilde A_\lambda{}^b{}_a) \Psi_b \\
  \nonumber [{\delta_1},{\delta_2}]\, \tilde A_\mu{}^b{}_a &=& v^\lambda\partial_\lambda  \tilde
 A_{\mu}{}^b{}_a +\tilde D_\mu(\tilde
\Lambda^b{}_a -v^\lambda\tilde A_\lambda{}^b{}_a )\;,
\end{eqnarray}
where $v^\lambda=-2i{\bar\epsilon_2}\Gamma^\lambda\epsilon_1$ and
$\tilde \Lambda{}^b{}_a=-2i{\bar \epsilon_2}\Gamma^{JK}\epsilon_1
X_c^J X_d^K f^{cdb}{}_{a}$, but only if the
following equations of motion are satisfied:
\begin{eqnarray}\label{EOMS}
\nonumber\Gamma^\mu D_\mu\Psi_a
+\frac{1}{2}\Gamma_{IJ}X^I_cX^J_d\Psi_bf^{cdb}{}_{a}&=&0\\
  \tilde F_{\mu\nu}{}^b{}_a
  +\varepsilon_{\mu\nu\lambda}(X^J_cD^\lambda X^J_d
+\frac{i}{2}\bar\Psi_c\Gamma^\lambda\Psi_d )f^{cdb}{}_{a}  &=& 0\;.
\end{eqnarray}
In this way we have found the fermion and gauge-field equations of motion. To find the equation of motion for the scalars, one takes the supersymmetry variation of the fermion equation above. The answer splits into two sets of terms, one that vanishes by virtue of the gauge field equation above and another whose vanishing implies the scalar equation of motion
\be
D^2X^I_a-\frac{i}{2}\bar\Psi_c\Gamma^I_{\ J}X^J_d\Psi_b f^{cdb}{}_a
   +\half X^J_bX^K_cX^I_eX^J_fX^K_g f^{bcd}{}_a f^{efg}{}_d= 0\ .
\ee

The free parts of the above equations of motion, obtained by setting all terms involving structure constants to zero, are respectively the massless Dirac equation, the equation of a flat gauge connection and the massless Klein-Gordon equation. The first and last are as expected, but the middle one is somewhat unusual -- it is not the equation of motion for a Yang-Mills gauge field, but rather the one that follows from a Chern-Simons action. Fortunately, it is just what we expected on grounds of conformal invariance.

It only remains to construct a lagrangian that gives rise to the full interacting equations of motion above. For this we need to introduce an inner-product or metric on the 3-algebra\footnote{For an alternative approach which does not utilise a metric see \cite{Gran:2008vi}.}
\be
\langle X,Y\rangle = h^{ab}X_aY_b\ .
\label{innerprod}
\ee
Requiring invariance of this inner-product under the gauge transformations
$\delta X_a=\tilde\Lambda^b{}_aX_b$, $\delta Y_a=\tilde\Lambda^b{}_aY_b$ implies that the structure constants with the last index raised by the metric $f^{abcd}=h^{de}f^{abc}{}_e$ are totally antisymmetric
\begin{equation}
f^{abcd}=f^{[abcd]}\ ,
\end{equation}
The lagrangian can now be written as
\be\label{action}
{\cal L} = -\frac{1}{2}D_\mu X^{aI}D^\mu X^{I}_{a}
+\frac{i}{2}\bar\Psi^a\Gamma^\mu D_\mu \Psi_a
+\frac{i}{4}\bar\Psi_b\Gamma_{IJ}X^I_cX^J_d\Psi_a f^{abcd}- V +{\cal L}_{CS}\ ,
\ee
with a sextic potential
\bea
V&=&\frac{1}{12}X^I_aX^J_bX^K_cX^I_eX^J_fX^K_g f^{abcd}f^{efg}{}_d\nn\\
&=&\frac{1}{12}\langle [X^I,X^J,X^k],[X^I,X^J,X^K] \rangle
\eea
and a ``twisted'' Chern-Simons term
\begin{equation}
{\cal L}_{CS}= \frac{1}{2}\varepsilon^{\mu\nu\lambda}\left(f^{abcd}A_{\mu
ab}\partial_\nu A_{\lambda cd} +\frac{2}{3}f^{cda}{}_gf^{efgb}
A_{\mu ab}A_{\nu cd}A_{\lambda ef}\right)\ ,
\end{equation}
Note that ${\cal L}_{CS}$ is written in
terms of a gauge field $A_{\mu ab}$ that differs from the ``physical''
gauge field we have previously encountered in the
supersymmetry transformations and equations of motion,
being related to it via
\be
\tilde A_{\mu}{}^b{}_a = A_{\mu cd}f^{cdb}{}_a\ .
\ee
In general, this equation cannot be inverted to determine $A$ in terms of $\tA$, but one can check that ${\cal L}_{CS}$ is invariant under shifts of
$A_{\mu ab}$ that leave $\tilde A_{\mu}{}^b{}_a$ invariant.  It is therefore
locally well-defined as a function of $\tilde A_{\mu}{}^b{}_a$.

It is not hard to check that the lagrangian is gauge invariant and supersymmetric up to a total derivative under the transformations (\ref{susygauged2}).  Note also that (\ref{action}) contains no free parameters, up to a rescaling of the structure constants.  In fact, given the presence of the Chern-Simons term, it is natural to expect the $f^{abcd}$ to be quantised and we will argue below that this is indeed the case.

The theory we have constructed is invariant under 16 supersymmetries and an $\SO(8)$ R-symmetry.  It is also conformally invariant at the classical level. These are all the continuous symmetries that are expected of multiple M2-branes.  Note that the Chern-Simons term naively breaks the parity that is expected to be a symmetry of the M2-brane worldvolume. However, we can make the lagrangian parity invariant if we assign an odd parity to $f^{abcd}$.  In particular, if we invert $x^2 \to -x^2$, we must then require that $X^I_a$ and $\tilde A^{\ a}_{\mu\ b}$ be parity-even for $\mu = 0,1$; $ \tilde A^{\ a}_{2\ b}$ and $f^{abcd}$ be parity-odd and $\Psi_a\to \Gamma_2\Psi_a$.  This assignment also implies that $A_{\mu ab}$ is parity-odd for $\mu=0,1$, while $A_{2 ab}$ is parity-even.

This would seem to be a complete success: We have a lagrangian with all the required symmetries for multiple M2-branes. One would expect the logical next step to be a determination of the possible consistent structure constants $f^{abc}{}_d$ which (following D-brane intuition) should be related to the number of coincident M2-branes. However at this stage we encounter a problem. If we assume that the metric $h^{ab}$ is positive definite, so that the kinetic and potential energies are all positive, then there turns out to be essentially a {\em unique} choice \cite{nagy-2007,Papadopoulos:2008sk,Gauntlett:2008uf} for $f^{abcd}$ that is totally anti-symmetric and that satisfies the fundamental identity, namely
\begin{equation}\label{onlyone}
f^{abcd} = \frac{2\pi}{k}\varepsilon^{abcd}\;,\qquad h^{ab}=\delta^{ab}\;,
\end{equation}
where $a,b,...=1,...,4$ and $k$ is a (for the moment, arbitrary) constant.

The uniqueness of the structure constants rules out the possibility that the lagrangian written above \cite{Bagger:2007jr} describes an arbitrary number of coincident M2-branes. Nevertheless, it is an interesting theory on its own. It provides the first example of an interacting lagrangian quantum field theory with maximal global supersymmetry that is not of Yang-Mills type.\footnote{One can alternatively arrive at this  theory starting from (gauged) $\mathcal N=8$ supergravity in 3d \cite{Nicolai:2001ac} and taking the global-supersymmetry limit \cite{Bergshoeff:2008ix,Bergshoeff:2008cz}.} Let us therefore study the theory in more detail.

The gauge algebra generated by $\tilde \Lambda^a{}_b$ is simply the space of all anti-symmetric $4\times 4$ matrices. This is of course $\mathfrak{so}(4)\simeq \su(2)\oplus \su(2)$. The split is realised by noting that the self-dual and anti-self-dual parts of $\tilde \Lambda{}^a{}_b$ commute with each other. Thus we write
\begin{equation}
\tilde A_\mu{}^a{}_b = \tilde A^+_\mu{}^a{}_b+\tilde A^-_\mu{}^a{}_b\;,
\end{equation}
where $\tilde A^\pm_\mu{}^a{}_b$ is the (anti)-self-dual part of $\tilde A_\mu{}^a{}_b$. Now the twisted Chern-Simons term can be written as \cite{Bagger:2007vi}
\begin{equation}\label{CSdual}
{\cal L}_{CS} = \frac{k}{8\pi}\epsilon^{\mu\nu\lambda} (\tilde A^+_\mu{}^a{}_b \partial_\nu \tilde A^+_{\lambda}{}^b{}_a
+\frac{2}{3} \tilde A^+_{\mu}{}^a{}_b \tilde A^+_{\nu}{}^b{}_c \tilde A^+_{\lambda}{}^c{}_a )-
\frac{k}{8\pi}\epsilon^{\mu\nu\lambda}( \tilde A^-_\mu{}^a{}_b \partial_\nu \tilde A^-_{\lambda}{}^b{}_a
+\frac{2}{3} \tilde A^-_{\mu}{}^a{}_b \tilde A^-_{\nu}{}^b{}_c \tilde A^-_{\lambda}{}^c{}_a )\;.
\end{equation}
The action of parity changes the sign of each of the two terms of ${\cal L}_{CS}$, and -- as we saw above -- flips the sign for $f^{abcd}$ which in our new notation amounts to swapping the two $\su(2)$ subalgebras. Combining the two Chern-Simons terms indeed leads to a parity-invariant
lagrangian \cite{Bandres:2008vf,VanRaamsdonk:2008ft}.

The (anti)self-duality constraint means that the independent gauge fields can be taken to be those whose indices $a,b$ take only the values $1,2,3$ and we relabel them $i,j$. Then we can further simplify the action by defining the hermitian $2\times 2$ matrices
\be
A^L_\mu=\half \epsilon_i^{~jk}\sigma_k\,{\tilde A}^+_\mu{}^i_{~j}\;,
\qquad A^R_\mu=\half \epsilon_i^{~jk}\sigma_k\,{\tilde A}^-_\mu{}^i_{~j}\;,
\ee
where $\sigma_k$ are the Pauli matrices. The gauge field action now reduces to the difference of two standard $\su(2)$ Chern-Simons actions, each of level $k$,
\be
\frac{k}{4\pi}\epsilon^{\mu\nu\lambda}{\rm Tr}\left[(A^L_\mu\partial_\nu A^L_\lambda - \frac{2i}{3}A^L_\mu A^L_\nu A^L_\lambda)-
(A^R_\mu\partial_\nu A^R_\lambda - \frac{2i}{3}A^R_\mu A^R_\nu A^R_\lambda)
\right]\;.
\label{CSdiff}
\ee
Moreover, the scalars can now be thought of as bi-fundamentals of the two $\su(2)$ gauge algebras. In this language they are denoted $X^I_{\alpha{\dot \beta}}$ with $\alpha,{\dot\beta} = 1,2$ and are defined in terms of $X_a, a=1,2,3,4$ by
\be
X^I_{\alpha{\dot\beta}} = \left(\half X^I_4 \one + \frac{i}{2}X^I_i\,\sigma^i\right)_{\alpha{\dot\beta}},\qquad i=1,2,3
\ee
where $\sigma^i$ are the Pauli matrices. As a consequence, they satisfy the reality condition
\be
X^{\dagger\,I}{}^{\alpha{\dot\beta}}=\epsilon^{\alpha{\dot\alpha}}
\epsilon^{\beta{\dot\beta}}X^I_{\beta{\dot\alpha}}\;.
\ee
The covariant derivative of \eref{covderblg} becomes
\be
D_\mu X^I=\del_\mu X^I-iA^L_\mu X^I + i X^I A^R_\mu
\label{bifcov}
\ee
and the sextic scalar self-interaction is just
\be
V(X)=\frac83\,\Tr\,\left( X^{[I}X^{J\dagger}X^{K]}
X^{K\dagger} X^J X^{I\dagger}\right)\;.
\ee

In the above discussion, the constant $k$ appears as an overall multiplicative coefficient, bearing the standard normalisation for the level of a Chern-Simons action. As such it is expected to be quantised in integers. To see this,
consider first a single $\su(n)$ gauge field $A_\mu$ and a Chern-Simons lagrangian
\begin{equation}
{\cal L}_{su(2)} = \frac{k}{4\pi}\epsilon^{\mu\nu\lambda}\,{\rm Tr}\,(A_\mu\partial_\nu A_\lambda - \frac{2i}{3}A_\mu A_\nu A_\lambda)\;,
\end{equation}
where ${\rm Tr}$ is the trace in the fundamental ($n\times n$) representation.
Under a large gauge transformation one has \cite{Deser:1981wh}
\begin{equation}
\int d^3 x{\cal L}_{su(n)} \to \int d^3 x{\cal L}_{su(n)}  + {2\pi k w}\;,
\end{equation}
where $w\in \mathbb Z$ is the winding number of the gauge transformation. In particular if we compactify spacetime to $S^3$ then a gauge transformation is a map from $S^3$ into  $\SU(n)$ which always contains a non-contractible 3-cycle. As usual, for the quantum theory to be well defined, we require that $\exp\left(i\int d^3 x{\cal L}_{su(n)}\right)$ remains  invariant under such a transformation. This fixes $k\in \mathbb Z$. The same result holds in our case with $n=2$ since both terms in \eref{CSdiff} are conventional Chern-Simons actions with the usual normalisation. As a result we also find $k \in \mathbb Z$.

To summarise, by exploiting all the desired symmetries we have found a lagrangian that appears to have the correct properties to describe multiple M2-branes.  Unfortunately, it is unique (up to the choice of the integer $k$, whose interpretation we will discuss below) and is thus unable to capture the dynamics of an arbitrary number of M2-branes. This issue will be addressed in the following section by relaxing the supersymmetry constraints of our theory.

\subsection{$\mathcal N=6$ 3-algebra theories}\label{Neq6}

It turns out that the most fruitful way to generalise the previous construction is to look for theories with less supersymmetry. In three dimensions it is possible to have field theories with $\mathcal N=8,6,5,4,3,2,1$ supersymmetry.\footnote{Indeed even $\mathcal N=0$ if one is so   inclined \cite{Gaiotto:2009mv}.} Since we have seen that $\mathcal N=8$ is very constrained and therefore likely to be of limited utility in studying M2-branes, the logical next step is to consider the case of $\mathcal N=6$. This is still a highly supersymmetric theory but as we will see, it is not constrained to have a fixed gauge group. In fact this direction leads to infinitely many interesting field theories, including the ABJM models \cite{Aharony:2008ug} that describe an arbitrary number of M2-branes.

The R-symmetry of an $\mathcal N=6$ superconformal field theory in 2+1d is $\SO(6)\simeq\SU(4)$. In fact we will find theories with $\SU(4)_R\times \U(1)_B$ global symmetry that can be thought of as a subgroup of the $\SO(8)$ R-symmetry of the ${\cal N}=8$ theory. The 12 supercharges transform in the $\mathbf 6$ of the $\SU(4)$, while the $\U(1)_B$ provides an additional global symmetry -- although it will eventually be gauged. The 8 transverse coordinates are grouped into four complex combinations that transform as the $\mathbf 4$ of $\SU(4)$.

Accordingly, we introduce four complex scalar fields $Z^A_a$, $A=1,2,3,4$, as well as their complex conjugates $\bZ_{A}^a$. The symmetries of the problem dictate that we must similarly group the fermions into sets of four complex 2-component spinors $\psi_{Aa}$, with their complex conjugates being denoted by\footnote{Thereby we abandon the $32\times 32$ notation of the previous section. In particular, the gamma-matrices  will henceforth be real $2\times 2$ matrices denoted $\gamma^\mu$ and satisfying $\gamma^0 \gamma^1 \gamma^2 =1$.} 
$\psi^{Aa}$.  A raised $A$ index indicates that the field is in the $\bf 4$ of $\SU(4)$; a lowered index transforms in the $\bar{\bf 4}$. We assign $Z^A_a$ and $\psi_{Aa}$ a $\U(1)_B$ charge of 1. Complex conjugation of fields raises or lowers the $A$ and $a$ indices and flips the sign of the $\U(1)_B$ charge.  The supersymmetry generators are denoted $\epsilon_{AB}$ and are antisymmetric under exchange of their indices. The reality condition $\epsilon^{AB} = \frac{1}{2} \varepsilon^{ABCD}\epsilon_{CD}$ ensures that they are in the $\bf 6$ of $\SU(4)$. Their $\U(1)_B$ charge is taken to vanish.

Having established the setup, one can follow the $\mathcal N=8$ discussion above to arrive at the form for the supersymmetry algebra that preserves the $\SU(4)$, $\U(1)_B$ and conformal symmetries. We will not go through the derivation here, but merely quote the result. Details can be found in Ref.~\cite{Bagger:2008se} where it is shown that the most general supersymmetry transformations are
\begin{eqnarray}\label{finalsusy}
\nonumber  \delta Z^A_d &=& i\bar\epsilon^{AB}\psi_{Bd} \\
\nonumber  \delta \psi_{Bd} &=& \gamma^\mu D_\mu Z^A_d\epsilon_{AB} +
  f^{ab}{}_{cd}Z^C_a Z^A_b \bZ_{C}^c \epsilon_{AB}+
  f^{ab}{}_{cd} Z^C_a Z^D_{b} \bZ_{B}^c\epsilon_{CD} \\
  \delta \tilde A_\mu{}^c{}_d &=&
-i\bar\epsilon_{AB}\gamma_\mu Z^A_a\psi^{Bb} f^{ca}{}_{bd} +
i\bar\epsilon^{AB}\gamma_\mu \bZ_{A}^b\psi_{Ba} f^{ca}{}_{bd} \;,
\end{eqnarray}
where $D_\mu Z^A_d= \del_\mu Z^A_d-\tA_\mu{}^c{}_d Z^A_c$ is a covariant derivative.

The above transformations close into translations and gauge transformations,
namely
\be
[\delta_1,\delta_2]\,Z^A_d=
v^\lambda D_\lambda Z^A_d
+ {\tilde\Lambda}^a{}_d Z^A_a\ ,
\label{compclose}
\ee
for the scalars (and similar expressions for the other fields),
where
\bea\label{symgen}
v^\lambda &=& \frac{i}{2}\bar\epsilon_2^{CD}\gamma^\lambda\epsilon_{1CD},\nn\\[2mm]
{\tilde \Lambda}^a{}_{d}&=&\Lambda^c{}_b f^{ab}{}_{cd},\quad
\Lambda^c{}_b = i(\bar\epsilon^{DE}_2\epsilon_{1CE}-
\bar\epsilon^{DE}_1\epsilon_{2CE})\,\bZ_{D}^c\, Z^C_b\;,
\label{compgauge}
\eea
provided that the fields satisfy the on-shell conditions
\begin{eqnarray}
 \gamma^\mu D_\mu\psi_{Cd} &=& f^{ab}{}_{cd} \psi_{Ca}
Z^D_b\bZ_{D}^c+2f^{ab}{}_{cd}\psi_{Da}Z^D_b\bZ_{C}^c-\varepsilon_{CDEF}f^{ab}{}_{cd}\psi^{Dc} Z^E_aZ^F_b.\nn\\[2mm]
\tilde F_{\mu\nu}{}^c{}_d &=&-
\varepsilon_{\mu\nu\lambda}\left(D^\lambda Z^A_a \bZ_{A}^b-
Z^A_aD^\lambda \bZ_{A}^b
-i\bar\psi^{Ab}\gamma^\lambda\psi_{Aa}\right)f^{ca}{}_{bd}\;.
\end{eqnarray}
As before, the scalar equations of motion can be obtained by performing a supersymmetry variation of the fermion equation and using the gauge field equation to eliminate part of the result.

The structure constants $f^{ab}{}_{cd} = -f^{ba}{}_{cd}$ define a new triple product
\begin{equation}
[T^a,T^b;\bar T_c] = f^{ab}{}_{cd}T^d\; ,
\end{equation}
which must satisfy the following fundamental identity:
\begin{equation}\label{complexFI}
f^{ef}{}_{gb}f^{cb}{}_{ad} +f^{fe}{}_{ab}f^{cb}{}_{gd}+
f^*_{ga}{}^{ fb} f^{ce}{}_{bd}+f^*_{ag}{}^{ eb} f^{cf}{}_{bd}=0\;.
\end{equation}
We see that in this case the triple product is linear and anti-symmetric in its first two entries, but complex anti-linear in the third.

Note that the structure constants  ${f^{ab}}_{cd}$ are now in general complex, and we have defined
\be
f^*_{ab}{}^{cd} = \left(f^{ab}{}_{cd}\right)^*\ .
\ee
Similarly, it is useful to define
\be
\Lambda^*_a{}^b = \left(\Lambda^a{}_b\right)^*\ .
\ee

Note that in this notation, which differs from \cite{Bagger:2008se} and was introduced in \cite{Nilsson:2008kq}, there are only unbarred upper and lower indices. They can be contracted, but that implies that the inner product (the analogue of \eref{innerprod}) is
\be
\langle \bar X, Y\rangle = \bar X^a Y_a\ .
\ee
This seems like  a special case,  equivalent to choosing
$h^{a{\bar b}}=\delta ^{a{\bar b}}$ on a complex manifold. One may  consider more general cases where $h^{a{\bar b}}\ne\delta ^{a{\bar b}}$  by changing the definition of complex conjugation. We will not consider such cases here.

In the special case that the structure constants are real, we can treat the third index on par with the first two (\ie consider it to be a raised index) and ask whether $f^{abc}{}_d$ is antisymmetric in $a,b,c$. When that is the case, we recover the supersymmetry transformations of the ${\cal N}=8$ theory.

Let us now construct an invariant lagrangian.  We have seen that
the supersymmetry algebra closes into a translation plus a gauge
transformation. By complex conjugating Eqs.~(\ref{compclose}), (\ref{compgauge}), we find that under gauge transformations
\begin{equation}
\delta_\Lambda\bZ_{A}^d =
{\tilde\Lambda}^*_{a}{}^d\,\bZ_A^a\;,
\label{cl}
\end{equation}
with $v^\lambda$ and ${\tilde\Lambda}^a{}_{d}$ given
in (\ref{symgen}) and
\be
{\tilde\Lambda}^*_a{}^b = \left(\tilde\Lambda^a{}_b\right)^*
=\Lambda^*{}_d{}^c{} f^*_{ac}{}^{db}\ .
\ee

To construct a
gauge-invariant lagrangian (or, for that matter, any gauge-invariant
observable) we need inner products to be gauge invariant, namely
$\delta_\Lambda (\bZ_{A}^a Z^A_a )=0$. This gives us
\be
\tilde\Lambda^*_b{}^a = -\tilde \Lambda^{a}{}_b\; .
\ee
In addition this requires that
\be
f^{ab}{}_{cd} = f^*_{cd}{}^{ab}\ .
\ee
This allows us to rewrite the fundamental identity as
\begin{equation}\label{complexFI2}
f^{ge}{}_{fd}f^{ab}{}_{cg} = f^{ae}{}_{fg}f^{gb}{}_{cd}+
f^{be}{}_{fg} f^{ag}{}_{cd}-f^*_{cf}{}^{eg} f^{ab}{}_{gd}\;.
\end{equation}

From these equations, we learn that the transformation parameters $\tilde \Lambda^a{}_b$
are elements of $\mathfrak u(n)$. The fundamental identity ensures that they form a Lie subalgebra of $\mathfrak u(n)$, \ie they are closed under ordinary matrix commutation.

The first term in (\ref{compclose}) contains a translation appearing as part of the covariant derivative $D_\mu Z^A_d
= \del_\mu Z^A_d-\tA_\mu{}^c{}_d Z^A_c$. The second piece of the covariant derivative is interpreted as a field-dependent gauge transformation with parameter
$\Lambda^c{}_d=-v^\mu \tilde A_\mu{}^c{}_d$. This implies that the gauge field also takes values in $\mathfrak u(n)$.

With these results, it is not hard to show that the following
lagrangian, invariant up to boundary terms, reproduces the
equations of motion:
\begin{eqnarray}\label{nsixlag}
\nonumber {\cal L} &=& - D^\mu \bZ_A^a D_\mu Z^A_a -
i\bar\psi^{Aa}\gamma^\mu D_\mu\psi_{Aa} -V+{\cal L}_{CS}\\[2mm]
&& -i f^{ab}{}_{cd}\,\bar\psi^{Ad} \psi_{Aa}\,
Z^B_b\bZ_{B}^c+
2if^{ab}{}_{cd}\,\bar\psi^{Ad}\psi_{Ba}\,Z^B_b\bZ_{A}^c\\[2mm]
\nonumber
&&+\frac{i}{2}\varepsilon_{ABCD}f^{ab}{}_{cd}\,\bar\psi^{Ad}\psi^{Bc}\,Z^C_aZ^D_b
-\frac{i}{2}\varepsilon^{ABCD}f^{cd}{}_{ab}\,\bar\psi_{Ac}\psi_{Bd}\,\bZ_{C}^a\bZ_{D}^b\;.
\end{eqnarray}
The potential is
\begin{equation}
V = \frac{2}{3}\,\Upsilon^{CD}_{Bd}\,\bar\Upsilon_{CD}^{Bd} \;,
\label{upsquare}
\end{equation}
with
\begin{equation}\label{upsilon}
\Upsilon^{CD}_{Bd} = f^{ab}{}_{cd}\,Z^C_aZ^D_b\bZ_{B}^c
-\frac{1}{2}\delta^C_Bf^{ab}{}_{cd}\,Z^E_aZ^D_b\bZ_{E}^c+\frac{1}{2}\delta^D_Bf^{ab}{}_{cd}\, Z^E_aZ^C_b\bZ_{E}^c\;.
\end{equation}
The twisted Chern-Simons term ${\cal L}_{CS}$ is given by
\begin{equation}\label{twistedABJM}
{\cal
L}_{CS}=\frac{1}{2}\varepsilon^{\mu\nu\lambda}\left(f^{ab}{}_{cd}\,
A_{\mu}{}^c{}_b\,\partial_\nu A_{\lambda}{}^d{}_a
+\frac{2}{3}f^{ac}{}_{dg}f^{ge}{}_{fb}\,
A_{\mu}{}^b{}_a\, A_{\nu}{}^d{}_c\,A_{\lambda}{}^f{}_e\right)\;.
\end{equation}
It satisfies
\begin{equation}
\frac{\delta{\cal L}_{CS}}{\delta A_{\lambda}{}^{a}{}_b}f^{ac}{}_{db} =
\frac{1}{2}\varepsilon^{\lambda\mu\nu}\tilde F_{\mu\nu}{}^{c}{}_d\;,
\end{equation}
up to integration by parts,
where $\tilde F_{\mu\nu}{}^a{}_b = -\partial_\mu \tilde
A_\nu{}^a{}_b+\partial_\nu \tilde A_\mu{}^a{}_b +  \tilde
A_\nu{}^a{}_e\tilde A_\mu{}^e{}_b- \tilde A_\mu{}^a{}_e\tilde
A_\nu{}^e{}_b $. Just as before, ${\cal L}_{CS}$ can be
viewed as a function of $\tilde A_\mu{}^c{}_d$
rather than $A_\mu{}^c{}_d$.

\subsection{From $\mathcal N=6$ 3-algebras to CS-matter theories}\label{from3altoCS}
The lagrangian constructed above can be given a more standard interpretation as a Chern-Simons matter theory, where the choice of 3-algebra determines the gauge group.  In this section we will show how to obtain ${\cal N} = 6$ Chern-Simons theories with gauge groups $\SU(m)\times \SU(n) \times \U(1)$ for $m\ne n$, $\SU(n)\times \SU(n)$, $\Sp(n) \times \U(1)$, and $\U(n)\times \U(n)$, all with matter in the bi-fundamental representation.

\subsubsection{Gauge group determined by 3-algebra}\label{gaugedet}

We start with what is perhaps the simplest 3-algebra, constructed from rectangular complex $m \times n$ matrices $X,Y,Z,$ as follows
\begin{equation}
\label{Neq6Mat}
[X,Y;Z] = - \frac{2\pi}{k}(XZ^\dag Y - YZ^\dag X)\;,
\end{equation}
where here $X^{\dagger}$ is the conjugate transpose of $X$.  The 3-algebra completely determines the gauge transformation of $X_{dl}$, where $d$ and $l$ are bifundamental indices, running from 1 to $m$ and $n$, respectively,
\begin{eqnarray}
\nonumber \delta X_{dl}& =& [X,Y;Z]_{dl} = f^{aibj}{}_{ckdl}\,\Lambda^{ck}{}_{bj}\,X_{ai} \\[1mm]
   & = &- \frac{2\pi}{k}(X_{dk}Z^{\dagger kb}Y_{bl} - Y_{dk}Z^{\dagger kb} X_{bl})\;.
\end{eqnarray}
This fixes the 3-algebra structure constants,
\begin{equation}\label{UcrossU}
f^{aibj}{}_{ckdl} = - \frac{2\pi}{k}(\delta ^a{}_d \delta ^b{}_c\delta ^i{}_k \delta ^j{}_l -  \delta ^a{}_c \delta ^b{}_d\delta ^i{}_l \delta ^j{}_k )\;.
\end{equation}
The $f^{aibj}{}_{ckdl}$ have the correct symmetries and satisfy the ${\cal N}=6$ fundamental identity.

It is a simple matter to determine the corresponding gauge group.  For the case at hand, we compute
\begin{equation}
\delta X _{dl} = \tilde\Lambda ^{ai}{}_{dl} X_{ai} = - \frac{2\pi}{k}\left(\delta^i{}_l \Lambda^{aj} {}_{dj}  - \delta^a{}_d \Lambda^{bi} {}_{bl}\right)X_{ai}\;.
\end{equation}
The matrix  $\tilde\Lambda^{ai} {}_{dl}$ has a nonvanishing trace for $m \neq n$ and a vanishing trace for $m = n$. Therefore the ${\cal N} = 6$ theory has $\SU(m) \times \SU(n) \times \U(1)$  gauge symmetry when $m \ne n$, and $\SU(n) \times \SU(n)$ otherwise.

A second choice of structure constants is given by
\begin{equation}
f^{ab}{}_{cd} = - \frac{2\pi}{k}(J^{ab} J_{cd} + (\delta^a{}_c \delta^b{}_d - \delta^a{}_d \delta^b{}_c ))\;,
\end{equation}
where $J^{ab}$ is the invariant anti-symmetric tensor of Sp($n$).  The structure constants also obey the fundamental identity and have the correct symmetries.  As above, the gauge symmetry can be determined from the gauge transformation on $X_d$,
\begin{equation}
\delta X _{d} = \tilde\Lambda ^{a}{}_{d} X_{a} = - \frac{2\pi}{k} [(\Lambda_d{}^a + \Lambda^a_d) - \delta^a{}_d \Lambda^b{}_b ] X_a\;.
\end{equation}
This transformation contains two parts:  The first is of the form $\delta' X_d = \tilde\Lambda'^a{}_d X_a$; the second is a phase.   It is not hard to check that $J_{ab}\Lambda'^b{}_c J^{cd} = \Lambda'^d{}_a $, so the gauge group is $\Sp(n) \times \U(1)$.

For the rest of the discussion, we will show how to lift the ${\cal N} = 6$   $\SU(2) \times \SU(2)$ theory to ${\cal N} = 8$ with the same gauge group, thus making a connection with Section~\ref{Neq8}. We first write the fields $Z^A _{\alpha \dot{\alpha}}$ in SO(4) notation \cite{Bagger:2010zq},
\begin{equation}
Z^A _d =  Z^A _{\alpha \dot{\alpha}}\bar{\sigma}_d ^{\dot{\alpha} \alpha}\;,
\end{equation}
where the $\bar{\sigma}_d ^{\dot{\alpha} \alpha}$ are the Pauli matrices of \cite{Wess:1992cp} (except taking $\sigma^0 \rightarrow i\sigma^0 = i\bar\sigma^0$ to make the gauge space Euclidean).  Because of the well-known identity
\begin{equation}
(\bar{\sigma}^a \sigma ^b \bar{\sigma}^c - \bar{\sigma}^c \sigma ^b \bar{\sigma}^a)^{\dot{\alpha}\alpha} = -2 \epsilon^{abcd}\bar{\sigma}_d ^{\dot{\alpha}\alpha}\;,
\end{equation}
the matrix representation of the SU(2) $\times$ SU(2) 3-algebra given in (\ref{Neq6Mat}) exactly reproduces the ${\cal N} = 8$ 3-algebra with\footnote{We absorb the constant of proportionality into $\epsilon^{abcd}$.}
$f^{abcd} =\epsilon^{abcd}$.

To find the full set of supersymmetry transformations, we start with the  ${\cal N} = 6 $ supersymmetry transformations presented above, parametrised by $\epsilon^{AB},$ and construct two additional supersymmetries, parametrised by a {\it complex} spinor $\eta$ of global $\U(1)_B$ charge $+2$.  It is a matter of algebra to find the full set of supersymmetry transformations
 \begin{eqnarray}\label{deltaMAT}
\nonumber \delta Z^A _d &=& i\bar{\epsilon}^{AD}\Psi_{Dd} + i \bar{\eta}\Psi^A _d \\[2mm]
 \nonumber    \delta \Psi_D ^d & = & \gamma ^{\mu}\epsilon_{AD} D_{\mu}Z^{Ad} + \gamma ^{\mu}\eta D_{\mu}\bar{Z}^d _D \\[1mm]
\nonumber                && +\ \epsilon ^{abcd} Z^A_a Z^B _b \bar{Z}_{Dc}   \epsilon_{AB} - \epsilon^{abcd}Z^A_a Z^B_b \bar{Z}_{Bc}\epsilon_{AD} \\[1mm]
&& -\ \epsilon^{abcd} Z^A_a \bar{Z}_{Ab}\bar{Z}_{Dc}\eta - \frac{1}{3} \epsilon_{ABCD}\epsilon^{abcd}\eta^* Z^A_a Z^B _b Z^C _c  \;,
\end{eqnarray}
where gauge indices can be moved up or down because the gauge group is $\SU(2) \times \SU(2) \simeq \SO(4)$.
Closing on the fermion gives
\begin{eqnarray}
  \nonumber [\delta_1 , \delta_2]\Psi_{Dd} & = & v^{\mu}D_{\mu}\Psi_{Dd} + \tilde\Lambda^a{}_{d}\Psi_{Da} \\[1mm]
     && +\ \frac{i}{2}\bar{\epsilon}^{CB} _{[2}\epsilon_{1]CD}E_{Bd} -\frac{i}{4}\bar{\epsilon}^{BE} _{2}\gamma ^{\mu}\epsilon_{1BE}\gamma_{\mu}E_{Dd} \\
 \nonumber    && +\ i \bar{\eta}_{[2}\epsilon_{1]CD}E^C _d - \frac{i}{2}(\bar{\eta} _{[2} \eta^* _{1]} + \bar{\eta}^* _{[2} \gamma^{\mu}\eta_{1]} \gamma_{\mu})E_{Dd}\;,
\end{eqnarray}
as required, where $E_{Dd}$ denotes the fermion equation of motion.  The same calculation also fixes the transformation of the gauge field
\begin{eqnarray}\label{deltaA}
\nonumber \delta \tilde{A}_{\mu}{}^{ad} &=& -i\epsilon^{abcd}\bar\epsilon_{BC}\gamma_{\mu}\Psi^B_b Z^C_c - i\epsilon^{abcd}\bar\epsilon^{BC}\gamma_{\mu}\Psi_{Bb}\bar{Z}_{Cc} \\[1mm]
&& +\ i\epsilon^{abcd} \bar\eta^* \gamma_{\mu}\Psi_{Bb}Z^B_c  + i\epsilon^{abcd}\bar\eta\gamma_{\mu}\Psi^B_b \bar{Z}_{Bc}\;.
\end{eqnarray}
Closing on $\tilde{A}_{\mu}{}^{ad}$ imposes the constraint on the gauge field strength.

The above supersymmetry transformations are manifestly $\SU(4) \times \U(1)_B$ covariant.  However, they must also be covariant under SO(8), the ${\cal N} = 8$ R-symmetry group.  As a check, therefore, one can compute their transformations under the twelve remaining generators of SO(8)/(SU(4) $\times$ U(1)$_B$), which we denote $g^{AB}$, with $\U(1)_B$ charge 2.  The transformations are
\begin{eqnarray}\label{GTT}
\nonumber \delta Z^A_a &=& g^{AB}\bar{Z}_{Ba} \\
\nonumber \delta \Psi_{Ba} &=& -\frac{1}{2}\epsilon_{BCDE}g^{DE}\Psi^C_a \\
 \delta \epsilon^{AB} &=& g^{AB}\eta^* + \frac{1}{2}\epsilon^{ABCD}g^*_{CD}\eta \\
\nonumber\delta \eta &=& -\frac{1}{2}g^{AB}\epsilon_{AB}\;,
\end{eqnarray}
consistent with the fact that $Z^A_a$, $ \Psi_{Bb}$ and $\epsilon^{AB}$ live in different SO(8) representations.  The transformations (\ref{GTT}) close into SU(4) $\times$ U(1) transformations, as required by the SO(8) algebra.  It can be shown that the supersymmetry transformations (\ref{deltaMAT}) and (\ref{deltaA}) are covariant under (\ref{GTT}), as they must be.  Thus, for the case of SO(4) gauge symmetry, the supersymmetry transformations (\ref{deltaMAT}) and (\ref{deltaA}) do indeed lift the ${\cal N} = 6$ theory to ${\cal N} = 8$.

\subsubsection{From $\mathcal N=6$ 3-algebras to $\U(n) \times \U(n)$ CS-matter theories}

We finally show how to extend ${\cal N} = 6$ theories with $\SU(n) \times \SU(n)$ gauge symmetry to $\U(n) \times \U(n)$.  We do this by gauging the global $\U(1)_B$ and requiring supersymmetry. Towards that end, we  introduce an abelian gauge field $B_\mu$ and redefine the covariant derivative $D_\mu$ to be
\begin{equation}\label{covder}
  D_\mu Z^A_a = \partial_\mu Z^A_a - \tilde A_{\mu}{}^b{}_a\,Z^A_b
  - i B_\mu\, \delta_a^b\, Z^A_b\;.
\end{equation}
Similar expressions hold for $D_\mu \psi_{Aa}$, $D_\mu{\bar Z}^a_A$ and $D_\mu\psi^{aA}$ with a flip in the sign of $\tA_\mu$ for fields with a lower $A$ index,\footnote{These transform in the $\bar{\bf 4}$ of $\SU(4)$.} and a flip in the sign of $B_\mu$ for fields with an upper $a$ index.\footnote{These have $\U(1)_B$ charge $-1$.}

Under the $\U(1)_B$ gauge
transformation we have
\begin{equation}
B_\mu \to B_\mu +  \partial_\mu\theta\ .
\end{equation}
Clearly, the action is now invariant under $\U(1)_B$ gauge transformations, so the full gauge symmetry is $\SU(n)\times \SU(n) \times \U(1)_B$.

Our next step is to make the lagrangian invariant under ${\cal N}=6$ supersymmetry. The transformations of $Z^A$, $\psi_A$ and
$\tilde A_\mu^a{}_b$ remain the same, except that the
covariant derivative now includes the $B_\mu$ gauge field.
We also need $\delta B_\mu$ which we simply take to be
\begin{equation}
\delta B_\mu = 0\ .
\end{equation}
Except for the covariant derivatives, the theory is the same as in \eref{nsixlag}, so the supersymmetry variation remains unchanged with
the exception of terms involving $[D_\mu,D_\nu]$, which now
includes a contribution from $G_{\mu\nu} =
\partial_\mu B_\nu - \partial_\nu B_\mu$.  Indeed, we find
\begin{eqnarray}\label{deltaL}
\nonumber \delta{\cal L}_{\SU(n)\times \SU(n)}^{\textrm gauged} &=&-\frac{1}{2}
G_{\mu\nu}\bar\epsilon_{AB}\gamma^{\mu\nu}\psi^{Aa}Z^B_a+\frac{1}{2}
G_{\mu\nu}\bar\epsilon^{AB}\gamma^{\mu\nu}\psi_{Aa}\bZ_B^a\\
&=& -\frac{1}{2}\varepsilon^{\mu\nu\lambda}G_{\mu\nu}\bar\epsilon_{AB}\gamma_{\lambda}\psi^{Aa}Z^B_a
+\frac{1}{2}\varepsilon^{\mu\nu\lambda}\bar\epsilon^{AB} G_{\mu\nu}\bar\epsilon\gamma_\lambda\psi_{Aa}\bZ_B^a\ ,
\end{eqnarray}
where we have used
$\gamma^{\mu\nu}=\varepsilon^{\mu\nu\lambda}\gamma_\lambda$. To
cancel this we introduce a new field $Q_\mu$ and a new term in the
lagrangian
\begin{equation}\label{L}
 {\cal L}_{{\U}(n)\times {\U}(n)} =  {\cal L}_{\SU(n)\times \SU(n)}^{\textrm gauged} +
 \frac{k'}{8\pi}
 \epsilon^{\mu\nu\lambda}G_{\mu\nu}Q_\lambda\;,
\end{equation}
where the first term on the right hand side includes the $B_\mu$ gauge field and $k'$ is an as-of-yet-undetermined real constant. Comparing with (\ref{deltaL}), we see that the complete lagrangian is supersymmetric if we take
\begin{equation}
  \delta Q_\lambda =
  \frac{4\pi}{k'}\bar\epsilon_{AB}\gamma_{\lambda}\psi^{Aa}Z^B_a-\frac{4\pi}{k'}\bar\epsilon^{AB} \gamma_\lambda\psi_{Aa}\bar
  Z_B^a\ .
\end{equation}

The supersymmetry transformation $\delta B_\mu=0$ implies
$[\delta_1,\delta_2]\,B_\mu=0$, whereas one would have expected the commutator to close onto translations and possible gauge transformations.
If, however, the equations of motion are $G_{\mu\nu}=0$, then it is consistent to say that, on-shell,
\begin{equation}
[\delta_1,\delta_2]\,B_\mu = v^\nu G_{\nu\mu}\qquad v^\nu=
\frac{i}{2}(\bar\epsilon_2^{CD}\gamma^\nu\epsilon^1_{CD})\ ,
\end{equation}
which is a combination of a translation and a $\U(1)_B$  gauge transformation.

We must also check the closure on $Q_\mu$. Let us first define the abelian field strength associated to $Q_\mu$ by
\be
H_{\mu\nu} =\partial_\mu Q_\nu -
\partial_\nu Q_\mu\ .
\ee
We find that
\begin{equation}
[\delta_1,\delta_2]\,Q_\mu = \frac{k'}{4\pi}v^\nu \varepsilon_{\mu\nu\lambda}(iZ^A_a D^\lambda \bZ^a_A -i
D^\lambda Z^A_a \bZ_A^a -\bar\psi^A_a\gamma^\lambda \psi_A^a) + D_\mu\Lambda\ ,
\end{equation}
where $\Lambda = (k'/4\pi) (\bar\epsilon_2^{AC}\epsilon_{1BC}-\bar\epsilon_1^{AC}\epsilon_{2BC} )\bar
Z^a_BZ^B_a$. Therefore, if the on-shell condition is
\begin{equation}
H_{\mu\nu} = -\frac{k'}{4\pi}\varepsilon_{\mu\nu\lambda}(iZ^A_a D^\lambda \bZ^a_A -i D^\lambda Z^A_a \bZ_A^a
-\bar\psi^A_a\gamma^\lambda \psi_A^a)\ ,
\end{equation}
we again find a translation plus a $\U(1)_Q \times\U(1)_B$ gauge
transformation
\begin{equation}
[\delta_1,\delta_2]\,Q_\mu = v^\nu H_{\nu \mu} + D_\mu\Lambda\ .
\end{equation}
Thus we see that $Q_\mu$, which started off life as a Lagrange multiplier for the constraint $G_{\mu\nu}=0$, naturally inherits a $\U(1)$ gauge symmetry of its own. The closure on the other fields remains unchanged from the $\SU(n) \times \SU(n)$ lagrangian, except that the connection now involves the $\U(1)_B$ gauge field.

If we write $B_\mu = A^L_\mu-A^R_\mu$ and $Q_\mu= A^L_\mu+A^R_\mu$, the new term in (\ref{L}) can be written in the following form, up to a total derivative:
\begin{equation}
{\cal L}_{\U(1)\times \U(1)\ CS} = \frac{k'}{4\pi} \epsilon^{\mu\nu\lambda}A^L_\mu\partial_\nu A^L_\lambda
-\frac{k'}{4\pi}\epsilon^{\mu\nu\lambda}A^R_\mu\partial_\nu A^R_\lambda\;,
\end{equation}
This is nothing but the Chern-Simon lagrangian for a $\U(1)\times\U(1)$ gauge theory.

We have therefore constructed a family of ${\cal N}=6$ Chern-Simons-matter lagrangians that have gauge fields in $\U(1)\times \SU(n)\times \U(1)\times \SU(n)$  and are parametrised by two numbers $k$ and $k'$, associated respectively to the $\SU(n)$ and $\U(1)$ factors. From the point of view of supersymmetry the levels $k$ and $k'$ are arbitrary and independent. Although $k$ must be an integer in the quantum theory, $k'$ need not be (indeed $k'$ can be absorbed into the definition of $Q_\lambda$); see for example Ref.~\cite{Dunne:1998qy}. (The possibility of choosing different levels for the $\SU(n)$ and $\U(1)$ factors was also pointed out in \cite{Aharony:2008ug}.)

With the special choice
\begin{equation}
k' = n k\ ,
\end{equation}
we see that the addition of the $\U(1)\times\U(1)$ Chern-Simons term simply converts the $\SU(n)\times \SU(n)$
level $(k,-k)$ Chern-Simons term ${\cal L}_{CS}$ with connection $\tilde A^a{}_b$ into a $\U(n)\times\U(n)$ level $(k,-k)$ Chern-Simons term with connection $\tilde
A_\mu^{L/R}+ iA^{L/R}_\mu$. In terms of $A^{L/R}_\mu$, the supersymmetry transformations are simply
\begin{equation}
\delta A^R_\lambda =\delta A^L_\lambda=
\frac{2\pi}{nk}\bar\epsilon_{AB}\gamma_{\lambda}\psi^{Aa}Z^B_a-\frac{2\pi}{nk}\bar\epsilon^{AB} \gamma_\lambda\psi_{Aa}\bar
Z_B^a\ .
\end{equation}

To summarise, we have used $\mathcal N=6$ 3-algebras to construct a variety of Chern-Simons-matter theories \cite{Lambert:2010ji}, a big step forward from the single $\mathcal N = 8 $ model of Section~\ref{Neq8}.  Their lagrangian is given by the set of equations (\ref{nsixlag})-(\ref{twistedABJM}).  One could look for further generalisations, including constructions with less supersymmetry, \eg ${\cal N}=5,4$.  We will come back to such theories in Chapter~\ref{chapter7}.  However, with the results in hand, we have what we need to understand multiple M2 branes.  Before we continue with their physical analysis, we provide a brief mathematical description of 3-algebras.

\subsection{Some mathematics of 3-algebras}

We have seen how Euclidean 3-algebras have been instrumental in the construction of three-dimensional  Chern-Simons-matter theories with ${\cal N}=8,6$ supersymmetry. Given their importance it is appropriate to pause our analysis, pertaining to their relation to M2-branes, and do them (partial) justice by providing a brief mathematical discussion of their properties.

Although perhaps novel to the mainstream string theory literature, 3-algebras have been studied in the mathematical and physical literature for more than 50 years.  They go by several names (Filipov algebras, ternary algebras, triple systems...). A selection of papers is given in
\cite{Jacobson,Lister,Faulkner,Filipov, Kasymov, Takhtajan:1993vr}.
 More recent and relevant discussions for our purposes can be found in \cite{deMedeiros:2008bf,
Cherkis:2008qr,Nilsson:2008kq,deMedeiros:2008zh,Cherkis:2008ha, deMedeiros:2009hf,Palmkvist:2009qq,deAzcarraga:2010mr,MendezEscobar:2010zq}.

At the most general level, a  3-algebra is simply a vector space ${\cal V}$ with a triple product
\begin{equation}
[\cdot,\cdot,\cdot]:{\cal V}\otimes {\cal V}\otimes  {\cal V} \to {\cal V}
\end{equation}
that is linear in each of the entries and satisfies a fundamental identity that generalises the concept of the Jacobi identity.  Although in the cases above we assumed that the triple product had various symmetry properties, this is not always required, and we do not require it in this section. Imposing symmetry properties restricts the 3-algebra and leads to Chern-Simons-matter lagrangians with different amounts of supersymmetry.

If the vector space ${\cal V}$ is real then we have a real 3-algebra. We can also introduce the notion of a complex 3-algebra by taking ${\cal V}$ to be a complex vector space and defining
\begin{equation}
[\cdot,\cdot;\cdot]:{\cal V}\otimes {\cal V}\otimes \bar{\cal V} \to {\cal V}\;,
\end{equation}
where $\bar {\cal V}$ is the complex (Hermitian) conjugate of ${\cal V}$. We also assume a similar map acting on the complex conjugate space
\begin{equation}
[\cdot,\cdot;\cdot]: \bar{\cal V}\otimes  \bar{\cal V}\otimes {\cal V} \to  \bar{\cal V}\;.
\end{equation}
(Note that we use the same notation for both maps since choosing which is which is easily determined by the elements on which it acts.)

Such maps preserve a ${\mathbb Z}_2$ grading where elements of $\cal V$ have charge 1 and those of $\bar{\cal V}$ have charge -1. In this case we require that the triple product be complex linear in the first two entries and anti-linear in the third entry. In addition, one can also introduce the notion of a quaternionic 3-algebra, but we will not discuss it here.

A complex 3-algebra can be viewed as a special case of a real 3-algebra and conversely a real 3-algebra is obtained from a complex 3-algebra by taking all the elements to be real (in cases where $\bar {\cal V}$ is naturally isomorphic to $\cal V$) and restricting the field associated to ${\cal V}$ to be $\mathbb R$. Thus in what follows we will only consider complex 3-algebras since the results automatically apply to real 3-algebras as well.

The key defining feature of a 3-algebra  is that the generalisation of the adjoint map should act as a derivation. In particular if we fix any two elements of $U \in {\cal V}$, $\bar V \in \bar {\cal V}$, then these induce  the linear map  $\varphi_{U,\bar V}:{\cal V}\to {\cal V}$ and $ \varphi_{U,\bar V}:\bar{\cal V}\to \bar{\cal V}$ defined by
\begin{equation}
\varphi_{U,\bar V}(X)=[X,U;\bar V]\qquad  \varphi_{U,\bar V}(\bar X)=-[\bar X,\bar V; U]\ .
\end{equation}
(Note the minus sign which is chosen so that in the real, totally anti-symmetric case, the two actions of  $\varphi_{U,\bar V}$ agree.)
We require that this map is a derivation in the sense that
\begin{equation}
 \varphi_{U,\bar V}([X,Y;\bar Z]) = [\varphi_{U,\bar V}(X),Y;\bar Z] + [X,\varphi_{U,\bar V}(Y);\bar Z] + [X,Y;  { \varphi_{U,\bar V}(\bar Z)}]\ ,
\end{equation}
or, equivalently,
\begin{equation}
 [[X,Y;\bar Z],U;\bar V] = [[X,U;\bar V],Y;\bar Z] + [X,[Y,U;\bar V];\bar Z] - [X,Y; { [\bar Z,\bar V,U]}]\ ,
\end{equation}
for all elements of $\cal V$. This is referred to as the fundamental identity and plays a role analogous to the Jacobi identity in Lie-algebras.

Since we are interested in physical theories, we also require that the vector space $\cal V$ admits an inner-product that we denote by $\langle\cdot,\cdot\rangle$ and take to be complex linear in the second entry and anti-linear in the first. This needs to be invariant with respect to the action of the map $\varphi_{U,\bar V}$ in the sense that $\langle \varphi_{U,\bar V}(\bar X),Y \rangle+\langle \bar X,\varphi_{U,\bar V}(Y) \rangle=0$ or
\begin{equation}
\langle {[\bar X,\bar V;U]},Y \rangle =\langle \bar X,[Y,U;\bar V]\rangle\  .
\end{equation}

Next we observe that a 3-algebra has a natural Lie-algebra associated to it. In particular, let $ {\cal G} \subset \mathrm{GL}(\cal V)$ be the vector space of all linear maps of $\cal V$ spanned by elements of the form $\varphi_{U,\bar V}$ for some pair $U\in \cal V$, $\bar V\in \bar \cal V$. Furthermore, we observe that the fundamental identity can be written as
\begin{equation}
[ \varphi_{U,\bar V},\varphi_{Y,\bar Z}](X) =  \varphi_{\varphi_{U,\bar V}(Y),Z}(X) - \varphi_{Y, \varphi_{U,\bar V}(\bar Z)}(X)\ ,
\end{equation}
and hence the commutator of two elements of $\cal G$ is contained in $\cal G$. Since
the composition of maps in $\mathrm{GL}({\cal V})$ is associative  the Jacobi identity is automatically satisfied. Thus ${\cal G}$ is a sub-Lie-algebra of $\mathrm{GL}({\cal V})$.

In the case that $[\cdot,\cdot;\cdot]$ is either symmetric or anti-symmetric in the first two entries, the inner-product $\langle\cdot,\cdot\rangle$ induces an invariant inner-product $(\cdot,\cdot)$  on ${\cal G}$:
\begin{equation}
(\varphi_{Y,\bar Z},\varphi_{U,\bar V}) =  \langle  \bar Z,[Y,U;\bar V] \rangle\; .
\end{equation}
The condition $[X,Y;\bar Z]=\pm [Y,X;\bar Z]$ implies that $(\varphi_{Y,\bar Z},\varphi_{U,\bar V})=(\varphi_{U,\bar V},\varphi_{Y,\bar Z})$ as required for a metric.
This metric is gauge invariant and non-degenerate (assuming the 3-algebra satisfies a certain semi-simple condition) but is not the usual Killing-form on a Lie-algebra. In particular it is not positive definite in general. This is clearly the case if $[X,Y;\bar Z]=- [Y,X;\bar Z]$ since then $(\varphi_{U,\bar V},\varphi_{U,\bar V})=0$. For example in the totally anti-symmetric 3-algebra (\ref{onlyone}) where ${\cal G} = \frak{so(4)}\cong \frak{su(2)}\oplus \frak{su(2)}$, one finds the inner-product $(\cdot,\cdot)$ acts as $+4\pi/k$ times the Killing form on one $\frak{su(2)}$ factor (self-dual gauge fields) and $-4\pi/k$ times the Killing form on the second (anti-self-dual gauge fields). This inner-product  appears, through its inverse, in the action through the Chern-Simons term and therefore  is not required to be positive definite.

The notable feature of the Lie algebras generated in this way from a 3-algebra is that they are not typically simple but have a product structure. Although this follows naturally from the 3-algebras, from the point of view of the gauge theory this is something of a surprise and helps to explain why such highly supersymmetric Chern-Simons gauge theories took so long to discover; the amount of supersymmetry largely depends on the choice of non-simple gauge group.

Stated another way, we see that $\cal V$ is the vector space of a representation of $\cal G$. In fact we can turn this around. Given any Lie-algebra $\cal G$ with invariant inner-product $(\cdot,\cdot)$ and a representation $R:{\cal G}\to {\cal V}$ with invariant inner-product $\langle\cdot,\cdot\rangle$, we can construct a triple product on $\cal V$ that satisfies the fundamental identity. To see this we construct the Faulkner map \cite{deMedeiros:2008zh}
\begin{equation}
\varphi:{\cal V}\times \bar {\cal V} \to {\cal G}\;,
\end{equation}
which is defined as follows. We first note that for any two elements $U\in {\cal V}$, $\bar V\in \bar {\cal V}$ we can construct an element $\varphi_{U,V}^*$ of the dual space ${\cal G}^*$ (\ie the space of linear maps from $\cal G$ to $\mathbb C$) by
\begin{equation}
\varphi^*(g)_{U,\bar V}= \langle \bar V,g(U)\rangle\;,
\end{equation}
where $g \in {\cal G}$.
However since $\cal G$ has an inner-product we can identify ${\cal G}^*$ with $\cal G$. In particular $\varphi^*_{U,V}$ can be realised by an element of $\varphi_{U,V}\in\cal G$ such that
\begin{equation}
\varphi^*(g)_{U,\bar V} = (\varphi_{U,\bar V},g) \ ,
\end{equation}
for all $g\in \cal G$. Thus we have constructed  $\varphi:{\cal V}\times{\cal V}\to {\cal G}$. Finally we observe that the Faulkner map defines a triple product on $\cal V$
\begin{equation}
[W,U;\bar V] = \varphi_{U,\bar V}(W)\;.
\end{equation}
By construction this map is linear in the first two entries and complex anti-linear in the third. Furthermore the fundamental identity is just the statement that the Faulkner map $\varphi_{U,\bar V}$ is equivariant:
\be
[g,\varphi_{U,\bar V}] = \varphi_{g(U),\bar V}+\varphi_{U,g(\bar V)}\ ,
\ee
where $g$ is an element of $\cal G$.

It is perhaps helpful now to be a little less mathematical   and illustrate the Faulkner construction using symbols more familiar to physicists. Suppose that we have a Lie-algebra $\cal G$ with generators $(T^r)^a{}_b$, $r=1,...,{\rm rank}({\cal G}) $ that act in some (typically reducible) representation  $\cal V$, where $a,b=1,...,{\rm dim}({\cal V})$. We further suppose that $\cal G$  and $\cal V$ have invariant, non-degenerate (but typically not positive definite) metrics $h_{rs}$ and $g^{ab}$. The Faulkner construction says that
\be
\varphi_{U,\bar V}(g)_r = h_{rs}(T^s)^a{}_bU_a\bar V_eg^{be}\ ,
\ee 
and the triple product structure constants are 
\be
f^{abc}{}_{d} =  (T^r)^b{}_e(T^s)^a{}_d h_{rs}g^{ce} \ .
\ee
In terms of irreducible representations, where we can write $h_{rs} = c_R\kappa^R_{rs}$ with $\kappa^R_{rs}$  the Killing form and $c_R$ a constant, we have
\be
f^{abc}{}_{d} =  \sum_R c_R(T^r)^b{}_e(T^s)^a{}_d \kappa^R_{rs}g^{ce} \ .
\ee

Furthermore, since a Lie-algebra always has the adjoint representation,  a 3-algebra is really an extension  of a Lie-algebra  to include additional, preferred, representations. Indeed, one can think of a Lie-algebra as a special case of a 3-algebra where the preferred representation is the adjoint. In this case the triple product is simply
\be
[X,Y,Z] = [[X,Y],Z]\ ,
\ee
where $[X,Y]$ is the Lie-bracket. One can check that, as a consequence of the Jacobi identity, $[[X,Y],Z]$ satisfies the fundamental identity.  

Thus 3-algebras can be viewed  as encoding the data of familiar Lie-algebra representation theory. They arise in Chern-Simons-matter theories since supersymmetry requires that the dynamical fields  sit in different representations from the  (non-dynamical) gauge fields -- which are, as always, in the adjoint. In particular, thinking in terms of 3-algebras enabled the discovery of new maximally supersymmetric gauge theories.
 This stands in contrast to more familiar Yang-Mills theories with dynamical gauge fields, where supersymmetry requires that matter fields be in the adjoint representation.


\section[The effective action of multiple M2-branes]{{\Large{\bf The effective action of multiple M2-branes}}}\label{chapter4}

We have thus far constructed a set of novel three-dimensional gauge theories  with $\mathcal N=8,6$ supersymmetry based on 3-algebras, while also providing some rationale on why these have the correct features to capture the low-energy dynamics of M2-brane configurations in M-theory. In this section we will  see explicitly how this connection arises and establish them as the gauge theories describing the CFT side of an $\mathrm{AdS}_4/\mathrm{CFT}_3$ duality.

\subsection{Brane derivation}\label{branederivation}

Following the construction of the ${\cal N}=8$ $\SU(2)\times\SU(2)$ BLG model and, as we will shortly see, its interpretation as describing two M2-branes \cite{Lambert:2008et,Distler:2008mk}, the ${\mathcal N}=6$ models with gauge group $\U(n)\times \U(n)$  were proposed by Aharony, Bergman, Jafferis and Maldacena (ABJM) in \cite{Aharony:2008ug}. One way in which the result can be derived is using brane constructions, from which one naturally obtains not only the above gauge theories but also the precise M-theory system that they describe. This is a rather lengthy, but ultimately insightful construction, which can be broken up into the following steps:
\begin{enumerate}
\item Construct a ``base'' type IIB brane configuration.

\item T-dualise to type IIA. Then lift to M-theory and take the near-horizon limit to obtain the candidate dual geometry on the gravity side.

\item Start again with the ``base'' type IIB brane configuration  and its associated low-energy theory.

\item Take the decoupling limit  and flow to the IR  to obtain an $\mathcal N= 6$   Chern-Simons-matter theory.

\end{enumerate}

\noindent
In what follows, we will explain each of these steps in detail.

\subsubsection{The ABJM brane construction}\label{breaneology}

We begin with the classic Hanany-Witten configuration \cite{Hanany:1996ie}. This is made up of intersecting D3, D5 and NS5-branes.  Consider $n$ D3-branes extended along the $\{x^0,x^1,x^2,x^6\}$ directions and suspended between two parallel NS5-branes that lie along $\{x^0,...,x^5\}$ and are separated by a finite distance $l$ along $x^6$. Because of the latter, the low-energy theory on the D3-branes reduces to a certain three-dimensional $\U(n)$ Yang-Mills gauge theory along $\{x^0,x^1,x^2\}$.

To determine this theory, notice that the NS5-branes impose boundary conditions on the worldvolume fields, reducing the supersymmetry by a factor of 2. In 2+1d, the  $\mathcal N=8$ vector multiplet decomposes into the sum of an $\mathcal N=4$ vector and hypermultiplet. It can be argued \cite{Hanany:1996ie} that the latter gets projected out, leading to a theory with 8 supersymmetries. The bosonic field content is then a three-dimensional gauge field along with three scalars that parametrise the fluctuations of the D3-branes along the NS5-brane directions $\{x^3, x^4, x^5\}$.

\begin{figure}[t]
\centering
\includegraphics[width=0.6\textwidth]{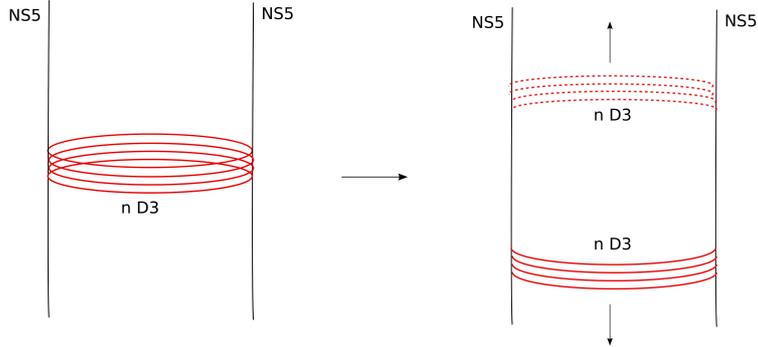}
\caption{The D3-brane segments can move independently.\label{figure1}}
\end{figure}

Let us now take the $x^6$ direction to be  compact with period $2\pi R $. The D3-branes are now chosen to wrap $x^6$ and the NS-branes are located at $x^6=0$ and  $x^6=\pi R$. The D3-branes are free to move up and down along the NS5-branes and furthermore they  can split up into two independent sets of $n$ D3-branes (see Fig.~\ref{figure1}), corresponding  to the segments $0<x^6<\pi R$ and $\pi R < x^6< 2\pi R $.  Thus the low-energy gauge group is  $\U(n)\times \U(n)$.

This time the field content evidently consists of two ${\cal N}=4$ $\U(n)$ vector multiplets, each containing a gauge field, 3 scalars and fermions in the adjoint of $\U(n)$. In addition we get a hypermultiplet in the bi-fundamental $(\bf{ n},\bf{\bar {n}})$ of $\U(n)\times \U(n)$ corresponding to the open strings that stretch between the two segments of D3-branes, as well as a hypermultiplet in the $(\bf{\bar {n}} ,\bf{n})$ for strings that stretch the opposite way. In terms of ${\cal N}=2$ language these give two chiral superfields ${\cal A}_i$, $i=1,2$ in the  $(\bf{ n},\bf{\bar {n}})$ and two more chiral superfields ${\cal B}_i$, $i=1,2$  in the  $(\bf{\bar {n}} ,\bf{n})$.

So far we just have a familiar $\U(n)\times \U(n)$ Yang-Mills theory in three dimensions, without any Chern-Simons terms. However, it is possible to obtain the latter by integrating out the massive fundamental fermions. This is due to the fact that three-dimensional Yang-Mills theories with fundamental fermions exhibit a parity anomaly at one loop. One can therefore integrate them out only at the expense of introducing a parity-violating Chern-Simons term. The Chern-Simons level receives a contribution of $\pm \frac{1}{2}sgn(m_f)$ for each fundamental or anti-fundamental fermion respectively, with $sgn(m_f)$ being the sign of the mass term \cite{1983PhRvL..51.2077N,1984PhRvD..29.2366R}.

Motivated by the above observation, we need to introduce some massive multiplets in the fundamental representation. To do this we add $k$ D5-branes along $\{x^0,x^1,x^2,x^3,x^4,x^9\}$ but sitting at $x^6=0$. These intersect the D3-branes and one of the NS-branes (the one at $x^6=0$), breaking a further half of the supersymmetry down to ${\cal N}=2$. This addition leads to $k$ chiral multiplets in the fundamental and anti-fundamental representation of each  $\U(n)$ factor, from the strings stretching between the D5-branes and each set of D3-branes.\footnote{Note that when the   two sets of D3-branes touch at $x^6=0$, the two types of 5-3 open strings can join up and form a   5-5 string. Thus if one is in the fundamental of $\U(n)$ then the other is in the   anti-fundamental.}

\begin{figure}[t]
\centering
\includegraphics[width=0.6\textwidth]{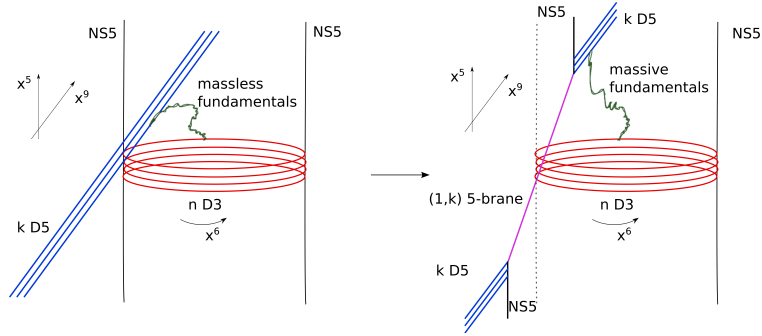}
\caption{Introduction of D5's and mass deformation.\label{figure2}}
\end{figure}

To induce a mass term for these fundamental chiral superfields, we deform the NS5-brane/D5-brane intersection into a $(p,q)$ 5-brane web \cite{Bergman:1999na}. In particular, we break up the intersection in the $x^9$ direction and replace it by a $(1,k)$ 5-brane along $\{x^0,x^1,x^2,x^2,x^3,x^4\}$ and at a particular angle $\theta$
in the $\{x^5,x^9\}$ plane (see Fig.~\ref{figure2}). Supersymmetry determines the angle $\theta$ that the $(1,k)$ brane makes with the NS5-brane \cite{Aharony:1997ju}. Because of the minimal coupling between  the charged matter fields and the vector multiplet
\be
\int d^4 \theta\; Q^\dagger e^V Q \;, \qquad \int d^4 \theta\; \tilde Q^\dagger e^{-V} \tilde Q\;,
\ee
a VEV for the vector multiplet scalar associated with the D5's results in a real mass  with opposite sign for the fundamental and anti-fundamental chiral superfields respectively. As a result, by integrating out the $3-5$ strings, one obtains a Chern-Simons term with level $k$ for the first $\U(n)$ factor and $-k$ for the second. Note that while each Chern-Simons term breaks parity independently, their combination does not as long as it is accompanied by  a simultaneous  exchange of the two gauge fields.

We have therefore managed to find a complicated brane configuration, the effective theory for which produces the desired Chern-Simons terms. The final step is to note that by rotating the $(1,k)$ five-brane relative to the NS5 sitting at $x^6 = \pi R$ by equal angles $\theta = \arctan(k)$ in the $\{x^3,x^7\}$, $\{x^4,x^8\}$ and $\{x^5,x^9\}$ planes, the supersymmetry is enhanced from $\mathcal N=2$ to $\mathcal N=3$ \cite{Kao:1995gf,Kitao:1998mf}. This completes the construction of our ``base'' IIB brane system, though we will return to it shortly to discuss how the supersymmetry gets enhanced beyond $\mathcal N=3$.

\subsubsection{From branes to  the ABJM geometry}\label{branestoABJM}

The above brane construction can be related to M-theory and M2-branes through a series of duality transformations. We first map to type IIA string theory by T-dualising along $x^6$. The D3-branes are now D2-branes along $\{x^0,x^1,x^2\}$. At the same time the NS5-brane is mapped to a Kaluza-Klein  monopole along $\{\tilde x^6,x^7,x^8,x^9\}$, where $\tilde x^6$ is the compact direction T-dual to $x^6$. On the other hand, the $(1,k)$ 5-brane gets mapped to a bound state of a Kaluza-Klein monopole along with $k$ units of D6-brane flux. Recall that this object lies along $\{x^0,x^1,x^2\}$ and also lies at a fixed angle $\theta$ in the $\{x^3,x^7\},\{x^4,x^8\},\{x^5,x^9\}$ planes.

The type IIA Kaluza-Klein monopoles that we have obtained are identical to the geometries which we reviewed around \eref{KKinone}, the only difference being that now the circle that plays a crucial role in constructing the monopole is $x^6$ rather than the M-theory direction. We have seen that they correspond to purely gravitational solutions  described by the metric \cite{Gibbons:1979zt,Aharony:2008ug}
\be\label{KKmonopole}
ds^2 = U dx^a dx_a + U^{-1} (d \phi +\omega^a dx_a)^2\;,
\ee
where $a= \{1,2,3\}$, $\phi= \phi + 2\pi$, $\pd_a \pd^a U = 0$ and $\pd_a \omega_b - \pd_b \omega_a =\epsilon_{abc} \pd^c U$, with all the indices raised and lowered with  the three-dimensional Euclidean metric.\footnote{There is a slight change in notation from Chapter \ref{chapter1}. Here the vector field will be denoted $\omega$ rather than $A$ and we will use explicit index notation in place of vector notation (\eg $\del_a$ instead of $\vec\nabla$).}

As previously explained in Chapter \ref{chapter1}, this geometry describes a nontrivial circle fibration over $\mathbb R^3$,\footnote{We will also use $\;\{|\vec{x}|,\vartheta, \varphi\}$ spherical coordinates for $\mathbb R^3$ in what follows.} where the circle shrinks to zero size at the origin. Moreover, requiring the absence of singularities imposes an $n\in \mathbb Z$ ``Dirac'' quantisation condition on the flux of $\omega$. A simple solution is when $U$ is a harmonic function with
\be\label{TN}
 U = U_{\infty}+ \frac{q}{2 | \vec x\; |}
\ee
where $U_{\infty}$ is a constant parametrising the size of the circle at infinity (we had earlier just set this to 1, but now it will become relevant), while  $\omega_\varphi = \frac{1}{2} q \cos \vartheta$. In the ``near core'' limit  ($|\vec x|\to 0$), $U\sim \frac{1}{2} q |\vec x|$ and via a coordinate change the geometry can be seen to reduce to a $\mathbb C^2/\mathbb Z_q$ orbifold.

It is now time to reap the rewards of our efforts by lifting the whole configuration to M-theory. By going to strong coupling we decompactify the $x^{10}$ spatial direction and the D2-branes turn into M2-branes along $\{x^0,x^1,x^2\}$. The Kaluza-Klein monopole associated with $\tilde x^6$  remains a Kaluza-Klein monopole in eleven dimensions. But recall that upon decompactification a pure D6-brane also turns into a Kaluza-Klein monopole associated with the M-theory circle, $x^{10}$. As a result, the initial $(1,k)$ five-brane in our type IIB configuration, which becomes a Kaluza-Klein monopole carrying $k$ units of D6-brane flux after T-duality, now becomes a ``tilted'' Kaluza-Klein monopole in which the circle is a linear combination of $\tilde x^6$ and $x^{10}$.

Therefore, the M-theory system constitutes exclusively of M2-branes extending in $\mathbb R^{2,1}$ and probing a nontrivial 4-complex-dimensional background given by the superposition of the two Kaluza-Klein monopoles. This type of transverse geometry has been investigated in the literature and goes by the name of ``toric hyper-K\"ahler,'' in general preserving six bulk supercharges \cite{Gauntlett:1997pk}.

The metric, which we will shortly write down, is a generalisation of the Kaluza-Klein monopole that was written down in \eqref{KKmonopole} above. In place of the single harmonic function $U$, we will now require a positive-definite $2\times 2$ matrix of harmonic functions $U_{ij}$ obeying
\be
U^{ij}\pd_{(i)}^a \pd_{(j)\,a} U_{kl} = 0\;.
\ee
where $\del^a_{(i)}\equiv \del/\del x^{(i)}_a$ and $U^{ij}$ is the matrix inverse of $U_{ij}$.
This condition can be shown to imply that each entry of $U_{ij}$ is harmonic.
Correspondingly there will be a $2\times 2$ matrix of vector fields $\omega_{ij}^a$ that are related to $U_{ij}$ via a generalisation of the usual relation between a vector field and a harmonic function,
\be
\partial^a_{(j)} \omega^b_{ki}-\partial^b_{(k)}\omega_{ji}^a = \epsilon^{abc}\partial_{(j)\,c}U_{ki}\;.
\label{curlom}
\ee
This relation is required for the metric to be hyper-K\"ahler.

Because the Kaluza-Klein monopoles are tilted with respect to each other, the two sets of coordinates
\be
\{x_a^{(1)}\}=(x^3,x^4,x^5),\qquad
\{x_a^{(2)}\}=(x^7,x^8,x^9)
\ee
play complementary roles in the metric. Moreover the metric has two angular coordinates $\phi_1 = \tilde x^6$ and $\phi_2 = x^{10}$, with period $2\pi$, that label the two Kaluza-Klein circles in the problem and together parametrise a 2-torus. The modulus of this 2-torus is
given by
\be
\label{torus}
\tau = -\frac{U_{12}}{U_{11}}+ i \frac{\sqrt {\det  U} }{U_{11}}\;.
\ee

With these preliminaries in place, the metric for the geometry under consideration can now be written as
\be
ds^2 = U_{ij} \;dx_a^{(i)} dx_a^{(j)} + U^{ij}(d\phi_i + \omega_{ik}^a dx_a^{(k)})(d\phi_j + \omega^b_{jl}dx^{(l)}_b)\;.
\ee
Note that \eref{curlom} implies the linear relation $\pd_{(i)\,a}U_{jk} - \pd_{(j)\,a}U_{ik} =0$. Since the equations for $U$ and $\omega$ are linear,  linear superpositions of simple solutions are also solutions to the supergravity equations, a feature that we will put to use below.

It is important to note that the modular parameter $\tau $ is acted upon by a set of $\SL(2,\mathbb R)$ fractional linear transformations which leave the torus invariant. Through \eref{torus} these induce a set of transformations on $U$, where  $U\to G^{T} U G$ with $G\in \SL(2,\mathbb R)$. It turns out that  a $G\in \SL(2,\mathbb Z)$ subgroup of the latter is a symmetry,  generating new solutions of the eleven-dimensional supergravity equations \cite{Gauntlett:1997pk}. In fact,  a more general transformation of $U$ with $G\in \mathrm{GL}(2,\mathbb Z)$ is {\it still} a symmetry of the theory, although this now leads to a change in the asymptotic shape of the torus.

This knowledge can be put to work  by allowing us to explicitly write down and study the geometry in which we are interested. The Kaluza-Klein monopole associated with the $x^{10}$ direction can easily be embedded in this eight-dimensional framework by choosing
\be\label{KK1}
U_1 = U_\infty + \twobytwo{h_1}{0}{0}{0}\;,\qquad h_1 = \frac{1}{2 |\vec x^{\,(1)} |}\;,\qquad U_\infty= \twobytwo{1}{0}{0}{1}\ .
\ee
It is a nice regular geometry (we have set $q$ of \eref{TN} to one). The other monopole, corresponding to what used to be the $(1,k)$ fivebrane in the IIB picture, is given by \cite{Aharony:2008ug}
\be\label{KK2}
U'_2 = U'_\infty + \twobytwo{ h_2}{k h_2}{k h_2}{k^2 h_2}\;,\qquad h_2 = \frac{1}{2 |\vec x^{\,(1)} + k \vec x^{\,(2)} |}\;,\qquad U'_\infty= \twobytwo{1}{0}{0}{1}\;.
\ee
This is also a simple geometry. To see this, apply a $\mathrm{GL}(2,\mathbb Z)$ transformation that acts as $U_2\to G^{T} U_2 G$, while simultaneously taking ${\vec x}^{\,(i)} \to {G^i}_{j} {\vec x}^{\,(j)}$ and $\phi_i \to \phi_j \,{G^{j}}_i$ to keep the line element invariant, with
\be
{G^i}_j = \twobytwo {1}{0}{-k^{-1}}{k^{-1}}\;.
\ee
Under this, \eref{KK2} is equivalent to
\be
U_2 = U_\infty + \twobytwo{ 0}{0}{0}{h_2}\;,\qquad h_2 = \frac{1}{2 |\vec x^{\,(2)}|}\;,\qquad U_\infty=k^{-2}\twobytwo{k^2+1}{-1}{-1}{1}\;,
\ee
with new angles $\{\phi_1',\phi_2'\} = \{\phi_1 -  \phi_2/k,\phi_2/k\}$
and modified periodicity $\{ -2 \pi/k,2 \pi/k\}$ respectively. The fact that $U_\infty\neq\one$ and that the periodicity of the circles parametrising the torus has changed is a result of the $\mathrm{GL}(2,\mathbb Z)$ transformation, which as we have mentioned above does not preserve the torus.

The final step is to combine \eqref{KK1} and \eqref{KK2} by linear superposition, such that the solution is given by $U = U_\infty + \textrm{diag}(h_1,h_2)$, with $h_{1,2} =\frac{1}{2}|\vec x^{\,(1),(2)}|^{-1}$. The nonzero elements of $\omega$ are then $({\omega_{\varphi}}_{\,11},
{\omega_{\varphi}}_{\,22}) = \frac{1}{2}(\cos{\vartheta_1},\cos{\vartheta_2})$ and the periodicity of the $\phi'$ angles is $\{ -2 \pi/k,2 \pi/k \}$ as already discussed. In the limit where both $\vec x^{\,(1)}$
and $\vec x^{\,(2)}$ become simultaneously small (the regime where the two Kaluza-Klein monopoles ``intersect'' and we are near both cores) we can neglect the contribution from $U_\infty$ to obtain the ``near-core'' metric
\be
ds^2 = \sum_{i =1,2}\Big( \frac{1}{2 |\vec x^{\,(i)}|} d\vec x^{\,(i)} \cdot d\vec x^{\,(i)} + 2 |\vec x^{\,(i)}|\, (d\phi'_i + \frac{1}{2}\cos{\vartheta_i}\, d\varphi_i)^2\Big)\;.
\label{ncore}
\ee
Writing $(d\vec x^{\,(i)})^2 = d|\vec x^{\,(i)} |^2 + |\vec x^{\,(i)} |^2 (d\vartheta_i^2+\sin^2\vartheta d\varphi_i^2) $, and through the change of variables $|\vec x^{\,(i)} | = \frac{1}{2}r_i^2$, we arrive at
\be
ds^2 = \sum_{i =1,2}\Big( dr_i^2  + r_i^2  (d\phi'_i + \frac{1}{2}\cos{\vartheta_i} d\varphi_i)^2+\frac{r_i^2}{4} (d\vartheta_i^2+\sin^2{\vartheta_i}\,d\varphi_i^2)\Big)\;.
\ee
Because of the $\{ -2 \pi/k,2 \pi/k \}$ identifications on the $\phi_i'$'s, this looks like two copies of $\mathbb R^4/\mathbb Z_k$ for a  parametrisation in terms of spherical coordinates, where the orbifold acts on the three-sphere and  $S^3/\mathbb Z_k$ is realised as the Hopf fibration $S^1/\mathbb Z_k\hookrightarrow S^3/\mathbb Z_k\stackrel{\pi}{\to}S^2$.

Alternatively, we can make a final change of coordinates
\bea
&&z^{1} = r_1 \cos\frac{\vartheta_1}{2} e^{-i\phi_1' -\frac{i}{2} \varphi_1 }
\qquad z^{2} = r_2 \cos\frac{\vartheta_2}{2} e^{i\phi_2' + \frac{i}{2}\varphi_2}\nn\\[2mm]
&&z^{3} = r_1 \sin\frac{\vartheta_1}{2} e^{-i\phi_1' + \frac{i}{2}\varphi_1 }
\qquad z^{4} = r_2 \sin\frac{\vartheta_2}{2} e^{i\phi_2' - \frac{i}{2}\varphi_2}
\;,
\eea
to obtain
\be
ds^2 = \sum_{A = 1}^4 |dz^A|^2\;.
\ee
We see that the identifications $\{\phi_1',\phi_2'\} \sim \{\phi_1',\phi_2'\} + \{ -2 \pi/k, 2 \pi/k \}$ can be simply expressed as  $z^A \sim e^{2 \pi i/k} z^A$
and we have arrived at a $\mathbb C^4/\mathbb Z_k$ orbifold of M-theory.

Thus, to summarise this section, our base system defined by a brane construction is dual to M2-branes transverse to a $\mathbb C^4/\mathbb Z_k$ orbifold geometry. This orbifold has been studied previously in Refs.~\cite{Halyo:1998pn} and \cite{Morrison:1998cs} and was discussed in the context of M2-branes in Ref.~\cite{Distler:2008mk}. For general $k\ge 3$ it is known to preserve ${\cal N}=6$ supersymmetry in the bulk. This can be demonstrated using the following argument: The element of $\mathrm{Spin}(1,10)$ corresponding to the $\mathbb Z_k$ rotations is
\begin{eqnarray}
e^{\pi(\Gamma_{34}+\Gamma_{56}+\Gamma_{78}+\Gamma_{910})/k}\
\end{eqnarray}
and the preserved supersymmetries must be left invariant by the action of this group element. For generic $k$ this implies
\begin{eqnarray}
(\Gamma_{34}+\Gamma_{56}+\Gamma_{78}+\Gamma_{910})\epsilon=0\;.
\end{eqnarray}
One way to solve this is to take $\Gamma_{3456}\epsilon =\Gamma_{78910}\epsilon = \epsilon$ and another way is
$\Gamma_{3478}\epsilon =\Gamma_{56910}\epsilon =  \epsilon$.  In both cases there are 8 components of $\epsilon$ that survive.
Naively this gives us $8+8=16$  preserved supersymmetries. However we have counted twice the supersymmetries that satisfy both conditions \ie\
$\Gamma_{3456}\epsilon = \Gamma_{78910}\epsilon=\Gamma_{3478}\epsilon=\Gamma_{56910}\epsilon=\epsilon$. There are 4 of these and thus we find $12$ independent supersymmetries. Note that these projectors imply $\Gamma_{012}\epsilon = \epsilon$ and therefore placing M2-branes at the fixed-point of the orbifold does not break any more supersymmetry, as expected by the brane construction. Thus the corresponding worldvolume theory should have ${\cal N}=6$ supersymmetry, which is what we have already found. Of course the $k=1,2$ cases are special. For $k=1$ there is no orbifold and for $k=2$ the orbifold is simply ${\mathbb C}^4/{\mathbb Z}_2$. Hence in these cases the background preserves $32$ and $16$ supersymmetries respectively.

\subsubsection{From branes to the ABJM gauge theory}

Having established the connection to M2-branes, let us now return to the gauge theory discussion. Even though at the end of Section~\ref{breaneology} we recovered a theory with two Chern-Simons terms with equal but opposite levels, we still have the presence of Yang-Mills kinetic terms. This is addressed by taking the limit in which we decouple gravity from the brane worldvolume ($\alpha'\to 0$) and then looking at the theory for very low energies. Since in three-dimensions the standard Yang-Mills kinetic term is an IR-irrelevant operator,\footnote{Three-dimensional Yang-Mills is weakly coupled in the UV and strongly coupled in the IR.} it drops out in this limit. One then arrives at a conformal Chern-Simons-matter gauge theory.

On general grounds the lagrangian of this theory must be captured by Eqs.~(\ref{nsixlag})-(\ref{twistedABJM}) for some choice of structure constants $f^{ab}{}_{cd}$.
As we have already seen, it is possible to re-express the $\mathcal N=6$ 3-algebra results of Section~\ref{Neq6} in terms of a product Lie-algebra with bi-fundamental matter, by considering a triple-product $[Z^A,Z^B;\bar Z_C] = -(2 \pi/k) (Z^A\bar Z_C Z^B -Z^B\bar Z_C Z^A)$. This re-expresses the twisted Chern-Simons term as a difference of two Chern-Simons terms with the same level and gauge group $\U(n)\times \U(n)$. Thus finally we have constructed the ABJM \cite{Aharony:2008ug} theory
\begin{eqnarray}\label{niceaction}
\nonumber {\cal L} &=& -{\rm Tr}(D^\mu \bar Z_A,D_\mu Z^A) -
i{\rm Tr}(\bar\psi^A,\gamma^\mu D_\mu\psi_A) -V+{\cal L}_{CS}\\[2mm]
&& +\frac{2\pi i}{k}{\rm Tr}(\bar\psi^A \psi_{A}\bar Z_{B}
Z^B - \bar\psi^AZ^B \bar Z_{B}
\psi_{A})-\frac{4\pi i}{k}{\rm Tr}(\bar\psi^A\psi_{B}\bar Z_{A}Z^B -\bar\psi^AZ^B\bar Z_{A}\psi_{B} )\\
\nonumber &&-\frac{2\pi i}{k}\varepsilon_{ABCD}{\rm Tr}(\bar\psi^A Z^C\psi^B Z^D) +\frac{2\pi i}{k}\varepsilon^{ABCD}{\rm Tr}(\bar
\psi_{A}\bar Z_{C}\psi_B\bar Z_D)\;,
\end{eqnarray}
where the sextic scalar  potential is
\begin{eqnarray}
  V = \frac{1}{3}{\rm Tr}\left(4Z^A\bar Z_AZ^B\bar Z_CZ^C\bar Z_B  - 4Z^A\bar Z_BZ^C\bar Z_AZ^BZ_C -Z^A\bar Z_AZ^B\bar Z_BZ^C\bar Z_C-\bar Z_A Z^A\bar Z_BZ^B\bar Z_C Z^C\right)\;.
  \end{eqnarray}
The complex scalars $Z^A$ and fermions $\psi_A$ transform in the bi-fundamental of the two gauge groups and also carry an R-symmetry index $A = 1,...,4$. The covariant derivatives act accordingly\footnote{Note that, in contrast to our notation in Chapter~\ref{chapter3}, we will work in ``physics'' conventions  with hermitian generators when dealing with CS-matter actions in the Lie algebra formulation.}
\begin{equation}
D_\mu Z^A = \partial_\mu Z^A - iA^L_\mu Z^A + i Z^A A^R_\mu\;.
\end{equation}
Finally, the piece
\begin{equation}\label{CSterm}
{\cal L}_{CS} = \frac{k}{4\pi}\varepsilon^{\mu\nu\lambda}\left({\rm Tr}( A^L_\mu\partial_\nu
A^L_\lambda - \frac{2}{3} i A^L_\mu A^L_\nu  A^L_\lambda) -{\rm Tr}(
A^R_\mu\partial_\nu   A^R_\lambda - \frac{2}{3}i  A^R_\mu A^R_\nu   A^R_\lambda)\right)\ ,
\end{equation}
encodes the Chern-Simons terms for the $\U(n)\times \U(n)$ gauge fields. Note that the Chern-Simons level $k$ plays the role of a (discrete) coupling constant in this theory.

We also collect the set of supersymmetry transformations that leave the above action invariant
\bea
\delta Z^A &=&i\bar\epsilon^{AB}\Psi_{B}, \label{4}\cr
\delta\psi_{B}&=&\gamma^\mu\epsilon_{AB} D_\mu Z^A+ \frac{2\pi}{k} (Z^C\bar Z_BZ^D-Z^D\bar Z_B Z^C) \epsilon_{CD}-\frac{2\pi}{k}(Z^A\bar Z_CZ^C-Z^C\bar Z_CZ^A)\epsilon_{AB}\cr
\delta  A_\mu^{L}&=& -\frac{2\pi}{k} (\bar \epsilon_{AB}\gamma_\mu Z^B\bar\psi^{A}-\bar \epsilon^{AB}\gamma_\mu \psi_A \bar Z_B)\cr
\delta A_\mu^R&=& -\frac{2\pi}{k}(\bar \epsilon_{AB}\gamma_\mu \bar\psi^{A}Z^B-\bar \epsilon^{AB}\gamma_\mu \bar Z_B\psi_A )\label{5}\;.
\eea

\subsection{The ABJM proposal for AdS$_4$/CFT$_3$}

We are finally in a position to see how the brane construction naturally leads to the formulation of an AdS/CFT  duality \cite{Maldacena:1997re,Aharony:1999ti} as proposed by ABJM \cite{Aharony:2008ug}. The superconformal $\U(n)\times\U(n)$ CS-matter gauge theory with $\mathcal N=6$ supersymmetry has a (discrete) gauge coupling $g=1/k$. This theory is weakly coupled for large $k$. For large $n$, it also admits an 't Hooft expansion in powers of $1/n^2$. The planar diagrams have an effective 't Hooft coupling $\lambda\equiv g n = n/k$, which can be kept small when $k \gg n$. The claim is that, as in the correspondence between $\mathcal N=4$ SYM and string theory  on $\textrm{AdS}_5\times S^5$, the ABJM theory is dual to the geometry arising from the near-horizon limit of $n$ M2-branes on a $\mathbb C^4/ \mathbb Z_k$ orbifold singularity, which will turn out to be $\textrm{AdS}_4\times S^7/\mathbb Z_k$. The CS level $k$ is identified with the rank of the $\mathbb Z_k$ orbifold.

In order to see how this comes about, let us look at the near-horizon geometry of M2-branes on $\mathbb C^4/\mathbb Z_k$, which we derived in a previous section, in more detail. Let us start with $k=1$, \ie no orbifold. The near-horizon limit for the geometry generated by $n$ M2's in $\mathbb C^4$ gives rise to $\mathrm{AdS}_4\times S^7$ in the presence of $n$ units of  4-form flux \cite{Maldacena:1997re,Aharony:1999ti}. As originally pointed out in Ref.~\cite{Nilsson:1984bj}, it is convenient to think of $S^7$ as a Hopf fibration: a fibre bundle whose one-dimensional fibre is a circle, $S^1$, and whose base is
the 3-complex-dimensional manifold $\cp^3$. This is denoted in the mathematics literature as $S^1\hookrightarrow S^7\stackrel{\pi}{\rightarrow} \cp^3$. Reducing M-theory along the fibre, one recovers type IIA string theory compactified on $\textrm{AdS}_4\times \cp^3$. Notice that the fibration defines a specific direction within $S^7$ to be the M-direction along which we compactify.

Interestingly, even though the M-theory description is maximally supersymmetric, this is not what the IIA description sees.  Following the recipe of Kaluza and Klein, zero modes cannot carry any momentum in the $S^1$ direction -- in other words, they must be invariant under this U(1). Since the  circle fibration is realised in a nontrivial manner, only some of the supercharges are invariant, and as a result the IIA theory ends up with $\mathcal N=6$ or $\mathcal N=0$ supersymmetry, depending on the orientation of the $S^7$ \cite{Nilsson:1984bj} (see also \cite{Duff:1997qz,Halyo:1998pn}). Here we will choose the $\mathcal N=6$ orientation, in order to match with the gauge theory result.

Now we can implement the orbifold action $z^A\to e^{2\pi i/k} z^A$ on the above construction. It reduces the $S^7$ factor in the near-horizon geometry to $S^7/{\mathbb Z}_k$.  This action commutes with, and therefore preserves, an $\SU(4)\times\U(1)$ subgroup of the $\SO(8)$ isometry group of $S^7$. The $\U(1)$ subgroup acts as a common phase on all the $z^A$, while under $\SU(4)$ the $z^A$ transform in the fundamental representation. As one would expect, $\SU(4)\times\U(1)$ is precisely the isometry group of $S^7/{\mathbb Z}_k$ for generic $k$. In terms of the Hopf fibration, ${\mathbb Z}_k$ acts only on the $S^1$ fibre, reducing it to $S^1/{\mathbb Z}_k$ which is simply a circle $k$ times smaller in circumference than the original one. The $\U(1)$ factor of the isometry group acts as a shift along the $S^1$ fibre leaving the base unaffected, while the $\SU(4)$ factor acts purely on the $\cp^3$ base.\footnote{The isometry group of $\cp^n$ is $\SU(n+1)$.}

To summarise, the gravity dual to the Chern-Simons-matter theory at level $k$ is $\mathrm{AdS}_4\times S^7/\mathbb Z_k$, with $S^1/\mathbb Z_k\hookrightarrow S^7/\mathbb Z_k\stackrel{\pi}{\rightarrow} \cp^3$. This also suggests the existence of a parameter regime  involving large $k$, in which the gravity description should more appropriately be thought of in terms of  IIA supergravity. Since the original spacetime preserved 16 supercharges, while the orbifold for generic $k$ preserves only 12, this is in line with usual AdS/CFT intuition which dictates that performing a quotient on the compact part of the geometry will lead to reduced supersymmetry. The supercharges preserved by the orbifold action are neutral under the $\U(1)$ and hence for $k>2$ the IIA and M-theory descriptions now both have the same amount of supersymmetry.

The metric of the near-horizon M-theory geometry has the form
\be\label{Mgeometry}
ds^2 = \frac{R^2}{4} ds^2_{\textrm{AdS}_4} + R^2 ds^2_{S^7/\mathbb Z_k}
\ee
with
\be
ds^2_{S^7/\mathbb Z_k} = \frac{1}{k^2}(d \phi + k \omega)^2 + ds^2_{\cp^3}\;,
\ee
where $\phi$ has period $2\pi$, $R = (2^5 \pi^2 k n)^\frac{1}{6}$ is the radius of the sphere in Planck units, and we also have $k n$ units of 4-form flux. The 1-form $\omega$ is related to the K\"ahler 2-form $J$ of $\cp^3$ by $d\omega=J$. The Fubini-Study $\cp^3$ metric and associated K\"ahler form are given in homogeneous coordinates by
\bea
ds^2_{\cp^3} &=& \frac{1}{|z|^4}( |z|^2 dz_i \bar dz_i - \bar z_i z_j dz_id\bar z_j )\cr
J &\sim& i d \Big(\frac{z_i}{|z|}\Big) \wedge d \Big(\frac{\bar z_i}{|z|}\Big)\;,
\eea
with $|z|^2 = z_i \bar z_i$.

The IIA geometry is closely related to the above via the usual reduction formulae. The string frame metric and dilaton are given in string units by
\bea
ds_{IIA}^2 &=& \frac{R^3}{k}\Big(\frac{1}{4}ds^2_{AdS_4} + ds^2_{\cp^3} \Big)\nn\\[2mm]
e^{2 \Phi} &=& \frac{R^3}{k^3} \sim\frac{1}{n^2}\Big(\frac{n}{k}\Big)^\frac{5}{2} \ ,
\eea
with $n$ units of four-form flux on $\mathrm{AdS_4}$ and $k$ units of two-form flux on a $\cp^1\subset\cp^3$.

Having established the geometry, let us see in which parameter regime each of the descriptions is valid. The most conservative statement of AdS/CFT would be that the planar sector of the large-$n$ ABJM theory should be dual to supergravity on $\textrm{AdS}_4\times S^7/\mathbb Z_k$. In principle, we expect  the supergravity description to be good when the 't Hooft coupling is large, $\lambda\gg 1$, or $k\ll n$. But when is the IIA description more appropriate than the M-theory one? For that one needs to have that the size of the circle in \eqref{Mgeometry} be  small, \ie $R\sim ( kn)^\frac{1}{6}\ll k$ or $k^5\gg n$. Hence, in the strong 't Hooft coupling regime there are {\it two} supergravity descriptions of the theory depending on whether $k^5\gg n$ (IIA) or  $k^5\ll n$ (11d SUGRA). This establishes all the theories involved in the conjecture and their respective regimes of validity.

The $\mathrm{AdS}_4/ \mathrm{CFT}_3$ duality proposal has passed a great number of nontrivial of tests. For example, the spectrum of supergravity fields is in complete agreement with the spectrum of chiral primary operators from the gauge theory side \cite{Klebanov:2008vq}. This crucially requires the inclusion of monopole operators, which are to be discussed in detail in Section~\ref{thooft}, after identifying the $\U(1)$ of the circle direction with $\U(1)_B$ of the gauge theory, \eqref{covder}. Moreover, at large $n$ the full superconformal index of the $\mathcal N=6$ theory exactly agrees with the index over supersymmetric gravitons in $\mathrm{AdS}_4\times S^7/\mathbb Z_k$ \cite{Bhattacharya:2008bja,Kim:2009wb}. Details of the various tests and successes of the $\mathrm{AdS}_4/ \mathrm{CFT}_3$ correspondence, as well as a discussions of properties like integrability, are beyond the scope of this work and the interested reader may wish to consult the reviews \cite{Klebanov:2009sg,Beisert:2010jr,Klose:2010ki} and references therein.

In summary, the planar sector of the  ABJM theory (valid for $n\ll k$, when the theory is weakly coupled) is dual to type IIA supergravity on $\mathrm{AdS}_4 \times \cp^3$ (valid for $k\ll n\ll k^5$) or eleven-dimensional supergravity on $\textrm{AdS}_4\times S^7/\mathbb Z_k$ (valid for $n\gg k^5$).

It is worth remarking that the ABJM theory  reproduces the expected scaling for the number of degrees of freedom for $n$ M2-branes: When the gauge theory is put on a thermal circle, the dual description is in terms of an AdS-black-hole geometry. The free energy of this black hole was estimated to scale as $n^{\frac{3}{2}}$ \cite{Klebanov:1996un} and this is easily extended to the $\mathbb Z_k$ orbifold case, for which it has been argued in Ref.~\cite{Aharony:2008ug} that the scaling with $n,k$ is $\sim n^{\frac{3}{2}}\sqrt{k}$. This formula nicely reconciles the $n^{\frac{3}{2}}$ scaling at fixed $k$ (in particular $k=1$) that was predicted long ago in \cite{Klebanov:1996un}, and the $n^2$ scaling for the free energy of a large-$n$ gauge theory at fixed 't Hooft coupling $n/k$ which requires that $k$ scale like $n$.

These results are expected to be recovered by the statistical entropy of the massless modes on the worldvolume theory. Recent results on localisation \cite{Pestun:2007rz} have allowed for this to be explicitly verified -- albeit for the free energy of the Euclideanised ABJM model \cite{Kapustin:2009kz,Jafferis:2010un} on $S^3$ (rather than $S^2\times S^1$). This can be reduced to a matrix model that has a strong coupling expansion, the leading term of which beautifully reproduces both the $n^{\frac{3}{2}}$ scaling behaviour at fixed $k$ and the numerical coefficient of the gravity calculation \cite{Drukker:2010nc}. For more details of this fascinating direction, the reader may consult Ref.~\cite{Marino:2011nm}.  Other multiple M2-brane and ABJM-related literature includes \cite{GarciadelMoral:2009uf,delMoral:2009pb, SheikhJabbari:2008wq,AliAkbari:2008rm,Buchbinder:2008vi,Ivanov:2009nr,Buchbinder:2009dc}.

\subsection{ABJ and discrete torsion}

There exists a generalisation, due to Aharony, Bergman and Jafferis (ABJ) \cite{Aharony:2008gk}, of the ABJM model to the case where the matter fields are $m\times n$ complex matrices in the bi-fundamental of $\U(m)\times \U(n)$. In this case $A^L$ and $A^R$ are $\U(m)$ and $\U(n)$ gauge fields. The form of the action is unchanged from the $\U(n)\times \U(n)$ case.\footnote{Indeed, one can also consider the ABJM model with $\SU(n)\times \SU(n)$ gauge groups. The coupling to bi-fundamental matter makes this a rather different theory with no generic spacetime interpretation. However, we will see in Section~\ref{relationton8} that for $n=2$ these models are simply the $\mathcal N=8$ theories and in some cases are dual to the M2-brane gauge theories.} Let us assume without loss of generality that $n<m$ and write $m=n+l$. Indeed, the action follows directly from the ${\cal N}=6$ 3-algebra theory.

To understand these theories we return to the brane construction given above. We can easily generalise it to the case where the initial D3 segments involve different numbers of branes $n$ and $m$, leading to $\U(m)\times\U(n)$ Chern-Simons-matter theories which describe M-theory configurations with $l=m-n$ units of discrete torsion for the background 4-form. For $m=n+l$ this has the interpretation of $n$ M2-branes along with $l$ fractional M2's stuck on the $\mathbb C^4/ \mathbb Z_k$ orbifold singularity. Starting from the same $\U(m)\times\U(n)$ configuration one can also place an $\mathrm{O3}$ orientifold plane parallel to the D3's, resulting in $\mathrm{O}(2m)\times\Sp(n) $ and $\mathrm{O}(2m+1)\times\Sp(n) $ theories corresponding to M2-branes on a $\mathbb C^4/\hat{ \mathbb D}_k$ singularity, where $\hat{ \mathbb D}_k$ is the binary dihedral group of order $4k$ \cite{Sethi:1998zk,Berkooz:1998sn,Aharony:2008gk}.

Since the discrete torsion $l$ is only defined modulo $k$, we see that for these models to match the supergravity we must make two conjectures:\footnote{Here the subscripts refer to the level of the corresponding Chern-Simons term in the lagrangian.}
\begin{itemize}
\item $\U(n+l)_k\times \U(n)_{-k}$ has no supersymmetric vacuum if $l>k$
\end{itemize}
and
\begin{itemize}
\item $\U(n+l)_k\times \U(n)_{-k}$ is dual to  $\U(n)_k\times \U(n+k-l)_{-k}$\;.
\end{itemize}
Note that both these conjectures are at strong coupling since $m/k= (n+l)/k$ cannot be made small for the models under consideration. We should also mention that in these models, the parity symmetry is typically broken as it maps $k\leftrightarrow -k$.

In the rest of this review, we will largely just concentrate on the ABJM models. However almost all of our discussion also applies to the ABJ models. But there are also interesting and subtle effects that arise in the ABJ models (\eg see \cite{Bak:2008vd,Bergman:2009zh,Kluson:2009tz,Caputa:2009ug,Evslin:2009pk,Bianchi:2011fc}).


\section[Analysis of the theory I: basics]{\Large{\bf {Analysis of  the theory I: basics}}}\label{chapter5}

We have  argued  that the ABJM theory encodes the dynamics of multiple M2-branes probing a $\mathbb C^4/\mathbb Z_k$ singularity. We  proceed to find gauge-theoretic evidence for this by analysing the vacuum moduli space of the theory. We next look at the theory expanded around a particular point in the moduli space, obtained by allowing one of the scalars to develop a large vacuum expectation value. This will lead, via a novel Higgs mechanism, to the theory being recast in terms of three-dimensional super Yang-Mills after the scalar gets eaten by the Chern-Simons gauge field. At the end of this chapter we clarify the relationship between the $\mathcal N = 6$ ABJM and $\mathcal N=8$ BLG theories.

\subsection{Vacuum moduli space}\label{vacmodsp}

The vacuum moduli space of a gauge theory is the space of vacua of the theory modulo gauge transformations. For D-branes in string theory, this is the space of vacua of supersymmetric Yang-Mills theory in the appropriate space-time dimension. Consider for example the theory on $n$ D3-branes, which is ${\cal N}=4$ supersymmetric Yang-Mills theory in 3+1d with gauge group $\U(n)$. This theory has six scalar fields $\Phi^i, i=1,\cdots,6$, all in the adjoint representation. The classical potential is $-\Tr [\Phi^i,\Phi^j]^2$ and is minimised by having all the scalars be diagonal matrices
\be
\Phi^i = {\rm diag}(x^i_1,x^i_2,\cdots x^i_n)\;.
\ee
Therefore the moduli space is naively $({\mathbb R}^6)^n$, but we must remember that the Weyl group ${\mathbb S}_n$ of $\U(n)$ permutes the eigenvalues. Quotienting by it, one finds the moduli space of D3-branes in flat space-time to be
\be
({\mathbb R}^6)^n/{\mathbb S}_n\equiv {\rm Sym}_n({\mathbb R}^6)\;.
\ee
This has a simple physical interpretation as the space of $n$ indistinguishable D3-branes, each one free to move in 6 transverse spatial dimensions. Because of the high degree of supersymmetry, this space does not receive quantum corrections.

One would like to understand the corresponding vacuum moduli space in the ABJM theory, which would provide a crucial test of the claim that it describes M2-branes. As we have seen, the structure of ABJM field theory is quite different from Yang-Mills, and moreover there is an additional ingredient: a ${\mathbb Z}_k$ orbifold. Therefore the moduli space needs to be computed and compared with that of $n$ indistinguishable M2-branes on the transverse space
$\mathbb C^4/ \mathbb Z_k$.

The vacuum moduli space was initially studied for BLG theory in \cite{VanRaamsdonk:2008ft,Lambert:2008et,Distler:2008mk} and these methods were then applied in \cite{Aharony:2008ug} to study ABJM theory. We review these developments below, in this order.

\subsubsection{Moduli space for BLG theory}\label{BLGmodulispace}

We start by reviewing the results of Refs.~\cite{Lambert:2008et,Distler:2008mk} on the moduli space of the ${\cal N}=8$ theories. Subsequent to these works, it was realised \cite{Lambert:2010ji} that there are actually two infinite families of such theories. Both have $k\in \mathbb Z$ but one has gauge group $\SU(2)\times \SU(2)$ while the other has gauge group $(\SU(2)\times \SU(2))/{\mathbb Z}_2$. Here we wish to consider both, as they share the same $\su(2)\oplus \su(2)$ lagrangian, so we present an updated version of the original analysis.

From the BLG sextic potential it is easy to see that the vacuum moduli space consists of 8 real scalars that are diagonal $2\times 2$ matrices. We combine them into four complex scalars $Z^A$ and write them as
\begin{equation}
Z^A = {\rm diag}(z^A_1, z^A_2)\ .
\end{equation}
Gauge transformations take $Z^A \rightarrow g_L Z^A g_R^\dagger$.  There are two such transformations that keep $Z^A$ diagonal.  The first is a discrete transformation, which up to conjugacy may be taken to be $g_L=g_R=i\sigma_1$.  This identifies the configurations
\be\label{z2}
g_{12}:\quad {\rm diag}(z^A_1, z^A_2) \ \cong\  {\rm diag}(z^A_2, z^A_1)  \;
\ee
and results in a $\mathbb Z_2$ quotient of the moduli space.

The second is a continuous $\U(1)$ gauge symmetry, with $g_L = g_R^\dagger = e^{\frac{i}{2}\theta_B\sigma_3}$. For fields in the moduli space, the $\su(2)\times \su(2)$ lagrangian reduces to
\begin{eqnarray}
{\cal L} = -D_\mu z_1^A D^\mu z_{1A} - D_\mu z_2^A D^\mu z_{2A} +\frac{k}{2\pi}\varepsilon^{\mu\nu\lambda}B_\mu\partial_\nu Q_\lambda  \;,
\end{eqnarray}
where $B_\mu = A^{3L}_\mu-A^{3R}_\mu$ gauges the $\U(1)$ symmetry, $Q_\mu = A^{3L}_\mu+A^{3R}_\mu$, and $D_\mu z^A_{1,2} = \partial_\mu z^A_{1,2} \mp iB_\mu z^A_{1,2}$. Note the  factor of $2$ in the Chern-Simons term that arises from taking the trace over $2\times 2$ matrices.

At this point we eliminate $Q_\mu$ in favour of its field strength $H_{\mu\nu} = \partial_\mu Q_\nu -\partial_\nu Q_\mu$.  We then treat $H_{\mu\nu}$ as an independent field, subject to the Bianchi identity $\epsilon^{\mu\nu\lambda}\partial_\mu H_{\nu\lambda} = 0$, which we impose via the Lagrange multiplier $\sigma$:
\be
{\cal L} = -D_\mu z_1^A D^\mu z_{1A} - D_\mu z_2^A D^\mu z_{2A} +\frac{k}{4\pi}\varepsilon^{\mu\nu\lambda}B_\mu H_{\nu\lambda} + \frac{1}{4\pi}\varepsilon^{\mu\nu\lambda}\sigma\, \partial_\mu H_{\nu\lambda}  \;.
\label{sigmult}
\ee
Integrating out $H_{\nu\lambda}$ leads to the identification $B_\mu=\del_\mu\sigma/k$. The Lagrange multiplier $\sigma$ can now be absorbed in the $z^A$ by the redefinition
\begin{eqnarray}
w^A_1 = e^{i\sigma/k}z^A_1\ ,\qquad w^A_2 = e^{-i\sigma/k}z^A_2\ ,
\end{eqnarray}
yielding the lagrangian
\begin{eqnarray}
{\cal L} = -\partial_\mu w_1^A \partial^\mu w_{1A} - \partial_\mu w_2^A \partial^\mu w_{2A} \ .
\end{eqnarray}

We will now show that the field $\sigma$ is periodic. This arises from the fact that the fluxes of $F_{\mu\nu}^{L/R}$, the field strengths of $A_\mu^{L/R}$,  satisfy the Dirac quantisation condition. To see this, consider  some field $\Psi$ that couples to a $\U(1)$ field $A_\mu$ through $D_\mu\Psi =\partial_\mu \Psi - iA_\mu \Psi$. Let us now carry out parallel transport of $\Psi$ over a closed path $\gamma$. The resulting field $\Psi_\gamma$ is related to the initial one $\Psi_0$ by a $\U(1)$ transformation
\begin{equation}
\Psi_\gamma = e^{i\oint_\gamma A}\Psi_0\;.
\end{equation}
Now using Stokes' theorem we have $\int_\gamma A = \int_D F$
where $D$ is a two-dimensional surface whose boundary is $\gamma$. Hence we may write
\be
\Psi_\gamma =e^{i\int_D F}\Psi_0\;.
\ee
However the choice of $D$ is not unique.  Given any two such choices $D$ and $D'$ we require that the phase, viewed as an element of the gauge group $\U(1)$, is the same. This implies that
\begin{equation}
e^{i\int_{D-D'} F}=1
\end{equation}
and hence $\int_\Sigma F \in 2\pi \mathbb{Z} $, where $\Sigma=D-D'$ is any closed surface.

Applying this to the field $H$, we have the quantisation condition
\be
\frac{1}{4\pi}\int_\Sigma H\in {\mathbb{Z}}\;,
\ee
where the extra factor of 2 comes from the fact that $H$ is the sum of two independent field strengths. This holds in the $\SU(2)\times \SU(2)$ case, while in the $(\SU(2)\times \SU(2))/{\mathbb Z}_2$ case the phase above must be equal to 1 only up to a ${\mathbb Z}_2$ action so the right hand side can be a half-integer or integer, which we denote by $\half {\mathbb Z}$.

Converting the integral of $H$ over a surface into an integral of $dH$ (in components, $\half \varepsilon^{\mu\nu\lambda}\partial_\mu H_{\nu\lambda}$) over the entire 3-volume,\footnote{For this manipulation it is best to temporarily continue to Euclidean 3-space.} we find
in the $\SU(2)\times \SU(2)$ case that
\begin{eqnarray}
\frac{1}{4\pi}\int \varepsilon^{\mu\nu\lambda}\partial_\mu H_{\nu\lambda} \in  2{\mathbb Z}\;,
\end{eqnarray}
whereas for $(\SU(2)\times \SU(2))/{\mathbb Z}_2$ we have
\begin{eqnarray}
 \frac{1}{4\pi}\int \varepsilon^{\mu\nu\lambda}\partial_\mu H_{\nu\lambda} \in  {\mathbb Z}\ .
\end{eqnarray}
From the lagrangian \eref{sigmult} it follows that $\sigma$ must have the periodicity $\sigma \sim \sigma +  \pi$ for $\SU(2)\times \SU(2)$ and $\sigma \sim \sigma + 2\pi$ for $(\SU(2)\times \SU(2))/{\mathbb Z}_2$.

The result is the identification
\begin{eqnarray}
g_{\SU(2)} = \left\{  \begin{array}{ccc}
z^A_1  \cong e^{ \pi i/k}z^A_1& z^A_2\cong e^{- \pi i/k}z^A_2 & \SU(2)\times \SU(2)\\
z^A_1 \cong  e^{2\pi i/k} z^A_1& z^A_2 \cong e^{- 2\pi i/k} z^A_2 & (\SU(2)\times \SU(2))/{\mathbb Z}_2\\
                      \end{array}\right.\;,
\end{eqnarray}
which corresponds to a $\mathbb Z_{2k}$ or  $\mathbb Z_{k}$ quotient of the moduli space respectively.

The identifications $g_{12}$ from \eqref{z2} and $g_{\SU(2)}$ above do not commute; they generate the dihedral group\footnote{The dihedral group of order $2m$ is given by $\mathbb D_{2m} =\mathbb  Z_2\ltimes \mathbb Z_m$.}  $\mathbb D_{4k}$ for $\SU(2)\times \SU(2)$ or $\mathbb D_{2k}$ for $(\SU(2)\times \SU(2))/{\mathbb Z}_2$. It follows that the moduli spaces of the two theories are:
\bea\label{modspace}
&&({\mathbb C}^4\times{\mathbb C}^4)/{\mathbb  D_{4k}}\quad{\rm for}\quad \SU(2)\times \SU(2)\nn\\
&&({\mathbb C}^4\times {\mathbb C}^4)/{\mathbb D_{2k}}\quad{\rm for}\quad (\SU(2)\times \SU(2))/{\mathbb Z}_2.
\eea
In general, these moduli spaces do not have an obvious space-time interpretation in terms of M2-branes. However, we will see in Section~\ref{connectingtoBJG} that such an interpretation can indeed be provided for the special values $k=1,2$ and 4.

\subsubsection{Moduli space for ABJM theory}

We now move on to consider the vacuum moduli space of ABJM theory. Here we must consider the minima of the potential \eref{upsquare}. Because the potential is a perfect square, the scalars must satisfy $\Upsilon^{CD}_B=0$. Contracting over $B$ and $D$ implies that $[Z^D,Z^C;\bar Z_D]=0$ and substituting back into $\Upsilon^{CD}_B=0$ shows that $[Z^C,Z^D;Z_B]=0$ or
\begin{equation}
Z^C Z_BZ^D - Z^D Z_BZ^C=0 \;,
\end{equation}
for all $B,C,D$. Clearly this is solved if all the $Z^A$ commute. Hence by a gauge transformation we can assume that
\begin{equation}
Z^A = {\rm diag}(z_1^A,...,z_n^A)\ .
\end{equation}
To see that this is the generic solution one can compute the mass matrix for the off-diagonal components and see that it is positive definite for generic vacua. However, as is familiar from D-brane theories, there are special points in the moduli space with enhanced gauge symmetry and extra massless states.

We must now quotient by the surviving gauge symmetries. In addition, unlike the case of D-branes, the vacuum is also invariant under continuous transformations generated by the $\U(1)^n$ Cartan subalgebra.
These gauge transformations are trivial in the adjoint representation but not in the bi-fundamental representation. In particular, each $z_i^A$ couples to a $\U(1)$ gauge field  $B_\mu^i = A^{Li}_\mu-A^{Ri}_\mu$ obtained from the diagonal components of the gauge fields:
\begin{equation}
A^L_\mu = {\rm diag}(A_\mu^{L1},...,A_\mu^{Ln}),\qquad A^R_\mu = {\rm diag}(A_\mu^{R1},...,A_\mu^{Rn})\ .
\end{equation}
The lagrangian for the vacuum moduli becomes
\begin{equation}
{\cal L} = -\frac{1}{2}\sum_{i=1}^n D_\mu z^A_i D^\mu z^i_A + \frac{k}{4\pi}\varepsilon^{\mu\nu\lambda}\sum_{i=1}^nB^i_{\mu}\partial_\nu Q^i_{\lambda}
\end{equation}
where $Q^i = A^{Li}_\mu +A^{Ri}_\mu $ and $D_\mu z_i^A = \partial_\mu z^A_i - iB^i_{\mu }z^A_i$ (no sum on $i$).
Note that the last term of the lagrangian above has an extra factor of $\half$ compared to the analogous term in BLG theory.

Now, just as we did for that case, we introduce 2-forms $H_i$ and Lagrange multipliers $\sigma_i$ which imply the Bianchi identities $dH_i=0$, from which locally $H_i$ can be written as $dQ_i$ for a set of 1-forms $Q_i$. Then the above lagrangian is equivalent to
\begin{equation}\label{LdH}
{\cal L} = -\frac{1}{2}\sum_i D_\mu z^A_i D^\mu z^i_A + \frac{k}{8\pi}\varepsilon^{\mu\nu\lambda}\sum_iB_{i\mu}H_{i\nu\lambda}
+\frac{1}{8\pi}\varepsilon^{\mu\nu\lambda}\sum_i \sigma_i \,\partial_\mu H_{i\nu\lambda}\;.
\end{equation}

In this lagrangian, the $H_i$ are independent fields. We can integrate them out, after performing an integration by parts in the last term, to find $B_i  = d\sigma_i/k$. Finally, $\sigma_i$ can be eliminated by defining the fields $w^A_i = e^{-i\sigma_i/k}z^A_i$ whereupon the action becomes
\begin{equation}
{\cal L} = -\frac{1}{2}\sum_i \partial_\mu w^A_i \partial^\mu w^i_A\;,
\end{equation}
Under the gauge transformation $B_i\to B_i + d\theta_i$, $z_i^A\to e^{i\theta_i}z_i^A$ we have $\sigma_i\to \sigma_i +k \theta_i$. Thus $w^A_i$ are gauge invariant coordinates on the moduli space.

As before, the field $H$ satisfies the condition
\begin{equation}
\frac{1}{8\pi}\varepsilon^{\mu\nu\lambda}\int  \partial_\mu H _{i\nu\lambda} \in {\mathbb Z}\ ,
\end{equation}
where the integral is over 3-space. It follows that in the lagrangian \eref{LdH}, the contribution of the last term to the path integral is periodic under shifts of $\sigma_i$ by $2\pi n$ for any integer $n$, or in other words we must identify $\sigma_i\sim \sigma_i + 2\pi$.

Returning to our gauge invariant variables $w_i^A$ we see that they are subject to the identification
\be
w_i^A\cong e^{2\pi i/k}w_i^A\;.
\ee
Each $w^i$ parametrises not ${\mathbb C}^4$ but rather the orbifold space ${\mathbb C}^4/{\mathbb Z}_k$.
The collection of all the $n$ $w^i$'s then naively parametrises the product space $\left({\mathbb C}^4/{\mathbb Z}_k\right)^n$. However at this point we again recall that we must quotient by the Weyl group, which is the symmetric group ${\mathbb S}_n$ and permutes the $n$ copies in the product. As a result the moduli space of ABJM theory is
\begin{equation}
{\cal M}_k = \left({\mathbb C}^4/{\mathbb Z}_k\right)^n/{\mathbb S}_n\equiv
{\rm Sym}_n\left({\mathbb C}^4/ {\mathbb Z}_k\right)\;.
\end{equation}
This has precisely the right form to be physically interpreted as the moduli space of $n$ indistinguishable M2-branes moving in a ${\mathbb C}^4/ {\mathbb Z}_k$ transverse space.

\subsection{A novel Higgs mechanism}\label{Higgsneq8}

Let us re-examine the BLG theory, namely the $\mathcal N=8$, $\frak{su}(2)\oplus \frak{su}(2)$ theory of Section~\ref{Neq8}. While we have not yet provided a definitive physical interpretation for it, it will be argued in Section~\ref{connectingtoBJG} that for the special values $k=1,2,4$ it describes a pair of M2-branes in ${\mathbb R}^8$ or ${\mathbb R}^8/{\mathbb Z}_2$. We now show \cite{Mukhi:2008ux} that upon giving a VEV to a scalar, it can be rewritten as maximally supersymmetric $\U(2)$ Yang-Mills theory in 2+1d  with infinitely many corrections. In the process the pair of non-propagating Chern-Simons fields of BLG theory ``eat up'' a scalar and give rise to a single {\em massless} propagating vector field. Thus on the Coulomb branch, BLG theory has a propagating Yang-Mills field. This provides a key relation between its 3-algebra structure and the more familiar Lie algebra structure of Yang-Mills theories. The above considerations will then be extended to the ABJM case, where some new features arise. The novel Higgs mechanism provides a useful check of these theories and tests detailed features including the somewhat baroque Chern-Simons structure.

\subsubsection{A simplified version}

We first present a simple example of the novel Higgs mechanism that does not involve supersymmetry or 3-algebras. It is a property of a certain class of Chern-Simons theories in 2+1d, particularly those with difference-type actions.

Consider the $\SU(N)_k\times \SU(N)_{-k}$ Chern-Simons theory
\be
L_{CS}=\frac{k}{4\pi}\,\Tr\left( A\wedge d A
+ \sfrac23  A\wedge  A\wedge  A -
{\tilde A}\wedge d{\tilde A} -
\sfrac23 {\tilde A}\wedge {\tilde A}\wedge {\tilde A}\right)\;,
\label{diffcs}
\ee
where $A=A^aT^a$ and $\Tr T^a T^b=-\half \delta^{ab}$.
We would like to induce a particular type of mass matrix via the Higgs mechanism. Such a term arises by choosing a Higgs field $\Phi$ in the {\em bi-fundamental} representation, for example the $(N,{\bar N})$ of $\SU(N)\times\SU(N)$, which transforms as
\be
\delta \Phi = -\Lambda \Phi + \Phi{\tilde \Lambda}\,.
\ee
The covariant derivative on the Higgs field is then
\be
D_\mu\Phi = \del_\mu \Phi +A_\mu\Phi - \Phi \tA_\mu\;.
\ee
For convenience we normalise the scalar kinetic term as
\be
\frac{k}{4\pi}\Tr (D_\mu\Phi^\dagger D^\mu\Phi)
\ee
where this trace is, formally, unrelated to that in the gauge field
action -- here it just sums over two pairs of repeated indices in the
fundamental representation, one pair being associated to each factor
of $\SU(N)\times \SU(N)$. This kinetic term gives rise to the
interaction:
\be
\frac{k}{4\pi}\Tr\,\Big|A_\mu\Phi - \Phi \tA_\mu\Big|^2\;.
\ee
With a Higgs VEV proportional to the identity, $\langle \Phi\rangle =
v\one$, the mass term is equal to
\be
\frac{k}{4\pi}v^2\Tr(A_\mu-\tA_\mu)^2\;,
\ee
where now the trace is over the Lie algebra of $\SU(N)$ after
identifying the two factors in $\SU(N)\times \SU(N)$.

It is convenient to go to a different basis of gauge fields by taking the linear combinations
\be
 B = \sfrac{1}{2}(A- {\tilde A}),\quad  C = \sfrac{1}{2}
(A + {\tilde A})\;.
\ee
In these variables, and with the mass term, the lagrangian is
\be
{\cal L}=\frac{k}{\pi} \Tr\left(B\wedge F^{(C)} + \sfrac{1}{3}B\wedge B\wedge
  B - v^2 B\wedge {}^*\! B
\right)\;,
\label{cublag}
\ee
where $F^{(C)}= d C +  C\wedge  C$ is the standard non-abelian field strength for the vector field $C_\mu$.
From this follows the equation of motion for $B$
\be
F^{(C)} + B\wedge B - 2 v^2\,{}^*\! B=0\;.
\ee
We see that the field $B$ is algebraic. However, because of the quadratic term, it cannot be eliminated in a straightforward fashion. Instead one can solve the above equation recursively, to get:
\be
\begin{split}
B&= -\frac{1}{2v^2}{}^*\! F^{(C)} -\frac{1}{2v^2}{}^*\! (B\wedge B)\\
&= -\frac{1}{2v^2}{}^*\! F^{(C)} -\frac{1}{8v^6}{}^*\!
({}^*\! F^{(C)}\wedge {}^*\! F^{(C)})
  + \cdots \;.
\end{split}
\label{recurs}
\ee
The terms in the ellipsis above contain all powers of $F^{(C)}$ and the orders in this expansion are counted by the parameter $1/v^2$.

We may now insert \eref{recurs} back into the lagrangian of
\eref{cublag} to find:
\be
{\cal L} = \frac{k}{\pi}\left(-\frac{1}{4v^2}F^{(C)}\wedge{}^*\! F^{(C)}
-\frac{1}{24v^6}{}^*\! F^{(C)}\wedge{}^*\! F^{(C)}\wedge{}^*\! F^{(C)}
+\cdots
\right)\;.
\ee
In this process, a pair of non-propagating Chern-Simons gauge fields have been replaced by a single {\em propagating, massless} Yang-Mills type gauge field. Its single polarisation was gained by ``eating" a component of the Higgs field. This is the novel Higgs mechanism \cite{Mukhi:2008ux}. The Yang-Mills coupling constant is $\sqrt{4\pi v^2/k}$.

Note, however, that there are still higher-order terms in $F^{(C)}$. Taking $v\to\infty$ allows us to ignore them, but then the Yang-Mills term becomes very strongly coupled. This can be avoided by simultaneously scaling $k\to\infty,v\to\infty$ keeping $k/v^2$ fixed \cite{Distler:2008mk}. In this latter limit the higher-order terms do drop out and the Yang-Mills coupling $\sim v/\sqrt{k}$ remains finite and can be chosen arbitrarily.

We now continue to describe the novel Higgs mechanism in BLG and ABJM theories where we will encounter both some subtleties -- and a nice physical interpretation for the effect.

\subsubsection{The Higgs mechanism for BLG theory}

Recall that the BLG lagrangian is
\begin{multline}
  \nn {\cal L} = -\half D_\mu X^{aI}D^\mu X_a^{I} + \frac{i}{2}\bar{\Psi}^a\Gamma^\mu D_\mu \Psi_a   +\frac{i}{4} f_{abcd} \bar{\Psi}^b \Gamma^{IJ}X^{cI}X^{dJ}\Psi^a \\ -\frac{1}{12}   \left(f_{abcd}X^{aI}X^{bJ} X^{cK}\right)
  \left(f_{efg}^{\phantom{efg}d}X^{eI}X^{fJ}X^{gK} \right)\\
  +\half\,\vep^{\mu\nu\lambda}\left( f_{abcd} A_\mu^{~ab}\del_\nu  A_{\lambda}^{~cd} + \frac23     f_{aef}^{~~~~~g}\,f_{bcdg}\, A_{\mu}^{~ab} A_{\nu}^{~cd} A_{\lambda}^{~ef}\right)
\end{multline}
where
\begin{equation}
\nn D_\mu X^{aI} = \del_\mu X^{aI} + f^{a}_{~~bcd}A_\mu^{cd}X^{bI}\;.
\end{equation}
The structure constants are given by the 4-index totally anti-symmetric symbol $f^{abcd} = f \veps^{abcd}$, with $a,b,c,d\in \{1,2,3,4\}$, and with the Chern-Simons coefficient quantised as $f= 2\pi/k$, where $k \in {\mathbb   Z}$. We also fix the Chern-Simons level to the value $k=1$ for the remainder of this section.

Consider the situation in which one of the transverse dimensions, say $X^{a(8)}$, develops a VEV. Because of $\SO(4)$ invariance it is possible to rotate the scalar field that gets a VEV to have only the component $X^{4(8)}$. Thus the four indices split into $a\in\{1,2,3\}$ plus $4$, and this amounts to considering $\langle X^{4(8)}\rangle = v$. Note that $\langle X^{4(8)}\rangle$ preserves supersymmetry as long as no other field has a VEV. To see this, consider the fermion supervariation \eqref{susygauged2}. The first term on the RHS is zero because the scalar VEV is constant while the gauge field VEV is zero. The second term vanishes because $X^{4(8)}$ can occur at most once and the other two scalar fields have vanishing VEV. Therefore the theory expanded about this scalar VEV has maximal supersymmetry.

Now let us examine the various terms in the lagrangian and show how
they reproduce the desired $\U(2)$ SYM theory. To begin with, consider
the sextic potential. Introduce the labels $A,B,C\in \{1,2,3\}$ as
well as $i,j,k\in \{1,2,...,7\}$. Then the potential is
\begin{equation}
\begin{split}
V(X) &=\frac{1}{12} \sum_{I,J,K=1}^8
\left(\vep_{abcd} \vep_{efg}^{\phantom{efg}d}X^{aI}X^{bJ}X^{cK}
  X^{eI}X^{fJ} X^{gK}\right)\\
&=
\frac{1}{2} \sum_{i<j}^7
\left(\vep_{abcd} \vep_{efg}^{\phantom{efg}d}X^{ai}X^{bj}X^{c(8)}
  X^{ei}X^{fj} X^{g(8)}\right)\\
&~~~+
\frac{1}{2} \sum_{i<j<k}^7
\left(\vep_{abcd} \vep_{efg}^{\phantom{efg}d}X^{ai}X^{bj}X^{ck}
  X^{ei}X^{fj} X^{gk}\right)
\\
&= \frac{1}{2}\, v^2 \sum_{i<j}^7
\left(\vep_{AB4 D}\vep_{EF4}^{\phantom{EF4}D}
X^{Ai}X^{Bj}X^{Ei}X^{Fj}\right)
+ v \;{\cal O}\left(X^5\right) + {\cal O}\left(X^6\right)\;.
\end{split}
\end{equation}
In the last line we have inserted the VEV $\langle
X^{4(8)}\rangle= v$, which leads to a term quartic in the
remaining $X$'s. Note that in this term, only $X^{Ai}$ appear where
$A\in \{1,2,3\}$ and $i\in \{1,2,...,7\}$. The terms of order $v
\mathcal O (X^5)$ and $\mathcal O(X^6)$ have not been written
explicitly because they decouple in the limit $v\to\infty$, which we will eventually take.

Using $\vep_{ABD4}\equiv\vep_{ABD}$ where the latter is the 3-index
totally anti-symmetric symbol and structure constant of an $\mathfrak{su}(2)$ Lie
algebra, we see that the quartic term becomes
\begin{equation}
\frac{1}{2}\, v^2 \sum_{i<j=1}^7
\left(\vep_{ABC}\vep_{EF}^{\phantom{EF}C}
X^{Ai}X^{Bj}X^{Ei}X^{Fj}\right)\;,
\end{equation}
which is precisely the quartic scalar interaction of maximally
supersymmetric $\SU(2)$ Yang-Mills theory in 2+1d.

Following the same procedure, it is easy to check that the 2-fermion,
2-scalar coupling reduces to the Yukawa coupling of 2+1d Yang-Mills, plus terms with two fermions and two scalars
\begin{equation}
\frac{i}{4} \vep_{abcd} \bar{\Psi}^b \Gamma^{IJ}X^{cI}X^{dJ}\Psi^a
= \frac{i}{2} v\,\vep_{ABC} \bar{\Psi}^B \Gamma_{i}X^{Ci}\Psi^A
+ {\cal O}\left(X^2\Psi^2\right)\;.
\end{equation}
We see that the only scalars and fermions appearing in the first term
(which will be the leading term in the limit of large VEV) are
$\Psi^A$ and $X^{Ai}$.

Since kinetic terms are unaffected by a scalar VEV, it only remains to
understand the gauge field terms including couplings of gauge fields
through covariant derivatives. On the face of it this should be the
major stumbling block, for the gauge field in the 3-algebra theory
only has Chern-Simons couplings while the D2-brane Yang-Mills theory
requires a dynamical gauge field.  As we are committed to make no additional
assumptions to account for the dynamical gauge field, we simply
work out the full content of the theory in the presence of the VEV of
the scalar field $X^{4(8)}$. As before, we will see that the Higgs mechanism
and the original Chern-Simons coupling conspire to
provide the desired dynamical gauge field with all the right
properties.

In view of our split of indices $a,b\in \{1,2,3,4\}$ into $A,B\in
\{1,2,3\}$ and $4$, it is natural to break up the gauge field
$A_\mu^{ab}$ into two parts
\be
A_\mu^{~A 4}\equiv A_\mu^{~A}\qquad \text{and}\qquad \half\vep^A_{~BC} A_\mu^{~BC}\equiv B_\mu^{~A}\;.
\ee
Each of these is a triplet of vector fields. We can now re-write the
two terms in the Chern-Simons action as follows
\begin{equation}
\begin{split}
\half\,\epsilon^{\mu\nu\lambda}
\vep_{abcd}A_\mu^{~ab}\del_\nu A_{\lambda}^{~cd}
&= 2\,\epsilon^{\mu\nu\lambda}\vep_{ABC} A_\mu^{~AB}\del_\nu
A_\lambda^{~C}=
4\,\epsilon^{\mu\nu\lambda}\,
B_\mu^{~A} \del_\nu A_{\lambda\,A}\\
\frac13\,\epsilon^{\mu\nu\lambda}\,\vep_{aef}^{~~~~~g}\,\vep_{bcdg}\,
A_\mu^{~ab}A_\nu^{~cd}A_\lambda^{~ef} &=
-4\, \epsilon^{\mu\nu\lambda}\,\vep_{ABC} B_\mu^{~A}A_\nu^{~B}A_\lambda^{~C}
-\frac43\,\epsilon^{\mu\nu\lambda}\,\vep_{ABC}
B_\mu^{~A}B_\nu^{~B}B_\lambda^{~C}\;.
\end{split}
\end{equation}
We also need to consider the couplings arising from the covariant
derivative on $X^{A(I)}$. We have
\begin{equation}
\label{covdev}
\begin{split}
D_\mu X^{AI}&=
\del_\mu X^{AI} + \vep^A_{~bcd}A_\mu^{~cd}X^{bI}\\
&= \del_\mu X^{AI} + 2\,\vep^A_{~BC}A_\mu^{~C}X^{BI}
+2\,B_\mu^{~A}X^{4(I)}
\end{split}
\end{equation}
and
\beq\label{covdev4} D_\mu X^{4I} = \del_\mu X^{4I}-2
B_{\mu A}X^{AI}\;.
\ee
 Inserting these in the lagrangian (but
ignoring fermions) and using the VEV $\langle
X^{4(8)}\rangle = v$, we find the following terms involving
$B_\mu^{~A}$\;
\begin{equation}
\begin{split}
  \mathcal L_{\rm kinetic} = -2 v^2 B_\mu^{~A}B^\mu_A -2
  &B^{~A}_\mu X^{4I}D'^\mu X_A^{I}-2 vB^{~A}_\mu
  D'^\mu X_A^{(8)}\\-2B_{\mu
    A}X^{AI}B^\mu_BX^{BI}
  &-2 B^A_{\mu }B^\mu_AX^{4I}X_{4I}+ 2 B^\mu_A X^{AI}
  \pd_\mu X^{4I} +... \;,
\end{split}
\end{equation}
where we have defined a new covariant derivative which depends only on
$A_\mu^A$\;
\be
D'_\mu X^{AI} = \pd_\mu X^{AI}-2 \vep^A_{\phantom{A}BC}A^B_\mu X^{CI}\;.
\ee
Notice that the first term looks like a mass for $B_\mu^{~A}$, as might be
expected from the Higgs mechanism, but we will see in a moment that $B_\mu^{~A}$ is not in the spectrum of the theory.

The terms involving $B_\mu^A$ that come from the gauge field
self-couplings are
\begin{equation}
\cL_{\rm CS}  = 2\,\epsilon^{\mu\nu\lambda}\,
B_\mu^{~A} F'_{\nu\lambda A}
-\frac43\,\epsilon^{\mu\nu\lambda}\,\vep_{ABC}
B_\mu^{~A}B_\nu^{~B}B_\lambda^{~C} +...\;,
\end{equation}
where we have also defined
\beq F_{\nu\lambda}^{'A} = \pd_\nu
A_\lambda^A - \pd_\lambda A^A_\nu - 2\vep^A_{\phantom{A}BC}A^B_\nu
A^C_\lambda\;.
\eeq
Thus $B_\mu^{~A}$ is an auxiliary field appearing without derivatives. It
can therefore be eliminated via its equation of motion. We can extract
the leading part of such solution by temporarily neglecting the
quadratic term in $B_\mu^A$ coming from the cubic self-interaction as
well as terms coming from higher interactions with
scalars. Later we will show that these  would have led to
higher-order contributions that are suppressed in the strong-coupling
limit.  We therefore consider the set of couplings
\be
\cL =  -2 v^2 B_\mu^{~A}B^\mu_A -2 vB^{~A}_\mu
  D'^\mu X_A^{(8)} + 2\,\epsilon^{\mu\nu\lambda}\,
  B_\mu^{~A} F'_{\nu\lambda A} + \hbox{higher order}
\ee
and find that
\begin{equation}
B_\mu^{~A} =
\frac{1}{2v^2}
\epsilon_\mu^{\phantom{\mu}\nu\lambda}\,F_{\nu\lambda}^{'A}
- \frac{1}{2v}D'_\mu X^{A(8)}\;.
\end{equation}

Thus one of our gauge fields, $B_\mu^{~A}$, has been set equal to the
field strength of the other gauge field $A_\mu^{~A}$ (plus other
terms). Eliminating $B_\mu^{~A}$ gives rise to a standard
Yang-Mills kinetic term for $A_\mu^{~A}$! This is the miracle
that promotes the Chern-Simons gauge field $A_\mu^{~A}$ into a
dynamical gauge field.

Continuing with the computation, the sum of the Chern-Simons gauge
field action and the scalar covariant kinetic terms becomes (up to a
total derivative)
\begin{equation}\label{kinetic-cs}
-\frac{1}{v^2} F^{'A}_{\mu\nu} F^{'\mu\nu}_A
- \frac{1}{2}\pd_\mu X^{4I}\pd^\mu X_4^{I}
- \frac{1}{2}D_\mu X^{Ai}D^\mu X_A^{i}
+\mathcal O (B X\pd X)+\mathcal O (B^2 X^2) + \mathcal O (B^3)\;.
\end{equation}
The redefinition
\beq
A\ra \frac{1}{2}A\;,
\eeq
leads to
\beq
 D'_\mu X^{AI}\ra D_\mu X^{AI} \equiv \pd_\mu X^{AI} -
 \vep^A_{~BC}A_\mu^B X^{CI}
\eeq
and
\beq
 F'^A_{\mu\nu}\ra \frac{1}{2}F^A_{\mu\nu}\equiv
\frac{1}{2}\left(\pd_\mu A^{~A}_\nu - \pd_\nu A^{~A}_\mu -
 \vep^A_{~BC}A^{~B}_\mu A^{~C}_\nu\right) \;.
\eeq
Thus \eref{kinetic-cs} finally becomes
\beq
\begin{split}
&-\frac{1}{4 v^2} F^A_{\mu\nu} F^{\mu\nu}_A
- \frac{1}{2}\pd_\mu
X^{4I}\pd^\mu X_4^{I}
- \frac{1}{2}D_\mu X^{Ai}D^\mu X_A^{i}+\frac{1}{v}\mathcal
O\left( X\pd X \left(F/v+ DX\right)\right)\\
&~~~~~+\frac{1}{v}\mathcal O \left(X^2\left(F/v+ DX\right)^2\right)
 + \frac{1}{v^3}\mathcal O
\left(\left(F/v+ DX\right)^3\right)\;.\\
\end{split}
\eeq
The terms in $B^A_\mu$ that we had neglected will lead to higher
interactions with increasingly higher powers of $(F/v+D X)$ in
the numerator and $v$ in the denominator.

For the fermions, we easily find that
\beq
\frac{i}{2}{\bar\Psi}^a \Gamma^\mu D_\mu \Psi_a \to
\frac{i}{2}{\bar\Psi}^A \Gamma^\mu D_\mu \Psi_A +
\frac{i}{2}{\bar\Psi}^4 \Gamma^\mu \del_\mu \Psi_4
+ \hbox{higher order}\;,
\eeq
where $D_\mu$ on the LHS is the 3-algebra covariant derivative while
$D_\mu$ on the right is the Yang-Mills covariant derivative.

The theory we have obtained now has an interacting  $\SU(2)$ Yang-Mills piece
 supplemented with some decoupled fields as well as a variety
of higher-order terms.\footnote{The original 3-algebra still makes its
presence in the higher-order terms, to be understood geometrically in the next section.} The action can be written in the
form
\beq
\cL= \cL_{\rm SU(2)} + \cL_{\rm U(1)}
\eeq
where
\beq
\cL_{\rm U(1)}= -\half\del_\mu X^{4I}\del^\mu X^{I}_{4}
+ \frac{i}{2}{\bar \Psi}^4 \Gamma^\mu\del_\mu\Psi_4\;.
\eeq
For the SU(2) part, we rescale the fields as $(X,\Psi)\to
(X/v, \Psi/v)$, to find the action
\beq
\label{lcoupled}
\cL_{\rm SU(2)}
= \frac{1}{v^2}\cL_{0} + \frac{1}{v^3}\cL_{1} + {\cal
O}\left(\frac{1}{v^4}\right) \;,
\eeq
where $\cL_0$ is the action of maximally supersymmetric 2+1d SU(2) Yang-Mills theory
\begin{equation}\label{D2}
\begin{split}
\mathcal L_0 = & -\frac{1}{4}F_{\mu\nu\,A}F^{\mu\nu\,A}-\frac{1}{2}D_\mu
X^{Ai}D^\mu X_A^{~i} + \frac{1}{4}
\left(\veps_{ABC}X^{Ai}X^{Bj}\right)
\left(\veps_{DE}^{~~~C}X^{Di}X^{Ej}\right)\\
& +\frac{i}{2}\bar
\Psi^A \slashed{D}\Psi_A + \frac{i}{2}\veps_{ABC}
{\bar\Psi}^A \Gamma^i X^{Bi}\Psi^C\;,
\end{split}
\end{equation}
with the field strength and covariant derivative defined as
\begin{equation}
F^A_{\mu\nu} = \partial_\mu A^A_\nu - \partial_\nu A^A_\mu -
\vep^A_{~BC}A^B_\mu A^C_\nu\quad \textrm{and}\quad D_\mu^{AB} =
\partial_\mu\delta^{AB} + \vep^{AB}_{~~C}A^C_\mu\;.
\end{equation}
In the above, $\cL_0,\cL_1,...$ are all completely independent of $v$. Taking the limit $v\to\infty$, only the $\cL_0$ term remains. The interacting part of the surviving theory is precisely $\SU(2)$ Yang-Mills, the low-energy theory on two D2-branes. Then, since scalar fields have canonical dimension $\half$, we can identify $v\equiv g_{YM}$; this is the correct mass dimension for the Yang-Mills coupling in 2+1d, and it is in agreement with the fact that this theory is weakly coupled in the UV and strongly coupled in the IR.

Note that $B_\mu^{~a}$ has disappeared from the theory, while $A_\mu^{~a}$ no longer has a Chern-Simons coupling but rather a full-fledged $\SU(2)$ Yang-Mills kinetic term. The fields that survive in the D2-brane action have  the correct couplings to the newly-dynamical gauge field. Note that the terms corresponding to the modes $X^{A(8)}$ have disappeared; they played the role of the Goldstone bosons that gave a mass to $B^A_\mu$, and at the end were transmuted via the Higgs mechanism and the Chern-Simons coupling into the single physical polarisation of $A_\mu^{~A}$. 

One might be alarmed at the fact that the original gauge symmetry $\SO(4)\simeq \SU(2)\times \SU(2)$ appears to have been Higgsed to $\SU(2)\times \U(1)$ by a VEV of a field in the $\bf 4$ of $\SO(4)$. That is not quite the case. The Higgs mechanism breaks $\SO(4)$ to $\SO(3)\simeq \SU(2)$ as it should, since the scalar $X^{4(8)}$ that develops the VEV breaks $\SU(2)\times \SU(2)$ to a diagonal $\SU(2)$.  However, several free scalars are left over and the $\U(1)$ gauge field is obtained by dualising one of them.

The final theory also contains 8 non-interacting scalars
$X^{4I}$.  Of these, $X^{4i}, i=1,2,...,7$ correspond to the
centre-of-mass modes for the D2 worldvolume theory. The last scalar
$X^{4(8)}$, the one which originally developed a VEV, can now be
dualised via an {\it abelian} duality to yield an extra $\U(1)$ gauge
field. The free abelian multiplet is completed by $\Psi^4$, so the
full gauge group is SU(2) $\times$ U(1).  Note that the entire
multiplet comes from a direction that was {\it not central} in
the original 3-algebra  (in the sense that it does not satisfy $[T^4,T^I,T^J]=0$ for all $I,J$).

When  $v = g_{YM}\to\infty$, the theory on the D2-branes becomes strongly coupled.\footnote{We remind the reader that at this stage we have fixed the Chern-Simons level, which is an  otherwise free parameter of the theory, to $k=1$. In the next section we will relax this   assumption.} As we have already seen, the physics of strongly coupled Yang-Mills in 2+1d is expected to be captured by M2-branes. Hence, in this limit of $\U(2)$ Yang-Mills one expects to recover the low-energy physics of 2 M2-branes in flat space \cite{Mukhi:2008ux}.

While the novel Higgs mechanism plays a specific role in the context of M2-branes, as described above, it occurs quite generically in a class of Chern-Simons field theories in 2+1 dimensions. It is closely associated to the well-known phenomenon of topological mass generation in 2+1d and stems from a conflict between diagonalisability of kinetic and mass terms that arise in Chern-Simons type theories \cite{Mukhi:2011jp}.

\subsubsection{The Higgs mechanism for ABJM theory}\label{Higgsbifund}

We next turn our attention to applying the above mechanism to the case of the full $\U(n)\times\U(n)$ ABJM theory \cite{Pang:2008hw,Li:2008ya,Honma:2008ef}. Here, however, the bifundamental nature of the matter fields makes the technical discussion slightly different; it further involves a subtlety relating to the treatment of the abelian parts of the matter and gauge fields \cite{Chu:2010fk}. For the sake of simplicity, we focus our attention on the bosonic part of the action. In this case we will also reintroduce the $k$-dependence. In fact, we will consider the Higgsing process in a limit where not only the VEV $v$ but also the Chern-Simons level is taken large, in such a way that $v/k\to\textrm{fixed}$.

Consider once again the ABJM lagrangian \eqref{niceaction}
\begin{eqnarray}
\nonumber {\cal L} &=& -{\rm Tr}(D^\mu \bar Z_A,D_\mu Z^A) +  \frac{k}{4\pi}\varepsilon^{\mu\nu\lambda}\left({\rm Tr}( A^L_\mu\partial_\nu
A^L_\lambda - \frac{2}{3} i A^L_\mu A^L_\nu  A^L_\lambda) -{\rm Tr}(
A^R_\mu\partial_\nu   A^R_\lambda - \frac{2}{3}i  A^R_\mu A^R_\nu   A^R_\lambda)\right)
 \\[2mm]
&& - \frac{1}{3}{\rm Tr}\left(4Z^A\bar Z_AZ^B\bar Z_CZ^C\bar Z_B  - 4Z^A\bar Z_BZ^C\bar Z_AZ^BZ_C -Z^A\bar Z_AZ^B\bar Z_BZ^C\bar Z_C-\bar Z_A Z^A\bar Z_BZ^B\bar Z_C Z^C\right)\nn\\
\end{eqnarray}
where
\begin{equation}
\hat D_\mu Z^A = \partial_\mu Z^A - iA^L_\mu Z + i Z^A A^R_\mu\;.
\end{equation}
We  would like to see what happens under a perturbation schematically of the form
\be
Z^A=v\delta^{A4}+z^A\;,
\ee
with $A=1,...,4$, or more precisely in terms of the real parts\footnote{Here the fields $X^A$ after Higgsing are $n\times n$ hermitian matrices which will be expanded in a basis consisting of the unit matrix and a hermitian set of $\SU(n)$ generators.}
\be
Z^A =  v \delta^{A4}\one_{N\times N}+\frac{1}{\sqrt 2}X^A  + i\frac{1}{\sqrt 2} X^{A+4}\;.
\ee

For the Higgsing it is appropriate to define
\be
A^+_\mu = \frac{1}{2}(A^L_\mu +  A^R_\mu) \;, \qquad A^-_\mu = \frac{1}{2}(A^L_\mu - A^R_\mu)\;,
\ee
which, observing that $A^L_\mu Z^A - Z^A A^R_\mu =[A^+_\mu , Z^A]+\{ A^-_\mu, Z^A\}$,  translates into
\bea
\hat D_\mu Z^A &=& D_\mu Z^A-i\{ A^-_\mu, Z^A\}\cr
D_\mu Z^A  &=& \partial_\mu Z^A - i[A^+_\mu , Z^A]\cr
F^+_{\mu\nu} &=& \partial_\mu A^+_\nu -i [A^+_\mu , A^+_\nu]\;.
\eea
Note that the abelian gauge fields do not appear in the covariant derivative $D_{\mu}$.
In terms of these new variables the Chern-Simons part of the lagrangian becomes
\be
S_{CS}= \int d^3x \frac{k}{2\pi}\epsilon^{\mu\nu\lambda}\Tr \Big( A^-_\mu F^+_{\nu\lambda}-\frac{2i}{3}A^-_\mu A^-_\nu A^-_\lambda\Big)\;.
\ee

The  fields are $n\times n$ matrices which  can be expanded in terms of a complete basis of $\U(n)$ generators as follows\footnote{Here we normalise the $\SU(n)$ generators as $\Tr(T^aT^b)=\delta^{ab}$, with $T^0 = \one_{n\times n}$.}
\bea
Z^A &=& Z^A_0T^0+iZ^A_aT^a\;, \cr
A^L_\mu&=& A^{L0}_{\mu}T^0 + A_\mu^{La}T^a\;,\cr
 A^R_\mu &=& A^{R0}_\mu T^0+  A_\mu^{Ra}T^a\,,
\eea
and subsequently, including the VEV and writing things in terms of real components,
\be
Z^A=\left(\frac{X_0^A}{\sqrt{2}}+v\delta^{A,4}\right)T^0+i\frac{X_0^{A+4}}{\sqrt{2}}T^0+i\frac{X^A_a}{\sqrt{2}}T^a-
\frac{X_a^{A+4}}{\sqrt{2}}T^a\,.
\ee
 As a result one gets for the covariant derivative
\bea
\hat D_\mu Z^A&=&\frac{\pd_\mu X_0^A}{\sqrt{2}}T^0-\frac{D_\mu X_a^{A+4}}{\sqrt{2}}T^a+\frac{i\pd_\mu X^{A+4}_0}{\sqrt{2}}
T^0+\frac{iD_\mu X_a^A}{\sqrt{2}}T^a\nn\\
&&
-2ivA_{\mu a} ^-T^a\delta^{A4}-i\sqrt{2}A_{\mu a}^-X_0^AT^a+\sqrt{2}A^-_{\mu a}X_0^{A+4}T^a-A_\mu^{-a}d_{abc}T^cX_b^{A+
4}\nn\\
&& +iA_\mu^{-a}d_{abc}T^c X_b^A -2ivA_{\mu 0} ^-T^0\delta^{A4}-i\sqrt{2}A_{\mu 0}^-X_0^AT^0+\sqrt{2}A^-_{\mu 0}X_0^{A+4}T^0\nn\\
&& +\sqrt{2}A_{\mu 0}^-X_a^AT^a+i\sqrt{2}A^-_{\mu 0}X_a^{A+4}T^a\,,\label{covaderi}
\eea
where  $[T^a,T^b]=i{f^{ab}}_cT^c, \{ T^a,T^b\}={d^{ab}}_cT^c$.
 Then we obtain
\bea
\Tr|\hat D_\mu Z^A|^2&=&N\frac{(\pd_\mu X_0^A)^2}{2}
+\left(\frac{(D_\mu X)_c^{A+4}}{\sqrt{2}}\right)^2+\left(2vA_{\mu c}^-\delta^{A4}-\frac{(D_\mu X)_c^A}{\sqrt{2}}\right)^2\cr
&&+N\left(\frac{1}{\sqrt 2}\pd_\mu X_0^{A+4}- 2 v A^-_{\mu 0}\delta^{A4}\right)^2+{\rm subleading}\;.
\eea
Adding the following two terms, which are equal to zero by the Bianchi identity,
\be
-\frac{k}{2\pi}\epsilon^{\mu\nu\lambda} \frac{1}{v}\frac{1}{2\sqrt  2} (D_\mu X)_a^4 F^{+a}_{\nu\lambda}-\frac{nk}{2\pi}\epsilon^{\mu\nu\lambda} \frac{1}{v}\frac{1}{2\sqrt  2} (\pd_\mu X_0^8) F^{+0}_{\nu\lambda}
\ee
and with the inclusion of the Chern-Simons terms, we get that the action becomes
\bea
S &=& \int d^3x \Bigg[\frac{k}{2\pi}\epsilon^{\mu\nu\lambda}\Tr \Big( A^-_\mu F^+_{\nu\lambda}-\frac{2i}{3}A^-_\mu A^-_\nu A^-_\lambda\Big)
- \Tr |\hat D_\mu Z^A|^2\Bigg]\nn\cr\\[0mm]
&=& \int d^3 x \Bigg[ \frac{k}{2\pi} \epsilon^{\mu\nu\lambda}  \left(A^-_{\mu a}- \frac{1}{2v}\frac{1}{\sqrt 2}
(D_\mu X)_a^4\right) F^{+a}_{\nu\lambda }+\frac{nk}{2\pi} \epsilon^{\mu\nu\lambda} \left(A^-_{\mu 0}- \frac{1}{2v}\frac{1}{\sqrt 2}
(\pd_\mu X)_0^8\right) F^{+0}_{\nu\lambda } \cr
&&\qquad\qquad- \left(2v A_{\mu a}^- -  \frac{1}{\sqrt 2}(D_\mu X)^4_a\right)^2-
n\left(\frac{1}{\sqrt 2}\pd_\mu X_0^{A+4}- 2 v A^-_{\mu 0}\delta^{A4}\right)^2\cr
&&\qquad\qquad -   \frac{1}{2}(D_\mu X)_a^{I'} (D^\mu X)_a^{I'} -\frac{1}{2}n \pd_\mu X_0^A \pd^\mu X_0^A+
\textrm{higher order}  \Bigg]\;,
\eea
where $I'=\{1,2,3,5,6,7,8\}$. The higher order terms also include a contribution proportional to  $(A_\mu^-)^3$. However, these terms are subleading in the limit $k,v\to\infty$ and can be ignored.

At this point we can perform a shift in the $A^-_{\mu a}$ and in the abelian component $A^-_{\mu 0}$ of the gauge field
\be
A^-_{\mu a} \to A^-_{\mu a}  +  \frac{1}{2v}\frac{1}{\sqrt 2} (D_\mu X)^4_a \qquad\textrm{and}\qquad A^-_{\mu 0} \to A^-_{\mu 0}  +  \frac{1 }{2v}\frac{1}{\sqrt 2} (\pd_\mu X^8_0)\, ,
\label{Higgseating}
\ee
which leads to
\bea\label{above}
S &= &\int d^3 x   \Big(\frac{k}{2\pi} \epsilon^{\mu\nu\lambda} (A^-_{\mu a} F^{a+}_{\nu\lambda }+N A^-_{\mu 0} F^{+}_{\nu\lambda \;0})
-4 v^2 A_{\mu a}^- A_a^{-\mu}-4 n v^2  A_{\mu 0}^- A_0^{-\mu}\nn\\
&&\qquad -  \frac{1}{2}(D_\mu X)_a^{I'} (D^\mu X)_a^{I'} -\frac{1}{2}n\pd_\mu X_0^{\tilde I'} \pd^\mu X_0^{\tilde I'} + \textrm{higher order}  \Big)\;,
\eea
where $\tilde I'=\{1,...,7\}$.

It is interesting to observe that in the above expression, both the $X_a^4$ and the $X^8_0$ components vanish to leading order.  These fields make up the Goldstone modes that render, respectively, $A^+_{\mu a}$ and $A^+_{\mu 0 }$ dynamical to give back a $\U(N)$ gauge field. Without the vanishing of the $X_0^8$ one would have ended up with excessive degrees of freedom. We will see shortly that this also has an interpretation in terms of the M2's moving in the orbifold geometry.

We can now integrate out both the abelian and non-abelian components of $A^-_{\mu }$ to obtain
\be\label{integrate}
A^-_\mu = \frac{k }{16\pi v^2}\epsilon_{\mu\nu\lambda} F^{+\nu\lambda} + \textrm{higher order}
\ee
and upon plugging into \eqref{above} this gives
\be
S = \int d^3 x  \Big[- \Tr \Big(\frac{k^2 }{32\pi^2 v^2} F^{+\mu\nu}F_{\mu\nu}^+\Big)  -  \frac{1}{2}(D_\mu X)_a^{I'} (D^\mu X)_a^{I'} -\frac{1}{2}n\pd_\mu X_0^{\tilde I'} \pd^\mu X_0^{\tilde I'} + \textrm{higher order}  \Big]\;.
\ee
Then using the definition
\be
\frac{k^2}{32\pi^2 v^2} = \frac{1}{4 g^2_{YM}} \label{gidentif}
\ee
and taking the limit $k,v\to\infty$, with $k/v = \textrm{fixed}$, the higher order terms drop out. Combining the remaining traceless part of $X_a^8$ with the  trace part of $X_0^4$, we find the bosonic kinetic terms for $\U(n)$, three-dimensional Yang-Mills theory. 

Regarding the bosonic potential terms, we observe that all terms scaling like $v^6,...,v^3$ vanish, so one is left with a potential that is of order $v^2/k^2\propto g^2_{YM}$, and hence fourth order in the scalar fields, as expected. The remaining terms are subleading in $v$ and vanish in the $v\rightarrow \infty$ limit. The surviving term in this limit is
\be
- V_6\to-\frac{ g_{YM}^2}{4}\Tr \Big([X^{I'},X^{J'}][X^{J'}, X^{I'}]\Big)\;.
\ee
In this way we have recovered the full bosonic content of three-dimensional $\U(n)$ Yang-Mills by Higgsing the $\U(n) \times \U(n)$ ABJM theory.

\subsubsection{Higgsing and large-$k$ compactification}

We now proceed to assign a spacetime interpretation to the field-theoretic mechanism that we have thus far described in the ABJM case. We have already  established in Section~\ref{branestoABJM} that the orbifold  $\mathbb Z_k$ acts as $Z^A\to e^{2\pi i /k} Z^A$ on the complex coordinates transverse to the M2-brane worldvolume. Setting $Z^{1,2,3}=0$ for simplicity reduces us to $\mathbb  C/\mathbb Z_{k}$ as $k\rightarrow \infty$, with\footnote{The trace parts of the field theory scalars are related to spacetime coordinates by  multiplication with a factor of $T_{M2}^{-\half}$.}
\be
 Z^4 \rightarrow  Z^4 e^{2\pi i/k}\simeq  Z^4\Big(1+2\pi i \frac{1}{k}+...\Big)
\simeq Z^4+2\pi i \frac{Z^4}{k}\;.
\ee
Expanding around $Z^4=v+i 0$ with $(v/k) T_{M2}^{-\half}\equiv R$, we see that
\be\label{orbifold}
Z^4 T_{M2}^{-\half} \rightarrow Z^4 T_{M2}^{-\half}+2\pi i R
\ee
should be an invariance of the theory, or by writing $Z^4=X^4+iX^8$, that $X^8$ is compactified with radius $R$. This is the radius of the M-theory circle.  By letting $(v/k) T_{M2}^{-\half} = R\to 0$, one recovers the theory of D2-branes of type IIA string theory in flat space.

In particular, since three-dimensional Yang-Mills involves seven (as opposed to eight) scalars, it is natural to expect that one of the Goldstone modes that  render the gauge fields dynamical in the ABJM theory should be precisely $X_0^8$, corresponding to the centre-of-mass motion of the branes in that direction. On the other hand, for the $\mathcal N=8$ BLG model, the scalar degree of freedom that disappeared was exactly the one that developed the VEV. For our particular choice of VEV,  this  corresponds to $X^4_0$ being singled out, as opposed to $X^8_0$, as implied by (\ref{orbifold}). This is a sign that the orbifold picture is not an appropriate dual description of the BLG model for generic values of $k$. We will explicitly see in  Section~\ref{relationton8} that this is indeed the case.

The relation between the novel Higgs mechanism and large-$k$ compactification  can also be understood as follows: $\mathbb C^4/\mathbb Z_k$ can be thought of as a cone over $S^7/\mathbb Z_k$. The orbifold action leads to an opening angle that shrinks like $1/k$. In the limit where $k \to \infty$, this opening angle approaches zero, so at some point infinitely far out on the moduli space the local geometry approaches that of a cylinder ${S}^7/\mathbb Z_k \times \mathbb R$, where the former is always realised as a Hopf fibration. However,  ${S}^7/\mathbb Z_k$ then involves a $\cp^3$ base of infinite volume, while the $S^1/\mathbb Z_k$ fibre has a finite, tunable radius  (which can be taken to be small) because of the action of the orbifold. Moreover, the nature of the fibration is locally trivial and the cylinder is really $\mathbb R^6 \times S^1_{\mathrm{small}} \times \mathbb R \equiv \mathbb R^7 \times S^1_{\mathrm{small}}$. The scaling limit $k\to\infty$, $v\to\infty$ with $g_{YM}\to\text{fixed and small}$, precisely takes the  M2-branes out into this cylindrical space, where they should behave like D2-branes in type IIA string theory. So at low energies we expect a finitely coupled $\U(n)$ Yang-Mills theory -- and that is exactly what we find \cite{Distler:2008mk,Chu:2010fk}.

It is worth mentioning that the discussion in the limit of large-order $\mathbb Z_{k}$ orbifolds bears a strong resemblance to the ideas introduced in \cite{Ganor:1996nf,Ganor:1997jx} and used in the deconstruction approach to M5-branes \cite{ArkaniHamed:2001ie,Mukhi:2002ck}. In those works the order of the orbifold grows large in a similar way and the D-branes are simultaneously moved far away from the fixed point, so that they effectively  end up propagating on a cylinder. It is important to note that, compared to the starting quiver gauge theory, the deconstructed theory is higher dimensional and has enhanced supersymmetry. Another interesting point is the implementation of the Higgs mechanism for the ABJM model coupled to $\mathcal N=6$ conformal supergravity, or ``topologically gauged ABJM theory'' \cite{Chu:2009gi}. The Higgsed theory \cite{Chu:2010fk} has broken conformal invariance and reduces to 3d ``chiral supergravity'' in the sense of \cite{Li:2008dq}. The latter has an AdS$_3$ vacuum and should also admit a CFT$_2$ boundary description. Hence, the Higgs mechanism relates $\textrm{AdS}_4/\textrm{CFT}_3$ to $\textrm{AdS}_3/\textrm{CFT}_2$ in something that might be called ``sequential AdS/CFT'' \cite{Nilsson:2012ky,Ohl:2012bk}.

\subsection{Relation of ABJM to BLG}\label{relationton8}

At this stage we can close the circle of ideas by asking the following question: What is the relation of the original BLG theory of Section~\ref{Neq8} to the ABJM models?  

We first remind that for $n=2$  we can chose a basis for the  3-algebra of $2\times 2$ matrices given by
\begin{equation}
T^a =
\left\{-\frac{i}{\sqrt{2}}\sigma_1,-\frac{i}{\sqrt{2}}\sigma_2,-\frac{i}{\sqrt{2}}\sigma_3,\frac{1}{\sqrt{2}}\one_{2
\times 2}\right\}\ ,
\end{equation}
where $a=1,2,3,4$ and $\sigma_i$ are the Hermitian Pauli matrices: $\sigma_i\sigma_j = \delta_{ij}
+i\epsilon_{ijk}\sigma^k$. In this basis the structure constants and metric of the triple product defined in (\ref{Neq6Mat}) are
\begin{equation}
f^{abcd} = \frac{2\pi}{k}\epsilon^{abcd} \qquad\textrm{and}\qquad {\rm Tr}(T_bT^a) = \delta^a_b\ .
\end{equation}
Thus $f^{abcd}$ is real and totally anti-symmetric. In fact one can check that the $n=2$, ${\cal N}=6$ lagrangian constructed from this 3-algebra is the ${\cal N}=8$ lagrangian with gauge symmetry $\SO(4)\simeq\SU(2)\times\SU(2)$ but written in complex notation: $Z^A = (X^A+iX^{A+4})/\sqrt{2}$, for $A=1,2,3,4$.  (See Section~\ref{gaugedet}.)

Here we need to reiterate a subtlety that will be important for our discussion: The lagrangians are defined only in terms of the data of the Lie algebra and not the gauge group. This data is encoded by the 3-algebra. However to define the quantum theory we need to specify the gauge group and the choice of the latter has an effect through the flux quantisation conditions. Since more than one group can have the same Lie algebra we see that more than one theory can be associated to a lagrangian.  To make this distinction clear in this section we refer to the lagrangian in terms of its Lie algebra, \eg\  $\uf(n)\oplus\uf(n)$ or $\su(n)\oplus \su(n)$, but we will refer to the theories in terms of their gauge group, \eg\  $\U(n)\times \U(n)$, $\SU(n)\times \SU(n)$ or $(\SU(n)\times \SU(n))/{\mathbb Z}_n$.

We have seen in Section~\ref{from3altoCS} that  the $\U(n)\times \U(n)$  ${\cal N}=6$ models can be derived from the $\SU(n)\times \SU(n)$ models by gauging the global $\U(1)$. We now describe how to go backwards: namely integrating out the $\U(1)$ gauge field of the ABJM  lagrangians leads to   $\su(n)\oplus \su(n)$ lagrangians, along with   a ${\mathbb Z}_k$ orbifold action on the fields. In the case with $n=2$ we will show that the ABJM  model can be related to the ${\cal N}=8$ lagrangian with an additional ${\mathbb Z}_k$ orbifold.  However there is global information that needs to be taken into account and this only works when  $k$ and $n$ are relatively prime \cite{Lambert:2010ji}. Let us next see how that happens.

\subsubsection{From $\frak u(n)\times \frak u(n)$ to $\su(n)\times \su(n)$ CS-matter theories}

To begin let us go back and rewrite the ABJM lagrangian as
\begin{equation}
 {\cal L}_{\uf(n)\times \uf(n)} =  {\cal L}_{\su(n)\times \su(n)}^{\textrm{gauged}} +
\frac{nk}{4\pi}
 \varepsilon^{\mu\nu\lambda}B_\mu\partial_\nu Q_\lambda\ .
\end{equation}
As in Section~\ref{BLGmodulispace}, we introduce a Lagrange multiplier term
\begin{equation}
 {\cal L}_{\uf(n)\times \uf(n)}=  {\cal L}_{\su(n)\times \su(n)}^{\textrm{gauged}} +
 \frac{nk}{8\pi}
 \varepsilon^{\mu\nu\lambda}B_\mu H_{\nu\lambda} +\frac{n}{8\pi} \sigma\varepsilon^{\mu\nu\lambda} \partial_{\mu}H_{\nu\lambda}\ .
\end{equation}
Integrating the last term by parts we find
\begin{equation}
 {\cal L}_{\uf(n)\oplus \uf(n)} =  {\cal L}_{\su(n)\oplus \su(n)}^{\textrm{gauged}} +
 \frac{nk}{8\pi}
 \varepsilon^{\mu\nu\lambda}B_\mu H_{\nu\lambda}
 - \frac{n}{8\pi}\varepsilon^{\mu\nu\lambda}\partial_\mu\sigma H_{\nu\lambda}\ .
\end{equation}
We can now integrate out $H_{\mu\nu}$ to see that
\begin{equation}
B_\mu = \frac{1}{k}\partial_\mu \sigma\ .
\end{equation}
Thus under a $\U(1)_B$ gauge transformation one has that
\begin{equation}\label{fix}
\sigma \to \sigma +  {k}\theta\ .
\end{equation}
Substituting back  we find that the $\uf(n)\oplus \uf(n)$ lagrangian is equivalent to the ${\su(n)\oplus
\su(n)}$ lagrangian with new variables:
\begin{equation}\label{actionsequal}
{\cal L}_{\uf(n)\oplus \uf(n)}(Z^A,\psi_A,\tilde A_\mu^a{}_b,B_\mu,Q_\mu) \cong {\cal
L}_{\su(n)\oplus\su(n)}(e^{i\sigma/k}Z^A,e^{i\sigma/k}\psi_A,\tilde A_\mu^a{}_b)\;.
\end{equation}
The variables $\hat Z^A=e^{i\sigma/k}Z^A$ and $\hat \psi_A=e^{i\sigma/k}\psi_A$ are $\U(1)_B$ gauge invariant.

Most of the steps that we have outlined do not rely on the global choice for gauge group. The exception to this is the last step  (\ref{actionsequal})  where the infinitesimal gauge transformation was exponentiated to a finite group element. Thus we should be careful with some global issues. In particular, although the Lie-algebra decomposes as $\uf(n) \simeq \su(n)\oplus \uf(1)$ it is not true that $\U(n) \cong\U(1)\times \SU(n)$. Rather one finds that  $\U(1)\times \SU(n)$ is an $n$-fold cover of $\U(n)$. To see this, we note that the group homomorphism  $\omega: \U(1)\times \SU(n)  \to  \U(n)$ defined by $\omega (e^{i\theta},g_{\SU(n)}) = e^{i\theta}g_{\SU(n)}$ covers  $\U(n)$ $n$-times. In fact, the determinant satisfies $\det (g_{\U(n)}) = e^{in\theta}$, but this only determines $\theta$ modulo $(2\pi/n) \mathbb Z$. Thus we have $\U(n) \cong(\U(1)\times \SU(n))/{\mathbb Z}_n$, where ${\mathbb Z}_n = {\rm Ker}(\omega)$ is the centre of $\U(n)$.

The upshot is that although (\ref{actionsequal}) links the lagrangian of the $\frak u(n)\times \frak u(n)$ theory to that of $\su(n)\times \su(n)$, the gauge group is $(\SU(n)\times \SU(n))/{\mathbb Z}_n$ and not $\SU(n)\times \SU(n) $.\footnote{One might have expected   $(\SU(n)/{\mathbb Z}_n)\times (\SU(n)/{\mathbb Z}_n)$ but only the relative ${\mathbb Z}_n$   factor acts non-trivially.} As we will now see this leads to modified flux quantisation rules that affect the physical interpretation.

We next need to determine the periodicity of $\sigma$ in  (\ref{actionsequal}),  which follows from a quantisation condition on the flux $H$.  In this case the standard Dirac condition  $\int_\Sigma F \in 2\pi\mathbb{Z} $  is modified. In particular, the gauge group is $(\U(1)\times \mathrm{SU}(n))/\mathbb{Z}_n$ and we need only require that $ \int_\Sigma F \in (2\pi/n)\mathbb{Z}$, {\it i.e.} the $\U(1)$ phases computed by two different paths $D$ and $D'$ must be equal modulo $\mathbb{Z}_n$.  Thus we see that the quantisation condition is
\begin{equation}\label{quantcon1}
\int  dF^{L/R} \in  \frac{2\pi}{n} \mathbb{Z}\ .
\end{equation}
As we have emphasised, this fractional flux quantisation condition arises because the gauge group is $(\SU(n)\times \SU(n))/\mathbb Z_n$ instead of $\SU(n)\times \SU(n)$, with $\mathbb{Z}_n$ the relative centre of the two $\SU(n)$ factors. Thus we refer to the resulting Chern-Simons matter theory as the $(\SU(n)\times \SU(n))/\mathbb Z_n$-theory. This is distinct from  a theory with the same ${\cal L}_{\su(n)\oplus\su(n)}$ lagrangian but global $\SU(n)\times \SU(n)$ gauge symmetry and no fractional flux quantisation, which we refer to as the $\SU(n)\times \SU(n)$-theory.

After integrating out $H$, we are left with the condition $B = d\sigma/k$. This is analogous to what we obtained for the moduli space calculation with the difference that now we are in the full theory, not just the moduli space. Locally, $F_L-F_R=dB$ vanishes so that $F_L$ and $F_R$ must have the same flux. Note that we do not require that $\sigma$ is globally defined so there can be a non-zero Wilson line for the gauge field $B$.  However, since $F_L-F_R=dB=0$ in any open set where $\sigma$ is single-valued, it follows that $F_L=F_R$ globally. This generalises the flux quantisation argument of \cite{Martelli:2008si} to allow for a non-vanishing but trivial gauge field and applies to the full theory, not just the moduli space (but only for the overall $\U(1)$ fluxes). Since $H = F_L+F_R$ we have
\begin{equation}\label{Hflux}
\int d H=\int \frac{1}{2}\epsilon^{\mu\nu\lambda}\partial_\mu H_{\nu\lambda} \in \frac{4\pi } {n}\mathbb{Z}
\end{equation}
and $\sigma$ has period $2\pi$. Note that since $e^{i\theta}$ is a $\U(n)$ transformation, $\theta$ also has period $2\pi$. Thus we can fix the $\U(1)_B$ symmetry using (\ref{fix}) and set $\sigma=0\ {\rm mod }\ 2\pi$. However, this periodicity imposes an additional identification on the $\U(1)$-invariant fields
\begin{equation}\label{U(1)b}
\hat Z^A \cong e^{2\pi  i/k}\hat Z^A\qquad \textrm{and} \qquad \hat \psi_A \cong
e^{2\pi  i/k}\hat \psi_A\ .
\end{equation}

We are therefore told that the $\U(n)\times \U(n)$ ABJM theory is equivalent to a $\mathbb{Z}_k$ identification on the $(\SU(n)\times \SU(n))/\mathbb Z_n$-theory. Note that the $\mathbb Z_n$ quotient arises here as the relative part of the two $\mathbb Z_n$ factors from $\U(n)\simeq (\U(1)\times \SU(n))/\mathbb Z_n$.

However we need to be careful since  there could be obstructions at a global level.  To look for the latter  it is insightful to compute the moduli  space of the  $(\SU(n)\times \SU(n))/\mathbb
Z_n$-theory along with a ${\mathbb Z}_k$ orbifold and compare it to  the  $\U(n)\times \U(n)$ result.

For a general $n$ the
vacuum moduli space is obtained by setting
\begin{equation}
Z^A = {\rm diag}(z_1^A,...,z_n^A)\ .
\end{equation}
If we consider gauge transformations of the form $g_L=g_R$ then $Z^A$ behaves as if it were in the adjoint of $\SU(n)$ and hence cannot tell the difference between the $\SU(n)$ and $\U(n)$ theories. The result is that the gauge transformations which preserve the form of $Z^A$ simply interchange the eigenvalues $z_i^A$ leading to the symmetric group acting on the $n$ M2-branes, just as is the case in D-brane theories.

Next we can consider transformations in the diagonal subgroup of $\SU(n)$ or $\U(n)$. These act to rotate the phases of the $z^A_i$, however in the $\SU(n)$-theory they only do so up to the constraint that the diagonal elements must have unit determinant. In the $\U(n)$-theory this is not the case and there are $n$ independent $\U(1)$'s, one for each $z_i^A$, and each of these $\U(1)$'s leads to a $\mathbb{Z}_k$ identification on the moduli space. Thus for $\U(n)$ we indeed see that we find $n$ commuting copies of $\mathbb{Z}_k$ along with the symmetric group acting on the $z_i^A$.

For the $\SU(n)$-theory, even including the $\mathbb{Z}_k$ action of $\U(1)_B$, this will not always be the case. In particular, note that since the determinant of the gauge transformations coming from $\SU(n)$ is always one we have, for an arbitrary element of the moduli space orbifold group,
\begin{equation}
{\rm det} (g_{\U(1)}^{l_B} g_{0}) = {\rm det} (g_{\U(1)}^{l_B} ) = e^{2\pi i nl_B/k}\ .
\end{equation}
Here $g_{0}$ represents a generic element of the moduli space orbifold group obtained in the $(\SU(n)\times \SU(n))/\mathbb Z_n$-theory. On the other hand, the moduli space orbifold group of the $\U(n)$-theory generated by $n$ independent $\U(1)$'s has
\begin{equation}
{\rm det} (g_{1}^{l_1}...g_n^{l_n}) =  e^{2\pi i(l_1+...+l_n)/k}\ .
\end{equation}
If these two theories are to give the same moduli space then we must be able to have $e^{2\pi i
(l_1+...+l_n)/k}=e^{2\pi i n l_B/k}$ for any possible combination of $l_i$'s. Thus we are required to solve
\begin{equation}\label{llb}
l= n\;l_B\ {\rm mod}\;k\ ,
\end{equation}
for $l_B$ as a function of $l,n,k$, where $l = l_1+...+l_n$ is arbitrary. Hence, if this equation can be solved for $l_B$ then $g_{0}=e^{-2\pi i l_B/k}g_{1}^{l_1}...g_n^{l_n}$ is an element of $\SU(n)$ and can arise from the vacuum moduli space quotient group of the $(\SU(n)\times \SU(n))/\mathbb Z_n$-theory.

We will now show that (\ref{llb}) has solutions for all $l$ if and only if $n$ and $k$ are co-prime. In general the solution is $l_B = (l-pk)/n$ for any $p\in \mathbb{Z}$; however we require that $l_B$ is an integer. It is clear that we may view $l,k$ and $p$ as elements of $\mathbb{Z}/\mathbb{Z}_n$ and we are therefore required to solve the following equation for $p$
\begin{equation}
l=pk \ {\rm mod}\ n\ .
\end{equation}
This always has solutions if the map $\varphi:p\mapsto pk$ is surjective on $\mathbb{Z}/\mathbb{Z}_n$. Since $\mathbb{Z}/\mathbb{Z}_n$ is a finite set this will be the case if and only if $\varphi$ is also injective. Thus we wish to show that $pk=p'k\ {\rm mod}\ n$ implies $p=p'$. This is equivalent to showing that $qk=0\ {\rm mod}\ n$ implies $q=0\ {\rm mod}\ n$. Now suppose that $qk=rn$. If $k$ and $n$ are co-prime then all the prime factors of $k$ must be in $r$ and all the prime factors of $n$ must be in $q$. Thus $q=0\ {\rm mod}\ n$. On the other hand if $k$ and $n$ have a common factor $d$ then we find a non-zero solution by taking $q =n/d$ and $r=k/d$. Thus $qk=0\ {\rm mod}\ n$ has no non-trivial solutions for $q$ if and only if $n$ and $k$ are co-prime.

This result can been restated as follows: Although locally $\U(n)\simeq \U(1)\times \SU(n)$, this is not true globally. Even though the lagrangian is defined by local information at the Lie-algebra level, the map we constructed, reducing the $\U(n)\times \U(n)$-theory to a $\mathbb{Z}_k$ quotient of the $(\SU(n)\times \SU(n))/\mathbb Z_n$-theory, involves finite gauge transformations and is therefore sensitive to global properties of $\U(n)$. The above discussion shows that the vacuum moduli space quotient group of the $\U(n)\times \U(n)$ theories is not of the form $\mathbb{Z}_k\times G_0$, where $G_0\subset\SU(n)$, unless $n$ and $k$ are relatively prime \cite{Lambert:2010ji}.

We have therefore shown that if $n$ and $k$ have a common factor then the vacuum moduli spaces for the two theories do not agree, as there is a global obstruction to mapping the $\U(n)\times \U(n)$-theory to a $\mathbb{Z}_k$ quotient of the $(\SU(n)\times \SU(n))/\mathbb Z_n$-theory. On the other hand, if $n$ and $k$ are co-prime then the vacuum moduli space calculated in the $(\SU(n)\times \SU(n))/\mathbb Z_n$-theory, along with the $\mathbb{Z}_k$ identification coming from $\U(1)_B$, agrees with the vacuum moduli space of the $\U(n)\times \U(n)$-theory. This suggest that there is no global obstruction and in these cases the $\U(n)\times \U(n)$ theories are $\mathbb{Z}_k$ quotients of the $(\SU(n)\times \SU(n))/\mathbb Z_n$ theories. Thus one can conjecture that:

\begin{itemize}
\item $\U(n)_k\times \U(n)_{-k}$ is equivalent to a ${\mathbb Z}_k$ quotient of $(\SU(n)_k\times \SU(n)_{-k})/{\mathbb Z}_n$  if $k$ and $n$ are relatively prime\;.
\end{itemize}

\subsubsection{Connecting to the BLG models}\label{connectingtoBJG}

We are finally in position to connect this discussion with the moduli space results for BLG theory, obtained in Section~\ref{BLGmodulispace}. Our analysis implies that for $n=2$ and $k$ odd, the ${\cal N}=6$ ABJM models can be viewed as ${\mathbb Z}_k$ orbifolds of the ${\cal N}=8$ model with gauge group $(\SU(2) \times \SU(2))/\mathbb Z_2$. Clearly, for $k=1$, the ABJM model for two M2-branes in $\mathbb R^8$ is precisely the ${\cal N}=8$ $(\SU(2)\times \SU(2))/{\mathbb   Z}_2$ model of \cite{Bagger:2006sk,Bagger:2007vi,Bagger:2007jr, Gustavsson:2007vu}.

One can also find a connection, which does not fit the above pattern, when the moduli space is $({\mathbb C}^4 \times {\mathbb C}^4) /\mathbb D_{8}$. This arises from (\ref{modspace}) for  $k=2$ in the $\SU(2)\times \SU(2)$ theory and for $k=4$ in the $(\SU(2)\times \SU(2))/{\mathbb Z}_2$ theory.  It can be identified with the moduli spaces of two M2-branes in ${\mathbb R}^8/{\mathbb   Z}_2$  by introducing
\begin{eqnarray}
r_1^A = z_1^A + z_2^A\ ,\qquad r_2^A = i(z_1^A - z_2^A)\;,
\end{eqnarray}
so that in the language of Section~\ref{BLGmodulispace},
\begin{eqnarray}
\nonumber  g_{12}:\quad  r^A_1 \cong r^A_1\ ,&&r^A_2 \cong -r^A_2 \\
g_{12}\; g^2_{\SU(2)}:\quad r^A_1 \cong -r^A_1\ , && r^A_2 \cong r^A_2 \\
 \nonumber  g_{\SU(2)}\; g_{12}:\quad r^A_1 \cong r^A_2 \ ,&& r^A_2 \cong r^A_1\;.
\end{eqnarray}
These are indeed the identifications expected for the moduli space $(\mathbb R^8/\mathbb Z_2 \times \mathbb
R^8/\mathbb Z_2)/\mathbb Z_2$ of two M2-branes located at a $\mathbb Z_2$ orbifold singularity of M-theory.

Hence we have seen that for $k=1$, the $(\SU(2)\times \SU(2))/{\mathbb Z}_2$ theory is precisely the ABJM model at level $k=1$. We have also seen that the $k=2$ $\SU(2)\times \SU(2)$ theory  \cite{Lambert:2008et} and the $k=4$ $(\SU(2)\times \SU(2))/{\mathbb Z}_2$ theory \cite{Bashkirov:2011pt} both have the correct moduli spaces to describe two M2-branes in ${\mathbb R}^8/{\mathbb Z}_2$. Indeed, there are two such theories expected, corresponding to the presence or absence of discrete torsion. Therefore it is natural to identify them with the $k=2$ $\U(2)\times \U(2)$ ABJM and $k=2$ $\U(2)\times \U(3)$ ABJ models respectively.

  In summary, the following ${\cal N}=6$ ABJ(M) theories are dual to ${\cal N}=8$ BLG models:
\begin{itemize}
\item $\U(2)\times \U(2)$ is dual to $(\SU(2)\times \SU(2))/{\mathbb Z}_2$, both at $k=1$;
\item $\U(2)\times \U(2)$ is dual to $\SU(2)\times \SU(2)$, both at $k=2$;
\item $\U(2)\times \U(3)$ at $k=2$ is dual to $(\SU(2)\times \SU(2))/{\mathbb Z}_2$ at $k=4$ .
\end{itemize}
These proposed dualities have also been  tested  non-trivially  by showing that their superconformal indices agree \cite{Bashkirov:2011pt}.


\section[Analysis of the theory II: advanced topics]{{\Large{\bf Analysis of the theory II: advanced topics}}}\label{chapter6}

 In this section we continue to analyse ABJM theory by investigating some of its most puzzling features. We start by focusing on the special role of momentum along the M-theory circle and its relation to ``monopole'' or `` 't Hooft'' operators. After this we discuss hidden symmetries arising at low values of the Chern-Simons level $k$. We then include couplings to Ramond-Ramond background fields in the ABJM lagrangian and discuss the ensuing mass-deformed version of the theory. Finally, we discuss the physical and geometric interpretation for the vacua of the mass-deformed ABJM theory in terms of dielectric M2-branes in M-theory.

\subsection{11D momentum, fluxes and 't Hooft operators}\label{fluxesandtHooft}

Let us look more carefully at some subtle features of the ABJM lagrangian.
For this we initially work with the $\U(1)\times\U(1)$ theory and decompose the complex scalars into their magnitude and phase:\footnote{These $\theta$'s should not be confused with the $\vartheta_i$ coordinates introduced above \eref{ncore}.} $Z^A =R^Ae^{i\theta^A}/\sqrt{2}$. Then the bosonic part of the lagrangian is
\begin{equation}
{\cal L} = - \frac{1}{2} \sum_{A=1}^4\partial_\mu R^A\partial^\mu R^A - \frac{1}{2}  \sum_{A=1}^4 (R^A)^2(\partial_\mu\theta^A-B_\mu)(\partial^\mu\theta^A-B^\mu) + \frac{k}{4\pi}\varepsilon^{\mu\nu\lambda}B_\mu\partial_\nu Q_\lambda\;,
\end{equation}
where as before we have defined $B_\mu = A^L_\mu  -A^R_\mu$ and $Q_\mu = A^L_\mu + A^R_\mu$. Under a $\U(1)_B$ transformation we have
$\theta^A\to \theta^A +\lambda$ and $B_\mu\to B_\mu + \partial_\mu\lambda$. Thus the centre-of-mass component $\theta = \sum_A\theta^A$ can be set to zero by a gauge transformation. As we have already seen, this direction (the common phase of the four complex coordinates) plays the role of the M-theory circle. This leads to a puzzle: where has the eleven-dimensional momentum gone?

To answer this we  compute the corresponding hamiltonian. The conjugate momenta are
\begin{eqnarray}
\nonumber \Pi_{R^A} &=& \partial_0 R^A\\
\Pi_{\theta^A} &=& (R^A)^2(\partial_0\theta^A - B_0)\\
\nonumber \Pi_{Q_j}& =& \frac{k}{4\pi}\epsilon^{ij}B_i \\
\nonumber \Pi_{Q_0}& =& \Pi_{B_0}=0\ ,
\end{eqnarray}
where $i,j=1,2$.
Thus the hamiltonian is
\begin{eqnarray}
H  &=  &\int d^2x   \left\{\frac{1}{2}\sum_A\Pi_{R_A}^2 + \frac{1}{2}\sum_A(R^A)^{-2}\Pi_{\theta^A}^2 +\frac{1}{2}\sum_A(\partial_iR^A)^2 + \frac{1}{2}\sum_A(R^A)^2(\partial_i\theta^A - B_i)^2\right. \\
\nonumber &&\qquad \left.  -\frac{k}{4\pi}F_{12}Q_0+ \left(\sum_A\Pi_{\theta^A}+\frac{k}{4\pi}H_{12}\right)B_0\right\}\;,
\end{eqnarray}
where $H = dQ = dA^L+dA^R$ and $F = dB=dA^L - dA^R$.

We see that, as is always the case in gauge theories, the Hamilton equations for $B_0$ and $Q_0$ impose constraints. In particular we have  that
\begin{equation}
F_{12}=0\ ,
\end{equation}
so that the magnetic fluxes of $A^L$ and $A^R$ are always equal. However, we also find
\begin{equation}\label{aflux}
 H_{12} = -\frac{4\pi}{k }\sum_A\Pi_{\theta^A}\;.
\end{equation}
Thus we have established that turning on momentum around the M-theory circle is equivalent to turning on $H=dA^L+dA^R$ magnetic flux.  Indeed, since $F^L$ and $F^R$ have quantised fluxes, we have $\int d^2 x\, F^L = \int d^2 x\, F^R \in 2\pi  {\mathbb Z}  $  and hence
$$
\int d^2x\, H_{12}   \in 4\pi {\mathbb Z}\ ,
$$
so that
$$
\sum_A P_{\theta^A} \in k {\mathbb Z}\;,
$$
where $P_{\theta^A} = \int d^2 x\, \Pi_{\theta^A}$ is the total momentum. This is consistent with the ${\mathbb Z}_k$ orbifold which projects out momentum modes that are not a multiple of $k$.

Let us now repeat this analysis for the general $\U(n)\times \U(n)$ ABJM model. The lagrangian is
\begin{eqnarray}
{\cal L} &=& -  \Tr (D_\mu Z^A D^\mu \bar Z_A) -i\Tr(\bar\psi^A\gamma^\mu D_\mu \psi)+{\cal L}_{Yukawa}- V \\[2mm]
\nonumber &&+ \frac{k}{4\pi}\varepsilon^{\mu\nu\lambda}\Tr \left(A^L_\mu\partial_\nu A^L_\lambda  - \frac{2i}{3}A_\mu^LA_\nu^LA^L_\lambda\right)- \frac{k}{4\pi}\varepsilon^{\mu\nu\lambda}\Tr \left(A^R_\mu\partial_\nu A^R_\lambda  - \frac{2i}{3}A_\mu^RA_\nu^RA^R_\lambda\right)\;,
\end{eqnarray}
where ${\cal L}_{Yukawa}$ represent the Yukawa-type terms of the form $\bar \psi ZZ\psi$. The hamiltonian is given  by
\begin{eqnarray}
H&=& \int d^2 x\   \Tr (\Pi_{Z^A}\Pi_{\bar Z_A})+\Tr (D_i Z^A D^i\bar Z_A)-{\cal L}_{Yukawa} +V\\
\nonumber &&+ \Tr  \left(iZ^A\Pi_{Z^A}-i\Pi_{\bar Z_A}\bar Z_A +i\psi_A\psi^A- \frac{k}{2\pi}F^L_{12}\right)A^L_0
+ \Tr  \left(i\bar Z_A\Pi_{\bar Z_A} -i\Pi_{Z^A}Z^A -i\psi^A\psi_A+ \frac{k}{2\pi}F^R_{12}\right)A^R_0\;.
\end{eqnarray}
Thus we find the constraints
\begin{eqnarray}\label{naflux}
\nonumber  \frac{k}{2\pi}F^L_{12} &=& iZ^A\Pi_{Z^A}-i\Pi_{\bar Z_A}\bar Z_A +i\psi_A\psi^A\\
  \frac{k}{2\pi}F^R_{12}&=&i\Pi_{Z^A}Z^A-i\bar Z_A\Pi_{\bar Z_A} +i\psi^A\psi_A \ .
\end{eqnarray}

Let us look at the massless centre-of-mass modes. As was noted in the moduli space analysis of the previous chapter, in a generic vacuum the $Z^A$ all commute. Thus we can write
\begin{equation}
Z^A =
\left(
\begin{array}{cccc}
 z^A_1 &   & &  \\
  &  z^A_2 &   &\\
  &   &\ddots   &\\
  & & &z^A_n
\end{array}
\right)\ ,
\end{equation}
and set the massive off-diagonal fields and fermions to zero. If we write $z^A_i = R^A_i e^{i\theta^A_i}/\sqrt{2}$, we find
\begin{equation}
F^L_{12} = F^R_{12} = -\frac{2\pi}{k}
\left(
\begin{array}{cccc}
 \sum_A \Pi_{\theta^A_1} &   & &  \\
  &   \sum_A \Pi_{\theta^A_2} &   &\\
  &   &\ddots   &\\
  & & & \sum_A \Pi_{\theta^A_n}
\end{array}
\right)\;,
\end{equation}
which generalises the previous $\U(1)\times \U(1)$ case to the centre-of-mass motion of each of the M2-branes.  Thus we see once again that to include momentum modes around the M-theory circle we must turn on magnetic fluxes. In particular for the centre-of-mass coordinates we must turn on magnetic fluxes in the Cartan subalgebra, but more generally for any component of the fields we can introduce a corresponding flux to give it M-theory momentum \cite{Lambert:2011eg,Aharony:2008ug}.

\subsubsection{Group theory analysis of 't Hooft operators}\label{thooft}

So far our discussion has been classical. However the relations (\ref{aflux}) and (\ref{naflux}) are constraints and as such we must also impose them in the quantum theory. Therefore we continue to identify flux quantisation with the momentum  around the M-theory circle. However in the quantum theory we will need to include operators that create or destroy units of momentum and hence flux.

Such an operator is called an 't Hooft operator (or sometimes a  monopole operator). They were first introduced into gauge theories in \cite{'tHooft:1977hy} and can be specified by saying that they create a given flux through closed surfaces around some insertion point. So in other words the 't Hooft operator is specified by giving the (Euclidean) spacetime point $x_0$ and flux -- or equivalently, the singular behaviour of the gauge field at that point (in Euclidean space)
\begin{equation}\label{singularity}
F = \star \frac{Q_M}{2}d\left(\frac{1}{|x-x_0|}\right)  + {\rm nonsingular}\;,
\end{equation}
where $Q_M\in \uf(n)\times \uf(n)$ is the magnetic flux, and is subject to the standard Dirac quantisation condition
\begin{equation}
e^{2\pi i Q_M}  = 1\ .
\end{equation}
These operators should be viewed as prescribing the behaviour of the fields at the insertion point  in the path integral and hence are local.

A famous result of Goddard, Nuyts and Olive (GNO) \cite{Goddard:1976qe} asserts that the solution to the Dirac quantisation condition is such that $Q_M $ is determined, up to a gauge transformation, by a dominant weight of the ``magnetic'' dual gauge group.\footnote{We will have to assume some knowledge of group theory concepts at this stage, which can be found \eg in \cite{O'Raifeartaigh:1986vq}.} This dual gauge group is more commonly referred to (especially in the mathematical literature) as the Langlands dual. Therefore, for a gauge group denoted by $G$, the Langlands dual will be denoted by $^{L}G$.  The Dynkin diagram and hence the Lie algebra of the dual gauge group is obtained by mapping the simple roots $\vec\alpha_i$ of the original gauge group to the ``co-roots'': $\vec \alpha^{\vee}_i \equiv 2\vec\alpha_i/|\vec\alpha_i|^2$. To obtain the actual dual group one notes that the weights obtained from the flux $Q_M$ are in general a subset of all possible weights, corresponding to a dual group that is a quotient of the universal simply connected group associated to the dual Lie algebra. For example, as we will see shortly, the Langlands dual to $\U(n)$ is $\U(n)$ but the Langlands dual to $\SU(n)$ is $\SU(n)/{\mathbb   Z}_n$.

Let us illustrate this in the case at hand and for gauge group $\U(n)$. By conjugation, which is simply the action of the gauge group, we can choose to have $Q_M$ in the $\U(1)^{n}$ Cartan subalgebra
\be\label{GNO}
Q_M = \vec q\cdot \vec H + q \one\  ,
\ee
where $\vec H$ generates the Cartan subalgebra of $\SU(n)$ and $\one$ is the abelian $\U(1)$ generator ({\it i.e.} the identity operator). We can
think of $Q_M$  as a diagonal $n\times n$ matrix  $Q_M = {\rm diag} (q_1,...,q_n)$  in the fundamental representation with highest weight $\vec\lambda^{1}$. The states in this representation are  given by the  orthogonal  basis
\begin{equation}\label{rootstring}
|\vec\mu^1\rangle = |\vec\lambda^1\rangle\ ,  \quad |\vec\mu^2\rangle = |\vec\lambda^1-\vec\alpha_1\rangle \ ,  \quad |\vec\mu^3\rangle = |\vec\lambda^1-\vec\alpha_1-\vec\alpha_2\rangle \ ,\quad \ldots \ , \quad |\vec\mu^n\rangle =  |\vec\lambda^1-\vec\alpha_1-...-\vec\alpha_{n-1}\rangle \ ,
\end{equation}
where $\vec\alpha_i$, $i=1,...,n-1$ are the simple roots of $\SU(n)$. In particular the diagonal components are\footnote{Note that we are using $i$ to indicate the range $i=1,...,n-1$ as well as $i=1,...,n$ to avoid introducing additional symbols. }
\begin{equation}\label{qiis}
q_i  = \langle\vec\mu^i|Q_M|\vec\mu^i\rangle  = \vec q\cdot \vec\mu^i+q ,\qquad  i=1,...,n
\end{equation}
and the quantisation condition is simply that $q_i\in {\mathbb Z}$, $i=1,...,n$.

Following GNO, $\vec q$ should be a weight of $^L\SU(n)$ so we start by writing $\vec q  = w_i\; ^L\vec \lambda^i$. The $^L\vec\lambda^i$, $i=1,...,n-1$ are called the ``fundamental'' weights of $^L\SU(n)$ and are by definition dual to the $\SU(n)$ roots $\vec\alpha_i$: $^L\vec\lambda^i\cdot\vec\alpha_j=\delta^i_j$.\footnote{The $\vec \lambda^i$ that we used in \eqref{rootstring} are the fundamental weights of $\SU(n)$ and dual to the co-roots.} Fundamental weights form a basis in which we can expand any weight of the group. To determine the coefficients $w_i$ in terms of the fluxes we note that, for $i=1,...,n-1$, we have $w_i = \vec q\cdot \vec\alpha_i$. On the other hand we see that, from the basis (\ref{rootstring}) and definition (\ref{qiis}),
\be
 w_i = \vec q\cdot \vec\alpha_i = \vec q\cdot (\vec\mu^i- \vec\mu^{i+1})=q_{i}-q_{i+1}\ ,\qquad i=1,...,n-1\ .
\ee
Thus we find
\be\label{qexpand1}
\vec q =(q_1-q_{2})\; ^L\vec\lambda^1+(q_2-q_{3})\; ^L\vec\lambda^2...+(q_{n-1}-q_{n})\; ^L\vec\lambda^{n-1}\  \;,
\ee
which is indeed a weight of $^L\SU(n)$ since the $q_i$'s are quantised. We also need to  specify the abelian $\U(1)$ charge $q$. To fix this we see that by tracing over \eref{GNO} we get $\sum_i^n \vec\mu^i=\vec 0$ and therefore
\be
q = \frac{1}{n}(q_1+...+q_n)\ .
\ee

The observation of GNO is that within the Cartan subalgebra we can still act with gauge symmetries (which are just conjugations) in the Weyl subgroup to order the diagonal components, $q_1\ge q_2 \ge ...\ge q_n$. In this case $\vec q$ is what is called a ``dominant'' weight of $^L\U(n)$. An important result in representation theory states that dominant weights are in one-to-one correspondence with finite dimensional irreducible representations of the group.  Hence, appropriate choices of fluxes $q_i$ fully characterise irreducible representations of $^L\U(n)$, with the coefficients of \eqref{qexpand1} playing the role of Dynkin labels.

In fact, for $\U(n)$ we have no restrictions on $Q_M$ and one can clearly arrange for any weight by choosing the $q_i$ appropriately. Moreover, for this particular choice of group and in our normalisations, $^L\vec \lambda^i = \vec \lambda^i$ and thus the Langlands dual to $\U(n)$ is $^L\U(n) = \U(n)$.

However, it is interesting to note that all weights need not always arise: Consider $\SU(n)$, where we must impose the constraint $q_1+...+q_n=0$.  In this case one can show that $\vec q$ can also be written as
\be\label{qexpand2}
\vec q = q_1\vec \alpha_1 +(q_1+q_2)\vec\alpha_2 +...+(q_1+q_2+...+q_{n-1})\vec\alpha_{n-1}
\ee
and $\vec q$ is therefore also a root of $\SU(n)$.\footnote{One can recover the expansion \eqref{qexpand1} from \eqref{qexpand2} by simply using the root inner products. These can in turn be read off from the Cartan matrix of $\SU(n)$.} Roots are weights of the adjoint representation, which in turn is blind to the centre of the group. This means that the only representations that appear are those of $\SU(n)/{\mathbb Z}_n$ and the Langlands dual to $\SU(n)$ is $^L\SU(n)=\SU(n)/{\mathbb Z}_n$.

Since dominant weights arise as the highest weights of finite dimensional irreducible representations, GNO also conjectured that monopoles come in representations of the dual group with highest weight $\vec q$.  To see how this works in a familiar physical example, consider 4-dimensional maximally supersymmetric $\SU(2)$ gauge theory. The perturbative spectrum contains gauge fields in the adjoint representation. The nonperturbative spectrum includes monopoles corresponding to a given flux. In particular the monopole and anti-monopole have $Q_M=\pm{\rm diag}(1,-1)$, but are both mapped by the GNO prescription to the single highest weight $2\vec\lambda^1$.  Nevertheless these monopoles are physically distinct, despite the fact that their fluxes can be mapped to each other by a gauge transformation, because the theory is also specified by the VEV of the scalar fields. In other words, the monopole is determined by a scalar field of the form
\be
\Phi = {\rm diag}(v,-v) + \frac{Q_M}{4\pi r}+\ldots
\ee
and thus the  gauge transformation that maps $Q_M\to -Q_M$ also changes the vacuum. If we ask that we keep the vacuum fixed then we can no longer use gauge transformations to map $Q_M$ into the form with $q_1\ge q_2 \ge ...\ge q_n$.
In this case one sees that  the monopole and anti-monopole, along with the zero-flux state,  form a representation of $\SU(2)/{\mathbb Z}_2$ with highest weight $2\vec\lambda^1$ corresponding to the adjoint representation.

Another example is the case of $\U(3)$ with highest weight $\vec q=2\vec\lambda^1$ and charge $q=2$, corresponding to the symmetric representation. The fluxes in this representation are easily found to be
\begin{equation}
\left(
\begin{array}{ccc}
2  & 0  & 0  \\
 0 & 0  & 0  \\
 0 & 0  & 0
\end{array}
\right)\ ,\left(
\begin{array}{ccc}
1  & 0  & 0  \\
 0 & 1  & 0  \\
 0 & 0  & 0
\end{array}
\right)\ ,\left(
\begin{array}{ccc}
0  & 0  & 0  \\
 0 & 2  & 0  \\
 0 & 0  & 0
\end{array}
\right)\ ,\left(
\begin{array}{ccc}
1  & 0  & 0  \\
 0 & 0  & 0  \\
 0 & 0  & 1
\end{array}
\right)\ ,\left(
\begin{array}{ccc}
0 & 0  & 0  \\
 0 & 1  & 0  \\
 0 & 0  & 1
\end{array}
\right)\ ,\left(
\begin{array}{ccc}
0 & 0  & 0  \\
 0 & 0  & 0  \\
 0 & 0  & 2
\end{array}
\right)\  .
\end{equation}
Although some of the fluxes can be mapped to each other using gauge transformations, they correspond to physically distinct monopoles in the gauge theory.   Such gauge transformations also act on the VEVs of the scalar fields, so they cannot be used to identify fluxes. Therefore in a generic vacuum, the fluxes within a representation are physically distinct.  This implies that the 't Hooft  operators assemble into multiplets that transform under the dual gauge group.  In the next section we will see that in a Chern-Simons gauge theory, the 't Hooft operators transform in representations of the dual gauge group that are indeed determined by the fluxes.

We also note that the second flux in this representation:
\be
\left(
\begin{array}{ccc}
1  & 0  & 0  \\
 0 & 1  & 0  \\
 0 & 0  & 0
\end{array}
\right)\ ,
\ee
is already ordered to $q_1\ge q_2\ge q_3$ and thus maps to the dominant weight $\vec\lambda^2$. One may try to view this as the highest weight of the anti-fundamental representation of $\SU(3)$. However since it has $\U(1)$ charge $2$ this cannot be viewed as  a representation of $\U(3)$ (but rather  $\SU(3)\times \U(1)$).

\subsubsection{'t Hooft operators in ABJM}

Let us now consider 't Hooft operators in Chern-Simons theories and in particular ABJM. More detailed discussions can be found in \cite{Borokhov:2002ib,Borokhov:2002cg,Borokhov:2003yu,Klebanov:2008vq,Hosomichi:2008ip, Benna:2009xd, Gustavsson:2009pm, Kwon:2009ar,Kim:2009ia, Kim:2010ac, Bashkirov:2010kz, Kapustin:2010xq, Martinec:2011}. In particular, let us take the Euclideanised theory and denote by ${\cal M}_{Q_M}(x)$ an 't Hooft operator which creates a flux
\be
\frac{1}{2\pi}\int_{S^2} F^L=\frac{1}{2\pi}\int_{S^2} F^R  =  Q_M \in \mathfrak{u}(1)^n\ ,
\ee
on arbitrarily small spheres surrounding the point $x$.  An alternative view of 't Hooft operators can be found in conformal field theory using the operator-state mapping. In particular, through conformally mapping ${\mathbb R}^3$ to $\mathbb R\times S^2$ one can replace the insertion of an 't Hooft operator at a point to the creation of a state at $t=-\infty$ which carries magnetic flux $Q_M$ through $S^2$.

Our first observation is that, because of the Chern-Simons term, a single 't Hooft operator is not invariant under  gauge transformations. Since the flux  lies in the Cartan subalgebra, the 't Hooft operator breaks the gauge group $G= \U(n)\times \U(n)$ down to the diagonal $\U(1)^{n}\times \U(1)^n$ subgroup.  To see how the 't Hooft operator transforms under this group,   consider an infinitesimal gauge  transformation generated by $\omega_L(x),\ \omega_R(x)$, which  we assume to vanish  at infinity. Although the flux $Q_M$ remains unchanged, because of the Chern-Simons term   we find
\bea\label{charges}
 {\cal M}_{Q_M}(x)  & \to & e^{(ik/2\pi)\, tr \int ( D\omega_L \wedge  F^L-D\omega_R\wedge F^R)}  {\cal M}_{Q_M}(x)\nonumber \\
& = &e^{i k\, tr((\omega_L(x)-\omega_R(x)) Q_M)}{\cal M}_{Q_M}(x)\ .
\eea
Thus  an 't Hooft operator that creates a given flux $Q_M$ is not gauge invariant. However, by taking several such 't Hooft operators together, we can create a multiplet of local operators that transforms under some representation of the gauge group.
To obtain the transformation of ${\cal M}_{Q_M}$ under the full $\U(n)\times \U(n)$ gauge group, we use the method of induced representations, based on the charges of 't Hooft operators under the $\U(1)^n\times \U(1)^n$ subgroup. Hence we see that, following the GNO map from charges to weights constructed above,  we can identify a multiplet of  't Hooft  operators with the various operators obtained from fluxes that appear in an irreducible representation with highest weight
\be
\vec \Lambda  = k\vec q\oplus-k\vec q\;.
\ee
 In particular, the $\U(1)^n\times \U(1)^n$ charges of the states in this representation agree with  (\ref{charges}), essentially by construction.  We note that the second factor is not a dominant weight but rather the negative of a dominant weight. We therefore identify the representation of the $\U(n)_R$ factor as the Cartan dual representation of $\U(n)_R$ (whose lowest weight is $-k\vec q$) and in order to avoid unnecessary notation we will not always write out the two factors of the highest weight.  Thus taking  the multiplet of operators corresponding to the fluxes found in the representation with highest weight  $k\vec q\oplus-k\vec q$ we obtain an 't Hooft operator
\be
{\cal M}_{\vec\Lambda}(x)\ ,
\ee
which transforms under gauge transformations in the representation of $\U(n)\times \U(n)$ with highest weight $k\vec q\oplus-k\vec q$.

Secondly, we want to consider {\it supersymmetric} 't Hooft operators. This means that the classical field configuration near the insertion point needs to preserve some fraction of the supersymmetry. As we will see, this means that in addition to a singularity in the gauge field, we must also require a singularity in the scalar fields. Assuming that the fermions vanish and that the scalar fields remain in the vacuum moduli space  - {\it i.e.} commuting but not necessarily constant - the supersymmetry variation is
$$
\delta \psi_A = \gamma^\mu D_\mu Z^B\epsilon_{BA}\ .
$$
To proceed we note that we are looking in the Euclidean regime. In such cases we do not need to look for real solutions. In particular if we expand $Z^A=(X^A+iY^{A})/\sqrt{2}$ in terms of real scalars then when we look for solutions in the Euclidean regime we no longer require that $X^A$, $Y^{A}$ be real. This means that we should not identify $Z^A$ and $Z_A$ as complex conjugates of each other but rather as independent fields. Similarly we no longer require that $\epsilon_{AB}^* = \epsilon^{AB}$, although we still impose $\epsilon^{AB} = \frac{1}{2}\varepsilon^{ABCD}\epsilon_{CD}$ \cite{Hosomichi:2008ip}.

With this in mind we can find supersymmetric configurations by taking a single scalar, say $Z^1$, and setting
$$
D_\mu {Z}^1=0\ ,
$$
but not $D_\mu {\bar Z}_{1}\ne 0$. This will then preserve the supersymmetries $\epsilon_{1A}$, $A=2,3,4$ {\it i.e.} half. Note that since we do not assume that $\epsilon_{AB}^*=\epsilon^{AB}$ the remaining supersymmetries $\epsilon_{AB}$, $A,B\ne 1$ and $\epsilon^{1A}$ are not related to $\epsilon_{1A}$ and hence can still be non-vanishing.

The equations of motion for the gauge field can now be written as
\begin{eqnarray}
\frac{k}{2\pi}\varepsilon_{\mu\nu\lambda} F^{\nu\lambda}_L &=& iZ^1D_\mu {\bar Z}_1  = D_\mu(iZ^1{\bar Z}_1) \\
\frac{k}{2\pi}\varepsilon_{\mu\nu\lambda} F^{\nu\lambda}_R &=& i D_\mu {\bar Z}_1Z^1  = D_\mu(i{\bar Z}_1Z^1)\nonumber\;.
\end{eqnarray}
Note that if $Z_1$ remains in the vacuum moduli space, \ie commuting,  then $F_L=F_R$.  In addition the Bianchi identity implies that $D_\mu D^\mu( Z^1 \bar Z_1)=0$. Thus a supersymmetric 't Hooft operator  is determined by an harmonic function which takes values in the Cartan-subalgebra and whose pole defines the magnetic charge $Q_M$, {\it c.f.} (\ref{singularity}):
\begin{equation}
Z^1{\bar Z}_1 = -\frac{2\pi i}{k}\frac{Q_M}{|x-x_0|}  + {\rm nonsingular}\ .
\end{equation}

The 't Hooft operators can be used to construct important observables in ABJM. As we have seen they are crucial for obtaining states with momentum around the M-theory circle. Indeed the circle action corresponds to $\U(1)_B$ rotations which have been gauged. Thus if we try to construct gauge invariant operators out of the local fields then they must be neutral with respect to $\U(1)_B$ and hence  have zero momentum around the M-theory circle. For example if we concentrate on the scalars then the only gauge invariant operators are analogues of spin-chains in the AdS$_5$/CFT$_4$ correspondence:
\begin{equation}
 \Tr (Z^A{\bar Z}_BZ^C....)\;.
\end{equation}
But these all carry zero $\U(1)_B$ charge and hence are invariant under rotations of the M-theory circle. However, we can rectify this by allowing operators such as
\begin{equation}
  \Tr ({\cal M} _{\vec\Lambda}Z^{A_1}Z^{A_2}Z^{A_3}....Z^{A_p})
\end{equation}
so long as $\cal M$ is an  't Hooft operator with $\U(1)_B$ charge $-p$ and is in the dual to the  $p^{th}$ symmetric representation of $\U(n)\times \U(n)$. This corresponds to highest weight vector
\be
\vec \Lambda= p\vec\lambda^{n-1}\ ,
\ee
and hence to $q_1=...=q_{n-1}=0$, $q_n=-p/k$. Note once again that this exists as long as $p$ is a multiple of $k$, {\it i.e.} it only allows for eleven-dimensional momenta that are multiples of $k$, in agreement with expectations from the ${\mathbb Z}_k$ orbifold projection.

\subsubsection{Abelian 't Hooft operators}

The 't Hooft operators that we have discussed might seem somewhat abstract in nature. So let us try to shed some light on these operators in the simpler abelian case.
In fact, we have already seen examples of abelian 't Hooft  operators -- the exponentials $e^{i \sigma(x)}$ --
in the moduli space computation of Section~\ref{vacmodsp}. Imagine now that we are working with $\U(1)^n\times \U(1)^n$ ABJM theory.

Let us first write down the total charges under the groups $\U(1)^i_B$ and $\U(1)^i_Q$ obtained by taking the difference and sum, respectively, of the gauge fields in the original groups
\bea
{\cal Q}^i_B &=& \frac{k}{4\pi}\int H^i_{12}+\Sigma_A\Pi_{\theta_i^A}\nn\\
{\cal Q}^i_Q &=& \frac{k}{4\pi}\int F^i_{12}\;.
\eea
Both of the total charges are constrained to be zero by gauge invariance. The second equation says the fluxes associated to $A^{Li},A^{Ri}$ are equal while the first equation equates the common flux to the charge coming from the matter current (which in turn we have identified with momentum in the M-direction).

Now recall from the analysis following \eref{LdH} that
the fields $\sigma_i$ obtained by dualising $H^i$ are shifted by $k\theta_i$ under the $i^{\rm th }$ $\U(1)_B$ subgroup. It follows that
\begin{equation}
{\cal M} _i= e^{i \sigma_i(x)}\ ,
\end{equation}
has charge $k$ under this subgroup. Furthermore inserting $p$ factors of $e^{i\sigma_i(x)}$ into the path integral
\begin{equation}
Z = \int [dz^A_i][dA^L_i][dA^R_i]e^{i\int d^3x {\cal L}}\;,
\end{equation}
with $\cal L$ given by (\ref{LdH}), is equivalent to shifting
\begin{equation}
 \frac{1}{8\pi}\int d^3y\, \varepsilon^{\mu\nu\lambda} \partial_\mu H _{i\nu\lambda}(y)\to \frac{1}{8\pi}\int d^3y \left( \varepsilon^{\mu\nu\lambda}  \partial_\mu H _{i\nu\lambda}(y) + 8\pi p  \delta(x-y)\right) \  ,
\end{equation}
\ie\ it has created $p$ units of flux in the  $i^{\rm th }$ $\U(1)_L$ and $\U(1)_R$ subgroups with the singularity located at $x$. The gauge invariant coordinate $w^A_i = e^{-i\sigma_i/k}z^A_i$ on the moduli space is then identified with ${\cal M}_i^{-1/k}z^A_i$.

In this simple abelian case we also see that the 't Hooft operator is just a Wilson line for the $i^{\rm th}$ $\U(1)_B$ gauge field
\begin{equation}
e^{i\sigma_i(x)} = e^{i\int_\gamma d\sigma_i} = e^{ik\int_\gamma B_i}\;,
\end{equation}
where we used the fact, derived after \eref{LdH}, that $B_{i\,\mu}= \del_\mu\sigma_i/k$. The integral is over a curve $\gamma$ that ends at the spacetime point $x$ with some some starting fixed reference point (that we could take to be at infinity). However such an interpretation is not possible in the non-abelian theory and   one must use the definition above in terms of singularities of the fields in the path integral.

In the abelian example we can also see some important quantum properties of 't Hooft operators that we expect to also be valid in the non-abelian case. One such property is that the 't Hooft operator is covariantly constant
\be
D_\mu e^{i\sigma_i} =  (i\partial_\mu \sigma_i - ikB_{\mu i} )e^{i\sigma_i} =0\ .
\ee
This is more generally true for any $e^{ip\sigma_i}$, an operator that has charge $kp$, since the covariant derivative changes accordingly. Because of the absence of singularities (see below) this can be thought of as the product of $p$ coincident 't Hooft operators.

A related point is that, in conformal field theory,  $e^{i\sigma_i}$ has dimension zero. This is clear classically since $B_{\mu i}=\partial_\mu\sigma_i/k$ should have dimension one, but one may wonder whether or not it holds in the quantum theory. After all, in perturbative string theory we are used to the notion that, due to normal ordering effects, the free boson vertex operator $e^{ikX}$ is not dimensionless but rather has conformal dimension $\alpha'k^2$. However in the present case, since the momentum conjugate to $B_{\mu}^i = A_\mu^{Li}-A^{Ri}_\mu$ is
$ \varepsilon^{0\mu\nu}(k/2\pi)Q^i_\nu = \varepsilon^{0\mu\nu}(k/2\pi) (A_\nu^{Li}+A^{Ri}_\nu)$, there is no normal ordering ambiguity in the definition of $e^{ip\sigma_i}$, and therefore it has dimension zero also in the quantum theory.

In fact this argument strongly suggests that 't Hooft operators have dimension zero in the full ABJM theory. To see this first consider a vacuum where the scalar field VEV's have been sent to infinity. Here the theory is purely abelian and as above the 't Hooft operators have dimension zero. Let us now allow for finite scalar VEV's. Since the scalar VEV's have conformal-dimension $
\frac{1}{2}$ in three-dimensions, the conformal dimension of 't Hooft operators cannot depend on them. Hence the 't Hooft operators remain dimension zero at finite values of the scalar VEV's in the full ABJM theory. For alternative and more detailed discussions on the dimensions and R-charges of 't Hooft operators in Chern-Simons gauge theories, see \cite{Benna:2009xd}.

It is important to remember that although 't Hooft operators have dimension zero, they can still have non-trivial OPE's with other local operators.
In the abelian case one finds that $e^{i\sigma_i}$ does not commute with $H_{\mu\nu i} = \partial _\mu Q_{\nu i}- \partial _\nu Q_{\mu i}$, leading to a non-vanishing OPE representing the creation of magnetic charge at the insertion point of $e^{i\sigma_i}$. In the non-abelian case   we should expect  't Hooft operators that are, roughly speaking, constructed from $A^L_\mu-A^R_\mu$,  to  have non-trivial OPEs  with local operators constructed from  $A^L_\mu+A^R_\mu$. Since such 't Hooft operators are not the identity,  these Chern-Simons models describe logarithmic conformal field theories and hence are not unitary. This is due to the indefinite metric associated to the gauge fields which, although non-unitary in a formal sense, still leads to a unitary physical theory.

\subsection{Hidden symmetries  at $k=1,2$}

The derivation of the ABJM model applies even when $k=1,2$. However, in that case the M2-branes are propagating in $\mathbb R^8$ or $\mathbb R^8/\mathbb Z_2$ so their worldvolume theory should preserve ${\mathcal N}=8$ supersymmetry. This is not manifest in the lagrangian formulation. On the other hand, the brane derivation of the ABJM model implies that the full ${\mathcal N}=8$ supersymmetry must somehow be present. The first step towards resolving this issue is to note that for $k=1,2$ the theory is strongly coupled. Therefore the quantum theory can in principle have quite different properties to the classical theory, including additional symmetries.\footnote{This is rather like the opposite of an anomaly, in the sense that the classical lagrangian does not have all the symmetries of the quantum theory.} Consequently, one can propose that the additional supersymmetries, even if not manifest in the lagrangian, are present in the quantum theory.

We start with the observation that the two extra supercurrents are charged under $\U(1)_B$. Naively, there are no gauge invariant local observables that can carry such a $\U(1)_B$ charge. Therefore, to make them gauge invariant, we must introduce 't Hooft operators.  Let us see how this works in the special case of $k=1,2$.

First note that $k=1,2$ the ABJM model should have  an $\SO(8)$ R-symmetry. There is a manifest $\SU(4)$ symmetry which at the quantum level is generated by the currents
\begin{equation}
J_\mu^A{}_B= \Tr(Z^AD_\mu \bar Z_B - D_\mu Z^A \bar Z_B + i\psi^A\gamma_\mu\psi_B)\;.
\end{equation}
To enhance this to $\SO(8)$ we need currents of the form $\Tr(Z^A D_\mu Z^B - D_\mu Z^A Z^B + i\varepsilon^{ABCD}\psi_C\gamma_\mu\psi_D)$ but these are not gauge invariant. However, we can rectify this by including 't Hooft  operators with $\U(1)_B$ charge $-2$ and which are in the symmetric representation (so that $J_\mu^{AB}=-J_\mu^{BA}$)
\begin{equation}
J_\mu^{AB}= \Tr(({\cal M}_{ 2 \vec\lambda^{n-1}} )(Z^A D_\mu Z^B - D_\mu Z^A Z^B + i\varepsilon^{ABCD}\psi_C\gamma_\mu\psi_D))\;,
\end{equation}
where  $2\vec\lambda^{n-1}$ is the highest weight of the  symmetric (anti-fundamental) representation.
This is obtained by taking the flux $q_1=q_2=....=q_{n-1}=0$, $q_{n} = -2/k$ and only exists precisely when $k=1,2$.

In addition there should be  an extra ${\cal N}=2$ supersymmetry current at $k=1,2$. In this case we can construct
 \begin{equation}
 \Tr({\cal M}_{2\vec\lambda^{n-1}} D_\mu Z^A \psi_A)\ ,
\end{equation}
which has all the desired properties.\footnote{For related treatments see \cite{Gustavsson:2009pm,Kwon:2009ar,Samtleben:2010eu}.}

Finally at $k=1$ there should be a current that generates translations along the M-theory circle. A natural candidate for this is
 \begin{equation}
 \Tr({\cal M}_{\vec\lambda^{n-1}} Z^A)\;,
\end{equation}
where  $\vec\lambda^{n-1}$ is the highest weight of the anti-fundamental with $\U(1)_B$ charge $-1$ and  can only arise at $k=1$ with $q_1=q_2=...=q_{n_1}=0$, $q_n = -1$.

Thus we see that there are candidate operators that can enhance supersymmetry and translational invariance exactly as expected on physical grounds when $k=1,2$. These operators necessarily involve 't Hooft operators in an important way. It is a different matter to rigorously prove that the operators constructed above actually achieve the desired result, a task we will not attempt here.

\subsection{Background fields}

For a single M2-brane propagating in an eleven-dimensional spacetime with coordinates $x^M$, the full non-linear effective action including fermions and $\kappa$-symmetry was discussed in Section~\ref{worldvol}. The bosonic part of the effective action is \cite{Bergshoeff:1987cm}
\begin{eqnarray}\label{BSTaction}
\nonumber S &=& -T_{M2}\int d^3\sigma  \sqrt{-{\rm det}(\partial_\mu
x^M\partial_\nu x^N g_{MN})}\\&&\hskip2cm  + \frac{T_{M2}}{3!}\int
d^3\sigma \, \epsilon^{\mu\nu\lambda}\partial_\mu x^M \partial_\nu
x^N\partial_\lambda x^P C_{MNP}\ .
\end{eqnarray}

So far the topic of this review has been on the non-abelian generalisation of the first term, in the decoupling limit $T_{M2}\to \infty$ while keeping $X^M  = \sqrt{T_{M2}}x^M$ finite,  corresponding to multiple M2-branes propagating in flat space (or orbifolds of flat space). We now switch gears and focus on the generalisation of the second term, namely the coupling of multiple M2-branes to the background M-theory gauge fields.

In the well studied case of D-branes, where the low energy effective theory is a maximally supersymmetric Yang-Mills gauge theory with fields in the adjoint representation, the appropriate generalisation was given by Myers \cite{Myers:1999ps}. In the case of multiple M2-branes, the scalar fields $X^I$ and fermions now take values in a 3-algebra which carries a bi-fundamental representation of the gauge group.  Here we will follow closely reference \cite{Lambert:2009qw} and we refer the reader there for more technical details. In addition we will restrict ourselves to terms which survive under $T_{M2}\to \infty$. For alternative discussions of the coupling of multiple M2-branes to background fields, including terms that do not survive this limit, see \cite{Li:2008eza,Ganjali:2009kt,Kim:2009nc,Allen:2011pm}.

\subsubsection{BLG  theory}

Let us first consider the maximally supersymmetric case.
Assuming that there is no metric dependence
we start with the most general form for a non-abelian pull-back of the background gauge fields
to the M2-brane worldvolume
\begin{eqnarray}\label{bgdcoupl}
\nonumber  S_{C} &=& \frac{1}{3!}\epsilon^{\mu\nu\lambda}\int
d^3 x \Big( a\,T_{M2} C_{\mu\nu\lambda}+3b\, C_{\mu IJ} \, {\rm Tr}
  (D_\nu X^I,D_\lambda X^J)  \\
\nonumber     && + 12c\, C_{\mu\nu IJKL} \, {\rm Tr}(D_\lambda
   X^I,[X^J,X^K,X^L])\\
   && + 12d\,C_{[\mu IJ}C_{\nu KL]} \, {\rm Tr}(D_\lambda
   X^I,[X^J,X^K,X^L])+\ldots
 \Big)\;,
\end{eqnarray}
where $a,b,c,d$ are dimensionless constants that we have included for generality. The ellipsis denotes terms that are proportional to negative powers of $T_{M2}$ and hence vanish in the limit $T_{M2}\to \infty$.

Let us make several comments. First note that we have allowed the possibility of higher powers of the background fields. In D-branes the Myers terms are linear in the Ramond-Ramond fields, but they also include non-linear couplings to the NS-NS 2-form. Since all these fields come from the M-theory 3-form or 6-form, this suggests that we allow for a non-linear dependence in the M2-brane action.

Also note that gauge invariance rules out any terms where the $C$-fields have an odd number of indices that are transverse to the M2-branes.  For $k=2,4$ this is expected from the spacetime interpretation, where the ${\cal N}=8$ theory with gauge group $\SU(2) \times \SU(2)$ or $(\SU(2)\times \SU(2))/{\mathbb Z}_2$, respectively, describes M2-branes on a ${\mathbb C}^4/{\mathbb Z}_2$ orbifold, as discussed in Section \ref{connectingtoBJG}.   For these cases
we must set to zero any components of $C_3$ or $C_6$ with an odd number of $I,J = 1,...,8$ indices.  Therefore, in what follows, we will restrict our attention to the case with $k=2$.  (Similar results hold for $k=1$.  However, in that case, {}'t Hooft operators play an important role in restoring translational invariance.)

The first term in \eref{bgdcoupl} is the ordinary coupling of an M2-brane to the background 3-form. Hence we should take $a=n$ for $n$ M2's. The second term leads to a non-Lorentz invariant modification of the effective three-dimensional kinetic energy. It is also present in the case of a single M2-brane action (\ref{BSTaction}), where we find $b=1$. We will assume the same to be true in the non-abelian theory.  The final term, proportional to $d$, in fact vanishes because ${\rm Tr}(D_\lambda X^{[I},[X^J,X^K,X^{L]}])= \frac{1}{4}\partial_\lambda{\rm Tr}(X^I,[X^J,X^K,X^L])$ which is symmetric under $I,J \leftrightarrow K,L$.
Note that we have allowed the M2-brane to couple to both the
3-form gauge field and  its electromagnetic 6-form dual, defined by
$G_4=dC_3$, $G_7=dC_6$ where $G_7$ is defined in (\ref{fdual}).

The equations of motion of eleven-dimensional supergravity imply that $dG_7=0$. However $G_7$ is not gauge invariant under $\delta C_3 = d\Lambda_2$. Thus $S_{C}$ is not obviously gauge invariant or even local as a functional of the eleven-dimensional gauge fields. We would like  to find an expression that is manifestly gauge invariant.

To discuss the gauge invariance under $\delta C_3=d\Lambda_2$, we
first integrate by parts and discard all boundary terms.  We find
\begin{eqnarray}\label{bgrcoupl2}
\nonumber  S_{C} &=& \frac{1}{3!}\epsilon^{\mu\nu\lambda}\int d^3 x\,
\Big(
n T_{M2} C_{\mu\nu\lambda}\\
&&+\frac{3}{2} G_{\mu\nu IJ} \, {\rm Tr}
  (X^I,D_\lambda X^J) -\frac{3}{2}C_{\mu IJ} \, {\rm Tr}
  ( X^I,\tilde F_{\nu\lambda}X^J)\\
\nonumber   && - c\, G_{\mu\nu\lambda IJKL} \, {\rm Tr}(
   X^I,[X^J,X^K,X^L])\Big)\ ,
\end{eqnarray}
where we have used the fact that $C_{\mu\nu I}$ and $C_{\mu\nu\lambda
IJK}$ have been projected out by the orbifold and hence $G_{\mu\nu
IJ} =2 \partial_{[\mu} C_{\nu ]IJ}$ and $G_{\mu\nu\lambda IJKL} = 3\partial_{[\mu}C_{\nu\lambda] IJKL}$.
We see that $S_C$ contains a coupling to the worldvolume gauge field strength $\tilde F_{\nu\lambda}$,
but this term is not invariant under the gauge transformation $\delta C_3 = d\Lambda_2$. However, it
can be cancelled by adding the term
\begin{equation}
S_{F}=\frac{1}{4}\epsilon^{\mu\nu\lambda}\int d^3 x \, {\rm Tr}(X^I,\tilde  F_{\mu\nu} X^J)
C_{\lambda
IJ}\
\end{equation}
to $S_{C}$. Such terms involving the worldvolume gauge field
strength also arise in the  action of multiple D-branes.

Next consider the terms on the third line of \eref{bgrcoupl2}. Although $G_7$ is not
gauge invariant, the combination $G_7+\frac{1}{2}C_3\wedge G_4$ is. Thus we also add the term
\begin{equation}
S_{CG}=-\frac{c}{2\cdot 3!}\epsilon^{\mu\nu\lambda}\int d^3 x \, {\rm Tr}(X^I,[X^J,X^K,X^L])(C_3\wedge G_4)_{\mu\nu\lambda IJKL}\
\end{equation}
and obtain a gauge invariant action.

To summarise, we find that the total flux terms are, in the limit $T_{M2}\to \infty$,
\begin{eqnarray}
\label{sflux0}
\nonumber   S_{flux} &=& S_C+S_F+S_{CG} \\
   &=& \frac{1}{3!}\epsilon^{\mu\nu\lambda}\int d^3
x\, \Big( nT_{M2} C_{\mu\nu\lambda}+\frac{3}{2} G_{\mu\nu IJ} \, {\rm Tr}
  (X^I,D_\lambda X^J)\\
\nonumber   && - c\,(G_7+\frac{1}{2}C_3\wedge G_4)_{\mu\nu\lambda
IJKL} \, {\rm Tr}(
   X^I,[X^J,X^K,X^L])\Big)\ .
\end{eqnarray}
We will argue later that $c=2$ by comparing the higher order terms in fluxes that are demanded by supersymmetry with those obtained because the supergravity background is no longer flat at quadratic order.

Now let us supersymmetrise (\ref{sflux0}) in a fixed (but gauge-invariant) background.\footnote{Similar calculations appear in \cite{Grana:2002tu,Marolf:2003vf,Camara:2003ku} where the flux-induced fermion masses on D-branes were obtained.}  Thus we consider a background in which
\begin{equation}\label{Sflux}
\mathcal{L}_{flux} = c\,\tilde G_{IJKL} \, {\rm Tr}(X^I,[X^J,X^K,X^L])\ ,
\end{equation}
where
\begin{eqnarray}
\nonumber \tilde G_{IJKL} &=&-
\frac{1}{3!}\epsilon^{\mu\nu\lambda}(G_7+\frac{1}{2}C_3\wedge G_{4})_{\mu\nu\lambda
IJKL}\\[2mm]
&=&
\frac{1}{4!}\epsilon_{IJKLMNPQ}G^{MNPQ}
\end{eqnarray}
and $G_{IJKL}$ is assumed to be constant.  After supersymmetrisation, one finds the lagrangian to be
    \begin{equation}
        \mathcal{L} = \mathcal{L}_{\mathcal{N}=8} + \mathcal{L}_{mass} + \mathcal{L}_{flux}\ ,
    \end{equation}
where $\mathcal{L}_{\mathcal{N}=8}$ is the lagrangian (\ref{action}) and
    \begin{align}
        \mathcal{L}_{mass} &= -\frac{1}{2}m^2\delta_{IJ} \, {\rm Tr} (X^{I} , X^{J} ) -\frac{ic}{16} \, {\rm Tr}\, ( \bar{\Psi} \Gamma^{IJKL} , \Psi )\, \tilde{G}_{IJKL}\ ,
    \end{align}
with
    \begin{equation}
m^2 = \frac{c^2}{32\cdot 4!} G^2
\end{equation}
and  $G^2 = G_{IJKL}G^{IJKL}$.

Next we need to modify the supersymmetry transformations $\delta \to \delta +\delta'$ to accommodate the flux terms. 
One finds that the required choice is
    \begin{align}
        \delta ' X^{I}_{a} &= 0 \nonumber \\
        \delta ' \tilde{A}_{\mu}{}^{b}{}_{a} &= 0 \\
        \delta ' \Psi_{a} &= \frac{c}{8} \Gamma^{IJKL} \Gamma^{M} \epsilon \,X^{M}_{a} \tilde{G}_{IJKL} \nonumber\ .
    \end{align}
Invariance  follows if  $\tilde{G}$ is self-dual and
    \begin{equation}\label{Gcon}
 G_{MN[IJ}
        G_{KL]}{}^{MN}=0\ .
    \end{equation}
The superalgebra can be shown to close on-shell.

We close by noting that setting
\begin{equation}
G = \mu (dx^3\wedge dx^4\wedge dx^5\wedge dx^6+dx^7\wedge dx^8\wedge dx^9\wedge dx^{10})
\end{equation}
leads to the mass-deformed lagrangian of \cite{Hosomichi:2008qk,Gomis:2008cv}. 
This is an interesting extension of the BLG theory that we will investigate shortly in its ABJM realisation.

\subsubsection{ABJM theory}

Let us now consider the more general case of ${\cal N}=6$
supersymmetry and in particular the ABJM \cite{Aharony:2008ug} and
ABJ \cite{Aharony:2008gk} models. Following the discussion of the previous section, we start with
\begin{eqnarray}
\nonumber   S_{C} &=& \frac{1}{3!}\epsilon^{\mu\nu\lambda}\int d^3 \, x\,
\Big(
nT_{M2} C_{\mu\nu\lambda}+\frac{3}{2} C_{\mu}{}^A{}_B \, {\rm Tr}
  (D_\nu{\bar Z}_A,D_\lambda Z^B)
+\frac{3}{2} C_{\mu}{}_A{}^B \, {\rm Tr}
  (D_\nu Z^A,D_\lambda \bar Z_B)\\[2mm]
&&+~\frac{3c}{2}C_{\mu\nu AB}{}^{CD} \, {\rm Tr}(
  [D_\lambda {\bar Z}_D,[Z^A,Z^B;{\bar Z}_C])
+\frac{3c}{2}C_{\mu\nu}{}^{AB}{}_{CD} \, {\rm Tr}(
  [D_\lambda Z^D,[\bar Z_A,\bar Z_B;Z^C])
\Big)\ .
\end{eqnarray}
Integrating by parts we again find a non-gauge invariant term proportional to $\epsilon^{\mu\nu\lambda }\tilde
F_{\nu\lambda}C_{\mu}{}^A{}_B$ which is canceled by adding
\begin{equation}
S_F = \frac{1}{8}\epsilon^{\mu\nu\lambda} \int d^3 x \, C_{\mu}{}^A{}_B \, {\rm Tr}
  ({\bar Z}_A,\tilde F_{\nu\lambda} Z^B)+C_{\mu}{}_A{}^B \, {\rm Tr}
  ({Z}^A,\tilde F_{\nu\lambda} \bar Z_B)\ .
\end{equation}
As was the case with the $\mathcal N=8$ theory, we  must also add
\begin{eqnarray}
S_{CG} &=& -\frac{c}{8\cdot 3!}\epsilon^{\mu\nu\lambda}\int d^3 x \, (C_3\wedge G_4)_{\mu\nu
AB}{}^{CD} \, {\rm Tr}(
  {\bar Z}_D,[Z^A,Z^B;{\bar Z}_C])
\end{eqnarray}
to ensure that the last term is gauge invariant. Thus in total we
have
\begin{eqnarray}
\nonumber   S_{flux} &=& S_C+S_F+S_{CG}\\
\nonumber &=&
\frac{1}{3!}\epsilon^{\mu\nu\lambda}\int d^3 x \,
\Big(
nT_{M2} C_{\mu\nu\lambda}\\&&+\frac{3}{4} G_{\mu\nu}{}^A{}_B \, {\rm Tr}
  ({\bar Z}_A,D_\lambda Z^B)+\frac{3}{4} G_{\mu\nu}{}_A{}^B \, {\rm Tr}
  ({Z}^A,D_\lambda \bar Z_B)\\
\nonumber  && -\frac{c}{4}(G_7+\frac{1}{2}C_3\wedge G_4)_{\mu\nu\lambda AB}{}^{CD}\, {\rm Tr}(
  [{\bar Z}_D,[Z^A,Z^B;{\bar Z}_C])
\Big) \, .
\end{eqnarray}
Continuing as before, we wish to supersymmetrise the action
    \begin{equation}\label{massdefimpl}
        \mathcal{L} = \mathcal{L}_{{\cal N}=6} + \mathcal{L}_{mass} + \mathcal{L}_{flux}\ ,
    \end{equation}
where ${\cal L}_{{\cal N}=6}$ is the ${\cal N}=6$ Chern-Simons-matter lagrangian     \eref{niceaction}.  We restrict to backgrounds where
\begin{equation}\label{Sflux2}
\mathcal{L}_{flux} = \frac{c}{4} \, {\rm Tr}( [\bar Z_D,[ Z^A, Z^B;\bar Z_C]) \tilde G_{AB}{}^{CD} \ ,
\end{equation}
with
\begin{eqnarray}
\nonumber \tilde G_{AB}{}^{CD} &=&-
\frac{1}{3!}\epsilon^{\mu\nu\lambda}(G_7+\frac{1}{2}C_3\wedge G_{4})_{\mu\nu\lambda
AB}{}^{CD}\\ &=&
\frac{1}{4}\epsilon_{ABEF}\epsilon^{CDGH}G^{EF}{}_{GH}\ .
\end{eqnarray}

We can supersymmetrise this term if we take $\mathcal{L}_{mass}$ to be
\begin{equation}\label{massterm}
\mathcal{L}_{mass} = - m^2 \, {\rm Tr}( \bar Z_A , Z^A ) + \frac{ic}{4} \, {\rm Tr}(\bar \psi^A , \psi_F ) \tilde G_{AE}{}^{EF}
\end{equation}
and include the following modification to the fermion supersymmetry variation
\begin{equation}
\delta' \psi_{Ad} =\frac{c}{4} \epsilon_{DF} Z^F_d \tilde G_{AE}{}^{ED}\ .
\end{equation}
We then find that supersymmetry requires
\begin{equation}\label{Gcon2}
\tilde{G}_{AE}{}^{EB} \tilde{G}_{BF}{}^{FC} = \frac{16 m^2}{c^2} \delta_A^C.
\end{equation}
It also restricts $\tilde{G}$ to have the form
\begin{equation}\label{Gid}
\tilde{G}_{AB}{}^{CD} = \frac{1}{2} \delta^C_B \tilde{G}_{AE}{}^{ED} - \frac{1}{2} \delta^C_A
\tilde{G}_{BE}{}^{ED} - \frac{1}{2} \delta^D_B \tilde{G}_{AE}{}^{EC} + \frac{1}{2} \delta^D_A
\tilde{G}_{BE}{}^{EC}\ ,
\end{equation}
with $\tilde{G}_{AE}{}^{EA}=0$. As a consequence, we  find that
\begin{equation}
m^2 = \frac{1}{32\cdot 4!}c^2 G^2\;,
\end{equation}
where $G^2 = 6G_{AB}{}^{CD} G^{AB}{}_{CD}=12 G_{AE}{}^{EB}G_{BF}{}^{FA}$.

Choosing $\tilde G_{AB}{}^{CD}$ to have the form (\ref{Gid}) with
\begin{equation}\label{ex2}
\tilde G_{AB}{}^{BC} = \left(
                  \begin{array}{cccc}
                    \mu & 0 &0&0\\
                    0 & \mu &0&0\\
                    0 & 0 &-\mu&0\\
                     0& 0 &0&-\mu\\
                  \end{array}
                \right)\ ,
\end{equation}
gives the mass-deformed ABJM lagrangian of \cite{Hosomichi:2008jb,Gomis:2008vc}.

 \subsubsection{Background curvature}

It is interesting to understand the physical origin of the mass-squared term in the  effective action \eref{massterm}, which is quadratic in the flux. Note that this term is a simple,  $\SO(8)$-invariant mass term for all the scalar fields. Furthermore it does not depend on any  non-abelian features of the theory. Therefore we can derive this term by simply considering a  single M2-brane and compute the unknown constant $c$.

We can understand the origin of this term as follows. We have just seen that it arises as a consequence of supersymmetry. For a single M2-brane, supersymmetry arises as a consequence of $\kappa$-symmetry and $\kappa$-symmetry is valid whenever an M2-brane is propagating in a background that satisfies the equations of motion of eleven-dimensional supergravity \cite{Bergshoeff:1987cm}.

The multiple M2-brane actions implicitly assume that the background is simply flat space or an orbifold thereof. However, the inclusion of  non-trivial flux implies that there is now a source for the eleven-dimensional metric, which is of order flux-squared. Thus for there to be $\kappa$-symmetry and hence supersymmetry it follows that the background must be curved. This in turn will lead to a potential in the effective action of an M2-brane. In particular given a 4-form flux $G_4$, the bosonic equations of eleven-dimensional supergravity are
\begin{eqnarray}
 \nonumber R_{mn} - \frac{1}{2}g_{mn}R &=& \frac{1}{2\cdot 3!}G_{mpqr}G_n{}^{pqr} - \frac{1}{4\cdot 4!}g_{mn}G^2 \\
   d\star G_4 -\frac{1}{2}G_4\wedge G_4 &=&0\ .
\end{eqnarray}
At lowest order in the fluxes we see that $g_{mn}=\eta_{mn}$ and $G_4$ is constant. However at second order there are source terms. To start, we assume that, at lowest order, only $G_{IJKL}$ is non-vanishing. To solve these equations we introduce a non-trivial metric of the form
\begin{equation}
g_{mn} = \left(
           \begin{array}{cc}
             e^{2\omega}\eta_{\mu\nu} & 0 \\
             0 & g_{IJ} \\
           \end{array}
         \right)\ ,
\end{equation}
where $\omega = \omega(x^I) = \omega( X^I/T_{M2}^{\frac{1}{2}})$ and
$g_{IJ} = g_{IJ}(x^I) = g_{IJ}( X^I/T_{M2}^{\frac{1}{2}})$.

Let us look at an M2-brane in this background. The first term in the action (\ref{BSTaction}) is
\begin{eqnarray}
\nonumber  S_1 &=& -T_{M2}\int d^3x  \sqrt{-{\rm det}(e^{2\omega}\eta_{\mu\nu}+\partial_\mu x^I\partial_\nu x^J g_{IJ})} \\
   &=&-T_{M2}\int d^3x \, e^{3\omega}\left(1 + \frac{1}{2}e^{-2\omega}\partial_\mu x^I\partial^\mu
   x^Jg_{IJ}+\ldots\right)\\
   \nonumber &=& -\int d^3x \left(T_{M2}e^{3\omega} + \frac{1}{2}e^{\omega}\partial_\mu X^I\partial^\mu
   X^Jg_{IJ}+\ldots\right) \ .
\end{eqnarray}
Next we note that, in the decoupling limit $T_{M2}\to \infty$, we
can expand
\begin{equation}
e^{2\omega(x)} = e^{2\omega(X^I/\sqrt{T_{M2}})} =
1+\frac{2}{T_{M2}}\omega_{IJ}X^IX^J+\ldots
\end{equation}
and
\begin{equation}
g_{IJ}(x) = g_{IJ}( X^I/\sqrt{T_{M2}}) = \delta_{IJ}+\ldots\ ,
\end{equation}
so that
\begin{equation}
S_1 =-\int d^3x \left(T_{M2} + 3\omega_{IJ}X^IX^J + \frac{1}{2}\partial_\mu X^I\partial^\mu
   X^J\delta_{IJ}+\ldots\right)\ ,
\end{equation}
where the ellipsis denotes terms that vanish as $T_{M2}\to \infty$. Thus we see that in the decoupling limit we
obtain the mass term for the scalars. Similar mass terms for M2-branes were also studied in
\cite{Skenderis:2003da}  for pp-waves.

To compute the warp-factor $\omega$ we can expand $g_{mn} =
\eta_{mn} + h_{mn}$, where $h_{mn}$ is second order in the fluxes,
and linearise the Einstein equation. If we impose the gauge
$\partial^m h_{mn} -\frac{1}{2}\partial_n h^p{}_p=0$, then Einstein's equations reduce to
\begin{equation}
\partial_I\partial^I e^{2\omega} = \frac{1}{3\cdot 4!} G^2
\end{equation}
and hence, to leading order in the fluxes,
\begin{equation}
e^{2\omega} = 1+\frac{1}{48\cdot4!}G^2 \delta_{IJ}x^Ix^J\ .
\end{equation}
Thus $S_1$ contributes the term
\begin{equation}\label{Sone}
S_1 = -\int d^3 x \, \frac{1}{32\cdot4!}G^2X^2
\end{equation}
to the potential.

Next we must look at the second term, the Wess-Zumino term, in (\ref{BSTaction})
\begin{equation}
S_2 = \frac{T_{M2}}{3!}\int d^3 x \, \epsilon^{\mu\nu\lambda} C_{\mu\nu\lambda}\ .
\end{equation}
Although we have assumed that $C_{\mu\nu\lambda}=0$ at leading order, the $C$-field equation of motion implies
that $G_{I \mu\nu\lambda }=\partial_ I C_{\mu\nu\lambda}$  is second order in $G_{IJKL}$. In particular if we
write $C_{\mu\nu\lambda} = C_0\epsilon_{\mu\nu\lambda}$ we find, assuming $G_{IJKL}$ is self-dual,  the
equation
\begin{equation}
\partial_I\partial^I C_0 = \frac{1}{2\cdot 4!}G^2\ .
\end{equation}
The solution is
\begin{equation}
C_0=\frac{1}{32\cdot 4!}G^2 \delta_{IJ}x^Ix^J\ .
\end{equation}
Thus we find that $S_2$ gives a second contribution to the scalar potential
\begin{equation}
S_2 = -\int d^3 x \, \frac{1}{32\cdot4!}G^2X^2\ .
\end{equation}
Note that this is equal to the scalar potential derived from $ S_1$ in \eref{Sone}. Therefore if we were to break supersymmetry and consider anti-M2-branes, where the sign of the Wess-Zumino term changes, we would not find a mass for the scalars.

In total we find the mass-squared
\begin{equation}
m^2= \frac{1}{8\cdot 4!}G^2 \ .
\end{equation}
Comparing with (\ref{Gcon}) we see that $c^2=4$, {\it e.g.} $c=2$.

\subsection{Dielectric membranes}\label{dielM2M5}

Having obtained the explicit form of the $\mathcal N=6$ supersymmetric lagrangian for the mass-deformed ABJM model, \eref{massdefimpl}, we can proceed to study the physics it describes. We recall that the undeformed ABJM action is given by the expression
\bea\label{abjmmass}
S_{\mathrm{ABJM}}&=&\int d^3 x\left[\frac{k}{4\pi}\epsilon^{\mu\nu\lambda}{\rm Tr}\left(A_\mu^{L}\pd_\nu A_\lambda^{L}+\frac{2i}{3}
A_\mu^{L} A_\nu^{L} A_\lambda^{L}-A_\mu^{R}\pd_\nu A_\lambda^{R}-\frac{2i}{3}
A^{R}_\mu A^{R}_\nu A^{R}_\lambda\right)\right.-{\rm Tr}\Big( D_\mu \bar Z_A D^\mu Z^A\Big) \nonumber\\
&&\left.+\frac{4\pi^2}{3k^2}{\rm Tr}\left(Z^A \bar Z_A Z^B \bar Z_B Z^C \bar Z_C+\bar Z_A Z^A \bar Z_B Z^B \bar Z_C Z^C+4 Z^A\bar Z_B Z^C\bar Z_A Z^B\bar Z_C-6 Z^A\bar Z_B Z^B \bar Z_A Z^C \bar Z_C\right)\right]\;,\nn\\
\label{abjmaction}
\eea
where on the first line we have the Chern-Simons gauge field and the matter kinetic terms, while on the second we have the sextic scalar potential. Focusing on the purely bosonic sector will prove enough for our purposes.

By splitting $Z^A=(R^\alpha,Q^{\alpha})$, where $\alpha = 1,2$, the mass deformation (\ref{ex2}) changes the potential to
\be\label{massdefpot}
V=|M^{ \alpha}|^2+|N^\alpha|^2\;,
\ee
where
\bea\label{interim}
M^{ \alpha}&=& \mu Q^{ \alpha} +\frac{2\pi}{k}(2Q^{[ \alpha }\bar Q_{\beta} Q^{\beta ]}+R^{\beta} \bar R_{\beta} Q^{\alpha}-Q^{\alpha} \bar R_\beta R^\beta
+2Q^{\beta} \bar R_\beta R^\alpha)\nonumber\\
N^\alpha&=&-\mu R^\alpha +\frac{2\pi}{k}(2R^{[\alpha}\bar R_\beta R^{\beta ]}+Q^\beta \bar Q_{\beta} R^\alpha-R^\alpha \bar Q_{\beta} Q^{\beta}
+2R^\beta \bar Q_{\beta} Q^{\alpha})\label{potterms}\;,
\eea
which in principle also involves a mass term for the fermions.  Note that the expressions in \eqref{interim} couple $R^\alpha$ with $Q^\alpha$ and break the $\SU(4)$ invariance.  Nevertheless, in the full scalar potential \eqref{massdefpot}, the terms that couple $R^\alpha$ and $Q^\alpha$ cancel out \cite{Gomis:2008vc}.  As a result, we will keep a different notation for their respective indices, with $R^\alpha$ and $Q^{\dot{\alpha}}$ for $\dot \alpha = 1,2$.  We conclude that the R-symmetry is broken down to the subgroup $\SU(2)\times\SU(2)\times\U(1)$.

\subsubsection{Vacua of the mass-deformed theory}

From the previous sections we see that the supersymmetric  vacua of the mass-deformed theory satisfy
\be
\frac{1}{2}M_B{}^C\epsilon_{CD}Z^D +\left([Z^C,Z^D;Z_B]+ [Z^E,Z^C;\bZ_E]\delta^D_B\right)\epsilon_{CD}=0\ ,
\ee
where $M_B{}^C$ has the form
\be
M_B{}^C = \left(
                  \begin{array}{cccc}
                    2\mu & 0 &0&0\\
                    0 & 2\mu &0&0\\
                    0 & 0 &-2\mu&0\\
                     0& 0 &0&-2\mu\\
                  \end{array}
                \right)\ .
\ee
For  maximally supersymmetric vacua we require that this is true   for all $\epsilon_{CD}=-\epsilon_{DC}$
\be\label{efg}
\frac{1}{4}M_B{}^C Z^D-\frac{1}{4}M_B{}^D Z^C + [Z^C,Z^D;Z_B]+\frac{1}{2}[Z^E,Z^C;\bZ_E]\delta^D_B-\frac{1}{2}[Z^E,Z^D;\bZ_E]\delta^C_B =0\ .
\ee
Taking the trace over $B,D$ implies that
\be
\frac{1}{2}M_B{}^CZ^B = [Z^D,Z^E;\bZ_E]\ .
\ee
Substituting back we find that \eref{efg}  is only  satisfied if $M_B{}^CZ^B=  2\mu Z^C$ or $M_B{}^CZ^B=-  2\mu Z^C$.

Thus the mass-deformed theory has two sets of ground states expressed in terms of the scalars $R^\alpha$ and $Q^{\dot{\alpha}}$. One set  corresponds to having $Q^{\dot \alpha}=0$ and $R^\alpha$ satisfying
\be\label{grvvvac}
R^\alpha = \frac{2\pi}{\mu k}\Big( R^\alpha \bar R_\beta R^\beta-R^\beta \bar R_\beta R^\alpha\Big)\;,
\ee
as can be easily seen from \eqref{massdefpot}-\eqref{interim}. This can be solved by the ansatz
\be\label{grvvsolution}
R^\alpha = f G^\alpha\;,
\ee
where $f^2 = \mu k /2 \pi$ and the $G^\alpha$'s are a set of complex, constant, $n\times n$ bi-fundamental matrices satisfying
\be\label{GRVV}
G^\alpha = G^\alpha \bar G_\beta G^\beta-G^\beta \bar G_\beta G^\alpha\;.
\ee
There exist irreducible solutions to the above equation, explicitly given by \cite{Gomis:2008vc}
\bea\label{BPSmatrices}
&& ( G^1)_{m,l }    = \sqrt { m- 1 } ~\delta_{m,l} \cr
&& ( G^2)_{m,l} = \sqrt { ( n-m ) } ~\delta_{ m+1 , l } \cr
&& (\bar G_1 )_{m,l} = \sqrt { m-1} ~\delta_{m,l} \cr
&& ( \bar G_2 )_{m,l} = \sqrt { (n-l ) } ~\delta_{ l+1 , m }\;.
\eea

Another set has $R^\alpha=0$, $Q^{\dot\alpha} = f G^{\dot\alpha}$, with the $G^{\dot\alpha}$'s satisfying once again (\ref{GRVV}). Moreover, one can easily construct reducible solutions using the above irreducible representations to form block diagonal matrices with block sizes that add up to $n$. It is also possible to construct reducible solutions where both $R^\alpha$ and $Q^{\dot   \alpha}$ are turned on, as long as the block components of $R^\alpha$ are zero when the respective ones of $Q^{\dot \alpha}$ are not, and vice-versa so that (\ref{GRVV}) is satisfied for each block \cite{Gomis:2008vc}.

What is the expected physical interpretation of these vacua in the context of M2-branes? By taking into consideration the background geometry that gives rise to the mass-deformed theory in the previous section, one would anticipate that we have described an M-theoretic version of the ``dielectric'' Myers effect \cite{Myers:1999ps}. That is, in the presence of the 4-form flux, the $n$ M2-branes are supposed to puff up into a fuzzy (or non-commutative) 3-sphere in the transverse 8-dimensional space, with the non-commutativity scale set by $1/n$. In the large-$n$ limit, the resulting configuration is an M2-M5 bound state and should also admit an equivalent interpretation in terms of a single M5-brane wrapping the $S^3$  \cite{Bena:2000zb}. In the following we will confirm this expectation.

\subsubsection{Geometric interpretation and Hopf fibration}

Let us examine how this picture emerges from the matrices \eqref{BPSmatrices} for our initial configurations with $R^\alpha= fG^\alpha $, $Q^{\dot \alpha} = 0$. At closer inspection, as seen from \eqref{BPSmatrices}, $G^1 = \bar G_1$ and one has three real degrees of freedom, as opposed to the four needed for the description of the expected 3-sphere. Moreover, the R-symmetry of the mass-deformed theory is only $\SU(2)\times\SU(2)\times\U(1)$ so the 3-sphere cannot be realised in the familiar $\SO(4)$-invariant way.

The key observation is that the following matrix combinations \cite{Nastase:2009ny}
\bea
{J^\alpha}_\beta = G^\alpha \bar G_\beta\qquad \textrm{and}\qquad \bar {J_\alpha}^\beta  = \bar G_\alpha G^\beta
\eea
are $n\times n$ adjoint matrices and $\U(2)$ symmetry generators. One can extract the $\SU(2)$ parts as follows
\bea\label{defining}
J_i&=&{(\tilde\s_i)^\alpha}_\beta G^\beta \bar  G_\alpha={(\tilde\s_i)^\alpha}_\beta{J^\beta}_\alpha\equiv{(\s_i)_\beta}^\alpha {J^\beta}_\alpha\;\cr
\bar{J}_i&=&{(\tilde\s_i)^\alpha}_\beta \bar G_\alpha G^\beta =  {(\tilde{\s}_i)^\alpha}_\beta {\bar{J}_\alpha\,}^\beta\equiv{(\sigma_i)_\beta}^\alpha{\bar{J}_\alpha\,}^\beta\;,
\eea
where $\tilde \sigma $ are the transpose of the Pauli matrices. The $J_i$ and $\bar J_i$'s then satisfy the $\SU(2)$ commutation relations
\be
[J_i, J_j] = 2i \epsilon_{ijk}J_k\qquad\textrm{and}\qquad [\bar J_i, \bar J_j] = 2i \epsilon_{ijk}\bar J_k\;.
\ee
Using these relations along with \eqref{GRVV}, one  finds that the $G^\alpha$, as well as all bi-fundamental fields, transform under the combined action
\be
J_i G^\alpha - G^\alpha \bar J_i = {(\tilde \sigma_i)^\alpha}_\beta G^\beta,
\ee
and as a result only a {\it single} diagonal $\SU(2)$ survives as a symmetry of the system.

In order to further analyse the geometry, one can use the well established fact that the algebra of large matrices, transforming in irreducible representations of a given symmetry group, approximates the algebra of functions on spaces with the same isometries. Or conversely, the matrix algebras can provide a finite-dimensional truncation/discretisation/quantisation of the continuous, ``classical'' geometry. Hence, one can define to leading order in the large-$n$ limit
\be
x_i \simeq \frac{J_i}{n}\qquad\textrm{and} \qquad \bar x_i \simeq \frac{\bar J_i}{n}\;,
\ee
which play the role of standard Euclidean coordinates on two,  at-first-sight-different, $S^2$'s.

One can similarly define
\be
g^\alpha \simeq\frac{G^\alpha}{\sqrt n}\qquad\textrm{and}\qquad g^*_\alpha \simeq \frac{\bar G_\alpha}{\sqrt n}
\ee
as some yet-to-be-understood {\it commuting} classical objects. In terms of the above definitions, the relations \eqref{defining} become
\bea\label{classical}
x_i &=&{(\tilde \sigma_i)^\alpha}_\beta g^\beta g^*_\alpha\cr
\bar x_i &=&{(\tilde \sigma_i)^\alpha}_\beta  g^*_\alpha g^\beta\;,
\eea
\ie in this limit $x_i \simeq \bar x_i$ and one has two versions of the same Euclidean coordinate on a single sphere. This is in line with our previous observation, stating that the solution  has only one $\SU(2)$ symmetry.

How does all this information fit together? The answer lies in recognising that \eqref{classical} is nothing but the expression for the familiar first Hopf map $S^3\stackrel{\pi}{\rightarrow} S^2$ from the unit 3-sphere to the unit 2-sphere. Note that in the above construction the 2-sphere coordinates $x_i, \bar x_i$ are invariant under multiplication of the classical $g^\alpha$'s (chosen such that $g^1 =  g_1^*$) by a $\U(1)$ phase. Using the latter, one could define some  $ \hat g^\alpha = e^{i \alpha(\vec x)} g^\alpha$ which would then describe a unit $S^3$ with $\hat g^\alpha \hat g^*_\alpha = 1$. However, in our case the $g^\alpha$'s are already defined modulo such a phase and they are just describing  a different parametrisation of the $S^2$ in terms of so-called Hopf spinors \cite{Hasebe:2004yp}.

It is interesting to note that in the same way that the $\SU(2)$ irreducible representations $J_i$ are ``fuzzy'' coordinates that ``discretise'' the classical 2-sphere coordinates, defined by $x_i x_i = 1$, the bifundamental matrices $G^\alpha$ ``discretise'' the classical Hopf spinors $ g^\alpha$. The latter are in fact equivalent to Killing spinors on $S^2$ and the $G^\alpha$'s can be thought of as ``fuzzy Killing spinors''. We refer the interested reader to \cite{Nastase:2009zu,Nastase:2010uy} for a detailed discussion of their properties.

\subsubsection{Brane interpretation}

The various pieces of our geometric analysis are now falling into place: It is clear that for a dielectric M5-brane to be emerging from this picture \`a la Myers, the $S^3$ that it is wrapping should be realised in terms of an $S^1\hookrightarrow S^3\stackrel{\pi}{\rightarrow} S^2$ Hopf fibration. However, in the ABJM model the M-theory direction is modded out by the $\mathbb Z_k$ orbifold action, which in turn implies that the Hopf fibration is instead $S^1/{\mathbb   Z}_k\hookrightarrow S^3/{\mathbb Z}_k\stackrel{\pi}{\rightarrow} S^2$. In the weak coupling limit, $k\to\infty$, the fibre shrinks and this is reflected by the fact that the vacuum solutions $G^\alpha$ only capture the $S^2$ base of the Hopf bundle.

As a result, the emerging dielectric brane is a D2-D4 bound state in type IIA on $\mathbb R^{2,1}\times S^2$, obtained from an M5 on $\mathbb R^{2,1}\times S^3/\mathbb Z_k$ in the $k\to\infty$ limit. This can also be verified by a small-fluctuation analysis around the irreducible vacua at large $n$ and leads to an abelian 5d worldvolume theory for the action of fluctuations \cite{Nastase:2009ny,Nastase:2009zu}. In turn, the latter also has an interpretation in terms of fluctuations around a D4-brane partially wrapping the (fuzzy) sphere with a worldvolume flux that provides the coupling to the D2-brane charge. To complete the lift to the full M5-brane description,  additional momentum modes along the M-theory circle must arise in a manner similar to the discussion in Section~\ref{fluxesandtHooft}, via $\U(1)_B$ fluxes in ABJM that give rise to {}'t~Hooft operators \cite{Lambert:2011eg}.

In order to further characterise the D2-D4 bound state, we note that there is a natural invariant that one can construct: First, we use $ Z^A = T_{M2}^\half z^A $ to convert the $Z^A$ kinetic term of the action (\ref{abjmmass})
\bea
S = -\int d^3 x \;\Tr( D_\mu Z^A D^\mu \bar Z_A)
\eea
to the physical form
\bea \label{physical}
S_{phys} = -T_{M2}  \int d^3 x \;\Tr( D_\mu z^A D^\mu \bar z_A )\;.
\eea
Now the $z^A$ are spacetime coordinates with dimensions of length and we can define a ``physical radius'' for the emerging sphere geometry
\bea\label{physrad}
R_{ph}^2 = { 2 \over n } \Tr( z^A \bar z_A) =8\pi^2 n  f^2 \ell_p^3\;.
\eea
This answer in terms of the M-theory constants can be further massaged through the structure of the Hopf fibration: In an appropriate parametrisation, for an $S^3$ radius $R_{ph}$, the fibre has radius $R_{ph}$, while the base $S^2$ has radius\footnote{See \eg the relevant geometric discussion in \cite{Ishii:2008tm}.} $\half R_{ph}$. Since the fibre plays the role of the M-theory circle, it is further modded out by the orbifold to give
\be
R_{11} = \frac{R_{ph}}{k} \;.
\ee
Then, using the M-theory -- type IIA relations $R_{11} = g_s \ell_s$, $\ell_p^3 = \ell_s^3 g_s$ and $f^2 =\mu k /2 \pi$, \eqref{physrad} becomes
\be
R_{ph}^2 = 4 \pi k n \mu R_{11} \ell_s^2 = 4 \pi n \mu R_{ph} \alpha'
\ee
and hence the radius of the $S^2$ is
\be
R_{ph} = 2n \mu \lambda\;,
\ee
 with $\lambda = 2 \pi \alpha'$, \ie for fixed $\mu$ it is simply a linear function in the size of the matrices $n$.

Even though all of the above discussion has primarily been for the case of the irreducible solutions that lead to a single higher dimensional brane description, the reducible solutions follow suit: Reducible representations of $m$ blocks with $n_1 + n_2 + \ldots + n_m = n$ correspond to concentric configurations of multi-centre D4's of different sizes. Of particular interest are the possibilities with $m$ copies of $n_m \times n_m$ equally sized blocks, where $m \,n_m = n$. Since in that case all radii have the same value and the branes are therefore coincident, one expects a worldvolume gauge symmetry enhancement $\U(1)^m\to\U(m)$. This provides a compelling starting point for studying multiple fivebranes in M-theory \cite{Lambert:2011eg}. Works in this direction include \cite{Gustavsson:2011af}.

It is important to add that our interpretation for the vacua of the mass-deformed ABJM theory can be confirmed by means of the gauge/gravity duality. The gravity solutions describing the M2-M5 bound state in $\mathbb C^4/\mathbb Z_k$, were found in \cite{Cheon:2011gv} and are given in terms of $\mathbb Z_k$ quotients of the smooth bubbling geometries of \cite{Bena:2004jw,Lin:2004nb}. The latter emerge as expected in the $k=1$ limit, preserve 16 bulk supercharges and are in one-to-one correspondence with partitions of $n$.\footnote{For $k=1$ the mass-deformed M2-brane theory is also related to the BFSS Matrix theory description of IIB string theory on the pp-wave \cite{Gomis:2008vc,Sethi:1997sw,Banks:1996my} and tiny graviton matrix theory \cite{SheikhJabbari:2004ik,SheikhJabbari:2005mf,Torabian:2007pk}.}  It can indeed be shown that the evaluation of the index for supersymmetric vacua from the gauge theory side at any $k$ reproduces exactly the counting expected from gravity, including the partitions of $n$ result for $k=1$ \cite{Kim:2010mr}.

\subsubsection{Fuzzy funnels revisited}

Finally, we can now go full circle and  reconsider the fuzzy funnel system of Basu-Harvey in the context of ABJM. Namely we consider $n$ M2-branes ending on an M5-brane. Here the M2-branes are in the $x^0,x^1,x^2$ plane and the M5-brane sits along $x^0,x^1,x^3,x^4,x^5,x^6$. The M2-branes preserve supersymmetries $\Gamma_{012}\epsilon=\epsilon$ whereas the M5-brane preserves $\Gamma_{013456}\epsilon=\epsilon$. Thus the common preserved supersymmetries satisfy $\Gamma_{2}\epsilon =  \Gamma_{3456}\epsilon$.
If we let $X^{I'}$, with $I'=\{3,4,5,6\}$, denote the fluctuations of the M2-branes that are tangent to the M5-brane, then we look for solutions where only these are non-vanishing and depend on $x^2$ -- the direction of the M2-branes that is orthogonal to the M5-brane. It will also be sufficient to set the gauge fields to zero.

Let us consider the BLG theory first for simplicity and look for $\frac{1}{2}$-BPS solutions. Here the condition  $\Gamma_{2}\epsilon =  \Gamma_{3456}\epsilon$ is equivalent to $ \Gamma^{I'J'K'}\epsilon =  \varepsilon^{I'J'K'L'}\Gamma_2\Gamma^{L'}\epsilon$. Thus the condition $\delta \Psi_a=0$ can be written as
\be
0=\left(\partial_2 X_a^{L'} - \frac{1}{3!}\varepsilon^{I'J'K'L'}f^{cdb}{}_a X_c^{I'}X_d^{J'}X_b^{K'}\right)\Gamma_2\Gamma^{L'} \epsilon\ .
\ee
From here we can read off the BPS equation \cite{Bagger:2006sk,Bagger:2007jr}
\be
\frac{d X^{I'}}{dx^2} = -\frac{1}{3!}\varepsilon^{I'J'K'L'} [X^{J'},X^{K'},X^{L'}]\ ,
\ee
which is essentially the Basu-Harvey equation \cite{Basu:2004ed}, in this case for just two M2-branes.

We can of course also do this for the ABJM theory \cite{Terashima:2008sy,Gomis:2008vc,Hanaki:2008cu,Nastase:2009ny}. In this case we need to set $Z^3=Z^4=0$. We then find (again assuming that the gauge fields vanish)
\bea
0= \gamma^2\partial_2 Z^\alpha \epsilon_{\alpha B} +
[Z^\gamma,Z^\alpha;\bZ_\gamma] \epsilon_{\alpha B}+
  [Z^\gamma,Z^\delta;\bZ_B] \epsilon_{\gamma\delta}\ ,
\eea
where $\alpha,\beta=1,2$. We can consider two cases. First $B=\beta'=3,4$ which gives
\be
0=\gamma^2\partial_2 Z^\alpha \epsilon_{\alpha \beta'} +
[Z^\gamma,Z^\alpha;\bZ_\gamma] \epsilon_{\alpha \beta'}\ .
\ee
This tells us that, assuming $\gamma^2\epsilon_{\alpha\beta'}=-\epsilon_{\alpha\beta'}$,
\be\label{bhabjm}
\frac{d Z^\alpha}{dx^2} = [Z^\gamma,Z^\alpha;\bZ_\gamma] = \frac{2\pi}{k}(Z^\gamma Z^\dag_\gamma Z^\alpha-  Z^\alpha Z^\dag_\gamma Z^\gamma )\ .
\ee
In the second case  $B=\beta$ and we find
\be\label{inbetween}
0= \gamma^2\partial_2 Z^\alpha \epsilon_{\alpha \beta} +
[Z^\gamma,Z^\alpha;\bZ_\gamma] \epsilon_{\alpha \beta}+
  [Z^\gamma,Z^\delta;\bZ_\beta] \epsilon_{\gamma\delta}\ ,
\ee
Next we note that since on $Z^1$ and $Z^2$ are non-vanishing, and $[Z^\gamma,Z^\delta;\bZ_\beta]$ is anti-symmetric in $\gamma,\delta$,
\bea
[Z^\gamma,Z^\delta;\bZ_\beta] \epsilon_{\gamma\delta} &=& 2 [Z^1,Z^2;\bZ_\beta] \epsilon_{12}\nonumber\\
&=& 2\varepsilon_{\beta\alpha}[Z^\gamma,Z^\alpha;\bZ_\gamma]\epsilon_{12}\\
&=& -2[Z^\gamma,Z^\alpha;\bZ_\gamma]\epsilon_{\alpha\beta}\ ,\nonumber
\eea
where $\varepsilon_{\beta\alpha}=-\varepsilon_{\alpha\beta}$ is the two-dimensional $\varepsilon$-symbol. Thus, given \eref{bhabjm}, \eref{inbetween} is satisfied if $\gamma^2\epsilon_{\alpha\beta}=\epsilon_{\alpha\beta}$ and half of the supersymmetries are preserved.

We are now just left with \eref{bhabjm}. To  solve this equation one can use the same bi-fundamental matrices as in our dielectric M2/M5 configuration, with the difference that the functional dependence of the solution is now in terms of the ``spike'' direction, $x^2$. Our ansatz is
\be
Z^\alpha = f(x^2) G^\alpha\ ,
\ee
where the $G^\alpha$ satisfy \eref{GRVV}. We then find the simple equation
\be
\frac {df}{dx^2} = -\frac{2\pi}{k}f^3\ ,
\ee
so that, ignoring the free translational zero-mode along $x^2$,
\be
f =  \sqrt{\frac{k}{2\pi}}\frac{2}{\sqrt{x^2}}\ .
\ee
As discussed in Section~\ref{BH} this reproduces the correct behaviour for both the radial profile and the energy, to account for the self-dual strings on the M5-brane worldvolume. Furthermore one can consider an M-theory version of the Nahm construction for self-dual strings \cite{Gustavsson:2008dy,Saemann:2010cp,Papageorgakis:2011xg,Palmer:2011vx}.


\section[Superconformal CS theories with reduced supersymmetry]{\Large{\bf Superconformal CS theories with reduced supersymmetry}}\label{chapter7}

In the previous chapters, we studied the ${\cal N} = 8$ and ${\cal N} = 6$ superconformal Chern-Simons theories in three spacetime dimensions.  We found that the most general such theories could be described in terms of 3-algebras.  For the case of ${\cal N} = 8$, the 3-algebra structure constants turned out to be real and totally antisymmetric,
\begin{equation}
\nn f^{abc}{}_d = f^{[abc]}{}_d, \qquad (f^{abc}{}_d)^* = f^{abc}{}_d.
\end{equation}
For the case of ${\cal N} = 6$, the constants were found to be complex, obeying
\begin{equation}
\nn f^{ab}{}_{cd} = -f^{ba}{}_{dc} = -f^{ab}{}_{cd}, \qquad (f^{ab}{}_{cd})^* = f^{cd}{}_{ab}\;.
\end{equation}
In each case the structure constants obey a fundamental identity, the analog of the Jacobi identity for an ordinary Lie algebra.  In this section we consider three-dimensional Chern-Simons theories with ${\cal N} = 5$ and ${\cal N} = 4$ superconformal symmetry \cite{Gaiotto:2008sd,Hosomichi:2008jd,Hosomichi:2008jb,Bergshoeff:2008bh,Aharony:2008gk}. We will see that they too are described by a set of 3-algebras \cite{deMedeiros:2008zh,deMedeiros:2009eq,Palmkvist:2009qq,Chen:2009cwa,Bagger:2010zq,Chen:2010xj,Kim:2010kq,Palmkvist:2011aw,Chen:2009ti,Chen:2011fe}. In this section we  closely follow the presentation of \cite{Bagger:2010zq}.

\subsection{Superconformal CS theories with ${\cal N} = 5$}\label{sec:level1} 

We start with the case of ${\cal N} = 5$.  For ${\cal N} = 8$ and ${\cal N} = 6$, the R-symmetry group is SO(8) and SO(6) $\simeq$ SU(4), respectively.  For ${\cal N} = 5$, the R-symmetry group is SO(5) $\simeq$ Sp(4).  Therefore we take the supersymmetry parameter $\epsilon_{AB}$ to be a spacetime spinor in the five-dimensional anti-symmetric tensor representation of Sp(4), with
\begin{equation}
\epsilon^{AB} = -\epsilon^{BA}, \qquad \epsilon^{AB}  \omega_{AB}= 0\;,
\end{equation}
where $A,B = 1,...,4$ and $\omega_{AB}$ is the invariant anti-symmetric tensor of Sp(4), with $\omega^{AB}\omega_{BC} = -\delta^A{}_C$ and $\omega^{AB} = (\omega_{AB})^*$.  The Sp(4) indices are raised and lowered using the antisymmetric tensors $\omega^{AB}$ and $\omega_{AB}$, respectively.  In particular, this implies
\begin{equation}
\epsilon_{AB} = \omega_{AC}\omega_{BD}\epsilon^{CD}\;,
\end{equation}
where $\epsilon_{AB} = (\epsilon^{AB})^*$.

For the ${\cal N} = 5$ theory, the matter fields are in the four-dimensional spinor representation of Sp(4).  The bosonic fields are scalars, which we write as $Z^A_a$, where $A=1,...,4$ and the index $a$ runs over the gauge group.  The fermionic fields are spacetime spinors, which we write as $\Psi_{Aa}$.  The fields obey reality conditions,
\begin{eqnarray}\label{constraint}
\nonumber (Z^A_a)^* &=& \bZ^a_A \ =\ -J^{ab}\omega_{AB}Z^B_b \\[1mm]
(\Psi_{Aa})^* &=& \Psi^{Aa}\ =\ -J^{ab}\omega^{AB}\Psi_{Bb}\;,
\end{eqnarray}
where $\omega_{AB}$ is the Sp(4) invariant tensor, and $J_{ab}$ is an invariant (anti-symmetric) tensor of the gauge group, with $J^{ab} J_{bc} = -\delta^a{}_c$.  The minus sign in the second term is chosen to render the constraint consistent with the ${\cal N} = 5$ supersymmetry transformations.  

With these conventions, the ${\cal N} = 5$ supersymmetry transformations take the following form
\begin{eqnarray}\label{transf1}
\nonumber \delta Z^A_d &=& i\bar{\epsilon}^{AD}\Psi_{Dd} \\[1mm]
\delta\Psi_{Dd} &=& \gamma^{\mu}\epsilon_{AD}D_{\mu}Z^A_d 
+ h^{abc}{}_d Z^A_aZ^B_bZ^C_c\epsilon_{AB}\omega_{DC} +\ j^{abc}{}_dZ^A_aZ^B_bZ^C_c \epsilon_{DC}\omega_{AB} \;,
\end{eqnarray} 
where the gauge-covariant derivative is given by
\begin{equation}
D_{\mu} Z^A _d = \partial_{\mu}Z^A _d - \tilde{A}_\mu{}^a{}_{d}Z^A _a \;.
\end{equation}
The tensors $h^{abcd}=\omega^{de}h^{abc}{}_e$ and $j^{abcd}=\omega^{de}j^{abc}{}_e$ are real with
\begin{eqnarray}
(h^{abcd})^* &=& h_{abcd}\ =\ \omega_{ae}\omega_{bf}\omega_{cg}\omega_{dh} h^{efgh} \nonumber\\[1mm]
(j^{abcd})^* &=& j_{abcd}\ =\ \omega_{ae}\omega_{bf}\omega_{cg}\omega_{dh} j^{efgh},
\end{eqnarray}
and, without loss of generality, they are anti-symmetric in their first two indices.  

Closing on the scalar, we find 
\begin{equation}
[\delta_1,\delta_2]Z^A_d = v^{\mu}D_{\mu}Z^A_d + \tilde{\Lambda}^a{}_d Z^A_a,
\end{equation}
with 
\begin{equation}\label{Neq5gauge}
\tilde{\Lambda}^a{}_d = i h^{abc}{}_d Z^B_bZ^C_c \omega_{DC} 
\bar{\epsilon}^{DF}_{[2}\epsilon_{1]BF}  \;,
\end{equation}
and
\begin{equation}
j^{abc}{}_d = \frac{1}{2} (h^{bca}{}_d - h^{acb}{}_d)\;.
\end{equation}
This implies
\begin{eqnarray}\label{transf2}
\nonumber \delta\Psi_{Dd} &=& \gamma^{\mu}\epsilon_{AD}D_{\mu}Z^A_d 
\ +\ h^{abc}{}_d Z^A_aZ^B_bZ^C_c\epsilon_{AB}\omega_{DC}\ -\ h^{acb}{}_d Z^A_aZ^B_bZ^C_c \epsilon_{DC}\omega_{AB}\;.
\end{eqnarray}
Closing on the fermion gives
\be
 [\delta_1,\delta_2]\Psi_{Dd} = v^\mu D_\mu \Psi_{Dd} +
\tilde\Lambda^a{}_d\Psi_{Da}-\ \frac{i}{2}\bar\epsilon_{[1}^{AC}\epsilon_{2]AD}E_{Cd} +\ \frac{i}{4}(\bar\epsilon^{AB}_1\gamma_\nu\epsilon_{2AB})\gamma^\nu
E_{Dd}\;, 
\ee
with the following fermion equation of motion:
\be
\nn E_{Dd} = \gamma^{\mu}D_{\mu}\Psi_{Dd} -\ h^{abc}{}_d(\Psi_{Dc}Z^A_aZ^B_b + \Psi_{Db}Z^A_aZ^B_c )\omega_{AB} +\ 2 h^{abc}{}_d (\Psi_{Ab}Z^A_aZ^C_c +
\Psi_{Ac}Z^A_aZ^C_b) \omega_{DC}\ =\ 0\;.
\ee
For these results to hold, the gauge field must transform as follows,
\begin{equation} \label{transf3}
\delta\tilde{A}_{\mu}{}^a{}_d \ =\  -i (h^{acb}{}_d + h^{abc}{}_d)
\omega^{BE}\bar{\epsilon}_{EC}\gamma_{\mu}\Psi_{Bb}Z^C_c \;.
\end{equation}

Closing on the gauge field imposes additional constraints:
\begin{eqnarray}
h^{abc}{}_{g}(h^{edg}{}_{f} + h^{egd}{}_{f})  Z^A_a Z^B_b Z^C_c Z^D_d \omega_{AD}\omega_{BC}
&=& 0 \nonumber \\
h^{abc}{}_{g}(h^{edg}{}_{f} + h^{egd}{}_{f})  Z^A_a Z^B_b Z^C_c Z^D_d \bar\xi_{AB[1} \gamma^\mu\xi_{2]CD} &=& 0\;.
\label{FIprecursor}
\end{eqnarray}
The ${\cal N}=5$ fundamental identity must be such that these constraints are satisfied.

Up to now, we have worked in complete generality.  To proceed further, we impose additional symmetries on the structure constants $h^{abc}{}_d$.  One choice is to take
\begin{equation}
h^{abe}{}_d J_{ce} = f^{ab}{}_{cd} = - f^{ba}{}_{cd} = - f^{ab}{}_{dc}  \;.
\end{equation}
The notation suggests that the $f^{ab}{}_{cd}$ are structure constants of the ${\cal N} = 6$ 3-algebra, and indeed the constraints (\ref{FIprecursor}) can be shown to be satisfied on account of the ${\cal N} = 6$ fundamental identity (\ref{complexFI}).

It is not hard to show that this ${\cal N}=5$ is just an ${\cal N}=5$ subalgebra of ${\cal N} = 6$.  We first use the constraint (\ref{constraint}) to eliminate $\omega_{AB}$ and $J_{ab}$ from the lagrangian and transformation laws.  We then remove the constraint so that $\bar{Z}^a_A$ and $\Psi^{Aa}$ become the {\it unconstrained} complex conjugates of $Z^A_a$ and $\Psi_{Aa}$.  With this interpretation, the transformations (\ref{transf1}) are precisely those of ${\cal N} = 6$ supersymmetry algebra, 
\begin{eqnarray}
\label{Neq6trans}
 \nonumber \delta Z^A _d &=& i\bar{\epsilon}^{AD}\Psi_{Dd}, \\[1mm]
 \delta\Psi_{Dd} &=& \gamma^{\mu}\epsilon_{AD}D_{\mu}Z^A _d 
+  f^{ab}{}_{cd}Z^A _a Z^B _b \bar{Z}_A^c \epsilon_{BD} \nonumber 
+ f^{ab}{}_{cd}Z^A _a Z^B_b \bar{Z}_D^c \epsilon_{AB} \nonumber \\[1mm]
\delta \tilde{A}_{\mu}{}^a {}_d &=& -i f^{ab}{}_{cd} (\bar{\epsilon}^{BC}\gamma _{\mu} \Psi_{Bb}\bar{Z}^c _C +
\bar{\epsilon}_{BC}\gamma_{\mu}\Psi^{Cc}Z^{B} _b )\;.
\end{eqnarray}
Indeed, the ``sixth" supersymmetry transformation, with $\epsilon_{AB} = -i \omega_{AB} \eta$, is explicitly broken by the constraint (\ref{constraint}).  When the constraint is removed, the full supersymmetry is restored.

A second choice for the structure constants is to take 
\begin{equation}
h^{abc}{}_{d} = g^{acb}{}_d - g^{bca}{}_d,
\end{equation}  
where
\begin{equation}
\label{symm1}
g^{acbd} = g^{cabd} = g^{bdac}.
\end{equation}  
This choice generates a set of ${\cal N} = 5$ theories that are not restrictions of ${\cal N} = 6$.  The conditions (\ref{FIprecursor}) are satisfied if
\begin{equation}
\label{symm2}
g^{(acb)d} = 0
\end{equation}  
and
\begin{equation}\label{Neq5FI}
J_{gj}(g^{afbg}g^{jchd} + g^{afgd}g^{hjbc} + g^{afhg}g^{jdbc} + g^{afgc}g^{bjhd}) = 0.
\end{equation}
Equation (\ref{Neq5FI}) is nothing but the ${\cal N} = 5$ fundamental identity.  

The ${\cal N} = 5$ supersymmetry transformations are found by substituting $g^{abc}{}_d$ for $h^{abc}{}_d$ in (\ref{transf1}), (\ref{transf2}) and (\ref{transf3}) \cite{Chen:2009cwa,Bagger:2010zq}
\begin{eqnarray}\label{n5transf}
\nonumber \delta Z^A_d &=& i\bar{\epsilon}^{AD}\Psi_{Dd} \\[1mm]
\nonumber \delta\Psi_{Dd} &=& \gamma^{\mu}\epsilon_{AD}D_{\mu}Z^A_d - g^{abc}{}_d Z^A_aZ^B_bZ^C_c \epsilon_{DB}\omega_{AC} +\ 2g^{abc}{}_d Z^A_aZ^B_bZ^C_c\epsilon_{AC}\omega_{DB} \nonumber\\[1mm]
\delta\tilde{A}_{\mu}{}^a{}_d &=&  3i g^{bca}{}_d
\omega^{BE}\bar{\epsilon}_{EC}\gamma_{\mu}\Psi_{Bb}Z^C_c .
\end{eqnarray} 
These transformations close into a translation and a gauge variation, with parameter
\begin{equation}
\tilde{\Lambda}^a{}_d = - \frac{3i}{2} g^{bca}{}_d Z^B_bZ^C_c \omega_{DC} 
\bar{\epsilon}^{DF}_{[2}\epsilon_{1]BF} .
\end{equation}
They also leave invariant the ${\cal N}=5$ lagrangian \cite{Chen:2009cwa}
\begin{eqnarray}
{\cal L} &=& -\,D^\mu \bar{Z}^a_A D_\mu Z^A_a -i \bar{\Psi}^{Aa}\gamma^\mu D_\mu \Psi_{Aa} - V + {\cal L}_{CS} \nonumber\\[3mm]
&& -\ 3i \,g^{acbd} \,\omega_{AB} \omega_{CD}\, ( Z^A_a Z^B_b \bar{\Psi}^C_c\Psi^D_d - 2
Z^A_a Z^D_b \bar{\Psi}^C_c\Psi^B_d )
\end{eqnarray} 
up to a total derivative, where
\begin{equation}
V = \frac{12}{5}\,\bar\Upsilon^{d}_{ABC}\Upsilon^{ABC}_d
\end{equation}
with
\begin{equation}
\Upsilon^{ABC}_d = g^{abc}{}_d \left(Z^A_a Z^B_b Z^C_c + \frac{1}{4} \omega^{BC} Z^A_a Z^D_b Z_{Dc}\right).
\end{equation}

\subsection{${\cal N} = 5$ gauge groups}\label{sec:level5}

In this section we construct ${\cal N} = 5$ gauge theories, built from the symmetric structure constants $g^{abc}{}_{d}$, with gauge transformations
\begin{equation}
\delta Z^A_d = \tilde{\Lambda}^a{}_d Z^A_a = g^{bca}{}_d \Lambda_{bc} Z^A_a.
\end{equation}
We will see that there are a host of such theories, including some with free parameters or exceptional gauge groups, in vivid contrast to ${\cal N} = 6$ or 8.

We start by constructing a set of $g^{abcd} = J^{de} g^{abc}{}_e$ that lead to an $\Sp(n) \times \SO(m)$ gauge group.  There are four combinations of the invariant tensors of Sp($n$) and SO($m$) that have the symmetries (\ref{symm1}):
\begin{eqnarray}
g_1^{aibjckdl} &=& (\delta^{ac}\delta^{bd} - \delta^{ad}\delta^{bc})J^{ij}J^{kl} \\
\nonumber g_2^{aibjckdl} &=& (J^{ik}J^{jl} + J^{jk}J^{il})\delta^{ab}\delta^{cd} \\
\nonumber g_3^{(\pm) aibjckdl} &=&(\delta^{ac}\delta^{bd} \pm \delta^{ad}\delta^{bc})(J^{ik}J^{jl} \pm J^{jk}J^{il}),
\end{eqnarray}
where $i,j,... = 1,...\ n$ are Sp($n$) indices, and $a,b,... = 1,...\ m$ are SO($m$).  However, there are only two linear combinations that satisfy (\ref{symm2}) and the fundamental identity (\ref{Neq5FI}):
\begin{eqnarray}\label{N5struct}
g^{aibjckdl} &=& - \frac{2\pi}{k} \left[ g_1^{aibjckdl} - g_2^{aibjckdl}  \right] \\[1mm]
\nonumber g^{aibjckdl} &=& - \frac{2\pi}{k} \left[ g_3^{(+)aibjckdl} + g_3^{(-)aibjckdl} \right].
\end{eqnarray} 

Let us look at the first case first.  The structure constants are 
\begin{equation}\label{Neq5g}
g^{aibjckdl} = - \frac{2\pi}{k} \left[(\delta^{ac}\delta^{bd} - \delta^{ad}\delta^{bc})J^{ij}J^{kl} - \delta^{ab}\delta^{cd}(J^{ik}J^{jl} + J^{jk}J^{il})\right].
\end{equation}
They give rise to the following gauge transformation:
\begin{eqnarray}
\delta Z^{Adl} &=&  - \frac{2\pi}{k}\left[(\delta^{ba}\delta^{cd} - \delta^{bd}\delta^{ca}) J^{jk}J^{il} - \delta^{bc}\delta^{ad}(J^{ji}J^{kl} + J^{ki}J^{jl}) \right]\Lambda_{bjck}  Z^A_{ai} \nonumber \\
&=  &  - \frac{2\pi}{k} \left[(\Lambda_{ajdk} - \Lambda_{djak}) J^{jk}J^{il} - \delta^{ad}(J^{ji}J^{kl} + J^{ki}J^{jl}) \Lambda_{bjbk} \right]
Z^A_{ai}.
\end{eqnarray}
The two terms are Sp($n$) and SO($m$) transformations, respectively, with matter in the fundamental representations of each \cite{Aharony:2008gk,Bergshoeff:2008bh,deMedeiros:2009eq}.

For the second case, the structure constants are simply
\begin{equation}\label{Neq5g2}
g^{aibjckdl} = - \frac{2\pi}{k}\left[J^{ik}J^{jl} \delta^{ac}\delta^{bd}  +  J^{il}J^{jk} \delta^{ad}\delta^{bc} \right].
\end{equation}
The indices are in standard direct product form, so the theory has gauge group Sp($mn$), with matter in the $mn$-dimensional fundamental representation.

For the special case of SO(4) $\times$ Sp(2) $\simeq$ SO(4) $\times$ SU(2), it is possible to add another term to the structure constants \cite{Bergshoeff:2008bh,deMedeiros:2009eq}:
\begin{equation}\label{SO(4)}
g^{aibjckdl} = - \frac{2\pi}{k}\left[ g_1^{aibjckdl} - g_2^{aibjckdl}  + \alpha \varepsilon^{abcd} J^{ij} J^{kl}\right].
\end{equation}
Here $\varepsilon^{abcd}$ is the totally antisymmetric SO(4)-invariant tensor.  The resulting $g^{aibjckdl}$ satisfy (\ref{symm2}) and the fundamental identity, for any choice of the parameter $\alpha$.  The gauge group is SO(4) $\times$ SU(2) for $\alpha \ne \infty$.  In the limit $\alpha \rightarrow \infty$, the gauge group is SO(4), and the resulting theory lifts to ${\cal N} = 6$ and $8$.

There are also two ``exceptional" theories with ${\cal N} = 5$.  The first arises from the tensor
\begin{equation}
\label{G2structure}
g^{aibjckdl} = - \frac{2\pi}{k} \left[ g_1^{aibjckdl} - g_2^{aibjckdl}  + \beta C^{abcd} J^{ij} J^{kl} \right],
\end{equation}
where $a,b,... = 1,...\ 7$ and $i,j,... = 1,2$ are SO(7) and SU(2) indices, respectively.  Here $C^{abcd}$ is the totally antisymmetric tensor that is dual to the octonionic structure constants\footnote{For a concise introduction to G$_2$, SO(7) and the octonions, as well as a host of useful identities, see Section 2 and Appendix A of \cite{Bilal:2001an}.}
 $C_{efg}$,
\begin{equation}
C^{abcd} = \frac{1}{3!}\varepsilon^{abcdefg}C_{efg}.
\end{equation}    
The tensor (\ref{G2structure}) satisfies (\ref{symm2}) and the fundamental identity for $\beta = 0$ or $\beta =  \frac{1}{2}$.  

When $\beta=0$, the $g^{aibjckdl}$ are just the Sp(2) $\times$ SO(7) structure constants discussed above.  When $\beta = \frac{1}{2}$, the gauge group is G$_2$ $\times$ SU(2).  In this case, the structure constants take the form
\be\label{G2g}
g^{aibjckdl} = - \frac{2\pi}{k} \Big[ \Big(\delta^{ac}\delta^{bd} - \delta^{ad}\delta^{bc} + \frac{1}{2}C^{abcd}\Big)J^{ij}J^{kl}  - \  \delta^{ab}\delta^{cd}(J^{ik}J^{jl}+J^{jk}J^{il }) \Big],
\ee
with $i,j,... = 1,2$.  The gauge transformation is then
\begin{eqnarray}
\nonumber \delta Z^{Adl} &=& g^{bjckaidl} \Lambda_{bjck} Z^A_{ai} \\[2mm]
&=&
- \frac{2\pi}{k} \Big[ J^{il} \Big(\delta^{ba}\delta^{cd} - \delta^{bd}\delta^{ca} + \frac{1}{2}C^{bcad}\Big)J^{jk} \Lambda_{bjck}-\ \delta^{ad}(J^{ji}J^{kl}+J^{ki}J^{jl }) \Lambda_{bjbk}  \Big].
\end{eqnarray}
The second term is clearly an SU(2) transformation.  The first is a G$_2 \subset$ SO(7) transformation, as can be seen by recognizing that the operator
\begin{equation}
{\cal P}^{abcd}_{14} = \frac{1}{3}\left(\delta^{ab}\delta^{cd} - \delta^{ac}\delta^{bd} + \frac{1}{2}C^{abcd}\right)
\end{equation}
is a projector from the adjoint $\bf 21$ of SO(7) to the adjoint $\bf 14$ of G$_2$, 
\begin{equation}
{\cal P}^{abcd}_{14}C_{bce} = 0.
\end{equation}
This proves that the gauge group is G$_2$ $\times$ SU(2), recovering the result found in \cite{Bergshoeff:2008bh,deMedeiros:2009eq}.

The second exceptional theory has SO(7) $\times$ SU(2) gauge symmetry with matter transforming in the {\it spinor} $\bf 8$ of SO(7) \cite{Bergshoeff:2008bh,deMedeiros:2009eq}.  The structure constants are
\begin{equation}
g^{aibjckdl} =- \frac{2\pi}{k} \left[ \delta^{ab}\delta^{cd}(J^{ik}J^{jl} + J^{jk}J^{il})-
\frac{1}{6} \Gamma^{ab}_{mn}\Gamma^{cd}_{mn}J^{ij}J^{kl} \right],
\end{equation}
where $a,b,... = 1,...\ 8$ and $i,j,... = 1,2$, and  $\Gamma^{ab}_{mn} = \frac{1}{2}(\Gamma_m\Gamma_n - \Gamma_n\Gamma_m)^{ab}$ is built from the $\SO(7)$ gamma matrices.  The structure constants have the correct symmetries and satisfy the fundamental identity.  The gauge transformations are
\begin{eqnarray}
\delta Z^{Adl} &=& g^{bjckaidl} \Lambda_{bjck} Z^A_{ai} \nonumber\\[1mm]
&=&- \frac{2\pi}{k} \Big[ \delta^{ad}(J^{ji}J^{kl} + J^{ki}J^{jl}) \Lambda_{bjbk} -\ \frac{1}{6} J^{il} \Gamma^{ad}_{mn}\Gamma^{bc}_{mn}J^{jk}\Lambda_{bjck} \Big]   Z^A_{ai}.
\end{eqnarray}
The gauge group is SO(7) $\times$ SU(2), with the matter fields transforming in the spinor representation of each.

In fact, the ${\cal N} = 5$ theories presented here are in one-one correspondence with the Lie superalgebras OSp($m|n$), D($2|1;\alpha$), G(3) and F(4) \cite{Gaiotto:2008sd,FigueroaO'Farrill:2009pa,deMedeiros:2009eq,Chen:2009cwa,Kim:2010kq}.  The 3-algebra structure constants can be built from the superalgebra structure constants as follows,
\begin{equation}
g^{abcd} = h_{mn} (\tau^{ma}{}_e J^{be})(\tau^{nc}{}_f J^{df})\;,
\end{equation}
where $h_{mn}$ is the invariant quadratic form on the algebra.  The $g^{abcd}$ obey the correct symmetries because $\tau^{ma}{}_e J^{be}=\tau^{mb}{}_e J^{ae}$.  They satisfy the ${\cal N}=5$ fundamental identity because of the $\tau^{ma}{}_b$ satisfy the superalgebra Jacobi identity.

\subsection{Lifting ${\cal N} = 5$ to ${\cal N} = 6$}

In this section, we show how to lift two theories with ${\cal N} = 5$ supersymmetry to ${\cal N} = 6$, along the lines of the lift from ${\cal N} = 6$ to ${\cal N} = 8$.  In particular, we lift the ${\cal N} = 5$ theories with Sp($n$) $\times$ SO(2) and SO(4) $\times$ SU(2) gauge symmetry to ${\cal N} = 6$ theories with Sp($n$) $\times$ U(1) and SO(4) gauge symmetry, respectively.  As we showed previously, the latter theory can then be lifted to ${\cal N} = 8$ \cite{Bagger:2010zq}.

To carry out the lifts, we first define unconstrained complex-conjugate scalar fields ${\cal Z}^A_a$ and $\bar{\cal Z}_A^a$, 
\begin{eqnarray}\label{calZ}
{\cal Z}^A_a &=& Z^A_{a1} + iZ^A_{a2} \nonumber \\[1mm]
 \bar{{\cal Z}}_A^a &=& \bar Z_A^{a1} - i\bar Z_A^{a2}.
\end{eqnarray}
Supersymmetry then requires that the superpartner $\Xi_{Aa}$ be defined as follows:
\begin{eqnarray}\label{calPsi}
\Xi_{Aa} &=& \Psi_{Aa1} + i\Psi_{Aa2} \nonumber \\[1mm]
\Xi^{*Aa} &=& \Psi^{Aa1} - i\Psi^{Aa2}.
\end{eqnarray}
The indices 1 and 2 refer to either SO(2) or SU(2), while $a$ refers to Sp($n$) or SO(4), respectively.  The definitions 
\begin{eqnarray}
\bar{Z}^{ai}_A &=& - \omega_{AB} J^{ab} \delta^{ij} Z^B_{bj} \nonumber\\
\Psi^{Aai} &=& - \omega^{AB} J^{ab} \delta^{ij} \Psi_{Bbj} 
\quad {\rm [for\ Sp(}n \rm) \times {\rm SO(2)]}
\end{eqnarray}
and
\begin{eqnarray}
\bar{Z}^{ai}_A &=& - \omega_{AB} \delta^{ab} \varepsilon^{ij} Z^B_{bj} \nonumber\\
\Psi^{Aai} &=& - \omega^{AB} \delta^{ab} \varepsilon^{ij} \Psi_{Bbj}
\quad {\rm [for\ SO(4) \times SU(2)]}
\end{eqnarray}
allow us to write the complex-conjugate expressions (\ref{calZ}) and (\ref{calPsi}) in terms of the original fields.  Note that this construction only works when one of the ${\cal N}=5$ gauge groups is SU(2) or SO(2).

We first consider the theory with Sp($n$) $\times$ SO(2) gauge symmetry, where $a,b,...=1,...\ n$ are Sp($n$) indices, and $i,j,... = 1,2$ are SO(2).  The conjugate scalar $\bar{{\cal Z}}_{A}^{a}$ takes the form
\begin{equation}
\bar{{\cal Z}}_{A}^{a} = -\omega_{AB} J^{ab}(Z^B_{b1}-i Z^B_{b2})\;, 
\end{equation}
and likewise for the conjugate spinor $\Xi^{*Aa}$.
With these definitions, the ${\cal N} = 5$ transformations, with 
\begin{equation}
g^{aibjckdl} =  \frac{4\pi}{3k} 
\left[(\delta^{ik}\delta^{jl} - \delta^{il}\delta^{jk})J^{ab}J^{cd} - \delta^{ij}\delta^{kl}(J^{ac}J^{bd} + J^{bc}J^{ad})\right],
\end{equation}
coincide with the ${\cal N} = 6$ transformations, with
\begin{equation}
f^{ab}{}_{cd} = - \frac{2\pi}{k} \left[J^{ab} J_{cd} + (\delta^a _c \delta^b _d - \delta^a _d \delta^b _c )\right],
\end{equation} 
for five of the six supersymmetries.

To find the sixth, we plug $\epsilon_{AB} \rightarrow -i \omega_{AB} \eta$ into (\ref{Neq6trans}) and collect terms.  After some calculation, we recover:
\begin{eqnarray}
\nonumber \delta Z^A_{dl} &=& - \omega^{AD} \bar\eta \Psi_{Ddl} \\[1mm]
\nonumber
\delta\Psi_{Ddl} &=& -i\gamma^{\mu}\omega_{AD} \eta D_{\mu}Z^A_{dl}  +\ i f^{ab}{}_{cd}(\omega_{AB}\omega_{CD} -\omega_{AC}\omega_{BD})\times (\varepsilon_{ik}\varepsilon_{jl} + \varepsilon_{jk}\varepsilon_{il} + i \delta_{ij}\varepsilon_{kl})
Z^A_{ai}Z^B_{bj}Z^{Cc}_{k}\, \eta \nonumber \\[1mm]
\delta \tilde{A}_{\mu}{}^{aidl} &=& i f^{abcd} (\bar{\eta}\gamma _{\mu} \Psi_{Bbj}Z^B_{ck}
-\bar{\eta}\gamma _{\mu} \Psi_{Bck}Z^B_{bj}) (\delta^{jk}\varepsilon^{il}+\varepsilon^{jk}\delta^{il})\;,
\end{eqnarray} 
where $\varepsilon^{ij}$ is the antisymmetric, invariant tensor of SO(2).  This is the extra supersymmetry transformation that lifts the ${\cal N} = 5$ theory with Sp($n$) $\times$ SO(2) gauge symmetry to the ${\cal N} = 6$ theory with Sp($n$) $\times$ U(1). 

Finally, we consider the ${\cal N} = 5$ theory with SO(4) $\times$ SU(2) gauge symmetry, with $g^{aibjckdl}$ given in (\ref{SO(4)}), in the limit $\alpha \rightarrow \infty$.  In this limit, the structure constants reduce to
\begin{equation}
g^{aibjckdl}  \rightarrow \alpha \varepsilon^{abcd} \varepsilon^{ij} \varepsilon^{kl},
\end{equation} 
where $a,b,...=1,...\ 4$ are SO(4) indices,  $i,j,... = 1,2$ are SU(2), and $\varepsilon^{ij}$ is the antisymmetric, invariant tensor of SU(2).  We first compute the gauge transformation.  Using (\ref{Neq5gauge}), we find
\begin{equation}
\label{limit}
\delta Z^{D}_{dl} \rightarrow \alpha \varepsilon^{abcd} \varepsilon^{jk} \varepsilon^{il} \Lambda_{bjck} Z^A_{ai}.
\end{equation}
This is a pure SO(4) gauge transformation (the SU(2) is not gauged in this limit).  Equation ({\ref{limit}) suggests that the SO(4) $\times$ SU(2) invariant ${\cal N}=5$ theory, in the $\alpha \rightarrow \infty$ limit, can be lifted to the SO(4) theory with ${\cal N}=6$ and 8.

We now construct the lift.  We first define the complex-conjugate scalars ${\cal Z}^{A}_{a}$ and $\bar{{\cal Z}}_{A}^{a}$.  For the case at hand, $\bar{{\cal Z}}_{A}^{a}$ is
\begin{equation}
\bar{{\cal Z}}_{Aa} = -i\omega_{AB}\delta_{ab} (Z^B_{b1}-i Z^B_{b2})
\end{equation}
and likewise for the spinor $\Xi^{*Aa}$.
As above, it possible to show that the ${\cal N} = 5$ transformations, with 
\begin{equation}
g^{aibjckdl} = \frac{4\pi}{3k} \varepsilon^{abcd} \varepsilon^{ij} \varepsilon^{kl},
\end{equation}
coincide with the ${\cal N} = 6$ transformations, with
\begin{equation}
f^{abcd} = - \frac{2\pi}{k} \varepsilon^{abcd},
\end{equation} 
for five of the six supersymmetries.

The sixth supersymmetry is derived in the same way as before.  Plugging $\epsilon_{AB} \rightarrow -i \omega_{AB} \eta$ into (\ref{Neq6trans}) and collecting terms, we find:
\begin{eqnarray}
\nonumber \delta Z^A_{dl} &=& - \omega^{AD} \bar\eta \Psi_{Ddl} \\[1mm]
\nonumber
\delta\Psi_{Ddl} &=& -i\gamma^{\mu}\omega_{AD} \eta D_{\mu}Z^A_{dl}  -\  \frac{4\pi}{k}\varepsilon^{abcd} \, \omega_{AB}\omega_{CD}\, \delta_{ik}\delta_{jl} \,
Z^A_{ai}Z^B_{bj}Z^C_{ck}\, \eta \nonumber \\[1mm]
\delta \tilde{A}_{\mu}{}^{aidl} &=& i \frac{4\pi}{k} \varepsilon^{abcd} \varepsilon^{il} \bar{\eta}\gamma _{\mu} \Psi_{Bbj} Z^{B}_{cj}\;.
\end{eqnarray}
Note that the interaction term explicitly breaks the global SU(2) symmetry.  The transformation is just what we need to lift the ${\cal N} = 5$ theory with $\SU(2)\times \SO(4)$ gauge symmetry, in the $\alpha \rightarrow \infty$ limit, to the ${\cal N} = 6$ theory with $\SO(4)$ gauge symmetry.  In Section~\ref{gaugedet}, we showed that this theory can again be lifted to ${\cal N} = 8$. 

\subsection{Superconformal CS theories with ${\cal N} = 4$}

In this section we use the results of the previous section to construct three-dimensional superconformal theories with ${\cal N}=4$  supersymmetry.  We exploit the fact that the ${\cal N}=4$ R-symmetry group is $\SO(4) \simeq \SU(2) \times \SU(2)$.  Following Gaiotto and Witten \cite{Gaiotto:2008sd}, we take the bosonic matter fields $Z^A_a$ to be in the ${\bf (2,1)}$ representation of $\SU(2) \times \SU(2)$, and the spinor fields $\Psi_{\dot A a}$ to be in the ${\bf (1,2)}$.  The notation is such that $A=1,2$ spans the spinor of the first SU(2), while $\dot A = 1,2$ spans the spinor of the second.  Indices are raised and lowered with the antisymmetric tensors $\varepsilon^{AB},$ $\varepsilon_{AB} = (\varepsilon^{AB})^*$ and $\varepsilon^{\dot A\dot B},$ $\varepsilon_{\dot A\dot B} = (\varepsilon^{\dot A\dot B})^*$, with $\varepsilon^{AB}\varepsilon_{BC} = - \delta^A{}_C$, and likewise for the dotted indices.\footnote{Note that the dotting of spinors has nothing to do with complex conjugation.}

For ${\cal N}=4$, the supersymmetry parameter is a vector of $\SO(4)$, or equivalently, in the ${\bf (2,2)}$ representation of $\SU(2) \times \SU(2)$.  Therefore we describe the supersymmetry parameter by a $2 \times 2$ matrix
\begin{equation}
\eta_{A \dot A} = 
\begin{pmatrix}
a & b \\
-b^* & a^*
\end{pmatrix}\;, \qquad
\eta^{A \dot A} = \varepsilon^{AB}\varepsilon^{\dot A \dot B}  \eta_{B \dot B}=
\begin{pmatrix}
a^* & b^* \\
-b & a
\end{pmatrix}\;. 
\end{equation}
With these conventions, the bosonic supersymmetry transformation takes the following simple form,
\begin{equation}
\delta Z^A_d = i\bar{\eta}^{A \dot A}\Psi_{\dot A d}\;.
\end{equation}
The index $d$ runs over the representation of the gauge group, exactly as in ${\cal N}=5$ and ${\cal N}=6$.  

The allowed gauge groups are determined by the 3-algebra structure constants.  There are essentially two choices, depending on whether the fields are real or complex.  For real fields, the gauge groups turn out to be those of ${\cal N}=5$, while for complex fields, they are those of ${\cal N}=6$.  

To see how this works, we first consider the case with complex fields.   We map the ${\cal N}=4$ fields into the ${\cal N}=6$ fields as follows:
\begin{eqnarray}
\begin{pmatrix}
Z^A_a \\
0
\end{pmatrix}
 \rightarrow 
\ Z_a^A 
\ &\quad&\,   
\begin{pmatrix}{\bar Z}_A^a \\
0
\end{pmatrix} 
\rightarrow\ {\bar Z}_A^a
\nonumber\\[1mm]
\begin{pmatrix}
0 \\
\Psi_{\dot A a}
\end{pmatrix}
\rightarrow \Psi_{Aa} 
 &&  
 \begin{pmatrix}
 0 \\
 \Psi^{\dot A a}
 \end{pmatrix}
\rightarrow \Psi^{Aa} \;,
\end{eqnarray}
where the R-symmetry indices run from 1 to 4 in the case of ${\cal N}=6$, and from 1 to 2 for ${\cal N}=4$.  In a similar fashion, we embed the ${\cal N}=4$ supersymmetry parameters in the ${\cal N}=6$ parameters as follows,
\begin{equation}\label{N4SUSY}
\begin{pmatrix}0&\eta^{A\dot B}\\-(\eta^{T})^{\dot A B}&0
\end{pmatrix}
\rightarrow\, \epsilon^{AB} \;,\qquad
\begin{pmatrix}0&\eta_{A\dot B}\\-(\eta^{T})_{\dot A B}&0
\end{pmatrix}
\rightarrow\, \epsilon_{AB} \;,
\end{equation}
which amounts to defining $\eta^{\dot A A} = - \eta^{A \dot A}$, and likewise for the lower indices.
With these conventions, it is not hard to extract the ${\cal N}=4$ supersymmetry transformations from (\ref{Neq6trans}),
\begin{eqnarray}
 \nonumber \delta Z^A_d &=& i\bar{\eta}^{A \dot D}\Psi_{\dot D d} \\[1mm]
 \delta\Psi_{\dot D d} &=& \gamma^{\mu}\eta_{A \dot D}D_{\mu}Z^A_d 
+  f^{ab}{}_{cd}Z^A_a Z^B_b \bar{Z}_A^c \eta_{B \dot D} \nonumber\\[1mm]
\delta \tilde{A}_{\mu}{}^a {}_d &=& -i f^{ab}{}_{cd} (\bar{\eta}^{C \dot B }\gamma_{\mu} \Psi_{\dot B b}\bar{Z}^c_C +
\bar{\eta}_{B \dot C}\gamma_{\mu}\Psi^{\dot C c}Z^{B}_b )\;.
\end{eqnarray}
These transformations close when the $f^{ab}{}_{cd}$ are the ${\cal N}=6$ structure constants.  In Chapter~\ref{chapter3} we found the allowed gauge groups to be $\SU(n) \times \SU(n)$, $\SU(n) \times \SU(m) \times \U(1)$ (when $n\ne m$), and $\Sp(n) \times \U(1)$.

When the fields are real, we proceed in a similar fashion.  We take the reality condition to be
\begin{eqnarray}\label{constraint2}
\nonumber (Z^A_a)^* &=& \bZ^a_A \ =\ -J^{ab}\varepsilon_{AB}Z^B_b \\[1mm]
(\Psi_{\dot Aa})^* &=& \Psi^{\dot Aa}\ =\ -J^{ab}\varepsilon^{\dot A \dot B}\Psi_{\dot B b}\;,
\end{eqnarray}
where $J^{ab}$ is the antisymmetric invariant tensor introduced previously.  As above, we extract the ${\cal N}=4$ supersymmetry transformations from those of ${\cal N}=5$.

To find the transformations we must embed the SU(2) invariant tensors into the invariant tensor of Sp(4).  We choose
\begin{equation}
\begin{pmatrix}
\varepsilon_{AB} & 0 \\
0 & \varepsilon_{\dot A \dot B}
\end{pmatrix}
\rightarrow\, \omega_{AB}\;,
\qquad
\begin{pmatrix}
\varepsilon^{AB} & 0 \\
0 & \varepsilon^{\dot A \dot B}
\end{pmatrix}
\rightarrow\, \omega^{AB}\;,
\end{equation}
where the index conventions are as before.  The supersymmetry parameter remains as in (\ref{N4SUSY}).  

With these conventions, the ${\cal N}=4$ supersymmetry transformations can read directly from (\ref{n5transf}).  We find
\begin{eqnarray}
\nonumber \delta Z^A_d &=& i\bar{\eta}^{A\dot D}\Psi_{\dot D d} \\[1mm]
\nonumber \delta\Psi_{\dot Dd} &=& \gamma^{\mu}\eta_{A\dot D}D_{\mu}Z^A_d + g^{abc}{}_d Z^A_aZ^B_bZ^C_c \eta_{B\dot D}\varepsilon_{AC} 
\nonumber\\[1mm]
\delta\tilde{A}_{\mu}{}^a{}_d &=&  -3i g^{bca}{}_d
\varepsilon^{\dot B\dot E}\bar{\eta}_{C\dot E}\gamma_{\mu}\Psi_{\dot B b}Z^C_c \;,
\end{eqnarray} 
where the $g^{abc}{}_d$ are the ${\cal N}=5$ structure constants.  The supersymmetry transformations close because the $g^{abc}{}_d$ enjoy the correct symmetries and obey the ${\cal N}=5$ fundamental identity.  The gauge groups are those of ${\cal N}=5$, namely $\Sp(n) \times \SO(m)$, $\SO(4) \times \Sp(2)$,  G$_2 \times \SU(2)$, and $\SO(7) \times \SU(2)$.

The Gaiotto-Witten models can be readily generalised by exploiting the fact that the ${\cal N}=4$ R-symmetry group contains two completely independent SU(2) factors \cite{Hosomichi:2008jd,Hosomichi:2008jb}.  To see how this works, we embed the ${\cal N}=4$ fields in the ${\cal N}=6$ fields as follows:
\begin{eqnarray}
\begin{pmatrix}
Z^A_a \\[1mm]
Z^{\dot A}_{\dot a}
\end{pmatrix}
\rightarrow \ Z_a^A 
\ &&\,   
\begin{pmatrix}{\bar Z}_A^a \\[1mm]
Z_{\dot A}^{\dot a}
\end{pmatrix} 
\rightarrow\ {\bar Z}_A^a 
\nonumber\\[1mm]
\begin{pmatrix}
\Psi_{A \dot a} \\[0mm]
\Psi_{\dot A a}
\end{pmatrix}
\rightarrow \Psi_{Aa} 
 &&   
 \begin{pmatrix}
  \Psi_{\dot A a} \\[0mm]
 \Psi^{\dot A a}
 \end{pmatrix}
\rightarrow \Psi^{Aa} 
 \;,
\end{eqnarray}
where the R-symmetry index runs as before, but now the gauge indices are free to run over different values for each of the SU(2) factors.  In the literature, the $Z^A_a$ are called hypermultiplets, while the $Z^{\dot A}_{\dot a}$ are sometimes called ``twisted'' hypermultiplets \cite{Hosomichi:2008jd}.

Formally, one can extract the ${\cal N}=4$ supersymmetry transformations from those of ${\cal N}=6$ and ${\cal N}=5$.  For complex fields, one finds
\begin{eqnarray}
 \nonumber \delta Z^A_d &=& i\bar{\eta}^{A \dot D}\Psi_{\dot D d} \\[1mm]
 \nonumber \delta Z^{\dot A}_{\dot d} &=& -i\bar{\eta}^{D \dot A}\Psi_{D \dot d} \\[1mm]
 \delta\Psi_{\dot D d} &=& \ \gamma^{\mu}\eta_{A \dot D}D_{\mu}Z^A_d 
+  f^{ab}{}_{cd}Z^A_a Z^B_b \bar{Z}_A^c \eta_{B \dot D}  +\ f^{\dot a b}{}_{\dot c d}Z^{\dot A}_{\dot a} Z^B_b \bar{Z}_{\dot A}^{\dot c} \eta_{B \dot D}  
- 2 f^{\dot a b}{}_{\dot c d}Z^{\dot A}_{\dot a} Z^B_b \bar{Z}_{\dot D}^{\dot c} \eta_{B\dot A}\nonumber \\[1mm]
 \delta\Psi_{D \dot d} &=& -\ \gamma^{\mu}\eta_{D\dot A}D_{\mu}Z^{\dot A}_{\dot d} 
-  f^{\dot a\dot b}{}_{\dot c\dot d}Z^{\dot A}_{\dot a} Z^{\dot B}_{\dot b} \bar{Z}_{\dot A}^{\dot c} \eta_{D \dot B} -\  f^{a \dot b}{}_{c \dot d}Z^A_a Z^{\dot B}_{\dot b} \bar{Z}_A^c \eta_{D \dot B}  
+ 2 f^{a \dot b}{}_{c\dot  d}Z^A_a Z^{\dot B}_{\dot b} \bar{Z}_{D}^{c} \eta_{A\dot B}\;,
\end{eqnarray}
and likewise for the gauge fields.  The structure constants are purely formal because the dotted and undotted gauge indices run over different values.  Nevertheless, it has been shown that the supersymmetry transformations close into the ${\cal N}=4$ algebra precisely when 
\begin{equation}
f^{ab}{}_{cd} = h_{mn} \tau^{ma}{}_c \tau^{nb}{}_d\;, \qquad f^{a\dot b}{}_{c\dot d} = f^{\dot b a}{}_{\dot d c} = h_{mn} \tau^{ma}{}_c \tau^{n\dot b}{}_{\dot d}\;, 
\end{equation}
where the $\tau^{ma}{}_c$ are structure constants of the Lie superalgebra OSp(2$|n$) or U($n|m$) (or its relatives SU($m|n$) and PSU($m|n$)), and $h_{mn}$ is the invariant quadratic form \cite{Chen:2010ib}.  The structure constants obey all the necessary identities because of the superalgebra Jacobi identities.  When the dotted and undotted indices are identified, the transformations describe the ${\cal N}=4$ subalgebra of ${\cal N}=6$.

A similar story holds when the fields are real.  The reality conditions are
\begin{eqnarray}
\nonumber (Z^A_a)^* &=& \bZ^a_A \ =\ -J^{ab}\varepsilon_{AB}Z^B_b \\[1mm]
 (Z^{\dot A}_{\dot a})^* &=& \bZ^{\dot a}_{\dot A} \ =\ -J^{\dot a\dot b}\varepsilon_{\dot A\dot B}Z^{\dot B}_{\dot b},
\end{eqnarray}
and likewise for $\Psi_{D\dot d}$ and $\Psi_{\dot D d}$.  Here $J^{ab}$ and $J^{\dot a\dot b}$ are antisymmetric tensors, possibly of different dimensions.  The supersymmetry transformations can be extracted from those of ${\cal N}=5$,
\begin{eqnarray}
 \nonumber \delta Z^A_d &=& i\bar{\eta}^{A \dot D}\Psi_{\dot D d} \\[1mm]
 \nonumber \delta Z^{\dot A}_{\dot d} &=& -i\bar{\eta}^{D\dot A}\Psi_{D \dot d} \\[1mm]
 \nonumber \delta\Psi_{\dot Dd} &=& \gamma^{\mu}\eta_{A\dot D}D_{\mu}Z^A_d 
+ g^{abc}{}_d Z^A_aZ^B_bZ^C_c \eta_{B\dot D}\varepsilon_{AC}-\ 3g^{\dot a\dot bc}{}_d Z^{\dot A}_{\dot a} Z^{\dot B}_{\dot b} Z^C_c\eta_{C \dot A}\omega_{\dot D\dot B} \nonumber\\[1mm]
   \nonumber \delta\Psi_{D\dot d} &=& -\gamma^{\mu}\eta_{D \dot A}D_{\mu}Z^{\dot A}_{\dot d}
 - g^{\dot a\dot b\dot c}{}_{\dot d} Z^{\dot A}_{\dot a}Z^{\dot B}_{\dot b}Z^{\dot C}_{\dot c} \eta_{D\dot B}\varepsilon_{\dot A\dot C} +\ 3g^{ab\dot c}{}_{\dot d} Z^A_a Z^B_b Z^{\dot C}_{\dot c}\eta_{A\dot C}\varepsilon_{DB} \;,
  \end{eqnarray}
and similarly for the gauge field.  As before, ${\cal N}=4$ closure occurs when 
\begin{equation}
g^{abcd} = h_{mn} J^{eb} J^{fd} \tau^{ma}{}_e \tau^{ncb}{}_f\;, \qquad g^{ab\dot c\dot d} = h_{mn} J^{eb} J^{\dot f \dot d} \tau^{ma}{}_e \tau^{n\dot c}{}_{\dot f}\;, 
\end{equation}
where the $\tau^{ma}{}_c$ are structure constants of the Lie superalgebra OSp($n|m$), or one of the exotics D($2|1;\alpha$), G(3) and F(4).  As before, the dotted and undotted indices can run over different representations of the superalgebra.

\begin{figure}[t]
\centering
\includegraphics[width=0.45\textwidth]{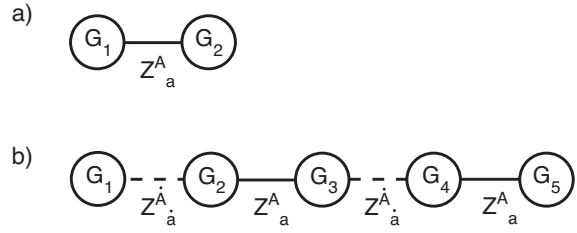}
\caption{
a)  A simple quiver from Gaiotto-Witten theory with hypermultiplets.  The gauge groups can be $G_1=\U(n_1)$, $G_2=\U(n_2)$ or $G_1=\Sp(n_1)$, $G_2=\SO(n_2)$.  b)  A longer quiver from a theory containing both hyper and twisted hypermultiplets.  The gauge groups can be $G_i=\U(n_i)$, $G_j=\U(n_j)$ or $G_i=\Sp(n_i)$, $G_{i+1}=\SO(n_{i+1})$.  The quiver can also be closed into a circle.\label{figure3}}
\end{figure}

The ${\cal N}=4$ construction gives rise to a host of models.  For Gaiotto-Witten theories, with just hypermultiplets, the story is relatively clear.  The only possible theories are those of ${\cal N}= 5$ and 6, with all matter fields in the same representation of the superalgebra gauge group.  The U($n|m$) and OSp($n|m$) theories can be described by a quiver diagram with a single link, as shown in Fig.~\ref{figure3}a.  The hypermultiplets are in the bifundamental representation, joining the two gauge groups.

For more general theories, containing both hyper and twisted hypermultiplets, the story is more interesting.  Since the dotted and undotted indices are independent, they can span different representations of the superalgebra.  For U($n|m$) and OSp($n|m$), one can exploit this fact to construct quiver theories, with the hyper and twisted hypermultiplets in bifundamental representations of the gauge groups \cite{Hosomichi:2008jd}.  In essence, the twisted hypers link together different Gaiotto-Witten theories, as shown in Fig.~\ref{figure3}b.  At its heart, this construction works because for ${\cal N}=4$, the hypers and twisted hypermultiplets transform independently under the two SU(2) factors of the R-symmetry group.


\section[Further 3-algebra directions]{{\Large{\bf Further 3-algebra directions}}}\label{chapter8}

In this review we have emphasized the 3-algebra approach to classifying supersymmetric gauge theories in three dimensions.  We found that it leads naturally to a description of multiple M2-brane systems. Even though one can recast the former in terms of conventional gauge theory language, the presence of 3-algebras is intriguing and one might wonder about their deeper connections to string and M-theory in general.

In this chapter we will briefly discuss some alternative and interesting applications of 3-algebras  less directly related to M2-branes. These will include 3-algebras with Lorentzian signature, their use in obtaining higher-derivative corrections to membrane theories, and the emergence of 3-algebras in a six-dimensional example.

\subsection{Lorentzian 3-algebras}\label{Lorentzian}

In Section~\ref{abelianM2D2} we saw how to transform the field theory for a single D2-brane into an M2-brane field theory. In the process, a vector field is exchanged for its dual scalar. Because this applies to a single brane, the operation is known as ``abelian duality.'' One may wonder whether the same process can be carried out starting with multiple D2-branes, and performing something like a ``non-abelian duality.''  This would perhaps provide an alternate route to finding multiple membrane field theories.

As we will show below, this is possible using an elegant generalisation of abelian duality. Moreover the resulting theory has a 3-algebra associated to it, and turns out to be precisely the $\mathcal N=8$ theory described in Section~\ref{Neq8} -- but with {\em Lorentzian} signature in field space. This change of signature evades the $\mathcal N=8$ uniqueness theorem, so the structure constants are no longer restricted as in \eref{onlyone} \cite{Gomis:2008uv,Ho:2008ei,Benvenuti:2008bt}.

Let us now describe this non-abelian duality, at first
in the usual $\alpha'\to 0$ limit of the multiple D2-brane action  \cite{Ezhuthachan:2008ch}.
We start with ${\cal N}=8$ supersymmetric Yang-Mills theory in
2+1d, based on any simple Lie algebra ${\cal G}$. Next, we introduce
two new adjoint fields, a vector $\boldB_{\mu}$ and a scalar
$\boldphi$. The non-abelian duality
transformation \cite{deWit:2004yr} is
\be
-\sfrac{1}{4 \gYM^2} \boldF^{\mu\nu}\boldF_{\mu\nu}
~~\rightarrow~~
\sfrac{1}{2}\epsilon^{\mu\nu\lambda}\boldB_\mu \boldF_{\nu\lambda}
-\sfrac{1}{2}\left(D_{\mu}\boldphi- \gYM \boldB_{\mu} \right)^2 \ ,
\ee
where $D_\mu$ is the covariant derivative with
respect to $\boldA$.

To prove that the right hand side of the above is equivalent to the left hand side, note the existence on the RHS of a new {\em noncompact abelian} gauge symmetry in addition to the usual gauge symmetry ${\cal G}$. The new symmetry acts as
\be
\delta \boldphi = \gYM \boldM,\qquad \delta \boldB_\mu = D_\mu \boldM \;,
\ee
where $\boldM(x)$ is an arbitrary matrix in the adjoint of $G$.  Now let us use this symmetry to set $\boldphi=0$. Then integrating out $\boldB_\mu$ gives the usual Yang-Mills kinetic term for $\boldF_{\mu\nu}$.

After the duality transformation, the action of the original ${\cal N} =8$
super-Yang-Mills theory becomes
\be
L = \Tr\Big(\sfrac{1}{2}
\epsilon^{\mu\nu\lambda}\boldB_{\mu}\boldF_{\nu\lambda}
- \sfrac{1}{2} \big(D_{\mu}\boldphi -\gYM \boldB_{\mu}\big)^2
-~~\sfrac{1}{2}D_\mu \boldX^i D^\mu \boldX^i
-\sfrac{\gYM^2}{4}[\boldX^i,\boldX^j]^2 + \hbox{fermions}\Big)\;.
\ee
This still only has  $\SO(7)$ invariance, while the expectation (as in
the abelian case) is to obtain $\SO(8)$ invariance in the end. To this
end, we rename $\boldphi$ as $\boldX^8$. By defining a constant
vector $\gYM^I = (0,\ldots,0,\gYM)$, we can unify all the scalar
kinetic terms as
\be
-\shalf \hD_\mu \boldX^I \hD^\mu \boldX^I
= -\shalf\left(\del_\mu \boldX^I -
[\boldA_\mu,\boldX^I] - \gYM^I \boldB_\mu\right)^2\;.
\ee

Let us now replace $\gYM^I$ by an arbitrary 8-vector of magnitude $||\gYM^I||=\gYM$. As a result the kinetic terms become formally invariant under an $\SO(8)$ that acts simultaneously on the fields and the coupling-constant vector. This is not a true symmetry of the theory but instead can be used to rotate $\gYM^I$ back to its original form. Therefore we so far have changed nothing from Yang-Mills.

Similarly the interaction term can be written in a
formally $\SO(8)$-invariant way
\be
\frac{\gYM^2}{4}[\boldX^i,\boldX^j]^2=
\frac{1}{12}\left(\gYM^I[\boldX^J,\boldX^K]+
\gYM^J[\boldX^K,\boldX^I]+\gYM^K[\boldX^I,\boldX^J]\right)^2\;.
\ee
Again, one can rotate the vector $\gYM^I$ by an $\SO(8)$ transformation
back to the form $(0,\ldots,0,\gYM)$ whereupon the interaction term
becomes that of the original Yang-Mills theory.

The final step is to promote the vector $\gYM^I$ to a new scalar
field. Introduce an 8-vector of new (gauge-singlet)
scalars $X_+^I$ and make the replacement
\be
\gYM^I \to X_+^I(x)\;.
\ee
This is legitimate if and only if $X_+^I(x)$ is rendered constant via
an equation of motion, in which case we recover the original theory
on-shell by writing $\langle X_+^I\rangle = \gYM^I$. Constancy of
$X^{I}_{+}$ is imposed by introducing a new set of abelian gauge
fields and scalars: $C_\mu^I,X_-^I$ and adding the following
constraint term to the lagrangian
\be
L_{C} = (C^{\mu}_{I}-\del^\mu X_-^I)\,\del_{\mu}X^{I}_{+} \;.
\ee
This lagrangian in turn has a shift symmetry
\be
\delta X_-^I = \lambda^I,\quad \delta C_\mu^I = \del_\mu \lambda^I\;,
\ee
which, since it acts as an abelian gauge symmetry on $C_\mu^I$,
removes the negative-norm states potentially associated to that
field.

We have thus ended up with the action
\bea\label{lorentzian3alg}
L \!\!\!&=&\!\!\!  \Tr\Big(\sfrac{1}{2}
\epsilon^{\mu\nu\lambda}\boldB_\mu \boldF_{\nu\lambda}
- \sfrac{1}{2} \hD_\mu \boldX^I\hD_\mu \boldX^I
 -~~ \sfrac{1}{12} \left( X^I_+[\boldX^J,\boldX^K] +
    X^J_+[\boldX^K,\boldX^I]
+ X^K_+[\boldX^I,\boldX^J]\right)^2\Big) \nn\\
&& +~~  (C^{\mu\,I}-
\del^\mu X_-^I) \del_\mu X^{I}_{+} + L_{\rm{gauge-fixing}}
+  \cL_{\rm{fermions}}\;.
\eea
As the notation suggests, $+$ and $-$ correspond to null directions in
field space, as we will next explain.

In fact the above action is a 3-algebra action with ${\cal N}=8$ supersymmetry but based on a Lorentzian-signature 3-algebra: The interactions depend on the triple product
\be
X^{IJK}\equiv  X^I_+[\boldX^J,\boldX^K] +
    X^J_+[\boldX^K,\boldX^I]
+ X^K_+[\boldX^I,\boldX^J]\;.
\ee
Thus the 3-algebra structure constants are
\be
f^{+abc}=f^{abc},\quad f^{-abc}=f^{+-ab}=f^{abcd}=0\;,
\ee
where $f^{abc}$ are the structure constants of the original Lie
algebra ${\cal G}$. A detailed study of 3-algebra theories with two or more time-like directions can be found in \cite{deMedeiros:2008bf,deMedeiros:2009hf}.

The above action, \eref{lorentzian3alg}, has manifest $\SO(8)$ invariance as well as ${\cal N}=8$ superconformal invariance \cite{Bandres:2008kj,Gomis:2008be}. However, both are spontaneously broken by giving a VEV $\langle X_+^I\rangle = \gYM^I$ and the theory reduces to ${\cal N}=8$ Yang-Mills with coupling $||\gYM^I||$.  The final theory has seven massless scalars, which can be thought of as the Goldstone bosons for the spontaneous breaking $\SO(8)\to \SO(7)$.

The derivation of 3-algebras via non-abelian duality is striking.  Unfortunately, in the end the theory so obtained seems to be just the original one re-written in a new way. To actually describe M2-branes we would need to find a way to take the VEV $\langle X_+^I\rangle\to\infty$ and this has not yet been understood.\footnote{A related discussion can be found in   \cite{Verlinde:2008di}. Interesting connections to ABJM theory have been investigated in \cite{Honma:2008jd,Antonyan:2008jf}.}

Another interesting application for a class of 3-algebras with $q+1$ time-like directions follows from the fact that the resulting BLG model can be identified with D$(2+q)$-branes on $T^{q}$ \cite{Ho:2009nk,Kobo:2009gz}. In particular, consider the 3-algebra with generators $(T^a_{\vec m},T^{+ },T^{+i},T^{- },T^{-}{}_j)$ where $i,j=1,..,q$, $\vec m \in {\mathbb Z}^{q}$ and totally anti-symmetric triple product whose non-vanishing components are
\bea
\nonumber [T^{+ },T^{+i},T^a_{\vec m}] &=& m^i T^a_{\vec m} \\
{}[T^{+}, T^a_{\vec m}, T^b_{\vec n}]  &=& m^i T^{-i} h^{ab} \delta_{\vec m, -\vec n}+ if^{ab}{}_c T^c_{\vec m+\vec n}\\
\nonumber [T^a_{\vec m},T^b_{\vec n},T^c_{\vec p}] &=& -if^{abc}T^{- }\delta_{\vec m+\vec n+\vec p,\vec 0}\ ,
\eea
where $f^{ab}{}_c$ are the structure constants of a Lie-algebra $\cal G$. This satisfies the fundamental identity. Furthermore an invariant inner-product is given by
\bea
\langle T^{+  },T^{- } \rangle &=& 1 \nonumber \\
\langle T^{+ i},T^{-}{}_j \rangle &=& \delta^i_j  \nn\\
\langle T^a_{\vec m},T^b_{\vec n} \rangle &=& h^{ab}\delta_{\vec m, -\vec n}\ ,
\eea
with all other terms vanishing and $h^{ab}$  the usual invariant  metric of $\cal G$. Expanding the fields in term of the generators one finds that again the components parallel to $T^{- },T^{-}_iS$ satisfy a shift symmetry that can be gauged to remove them as physical fields \cite{Bandres:2008kj,Gomis:2008be}. As a result the components of the fields parallel to $T^{+ },T^{+i}$ are set to constants. The remaining physical components parallel to $T^a_{\vec m}$ can then be interpreted as the Fourier modes of the fields of a D$(2+q)$-brane with Lie-algebra $\cal G$ wrapped on $T^q$.

\subsection{Higher-derivative corrections}

It is natural to ask if higher-derivative corrections to M2-brane actions (governed by the expansion parameter $\ell_p$) can be written down. For the abelian case the full higher-derivative M2-brane theory was written down in Chapter~\ref{chapter1}, in the DBI approximation. For the non-abelian case, one can no longer work to all orders in $\alpha'$ because the starting point, a non-abelian analogue of DBI, is still not known for multiple D2-branes. One approach would be to extend the duality transform of Section \ref{Lorentzian} by incorporating $\alpha'$ corrections. Indeed it has been shown \cite{Alishahiha:2008rs} that one can extend the non-abelian duality above to convert the multiple D2-brane field theory with leading $\alpha'$ corrections into an $\SO(8)$-invariant form for the leading higher-derivative corrections to multiple M2-branes.\footnote{See   also \cite{Iengo:2008cq} for an alternative proposal.}

Subsequently, the leading higher-derivative corrections to the $\mathcal N=8$ theory were calculated for both choices of 3-algebra signature  \cite{Ezhuthachan:2009sr}, using the novel Higgs mechanism of Ref.~\cite{Mukhi:2008ux}. The result, which we review below, strongly suggests that not just the leading term but also the higher-derivative corrections to M2-brane actions are governed by 3-algebras, reaffirming the relevance of this  mathematical structure to M2-branes.

The strategy of Refs. \cite{Alishahiha:2008rs,Ezhuthachan:2009sr} is to assume that $\ell_{p}$ corrections admit an organisation in terms of the 3-algebra product. Therefore one starts with the ansatz that the leading $\ell_{p}$ corrections take the most general form that can arise using 3-algebra ``building blocks,'' but with arbitrary coefficients. One then uses the novel Higgs mechanism to uniquely determine the value of these coefficients by matching to the leading $\alpha'$ corrections in the low-energy theory of two D2-branes. As explained in the introduction, these corrections are $\cO (\ell_{p}^3)$ in M-theory and $\cO(\alpha'^2)$ for the corresponding D2-branes in string theory. Following Ref.~\cite{Ezhuthachan:2009sr}, we first carry out this derivation for the $\SU(2)\times \SU(2)$ BLG theory and then briefly exhibit how it works for the Lorentzian 3-algebra theory.

\subsubsection{Bosonic part of the $\SU(2)\times\SU(2)$ theory}
We concentrate on the bosonic content of the theory. Our ansatz for the
BLG theory will contain all the terms built out of
3-algebra ``blocks'' that are gauge/Lorentz invariant, dimension six and
lead to expressions contained in the D2-brane effective action upon
Higgsing.  However some adjustments must be made for the fact that,
unlike for the D2-brane theory, our fields $X^I$ and the corresponding
triple-product
\begin{equation}
\label{xijkdef}
 [X^I, X^{J \dagger},X^K] = \sfrac{1}{3}\Big(X^{[I}
X^{J]\dagger}X^K -  X^{[I} X^{K\dagger} X^{J]} +X^K X^{[I\dagger}
X^{J]}\Big)
\end{equation}
are complex in the bi-fundamental formulation of
Ref.~\,\cite{VanRaamsdonk:2008ft}. As a result we first need to
re-examine the definition of symmetrised trace. We propose that this
definition be extended, for bi-fundamentals, to a symmetrisation of
the objects while keeping the daggers in their original
place. Explicitly
\be
\STr(AB^\dagger C D^\dagger)=\sfrac{1}{12}\Tr\Big[
A\big(
B^\dagger C D^\dagger +
B^\dagger D C^\dagger +
C^\dagger D B^\dagger +
C^\dagger B D^\dagger +
D^\dagger B C^\dagger +
D^\dagger C B^\dagger\big) ~+ ~\textrm{h.c.}~ \Big]\;.
\ee
Note that this reduces to the conventional definition for hermitian
fields, for which adding the complex conjugate is not necessary.

There is one simplification in the BLG theory that should be
noted at this stage. Because of the low rank of the gauge group,
$\SU(2)\times \SU(2)$, the following three $(X^{IJK})^4$ terms
are proportional to each other:
\bea\label{euclidid}
\STr\,\Big[ X^{IJK}X^{IJL\dagger}X^{MNK}X^{MNL\dagger}\Big]&=&
2 \,\STr\, \Big[X^{IJM}X^{KLM\dagger}X^{IKN}X^{JLN\dagger}\Big]\nn\\
 &=& \sfrac13 \,\STr\,\Big[
X^{IJK}X^{IJK\dagger}X^{LMN}X^{LMN\dagger}\Big]\;,
\eea
where
\be
X^{IJK} = X^{[I}X^{J\dagger} X^{K]}
\ee
Using this, we can write down the following general
ansatz for the ${\cal O}(\ell_p^3)$ corrections to the
BLG-theory
\bea
\label{3BIundetermined}
(D X)^4: && k^2\, \STr\, \Big[{\bf a}\,
D^{\mu}X^{I}\, D_{\mu}X^{J\dagger}\,
D^\nu X^J\,  D_\nu X^{I\dagger}+
{\bf b}\,  D^{\mu}X^{I}\, D_{\mu}X^{I\dagger}\,
D^{\nu}X^{J}\,D_{\nu}X^{J\dagger}\Big]\nn\\
X^{IJK} (D X)^3 :&& k^2\,
\varepsilon^{\mu\nu\lambda}\,\STr\,\Big[{\bf c}\,
X^{IJK} D_{\mu}X^{I\dagger}
D_{\nu}X^{J}\,D_{\lambda}X^{K\dagger} \Big]
\nn\\
(X^{IJK})^2 (D X)^2 :&& k^2\,
\STr\,\Big[{\bf d}\,X^{IJK}\,X^{IJK\dagger}\, D_{\mu} X^{L}\,
D^{\mu}X^{L\dagger} +
{\bf e} \, X^{IJK}\,X^{IJL\dagger}\,D_{\mu} X^{K}\,
D^{\mu}X^{L\dagger}\Big]\nn\\
(X^{IJK})^4 :&& k^2\,
\STr\,\Big[
{\bf f}\, \, X^{IJK}\,X^{IJK\dagger}\,X^{LMN}\,X^{LMN\dagger}\Big]\;,
\eea
where ${\bf a},{\bf b},{\bf c},{\bf d},{\bf e},{\bf f}$ are
constants which we will determine. The sum of all terms above will be
denoted $\Delta\cL$.

Note the absence of pure gauge field terms in \eref{3BIundetermined}. Higher dimension combinations of CS terms would break invariance under large gauge transformations. Higher powers of the field strength would explicitly break supersymmetry, which is expected to remain maximal in the $\ell_{p}$ expansion.

The next step is to Higgs the terms in
\eref{3BIundetermined} and compare them with the derivative-corrected
D2-brane theory, following our treatment of Section~\ref{Higgsneq8}. It turns out that  one can summarise the effect of the Higgsing  through 
a set of substitution rules. For the bosonic fields, they are\footnote{We have put adjoint fields in boldface. Also, by abuse of notation we have used the symbol $D_\mu$ on the LHS for the covariant derivative of bi-fundamental fields, as defined in \eref{bifcov}, while on the RHS it is the covariant derivative on adjoint fields. The distinction should be clear from the context.}
\bea\label{rules}
&& \!\!\!\! D^\mu X^8 \to \sfrac{1}{v}\boldf^\mu\;,~\quad
D^\mu X^i \to \sfrac{1}{v}D^\mu\boldX^i\;,~~\quad
 X^{ij8} \to -\sfrac{1}{4v}  \boldX^{ij}\;,\quad
X^{ijk}\to {\cal O}\Big(\sfrac{1}{v^3}\Big)\;
\nn\\
&& \!\!\!\!D^\mu X^{8\dagger}
\to -\sfrac{1}{v}\boldf^\mu\;, ~
D^\mu X^{i\dagger} \to -\sfrac{1}{v}D^\mu \boldX^i\;,
\!\!\quad
X^{ij8\dagger}\to \sfrac{1}{4v}  \boldX^{ij}\;,\quad
X^{ijk\dagger}\to {\cal O}\Big(\sfrac{1}{v^3}\Big)\;,
\eea
where $\boldf^\mu =\frac{1}{2}\varepsilon^{\mu\nu\lambda}\boldF_{\nu\lambda}$ and $\boldX^{ij}=[\boldX^i,\boldX^j]$. In principle, these rules could be modified once higher-derivative corrections are included. However, as shown in Ref.~\cite{Ezhuthachan:2009sr}, which the reader should consult for more details, these rules in fact turn out to need no modification.

Through the substitutions \eref{rules} the various terms in the bosonic action become
\bea\label{semifinal}
S^b_{\bf a} &=& {\bf a}\; \left(\sfrac{k}{v^2}\right)^2
\int d^3 x\;
\STr\left[D^\mu \boldX^{i} D_\mu \boldX^j D^\nu \boldX^{i} D_\nu \boldX^j+
2 D^\mu \boldX^{i} D_\nu  \boldX^i \boldf^\mu \boldf_\nu +
\boldf^\mu \boldf_\mu \boldf^\nu \boldf_\nu\right]\cr
S^b_{\bf b} &=& {\bf b} \;
\left(\sfrac{k}{v^2}\right)^2
\int d^3 x\;
\STr\left[D^\mu \boldX^{i} D_\mu \boldX^i D^\nu \boldX^{j} D_\nu \boldX^j+
2 D^\mu \boldX^{i} D_\mu  \boldX^i \boldf^\nu \boldf_\nu +
\boldf^\mu \boldf_\mu \boldf^\nu  \boldf_\nu\right]\cr
S^b_{\bf c}&=&  {\bf c} \;\left(\sfrac{k}{v^2}\right)^2
\int d^3 x\;
\STr\left[\sfrac{3}{4} \varepsilon^{\mu\nu\lambda}  D_\mu \boldX^{i}
\boldf_\nu D_\lambda \boldX^j  \boldX^{ij} \right]\cr
S^b_{\bf d} &=& {\bf d} \;
\left(\sfrac{k}{v^2}\right)^2
\int d^3 x\;
\STr\left[\sfrac{3}{16}D^\mu \boldX^{i} D_\mu \boldX^i
\boldX^{jk}\boldX^{jk}
+\sfrac{3}{16} \boldf^\mu \boldf_\mu \boldX^{ij}\boldX^{ij}\right]\cr
S^b_{\bf e} &=& {\bf e} \;
\left(\sfrac{k}{v^2}\right)^2
\int d^3 x\;
\STr\left[\sfrac{1}{8}D^\mu \boldX^{i} \boldX^{ij}\boldX^{kj}
D_\mu \boldX^k +
\sfrac{1}{16} \boldf^\mu \boldf_\mu \boldX^{ij}\boldX^{ij}
\right]\cr
S^b_{\bf f}&=& {\bf f} \;
\left(\sfrac{k}{v^2}\right)^2
\int d^3 x\;
\STr\left[\sfrac{9}{256} \boldX^{ij}\boldX^{ji}\boldX^{kl}
\boldX^{lk} \right]
\eea
plus terms in $\cO(1/v)$, where we are using $\boldX^{ij}=
[\boldX^i,\boldX^j]$.
Note that terms involving $\boldX^8$ are absent. This is as it should be,  since these Goldstone degrees of freedom need to disappear from
the action. Putting back the factor $\ell_p^3$ in the above terms and using
\be
(2 \pi)^2 \ell_{p}^3\left({k \over 2\pi v^2}\right)^2 =
{(2 \pi \alpha')^2 \over g_{YM}^2}
\ee
it is now straightforward to compare with the appropriate terms coming
from the D2-brane theory.

The precise form of the low-energy effective action for multiple parallel D-branes is still not known to all orders. However, up to order $\alpha'^2$ it has been explicitly obtained using open string scattering amplitude calculations\footnote{See \eg \cite{Bergshoeff:2001dc,Cederwall:2001td,Cederwall:2002df} and references therein.} and the result agrees with Tseytlin's proposal for a DBI action with a symmetrised prescription for the trace \cite{Tseytlin:1997csa}. Starting from D9-branes, the prescription requires symmetrisation over the gauge field strengths. For lower dimensional branes, T-duality requires that this carries on to scalar covariant derivatives and scalar commutators \cite{Taylor:1999pr,Myers:1999ps}. This proposal fails at order $\alpha'^4$ \cite{Hashimoto:1997gm} but is good enough for our purposes.

The form of the relevant action for two D2-branes is given at this
order by an appropriately modified, dimensionally reduced version of
the D9-brane answer provided in\footnote{Note
that the coefficients here are twice their value given in
\cite{Bergshoeff:2001dc}
because the normalisation of the
trace used there is $\Tr\,( T^a T^b)=\delta^{ab}$ while we
consistently use $\Tr\, (\sigma^a \sigma^b)=2 \delta^{ab}$.}
 \cite{Bergshoeff:2001dc}:
\bea
\label{bosonicDBI}
S_{\alpha'^2}^{b} &= &{(\alp)^2 \over g_{YM}^2}\int d^3x\;  \STr \Big[
  \sfrac{1}{4}\boldF_{\mu\nu}\boldF^{\nu\rho}\boldF_{\rho\sigma}\boldF^{\sigma\mu}  - \sfrac{1}{16}
    \boldF^{\mu\nu}\boldF_{\mu\nu} \boldF^{\rho\sigma}\boldF_{\rho\sigma} -
    \sfrac{1}{4}D_\mu   \boldX^{i} D^\mu  \boldX^i D_\nu  \boldX^{j} D^\nu \boldX^j\nn\\[2mm]
 &&+~  \sfrac{1}{2}D_\mu  \boldX^{i} D^\nu \boldX^i D_\nu \boldX^{j} D^\mu \boldX^j +
\sfrac{1}{4}  \boldX^{ij} \boldX^{jk} \boldX^{kl} \boldX^{li}  -
    \sfrac{1}{16} \boldX^{ij} \boldX^{ij} \boldX^{kl} \boldX^{kl}\nn\\[2mm]
 &&-~ \boldF_{\mu\nu}\boldF^{\nu\rho} D_\rho  \boldX^{i} D^\mu \boldX^i
  - \sfrac{1}{4} \boldF_{\mu\nu}\boldF^{\mu\nu}D_\rho \boldX^{i}
D^\rho  \boldX^i  -\sfrac{1}{8}\boldF_{\mu\nu}\boldF^{\mu\nu}  \boldX^{kl} \boldX^{kl}  \nn\\[2mm]
 && -~\sfrac{1}{4} D_\mu \boldX^{i} D^\mu \boldX^i \boldX^{kl} \boldX^{kl}
 - \boldX^{ij} \boldX^{jk}D^\mu \boldX^{k} D_\mu \boldX^i
-\boldF_{\mu\nu}D^\nu \boldX^{i} D^\mu \boldX^{j} \boldX^{ij} \Big].
\eea
Note that for $\U(2)$, one has the additional simplification:
\be
\STr\, \Big[\boldX^{ij}\boldX^{jk}\boldX^{kl}\boldX^{li} \Big] = \sfrac{1}{2}\STr\,
\Big[\boldX^{ij}\boldX^{ij}\boldX^{kl}\boldX^{kl} \Big]\;.
\ee
 It is then straightforward to compare the coefficients for all of these terms to
finally obtain\footnote{We note a sign difference in the value of the ${\bf c}$ coefficient compared to Ref.~\cite{Ezhuthachan:2009sr}.}
\be\label{bosoniccoeff}
\begin{split}
& {\bf a } = \sfrac{1}{2} \;,\quad {\bf b } =-\sfrac{1}{4}\;,\quad
{\bf c }  =\sfrac{4}{3} \;,\\
&{\bf d } = - \sfrac{4}{3}\;,\quad  {\bf e } = 8  \;,\quad {\bf f } =\sfrac{16}{9}\;.
\end{split}
\ee

It is important to note that the fixing of coefficients by the above comparison is nontrivial. There are 3-algebra terms of \eref{semifinal} that, after Higgsing, give rise to terms in the D2 action \eref{bosonicDBI} that come from different index contractions (that is, ultimately, different index contractions of the D9-brane theory before dimensional reduction). Also in some places, two terms in the 3-algebra theory lead to the same term in the D2 action. Hence, it was not obvious at the outset that there would be any values of the coefficients in the above expression that would lead to the D2 theory upon Higgsing. The fact that we find a consistent and unique set of coefficients is therefore very satisfying.

The Higgsing of the fermion terms follows the above discussion closely and for this reason we will not review it here.

\subsubsection{The four-derivative corrections in 3-algebra form}

In this section we will re-cast our results in 3-algebra language. There are several important reasons to do so: One is that we will uncover some new properties of 3-algebras, arising from the fact that at order $\ell_p^3$ we encounter traces of as many as four 3-algebra generators for the first time.
Another  is that corrections of order $\ell_p^3$ are already
known
\cite{Alishahiha:2008rs,Iengo:2008cq}  for the special case of
Lorentzian 3-algebras. By re-writing the derivative corrections of
$\SU(2)\times \SU(2)$ BLG theory in terms of 3-algebra quantities, we will be able to compare them with the results of
Refs.\,\cite{Alishahiha:2008rs,Iengo:2008cq}. Indeed, it is natural to
hope that all BLG theories (including both $\SU(2)\times \SU(2)$ and Lorentzian sub-classes) originate from a common 3-algebra formulation, even
though they were obtained using completely different procedures. As we
now have all the necessary data for determining what that formulation
is, we will compare the two classes of theories explicitly. After
dealing with some issues of normalisation we will find that there is
indeed complete agreement.

Yet another reason to re-express our results in 3-algebra language is to
open the possibility of extending this investigation to the ${\cal
N}=6$ 3-algebras of Refs.\,\cite{Bagger:2008se,Schnabl:2008wj} which
encode, among other things, the ABJM field theory. In the final
section we will make some general comments on how this might be
done.

We have obtained the four-derivative action in
bi-fundamental notation and we now want to express it in 3-algebra
form. For this purpose we will make use of the dictionary between the two languages that we described at the end of Section \ref{Neq8}.
Additionally, we have to deal with evaluating the symmetrised
trace of four 3-algebra generators. Symmetry restricts its form to be
\be\label{4Ts}
\STr\Big( T^a T^b T^c T^d \Big) = m\;h^{(ab}h^{cd)}\;,
\ee
where $m$ is an as yet undetermined numerical coefficient. However, the
Lorentzian 3-algebras can help us determine the latter as
follows: Lorentzian 3-algebras include a set of generators
corresponding to a compact subgroup of the theory's whole symmetry
group. One is then free to choose them as the generators of any
semi-simple Lie algebra, \eg SU(2). In turn, tracing over the latter
leads to a flat Euclidean block in the 3-algebra metric, $h^{ij} =
\delta^{ij}$. In any four-derivative Lorentzian 3-algebra action there
will be terms with components for which the generators in \eref{4Ts}
run over this subset. In that case, and once again taking into
consideration the appropriate definition of the trace, one can
explicitly evaluate the following expression for the particular case
of SU(2)
\be
\STr\Big( T^i T^j T^k T^l \Big) = 2 \;
\STr\Big( \sfrac{\sigma^i}{2}\sfrac{\sigma^j}{2}
\sfrac{\sigma^k}{2}\sfrac{\sigma^l}{2} \Big) =
\sfrac{1}{4}\; \delta^{(ij}\delta^{kl)}.
\ee
This fixes $m=\sfrac{1}{4}$.

Equipped with the above fact, we can finally rewrite our results and
obtain the leading derivative corrections to the bosonic part of the $\SU(2)\times\SU(2)$ BLG theory in 3-algebra form
\bea\label{A4algebrabosonic}
S^{b}_{\ell_{p}^3} &= & (2\pi)^2\ell_{p}^3\int d^3 x \;\STr\Big[ \sfrac14\,
D^\mu X^I D_\mu X^J
D^\nu X^J D_\nu X^I
-\sfrac18 D^\mu X^I D_\mu X^I D^\nu X^J D_\nu X^J\nn\\[1mm]
 &&\qquad+~\sfrac{1}{6}  \,\varepsilon^{\mu\nu\lambda}\,
 X^{IJK} D_\mu X^I
D_\nu X^J D_\lambda X^K\nn\\[1mm]
 &&\qquad +~\sfrac{1}{4} \,X^{IJK}X^{IJL}
D^\mu X^K D_\mu X^L
-\sfrac{1}{24} X^{IJK} X^{IJK} D^\mu X^L D_\mu X^L
\nn\\[1mm]
 &&\qquad+~\sfrac{1}{288} \,X^{IJK}X^{IJK} X^{LMN}X^{LMN}\Big]\;,
\eea
where now
\be
X^{IJK} = [X^I,X^J,X^K]\;.
\ee

\subsubsection{Derivative corrections in the Lorentzian theory}

In Ref.~\,\cite{Alishahiha:2008rs} the equivalent four derivative terms
were constructively obtained for Lorentzian 3-algebra theories and it
was conjectured there that the $\SU(2)\times\SU(2)$-theory should also be
expressed in the terms of the same 3-algebra structures at four
derivative order. We will now verify this conjecture.

Let us start by quoting the result found there for the higher-derivative corrections to Lorentzian 3-algebra theories. To avoid confusion with the Euclidean signature theory we have been discussing so far, we will henceforth denote all Lorentzian 3-algebra variables with a hat symbol on top. Accordingly, our notation for the field variables is that the eight adjoint scalars are denoted $\hX^I$, the fermions $\hlambda$, the sixteen gauge-singlet scalars and fermions $\hX^I_\pm,\hlambda_\pm$ and the pair of gauge fields is $\hA_\mu,\hB_\mu$.

As we saw in Section \ref{Lorentzian}, due to constraints the fields $\hX^I_-,\hlambda_-$ decouple and the fields $\hX^I_+,\hlambda_+$ are fixed to be a constant and zero, respectively. It was shown in Ref.~\,\cite{Alishahiha:2008rs} that the bosonic part of the $\ell_p^3$ correction can be written entirely in terms of the building blocks
\bea
\hD_\mu \hX^I &=& \del_\mu \hX^I - [\hA_\mu,\hX^I] - \hB_\mu \hX^I_+
\nn\\[1mm]
\hX^{IJK} &=& \hX^I_+[\hX^J,\hX^K] + \hX^J_+[\hX^K,\hX^I]
+ \hX^K_+[\hX^I,\hX^J]\;.
\eea
To simplify formulae, we present the results in symmetrised-trace form. Then Eq.(3.14) of Ref. \cite{Alishahiha:2008rs} is the sum of the following four 
terms\footnote{We have corrected a few of the
coefficients.} (we only write
the ${\cal O}(\ell_p^3)$ corrections, dropping the lowest-order terms)
\bea
\label{Lorentzianboslead}
(\hD \hX)^4: &&
~~\,\sfrac{1}{4}\, \STr\,\Big(\hD^\mu \hX^I \hD_\mu \hX^J
\hD^\nu \hX^J \hD_\nu \hX^I
-\shalf \hD^\mu \hX^I \hD_\mu \hX^I \hD^\nu \hX^J \hD_\nu \hX^J\Big)\nn\\[1mm]
\hX^{IJK} (\hD \hX)^3: &&
~~\,\sfrac{1}{6}\,\varepsilon^{\mu\nu\lambda}\,
\STr\,\Big( \hX^{IJK}\hD_\mu \hX^I
\hD_\nu \hX^J\hD_\lambda \hX^K\Big)\nn\\[1mm]
(\hX^{IJK})^2 (\hD \hX)^2: &&~~\, \sfrac{1}{4}\, \STr\,\Big(\hX^{IJK}\hX^{IJL}
\hD^\mu \hX^K\hD_\mu \hX^L
- \sfrac{1}{6} \hX^{IJK} \hX^{IJK} \hD^\mu \hX^L \hD_\mu \hX^L
\Big)
\nn\\[1mm]
(\hX^{IJK})^4: &&\,
~~\sfrac{1}{24}\,\STr\,\Big(\hX^{IJM}\hX^{KLM}\hX^{IKN}\hX^{JLN}-
\sfrac{1}{12} \hX^{IJK}\hX^{IJK} \hX^{LMN}\hX^{LMN}\Big)\;.
\eea
Here, the trace is defined using $\Tr\,( T^a T^b)= \delta^{ab}$ where
$a,b$ are adjoint Lie algebra indices.

Note that the above expression involves all possible terms one can
write down at this order using $\hD_\mu \hX^I$ and $\hX^{IJK}$ as
building blocks, with one apparent exception: The $(\hX^{IJK})^4$
terms could have contained one more distinct index contraction, namely
the one with $\hX^{IJK}\hX^{IJL}\hX^{MNK}\hX^{MNL}$. However, it is easy to
demonstrate the identity
\be
\label{xfourid}
\STr\,\Big(\hX^{IJK}\hX^{IJL}\hX^{MNK}\hX^{MNL}\Big) =  \STr\, \Big( \sfrac43
\hX^{IJM}\hX^{KLM}\hX^{IKN}\hX^{JLN}
+ \sfrac19  \hX^{IJK}\hX^{IJK}
\hX^{LMN}\hX^{LMN}\Big)\;,
\ee
as a result of which only two of the three possible ${\cal
O}(\hX^{IJK})^4$ terms are independent.

\subsubsection{Universal answer for the BLG theory}

We can now recover a universal answer for the four-derivative action to BLG theory for general 3-algebras. A reasonable guess would be to see whether \eref{A4algebrabosonic} provides the answer by simply replacing the $\SU(2)\times \SU(2)$ structure constants and metric with their Lorentzian counterparts inside the expressions. One then finds that all terms and coefficients in \eref{Lorentzianboslead} can be readily obtained except for $\cO(\hat X^{IJK})^4$. This discrepancy is easily traced back to the difference between the identities obeyed by quartic powers of triple-products in the two cases and is resolved by noticing that \eref{euclidid} is actually a special case of \eref{xfourid}, due to the particularly simple nature of the $\SU(2)\times\SU(2)$ structure constants $\varepsilon^{abcd}$. Therefore, at least within the class of BLG theories we are considering, we may assume that \eref{xfourid} holds in general, thereby dropping the hat in this equation.

This raises the interesting question, which to our knowledge has not yet been resolved, of whether this identity is also obeyed by other indefinite-signature BLG theories, notably those with multiple time-like directions as discussed in \cite{deMedeiros:2008bf,Ho:2009nk,deMedeiros:2009hf}.  If the answer turns out to be in the affirmative, we would have found a new relation for quartic products of structure constants that holds for a generic $\cN=8$ 3-algebra.

With these observations we can at last write a common expression for
both $\SU(2)\times\SU(2)$ and Lorentzian BLG theories
\bea\label{3algebrabosonic}
S^b_{\textrm{BLG},\ell_{p}^3} &= & \ell_{p}^3\int d^3 x \;\STr\Big[ \sfrac14\,
\Big( D^\mu X^I  D_\mu X^J
 D^\nu X^J  D_\nu X^I
-\shalf  D^\mu X^I  D_\mu X^I  D^\nu X^J  D_\nu X^J\Big)\nn\\[1mm]
 &&\qquad+~\sfrac{1}{6}  \,\varepsilon^{\mu\nu\lambda}\,
\Big( X^{IJK} D_\mu X^I
 D_\nu X^J D_\lambda X^K\Big)\nn\\[1mm]
 &&\qquad+~ \sfrac{1}{4} \,\Big(X^{IJK}X^{IJL}
 D^\mu X^K D_\mu X^L
- \sfrac{1}{6} X^{IJK} X^{IJK}  D^\mu X^L  D_\mu X^L
\Big)
\nn\\[1mm]
 &&\qquad+~\sfrac{1}{24} \,\Big(X^{IJM}X^{KLM}X^{IKN}X^{JLN}-
\sfrac{1}{12} X^{IJK}X^{IJK} X^{LMN}X^{LMN}\Big)\Big]\;.
\eea
It is very satisfactory that one can obtain the precise coefficients of \eref{bosoniccoeff} as well as \eref{Lorentzianboslead} from this expression upon specifying the 3-algebra.

In a similar manner, one can write down corrections for the fermion terms in 3-algebra form, the details of which are presented in Ref. \cite{Ezhuthachan:2009sr}. The resulting theory is expected to be supersymmetric, and some initial results in this direction appeared in \cite{Low:2010ie}. The full set of next-to-leading-order corrected supersymmetry transformations (to lowest order in the fermions) that leave the $\SU(2)\times\SU(2)$ BLG lagrangian invariant were presented in \cite{Richmond:2012by}. Computing the derivative corrections to the $\cN=6$ ABJM theory is an interesting and important  problem that remains open at the time of writing.

 \subsection{Applications to M5-branes}

We now switch gears and discuss an application of 3-algebras to the theory of M5 branes.  We have already seen several times how one can attempt to make a connection between M2 and M5-brane theories through M2$\perp$M5 funnels and ``dielectric'' configurations. This is because, compared to M2-brane systems, the formulation of an M5-brane theory is difficult at best: Even for the case of a single fivebrane it does not seem possible to write down a six-dimensional action with conformal symmetry because of the self-duality of the three-form field-strength \cite{Witten:1996hc}. In addition, the theory of multiple M5-branes is given by a conformal field theory in six-dimensions with  mutually local electric and magnetic states  and no coupling constant. All of these features are difficult to reconcile with  a lagrangian description.\footnote{However,  note that there exist proposals for lagrangian descriptions which relax some of the original assumptions and involve sacrificing manifest 6d Lorentz invariance \cite{Perry:1996mk,Aganagic:1997zq}, introducing a non-dynamical auxiliary scalar field \cite{Pasti:1997gx,Bandos:1997ui} or imposing the self-duality condition directly at the level of the quantum theory \cite{Cederwall:1997gg}.} However, the covariant equations of motion for the abelian M5-brane have been known for some time \cite{Howe:1983fr,Howe:1996yn,Howe:1997fb}.

In this section we investigate a potential direct relation between 3-algebras and multiple M-theory fivebranes by studying the equations of motion of a non-abelian $(2,0)$ tensor multiplet. Starting with the set of supersymmetry transformations for the abelian M5-brane, one can write an ansatz for a non-abelian generalisation. However, apart from the expected non-abelian versions of the scalars, fermions and the anti-symmetric three-form field strength, it turns out that one needs to also introduce a gauge field as well as a non-propagating vector field that transforms non-trivially under the non-abelian gauge symmetry and has a negative scaling dimension. Curiously, the ansatz involves ``structure constants'' with four indices that can be associated to a 3-algebra \cite{Lambert:2010wm}.

\subsubsection{A non-abelian $(2,0)$ tensor multiplet}

We start by giving the covariant supersymmetry transformations of a free six-dimensional $(2,0)$ tensor multiplet
\bea\label{Habelian}
\delta X^I &=& i \bar \epsilon \Gamma^I \Psi\cr
\delta  \Psi &=& \Gamma^\mu \Gamma_I \pd_\mu X^I \epsilon + \frac{1}{  3!}\frac{1}{2} \Gamma^{\mu\nu\lambda}H_{\mu\nu\lambda}\epsilon\; \cr
\delta H_{\mu\nu\lambda} &=& 3i \bar \epsilon \Gamma_{[\mu\nu}\partial_{\lambda]} \Psi \; ,
\eea
where $\mu = 0,...,5$, $I = 6,...,10$ and $H_{\mu\nu\lambda} = 3\partial_{[\mu}B_{\nu\lambda]}$ is self-dual. The supersymmetry generator $\epsilon$ is chiral: $\Gamma_{012345}\epsilon=\epsilon$ and the fermions $\Psi$ are anti-chiral: $\Gamma_{012345}\Psi=-\Psi$.
This algebra closes on-shell, with  equations of motion
\be
\Gamma^\mu \pd_\mu \Psi = 0 \;,\quad \pd_\mu\pd^\mu X^I =0\;, \quad\pd_{[\mu}H_{\nu\lambda\rho]}=
0\; .
\ee
We note that, from the point of view of supersymmetry, it is sufficient to write the algebra purely in terms of $H_{\mu\nu\lambda}$, and not mention $B_{\mu\nu}$.

We wish to try and generalise this algebra to allow for non-abelian fields and interactions. To this end we again assume all fields take values in some vector space with a basis $T_a$, so that $X^I = X^I_aT^a$, etc, and promote the derivatives to suitable covariant derivatives
\bea\label{deriv}
D_\mu X^I_a = \pd_\mu X^I_a - \tilde A_\mu^b{}_aX^I_b\;\ ,
\eea
where $\tilde A_\mu^b{}_a$ is a gauge field. We wish to have a system of equations in six-dimensions with $(2,0)$ supersymmetry and an $\SO(5)$ R-symmetry.

In order to obtain a term analogous to the $[X^I,X^J]$ for $\delta\Psi$ in (\ref{Habelian}), we need to introduce a $\Gamma_\mu$ matrix to account for the fact that $\epsilon$ and $\Psi$ have opposite chirality. A natural guess is to  propose the existence of a new field $C^\mu_a$. Starting from a suitably general possibility one then finds that the following ansatz works
\bea\label{ansatz1}
\delta X^I_a &=& i \bar \epsilon \Gamma^I \Psi_A\cr
\delta \Psi_a &=& \Gamma^\mu \Gamma^I D_\mu X_a^I \epsilon + \frac{1}{ 3!}\frac{1}{2} \Gamma_{\mu\nu\lambda}H_a^{\mu\nu\lambda}\epsilon- \frac{1}{2}\Gamma_\lambda \Gamma^{IJ} C^\lambda_b X^I_c X^J_d {f^{cdb}}_a\epsilon\cr
\delta H_{\mu\nu\lambda\; a} &=& 3 i \bar \epsilon \Gamma_{[\mu\nu}D_{\lambda]} \Psi_a +  i\bar \epsilon \Gamma^I \Gamma_{\mu\nu\lambda\kappa}C^\kappa_{b} X^I_c \Psi_d{f^{cdb}}_a \cr
\delta  \tilde A_{\mu\;a}^{b} &=& i \bar \epsilon \Gamma_{\mu\lambda} C^\lambda_c \Psi_d {f^{cdb}}_a\cr
\delta C^\mu_a  &=& 0\;.
\eea
Here again we see the appearance of 3-algebra-like $f^{cdb}{}_a$   ``structure'' constants.  As with the abelian case we also impose self-duality on the 3-form, $H_{\mu\nu\lambda a} = \frac{1}{3!}\epsilon_{\mu\nu\lambda\tau\sigma\rho}H^{\tau\sigma\rho}{}_a$.

The closure computation works in spirit much like the cases in Chapter~\ref{chapter3}. To summarise the results,\footnote{Full details of the calculation can be found in  \cite{Lambert:2010wm}.} one finds that closure of (\ref{ansatz1}) is on-shell and subject to the equations of motion
\bea\label{eq1}
D^2 X_a^I  &=&\frac{i}{2}\bar\Psi_CC^\nu_B\Gamma_\nu\Gamma^I \Psi_d f^{cdb}{}_a + C^\nu_b C_{\nu g} X^J_cX^J_eX^I_f f^{efg}{}_{d}f^{cdb}{}_a
 \nn\\[2mm]
 D_{[\mu}H_{\nu\lambda\rho]\;a}  &=&-\frac{1}{4}\epsilon_{\mu\nu\lambda\rho\sigma\tau}C^\sigma_b X^I_cD^\tau X^I_df^{cdb}{}_a - \frac{i}{8}\epsilon_{\mu\nu\lambda\rho\sigma\tau}C^\sigma_b \bar\Psi_c\Gamma^\tau \Psi_df^{cdb}{}_a  \nn\\[2mm]
\Gamma^\mu D_\mu\Psi_a  &=& -X^I_cC^\nu_b\Gamma_\nu\Gamma^I\Psi_d f^{cdb}{}_a  \nn\\[2mm]
\tilde F_{\mu\nu}{}^b{}_a  &=& -C^\lambda_cH_{\mu\nu\lambda\; d}f^{cdb}{}_a\;,
\eea
as well as the conditions
\bea\label{eq2}
C^\rho_cD_\rho X^I_D f^{cdb}{}_a = 0\;, &\quad& D_\mu C^\nu_a  = 0 \nn\\[2mm]
 C^\rho_cD_\rho \Psi_D f^{cdb}{}_a = 0\;, && C^\mu_cC^\nu_df^{bcd}{}_a = 0\nn\\[2mm]
C^\rho_cD_\rho H_{\mu\nu\lambda\;a} f^{cdb}{}_a = 0\;.&&
\eea
Furthermore one finds that the structure constants are   anti-symmetric: $f^{abc}{}_d= f^{[abc]}{}_d$ and obey the fundamental identity: $f^{[abc}{}_ef^{d]ef}{}_g=0$. These are precisely the structure constants for a real 3-algebra. We additionally need to endow the 3-algebra with an inner-product $\Tr (T^a, T^b) = h^{ab}$ with which one can construct gauge-invariant quantities. This in turn implies that $f^{abcd} = h^{de}f^{abc}{}_e$ is anti-symmetric in $c,d$ and hence anti-symmetric in all of $a,b,c,d$. 

The consistency of the above set of equations with respect to their scaling dimensions gives
\bea
[H] = [X]+1 \;,\qquad &&\qquad [\tilde A] = 1\;,\qquad\qquad [C] = 1-[X]\cr
[\epsilon] = -\sfrac{1}{2}\;,\qquad &&\qquad [\Psi] = [X] +\sfrac{1}{2}\;,\qquad\qquad [X] \;,
\eea
so one could still make this work with a set of noncanonical assignments that are  related to the choice of $[X]$. However the canonical choice is
$[X]=2,[H]=3,[\Psi]=\frac{5}{2},[C]=-1$. In particular, we see that the new field $C^\mu_a$ has scaling dimension $-1$. The theory does not have any a priori  dimensional or dimensionless  parameters. Therefore if we compactify it on a circle of radius $R$, we expect the expectation value of $C^\mu_a$ to be proportional to $R$, purely on dimensional grounds.

What is the physical content of the above equations? One sees immediately from (\ref{eq2}) that the fields cannot depend on the coordinate that is parallel to $C^\mu_a$. Thus the system is more of a five-dimensional theory than a six-dimensional one. However this is not entirely so. One can compute the six-dimensional energy-momentum tensor \cite{Lambert:2011gb}
\bea
\nonumber
T_{\mu \nu} &=& D_\mu X^I_a D_\nu X^{Ia} - \frac{1}{2} \eta_{\mu \nu} D_\lambda X^I_a D^\lambda X^{Ia} \\
\nonumber
	&&+ \frac{1}{4} \eta_{\mu \nu} C^\lambda_b X^I_a X^J_c C_{\lambda g} X^I_f X^J_e f^{cdba} f^{efg}{}_d + \frac{1}{4} H_{\mu \lambda \rho \; a} H_{\nu}{}^{\lambda \rho \; a} \\
	&&- \frac{i}{2} \bar{\Psi}_a \Gamma_\mu D_\nu \Psi^a + \frac{i}{2} \eta_{\mu \nu} \bar{\Psi}_a \Gamma^\lambda D_\lambda \Psi^a - \frac{i}{2} \eta_{\mu \nu} \bar{\Psi}_a C^\lambda_b X^I_c \Gamma_\lambda \Gamma^I \Psi_d f^{abcd} \,
\eea
of this theory and find that it does carry all six momenta \cite{Lambert:2011gb}. In particular one finds that the momentum of associated to the missing coordinate parallel to $C^\mu_a$ is given by the instanton number of the gauge fields over the purely spatial submanifold. Since this is discrete, we see that one can in principle interpret the system as applying to a case where one dimension has been compactified on a circle.

\subsubsection{Relation to five-dimensional SYM and DLCQ}

As we have already seen in Section~\ref{Lorentzian}, 3-algebras can be classified according to the signature of the metric in group space. We next  investigate the vacuum solutions of our six-dimensional equations for both Lorentzian and Euclidean  possibilities.
Let us begin with a Lorentzian-signature 3-algebra $T^a=\{T^+,T^-,T^A\}$ with structure constants given by
\be\label{lorentzian}
{f^{+AB}}_C = {f^{AB}}_C\;,\qquad {f^{ABC}}_- = f^{ABC}\;,
\ee
where ${f^{AB}}_C$ are the structure constants of the Lie algebra $\mathcal G$ and all remaining components of ${f^{abc}}_d$ vanishing. We  look for vacua of this theory in the particular case of $\mathcal G = \mathfrak {su}(N)$ by  expanding around the point
\be\label{vev}
\langle C^\lambda_A\rangle = g \delta_5^\lambda \delta^+_A \;,
\ee
while all other fields are set to zero. One then has from the fourth line of (\ref{eq1}) that
\be\label{FvH}
\tilde F_{\alpha\beta}{}^B{}_A =  -g H_{\alpha\beta 5\; D}f^{DB}{}_A\;,
\ee
with  $\mu =\{0,1,2,..,5\}$, $\alpha=\{0,1,2,3,4\}$ and all other components of $\tilde F_{\mu\nu}{}^A{}_B$ vanishing. As a result, the latter correspond to flat connections that can be set to zero up to  gauge transformations, while the second equation in (\ref{eq2})  reduces to $\pd_\mu g = 0$, rendering $g$ constant.

The rest of (\ref{eq1})-\eqref{eq2}  become:
\bea\label{YMeom}
0 &=&\tilde  D^\alpha \tilde  D_\alpha X_A^I - g \frac{i}{2}\bar\Psi_c\Gamma_5\Gamma^I \Psi_d f^{CD}{}_A -  g^2 X^J_CX^J_EX^I_F f^{EF}{}_{D}f^{CD}{}_A \nn\\
0 &=& \tilde  D_{[\alpha}H_{\beta\gamma] 5\;A} \nn\\
0 &=& \tilde  D^\alpha H_{\alpha\beta 5\; A}+\frac{1}{2} g f^{CD}{}_A ( X^I_C \tilde  D_\beta X^I_D + \frac{i}{2} \bar\Psi_C\Gamma_\beta \Psi_D )  \nn\\
0 &=& \Gamma^\mu \tilde  D_\mu\Psi_A+ g X^I_C\Gamma_5 \Gamma^I\Psi_Df^{CD}{}_A \nn\\
0 &=& \pd_5 X^I_D = \pd_5 \Psi_D = \pd_5 H_{\mu\nu\lambda\;D} \;,
\eea
where $\tilde  D_\alpha X^I_A= \pd_\alpha X^I_A - \tilde A_{\alpha}{}^B{}_A X^I_B$, while one also has from (\ref{ansatz1}) that
\bea\label{YMsusy}
\delta X^I_A &=& i \bar \epsilon \Gamma^I \Psi_A \nn\\
\delta \Psi_A &=& \Gamma^\alpha \Gamma^I \tilde  D_\alpha X_A^I \epsilon + \frac{1}{2} \Gamma_{\alpha\beta} \Gamma_5H_A^{\alpha\beta 5}\epsilon- \frac{1}{2}\Gamma_5 \Gamma^{IJ}  X^I_C X^J_D {f^{CD}}_A\epsilon \nn\\
\delta  \tilde A_{\alpha \;A}^{\;B} &=& i \bar \epsilon \Gamma_{\alpha}\Gamma_5  \Psi_d {f^{DB}}_A\;.
\eea
We immediately see that with the identifications
\be
g= g_{YM}^2\;, \qquad H_{\alpha\beta5}^A = -\frac{1}{g_{YM}^2} F_{\alpha\beta}^A\;, \qquad \tilde A_{\alpha\;A}^{\;B} =  A_{\alpha\;C} {f^{CD}}_A\;\ ,
\ee
we recover the equations of motion, Bianchi identity and supersymmetry transformations of five-dimensional $\SU(n)$ super-Yang-Mills  theory. In particular since $g$ has scaling dimension $-1$, we see that $g_{YM}$ also has the correct scaling dimension. Furthermore the fundamental identity reduces to the Jacobi identity for the structure constants of $\mathfrak{su}(n)$. Hence the off-shell $\SO(5,1)$ Lorentz and conformal symmetries are  spontaneously broken to an $\SO(4,1)$ Lorentz invariance.

However, we also have the additional equations
\bea\label{20eom}
0 &=&\pd^\mu \pd_\mu X_\pm^I  \nn\\
0 &=& \pd_{[\mu}H_{\nu\lambda\rho]\;\pm} \nn\\
0 &=& \Gamma^\mu \pd_\mu\Psi_\pm \;,
\eea
with transformations
\bea\label{20susy}
\delta X^I_\pm &=& i \bar \epsilon \Gamma^I \Psi_\pm\nn\\[2mm]
\delta \Psi_\pm &=& \Gamma^\mu \Gamma^I \pd_\mu X_\pm^I \epsilon + \frac{1}{ 3!}\frac{1}{2} \Gamma_{\mu\nu\lambda}H_\pm^{\mu\nu\lambda}\epsilon\nn\\[2mm]
\delta H_{\mu\nu\lambda\; \pm} &=& 3 i \bar \epsilon \Gamma_{[\mu\nu}\pd_{\lambda]} \Psi_\pm \;.
\eea
These comprise two free, abelian $(2,0)$ multiplets in six dimensions.

Thus for the choice of a Lorentzian 3-algebra, the vacua of the theory  correspond to the ones for five-dimensional super-Yang-Mills along with two free, abelian $(2,0)$ multiplets which are genuinely six-dimensional.  Presumably one must be gauged away in order to have a well-defined system of equations with positive definite energy. How about the case of a Euclidean 3-algebra? It turns out that this behaves in a qualitatively similar manner. On the other hand, taking 3-algebras with more than one timelike direction has been shown to lead to descriptions of various other $p$-branes in string theory \cite{Honma:2011br,Kawamoto:2011ab} in a manner similar to the BLG case above.

One can also study this system of equations when the reduction is performed on a time-like or null direction. For the latter case we introduce lightcone coordinates $x^\mu = \{x^+,x^-,x^i\}$ and take $C^\mu_a = g\delta^\mu_+\delta^+_a$, in which case the constrains (\ref{eq2}) lead to supersymmetric system where the fields depend on 4 space ($x^i$) and one null dimension $x^-$. The equations that follow from (\ref{eq1}) in this case are rather novel, for example the scalar potential vanishes, and yet they are invariant under 16 supersymmetries and an $\SO(5)$ R-symmetry.
It was shown in \cite{Lambert:2011gb} that these equations can be reduced to one-dimensional evolution on an instanton moduli space, where $x^-$ plays the role of time. Time evolution is then generated by the conserved momentum $T^{--}$ and this leads to geodesic motion on moduli space, modified by the inclusion of a potential and background gauge field when the scalars have non-vanishing VEVs. This system can then be quantised and this leads to the DLCQ description of the $(2,0)$ theory given by \cite{Aharony:1997th,Aharony:1997an}.

We therefore see that with the help of 3-algebras, it is possible to go from a conventional description of five-dimensional super-Yang-Mills, the low-energy theory on the D4-brane worldvolume, to an equivalent 3-algebraic version with off-shell $\SO(5,1)$ and conformal symmetries, as was also the case for D2-branes in Section~\ref{Lorentzian}. Furthermore, this approach  naturally includes the DLCQ quantisation of the M5-brane. Thus the system (\ref{ansatz1})-(\ref{eq2})   seems capable of describing M5-branes in the case that one dimension is compactified on a circle. Hopefully in this way new light can be shed on M5-branes by re-formulating D4-branes in terms of a $(2,0)$ system. This might  be pertinent given the recent conjectures stating that the $(2,0)$ theory should be {\it defined} as the strong-coupling limit of five-dimensional super-Yang-Mills \cite{Douglas:2010iu,Lambert:2010iw}.

In any case, these results should be viewed as exploratory in terms of applications to M5-branes. Even if we had achieved complete success in writing down a fully six-dimensional system of equations it would still not be enough to define the quantum theory without also giving a lagrangian or some quantisation prescription.  Nevertheless it is of interest to try and see what structures might be at play. The role of 3-algebras, and in particular totally anti-symmetric Lie 3-algebras, was not an assumption but rather emerged through the demands of supersymmetry.
It is tempting to note these 3-algebra structures seem related to 2-groups and 2-Lie algebras which have arisen in the mathematical literature (\eg see \cite{Martins:2010ry,Baez:2010ya,Hofman:2002ey} and references therein).

We should also mention that the M5-brane has been associated more directly with the M2-brane theories of Chapter~\ref{chapter3}, where the 3-algebra is taken to be the Nambu bracket associated to a 3-manifold $\Sigma$ \cite{Ho:2008nn,Ho:2008ve,Bandos:2008fr,Pasti:2009xc,Pasti:2011zz,Park:2008qe,Gustavsson:2010nc}. There have also been other   approaches to the M5-brane that we  have not been able to review here \cite{ArkaniHamed:2001ie,Ho:2011ni,Chu:2011fd,Samtleben:2011fj,Fiorenza:2012tb,Chu:2012um}.


\section{Closing remarks}\label{chapter9}

In this review we have attempted to explain some of the key developments regarding membranes in M-theory over the last five or six years.  These largely concern the formulation of 2+1 dimensional quantum field theories with extended superconformal invariance that describe multiple M2-branes. Conformal Chern-Simons gauge theories form an essential part of such field theories. 

In contrast to Yang-Mills gauge theories, the amount of supersymmetry of a  Chern-Simons gauge theory is largely controlled by the choice of the non-simple  gauge group. Furthermore the matter fields do not sit in the same, adjoint,  representation as the gauge fields. We have seen that a key property of multiple membrane theories is the central role played by the mathematical structure of 3-algebras, which are generalisations of the usual Lie algebras that define the more familiar Yang-Mills theories.
Specifying a 3-algebra is equivalent to giving a Lie-algebra along with a preferred representation. The symmetry properties of the 3-algebra are relatively directly related to supersymmetry and they explain the seemingly odd choices of gauge group that are required by extended supersymmetry.   

Another key aspect, which  enabled the analysis of these theories at a remarkable level of detail, is that the quantised Chern-Simons level $k$ defines the coupling constant $1/k$ of the theory, so that in the limit of large $k$ the theory becomes weakly coupled. From the bulk side $k$ is associated to the rank of an orbifold group, so one is really considering M2-branes propagating in a family of different backgrounds labeled by an integer, which at $k=1$ reduces to the flat, trivial background. On the other hand, in the limit of large $k$ there exists a duality between these field theories and AdS$_4$ backgrounds in type IIA string theory/M-theory, constituting a new and tractable example of AdS/CFT. These insights have made it possible to break fresh ground in recent years in a subject that dates back over a decade and a half.

Naturally, the angle through which these developments were presented was influenced by the authors' own contributions and interests. Several important developments in this area have been omitted from this review, a  significant one being the study of integrability in the AdS/CFT correspondence, of which AdS$_4/$CFT$_3$ forms an important recent class of examples with the CFT in question being one of the theories we have described here. This is a subject on its own, with its own language, motivations, features and results. For a review of M2-branes and  AdS/CFT see \cite{Klebanov:2009sg}. Furthermore we refer the reader to the overview \cite{Beisert:2010jr} of integrability in string theory, and more particularly to Ref.~\cite{Klose:2010ki} which is devoted to integrability in AdS$_4/$CFT$_3$.

Using the results that we have covered as a starting point, the most urgent area of investigation is clearly the dynamics of multiple M5-branes. Here we have surveyed some recent progress in this direction but it is likely that much more will come in the near future. There are of course many other open questions within the vast and beautiful structure of M-theory; we hope that their resolution will continue to benefit both mathematics and physics.

\section*{Acknowledgements}

The authors would like to thank M.~Alishahiha, G.~Bruhn, X.~Chu, J.~Distler, B.~Ezhuthachan, H.~Nastase, B.E.W.~Nilsson, S.~Ramgoolam, P.~Richmond, C.~S\"amann, M.~Schmidt-Sommerfeld, D.~Tong and M.~Van Raamsdonk for collaboration on various parts of the results presented here. JB acknowledges support from the U.S. National Science Foundation, grant NSF-PHY-0910467. NL was supported in part by STFC grant ST/G000395/1. SM is grateful to Trinity College, Cambridge, for a Visiting Fellow Commoner position during the Lent Term, 2012. He would like to thank the TH group at CERN for hospitality during the initial stages of the writing of this review, and the Isaac Newton Institute for Mathematical Sciences for hospitality at the final stage. The generous support of the people of India has, as always, been invaluable. CP is supported by the U.S. Department of Energy under grant DE-FG02-96ER40959. He would like to thank the Isaac Newton Institute for Mathematical Sciences for hospitality at the final stages of this work, as well as acknowledge the Department of Mathematics at King's College London, where a significant part of it was undertaken.


\bibliographystyle{elsarticle-num.bst}
\bibliography{M2review-v26.bib}

\end{document}